%% file: main.tex
\documentclass[krantz1]{krantz} 
\usepackage{fixltx2e,fix-cm}
\usepackage{amssymb}
\usepackage{amsmath}
\usepackage{graphicx}
\usepackage{subfigure}
\usepackage{makeidx}
\usepackage{multicol}
\usepackage{comment}

\usepackage{amssymb}
\usepackage{amsmath,amsthm}
\usepackage{graphics}

\usepackage{graphicx}
\usepackage{epsfig}
\usepackage{makeidx}
\usepackage{kotex}

\usepackage{pdflscape}

\usepackage{amsmath}
\usepackage{amssymb}
\usepackage{setspace}
\usepackage{wasysym}
\usepackage[authoryear]{natbib}

\usepackage{amsmath,amsthm,amssymb}
\usepackage{mathrsfs}
\usepackage{indentfirst}
\usepackage{relsize}
\usepackage{epsfig, epsfig, graphics, lscape, amsbsy, amstext, natbib}
\usepackage{multirow}
\usepackage{amsthm}
\usepackage{makecell}
\usepackage{authblk}
\usepackage{blindtext}

\usepackage{makeidx,epsfig,lscape}
\usepackage{mathptmx}
\usepackage{latexsym}
\usepackage{amsmath}

\usepackage{amssymb}
\usepackage{amsthm}
\usepackage{graphics}

\usepackage{graphicx}
\usepackage{epsfig}
\usepackage{makeidx}

\usepackage{bm, color, colortbl}
\usepackage{natbib}
\usepackage{chngcntr}
\usepackage{multirow}

\makeindex
\usepackage[T1]{fontenc}
\usepackage{lmodern}

\usepackage{fancyvrb}

\DefineVerbatimEnvironment{Highlighting}{Verbatim}{commandchars=\\\{\}}
\usepackage{framed}
\definecolor{shadecolor}{RGB}{248,248,248}

\usepackage{graphicx,grffile}

\newtheorem{definition}{Definition}[chapter]

\newtheorem{example}{Example}[chapter]
\newtheorem{theorem}{Theorem}[chapter]

\newtheorem{lemma}{Lemma}[chapter]

\newtheorem{remark}{Remark}[chapter]

\newcommand{\bx}{\mathbf{x}}

\newcommand{\by}{\mathbf{y}}
\newcommand{\bz}{\mathbf{z}}

\newcommand{\thetahat}{\hat{\theta}}

\newcommand{\blambda}{{\mbox{\boldmath$\lambda$}}}

\newcommand{\bomega}{{\mbox{\boldmath$\omega$}}}

\newcommand{\bbeta}{{\mbox{\boldmath$\beta$}}}

\newcommand{\bgamma}{{\mbox{\boldmath$\gamma$}}}
\newcommand{\indep}{ \stackrel{indep}{\sim} }
\newcommand{\iid}{ \stackrel{i.i.d.}{\sim} }
\newcommand{\PPS}{ {\rm PPS} }

\newcommand{\var}{\mbox{V}}
\newcommand{\cov}{\mbox{Cov}}

\newcommand{\eopen}{\begin{enumerate}}
\newcommand{\eclose}{\end{enumerate}}

\newcommand{\Ybar}{\bar{Y}}

\newcommand{\sumiN}{\sum_{i=1}^{N}}
\newcommand{\sumjM}{\sum_{j=1}^{M}}

\begin{document}




\setcounter{page}{7} 

\include{frontmatter/title}

\include{frontmatter/preface}
\tableofcontents

\mainmatter


\include{chapters/chapter1}

\include{chapters/chapter2} 

\include{chapters/chapter3}

\include{chapters/chapter4}

\include{chapters/chapter5}

\include{chapters/chapter6}

\include{chapters/chapter7}

\include{chapters/chapter8}

\include{chapters/chapter9}
\include{chapters/chapter10}

\include{chapters/chapter11}
\include{chapters/chapter12}

\include{chapters/chapter13}

\include{chapters/chapter14}

\include{chapters/chapter15}

\bibliographystyle{apalike}

\bibliography{ref}


\printindex

\end{document}

%% file: frontmatter/title.tex
\begin{landscape} 

\title{Statistics in Survey Sampling} 
\author{Jae Kwang Kim}
\maketitle

\end{landscape}

%% file: frontmatter/preface.tex
 \chapter*{Preface} 

 This textbook on survey sampling has its origins in a set of lecture notes prepared for a course on survey sampling at Iowa State University. Over the years, these notes have been refined and expanded into the comprehensive
volume you now hold. It is designed to serve both as an introductory text for students and as a reference for practitioners in the field.

Survey sampling is a crucial aspect of statistical practice, bridging the gap between theoretical statistics and practical data collection. This textbook aims
to provide readers with a thorough understanding of both the classical and modern methodologies in survey sampling. We cover fundamental principles such as simple random sampling, stratified sampling, and cluster sampling, as
well as advanced topics including model assisted estimation, two-phase sampling, nonresponse adjustments, and imputation.

The content of this book is structured to guide the reader from basic
concepts to more sophisticated techniques. We begin with an introduction
to the basic ideas and terminology in survey sampling, followed by detailed
discussions on various sampling designs and estimation procedures.
In recent years, the field of survey sampling has seen significant advancements, particularly in response to the challenges posed by new types of data and evolving technological capabilities. This textbook addresses these developments, providing insights into contemporary issues such as analytic inference
and analysis of voluntary samples.
Our goal is to equip readers with the knowledge and tools needed to design
and analyze surveys effectively. Whether you are a student new to the field or
a seasoned researcher looking to update your skills, this book offers valuable
resources to support your work.

I would like to express my gratitude to the students at Iowa State University, whose questions and feedback have greatly contributed to the development of these materials. Additionally, I am thankful for the support and
encouragement of my colleagues, whose expertise and collaboration have been
invaluable.

%% file: chapters/chapter1.tex
\setcounter{chapter}{0} 
\chapter{Introduction 
}

\section{Introduction}

Suppose that we are interested in obtaining the employment rate of the US population in a certain time. In this case, we can think of two ways to obtain the employment rate. One is to measure the employment status for every individual in the population and then take the mean of the measurement in the population. The other is to sample a subset of the population and then use the average employment status of the sample as an estimate for the employment status of the population. The former method is called census, and the latter is called sample survey. Roughly speaking, the census may obtain an accurate figure of the true employment rate, but the cost for census can be enormous. On the other hand, the sample survey can reduce the cost greatly, but the sample estimate can be quite different from the true population values. 

In this simple example, the true employment rate of the target population is called a parameter. From the sample survey, we obtain an estimate of the parameter. There are many different ways of obtaining samples from the same population. Thus, finding the best sample (and the corresponding estimator) is one of the main goals of survey sampling. 

Sample survey has the following advantages over the census:
\begin{enumerate}
\item It reduces the cost significantly. 
\item It takes much less time. Thus, we can obtain information about the population in a timely manner. 
\item Because we employ more trained interviewers in survey sampling, it can produce more accurate estimates. 
\item Sometimes, it is the only way to get information about the target
population. 
\end{enumerate}

The third point may sound strange, but it is often true in practice because the census may have larger non-sampling errors than the sample surveys. Because the sample survey is based only on a subset of the finite population, there is always a danger of failing to represent the population properly. The representativeness of a sample is a concept that roughly describes whether the sample can be treated as a miniature of the finite population \citep{kruskal1979}.
A representative sample can give the most of the pictures of the finite population in a cost-effective way. When we use a subset of the population to make inferences about the population, the sample estimate can be different from the parameter to some extent. 
 The sampling error of an estimator is the difference between the estimator and the parameter.  We can reduce the sampling error in two aspects. One is to use a suitable sampling rule and the other is to use enough sample size. Note that the sampling error of an estimator $\hat{\theta}$ can be written as
\begin{eqnarray}
\hat{\theta}- \theta &=& \{ \hat{\theta} - E ( \hat{\theta}
) \} +  \{  E ( \hat{\theta} )- \theta
\}. \label{error}
\end{eqnarray}
The first term, 
 $ \hat{\theta} - E( \hat{\theta})$, can be called the variation of $\hat{\theta}$ and the second term, $E(\hat{\theta}) - \theta$, is called the bias of $\hat{\theta}$. Increasing the sample size will reduce the absolute value of the variation part and using a sampling rule that guarantees unbiasedness will make the second part equal to zero.

The  accuracy of an estimator is often measured by mean squared errors, which is give by 
 \begin{eqnarray}
MSE ( \hat{\theta}) &\equiv& E\left\{  (
\hat{\theta} - \theta )^2 \right\} \notag \\
&=& \left\{  Bias ( \hat{\theta}) \right\}^2 + V (
\hat{\theta}). \label{mse}
\end{eqnarray}
By employing a probability sampling design, we  can make $Bias ( \hat{\theta})=0$. By increasing the sample size, we can make the variance smaller. Probability sampling will be introduced in the next section. 
\index{Mean Squared Error} 

\section{Probability sampling}

Probability sampling refers to a set of sampling methods in which the selection probability in each element in the population is known and strictly positive. \index{probability sampling} For example, simple random sampling assigns equal probability of selection to each element of the population. In this case, the selected sample is obtained objectively and does not involve subjective human selection. There are many different ways of obtaining probability samples, and the choice depends on the objective of the study, the cost, and the prior information available in the sampling frame. The following simple example will illustrate the basic idea. 


\begin{table}[htb]
\caption{Farm acreage and crop yield for Example \ref{example:1-1}}
\begin{center}
\begin{tabular}{c|c c}
\hline
    ID  & Acreage   & Yield ($y$)   \\
   \cline{1-3}
   1 & 4 & 1 \\
   \cline{1-3}
   2 & 6 & 3 \\
   \cline{1-3}
   3 & 6 & 5 \\
   \cline{1-3}
   4 & 20 & 15 \\
 \hline
\end{tabular}
\end{center}
\label{1-table1}
\end{table}

\begin{example}
\label{example:1-1}
Suppose that there are four farms in a town.   
The finite population consists of 4 farms in Table \ref{1-table1} and we are interested in estimating the average crop yield in the population. Here, the farm acreage is the auxiliary information that is available in advance and the crop yield is the item of interest in the study. Suppose further that,  due to the cost restriction, we want to select only $n=2$ farms and estimate the average crop yield from the sample. In this case, there are six ways to select a sample of size $n=2$ from the finite population of size $N=4$. Among the six possible samples, we select one sample and measure the crop yields from the sample. In probability sampling, we can select one sample randomly from the six possible samples. Here, ``random'' sampling means that the sample selection is determined solely by the random number generated for the sample selection, without involving any subjective human decision. To select a sample randomly, the selection probability is assigned in advance for each possible sample. Once the selection probability is assigned to each sample, the final sample is selected using a random number generated from a random-number-generating mechanism.

When selecting a sample from a finite population, there is a finite number of possible samples, and the probability distribution of the sample can be treated as a probability distribution for a discrete random variable. In this case, the probability distribution of a sample consists of all possible samples and their selection probabilities. The selection probabilities must be sum to one. Simple random sampling refers to the sampling mechanism where the selection probabilities are all equal for samples with the same sample size. 
\index{simple random sampling} Table \ref{1-table2} shows the probability distribution of the sample under simple random sampling using the example in Table \ref{1-table1}. Since there are 6 possible samples of size $n=2$ from the population of size $N=4$, the selection probabilities are all equal to $1/6$.  The probability distribution of the sample is simply called the sample distribution. \index{sample distribution}

\begin{table}[htb]
\caption{Sample distribution under simple random sampling }
\begin{center}
\begin{tabular}{c|c}
\hline
   Sample ID  &  Selection probability   \\
   \cline{1-2}
   1, 2 & 1/6 \\
   \cline{1-2}
   1, 3 & 1/6 \\
   \cline{1-2}
   1, 4 & 1/6 \\
   \cline{1-2}
   2, 3 & 1/6 \\
   \cline{1-2}
   2, 4 & 1/6 \\
   \cline{1-2}
   3, 4 & 1/6\\
 \hline
\end{tabular}
\end{center}
\label{1-table2}
\end{table}

Determining the sample distribution is equivalent to determining the sampling mechanism, the probability mechanism for sample selection. \index{sampling mechanism} Sampling design refers to the process of determining the sampling mechanism. Form the sample distribution, we can use a random number to select a sample. For example, in Table \ref{1-table2}, if the random number is 0.4 then $\{1,4\}$ is selected because 0.4 is greater than 2/6 and less than 3/6.

Note that we can compute an estimator from the sample observations. For example,  sample mean can be computed from each sample.   
Thus, we can  the sampling distribution of the sample mean  from the sample distribution. Probability distribution of an estimator consists of all possible values of the estimator and their  probabilities.  The possible value of an estimator can be computed from the observed values in the sample. Thus, sampling distribution of an estimator is induced from the sample distribution, the probability distribution of the sample. Table \ref{1-table3} presents the sampling distribution of the sample mean under the simple random sampling given by Table \ref{1-table2}. From the sampling distribution table, the sample mean $\bar{y}$ is distributed as a discrete random variable taking values on 2,3,4,8,9,10, with equal probability. Thus, we have $E(\bar{y}) = 6$ which is equal to the population mean. Under the sampling  distribution in Table  \ref{1-table3}, the expectation of the sample mean is equal to the population mean. That is, sample mean is unbiased for the population mean under simple random sampling.

\begin{table}[htb]
\caption{Sampling distribution of sample mean under simple random sampling }
\begin{center}
\begin{tabular}{c|c|c|c}
\hline
   Sample  ID  & $y$ value  & Sample mean ($\bar{y}$) & Selection probability \\
   \cline{1-4}
   1, 2 & 1, 3 & 2 & 1/6 \\
   \cline{1-4}
   1, 3 & 1, 5 & 3 & 1/6 \\
   \cline{1-4}
   1, 4 & 1, 15 & 8 & 1/6 \\
   \cline{1-4}
   2, 3 & 3, 5 & 4 & 1/6  \\
   \cline{1-4}
   2, 4 & 3, 15 & 9 & 1/6 \\
   \cline{1-4}
   3, 4 & 5, 15 & 10 & 1/6\\
 \hline
\end{tabular}
\end{center}
\label{1-table3}
\end{table}
\end{example}

Unbiasedness of an estimator is a desirable property but it does not mean that your estimator in a particular sample is close to the true parameter. In the example above, if sample $\{1,2\}$ is selected, your estimate of the population mean is 2, which is much smaller than the truth (=6). That is, unbiasedness does not imply accuracy. 

How to improve the accuracy of an estimator? One way is to increase the sample size. By increasing the sample size, the variance is reduced. If the estimator is unbiased, smaller variance means smaller mean squared error and it means high accuracy. Another way of improving accuracy is to use more efficient sampling designs. Making an efficient sampling design by properly incorporating the auxiliary information of the sampling frame is one of the main topics in survey sampling. Example 1.2 will provide an illustration for an efficient sampling design.

One thing to note from the above example is that the basis for inference is with respect to the sampling mechanism and has nothing to do with the actual value of $y_i$ in the population. That is, the reference distribution in computing the expectation of an estimator is the probability mechanism generated by a repeated application of the sampling design.  This approach is called the design-based approach \index{design-based approach} because the sampling design is used mainly to make statistical inferences about the population, treating $y_i$ as fixed. On the other hand, the model-based approach refers to an alternative approach to inferring a finite population by making model assumptions about $y_i$ in the population. In survey sampling, the design-based approach is more popular because it does not rely on model assumptions, which can be difficult to verify in practice. 
Also, design-based approach avoids the danger of human subjectivity in sample selection, which is an important issue in government official statistics.

\begin{example}
Consider the simple farm example in Example \ref{1-table1} and consider the following alternative sampling mechanism. 
\begin{table}[htb]
\caption{An alternative sample distribution}
\begin{center}
\begin{tabular}{c|c}
\hline
   Sample ID  &  Selection probability   \\
   \cline{1-2}
   1, 4 & 1/3 \\
   \cline{1-2}
   2, 4 & 1/3 \\
   \cline{1-2}
   3, 4 & 1/3\\
 \hline
\end{tabular}
\end{center}
\label{1-table4}
\end{table}

In this new sampling design, farm unit 4 is selected with certainty, and one of the other three farms is selected with equal probability. In this case, an unbiased estimator of the population mean is constructed by 
$$ \hat{\theta} = \frac{3}{4} \sum_{i=1}^3 I_i y_i + \frac{1}{4} y_i $$
where $I_i=1$ if unit $i$ is selected in the sample and $I_i=0$ otherwise. Note that each sample has observation $\{y_k, y_4\}, k=1,2,3$ and the mean estimator takes the form of $(3 y_k + y_4)/4$ because one selected $y_k$ represents three elements $\{y_1, y_2, y_3\}$ and $y_4$ represents itself only. This is an example of the Horvitz-Thompson estimator that we shall study in Chapter 2. 

\begin{table}[htb]
\caption{Sampling distribution of the mean estimator under the alternative sampling design }
\begin{center}
\begin{tabular}{c|c|c|c}
\hline
   Sample  ID  & $y$ value  & Mean estimator ($\hat{\theta}$) & selection probability  \\
   \cline{1-4}
   1, 4 & 1, 15 & 4.5 & 1/3 \\
   \cline{1-4}
   2, 4 & 3, 15 & 6 & 1/3 \\
   \cline{1-4}
   3, 4 & 5, 15 & 7.5 & 1/3 \\
 \hline
\end{tabular}
\end{center}
\label{1-table5}
\end{table}

Table \ref{1-table5} shows the sampling distribution of the mean estimator $\hat{\theta}$ under the alternative sampling design in Table \ref{1-table4}.  The new sampling design still provides unbiased estimation as the expectation of the mean estimator is equal to 6. However, the variance of the mean estimator is now 1.5, which is much lower than the variance 9.67 under simple random sampling.  
\end{example}

As can be seen in Table  \ref{1-table5},  the alternative sampling design significantly reduces the sampling variance without increasing the sample size. Efficient sampling design can improve the accuracy of an estimator without increasing the sample size. 

\section{Basic Procedures  for survey sampling}

Survey sampling is developed to collect information about characteristics of interest from a subset of the population to build a database for analytical or descriptive purposes. 
 The manual ``Survey Methods and Practice'', published by Statistics Canada (2003), provides an excellent summary of the basic procedures for sample surveys in government agencies. The life of a survey can be broken down into several phases. The first is the planning phase, which is followed by the design and development phase, and then the implementation phase. Finally, the entire survey process is reviewed and evaluated. The phases of the life of a survey are described below. 
 \subsection{Survey planning} 
 
 Survey planning is the first step in any survey sampling procedure. In the preliminary planning stage, those who are considering a sample survey should formulate the objectives of the proposed survey. The objective should include the specification of the information to be gathered and the determination of the target population,  to which the findings of the survey will be extrapolated. Once a survey proposal is formulated, it is important to determine whether a new survey is necessary, considering costs and other practical constraints in conducting a new survey. Sometimes, the goal can be achieved by adding equations to an existing survey's questionnaire or by redesigning an existing survey. 
 
 If it is decided to conduct a new survey, the planning team may proceed to formulate the statement of objectives and develop some appreciation of the frame options, the general sample size, the precision requirement, the data collection options and the cost. More discussion of the items in the Statement of Objectives is beyond the scope of this textbook. See Chapter 2 of \cite{statcan2003} for more details.

The selection of a sampling frame is also an important part in survey planning. The sampling frame ultimately defines the population to be surveyed, which may be different from the target population to which the Statement of Objective refers. In order to define the target population, the statistical agency begins with a conceptual population for which no actual list may exist. For example, the conceptual population may be all farmers. To define the target population, ``farmers'' must be defined. The target population is the population for which information is desired to apply. On the other hand, the survey population is the population that is \emph{actually} covered by the survey. 
The survey frame (or sampling frame) is a realized list of the survey population. For example, in the survey of household and expenditures in the US, the target population is the entire resident population of the US on a particular reference date. However, in reality, we do not have addresses for those living in institutions or living without fixed addresses. Thus, the survey population excludes those residents in the US on the same reference date. 

\subsection{Design and development} 

A subset of the survey population is selected from the sampling frame to collect the information we are interested in. Probability sampling, which is the main topic of this book, will be used to obtain a representative sample of the population. There are several different ways to select a probability sample. The choice of the sampling design depends on several factors such as the availability of the sampling frame, the heterogeneity of the study variable, the target precision and the cost of the survey measurements. For a given population, a balance of sampling error with cost and timeliness is achieved through the choice of design and sample size.  A more efficient estimation procedure can also be developed by incorporating the auxiliary information available throughout the finite population.

In addition to the choice of the sampling design and the estimation procedure, there are also choices of measurements to be taken and the procedures for taking these measurements. The subject matter persons, who will be users of the survey data, should provide the primary input to specify the measurements that are needed to meet the objective of the study. Once the measurements are specified, the measurement experts (survey methodologists or psychologists) begin designing the questionnaires or forms to be used to obtain the data from the sample individuals.  Survey questionnaires may undergo some pilot survey and revision before being used in the main survey.

\subsection{Implementation} 

Having ensured that all systems are in place, the survey can now be launched. This is the implementation phase.  Interviewers are trained, the sample is selected, and information is collected, all in a manner established during the development phase. Following these activities, data processing begins. Processing activities include data capture, coding, editing, and imputation. The result is a well-structured and complete data set from which it is possible to produce required tabulations and to analyze survey
results. The confidentiality of these results is then checked and distributed. At each step, data quality
must be measured and monitored using methods designed and developed in the previous phase.

\subsection{Survey evaluation} 

Once the survey is over, the entire process can be documented and evaluated. This involves assessments
of the methods used, as well as evaluations of operational effectiveness and cost performance. These
evaluations serve as a test of the suitability of technical practices. They also serve to improve and
guide the implementation of specific concepts or components of methodology and operations, within and
between surveys.

\section{Survey errors} 

Estimates obtained from a sample survey can suffer from several sources of errors. The errors of an estimator can be classified into two categories. One is the sampling error, the error due to selecting only a subset of the population, and the other is non-sampling error, the error that can be obtained even under census. Non-sampling errors consist of coverage error, response error, measurement error, and processing error, etc. The coverage error occurs when the sampling frame is incomplete in the sense that it does not fully cover the target population. Response error occurs when the sampled element does not respond to the survey.  Measurement error occurs when there is a discrepancy between the actual value and the reported value for certain study items. Measurement error exists when there is misunderstanding of the survey questions, or due to a false answer, or due to interviewer effect. Processing error refers to the errors that occur when transferring the survey answers to the computing process system. 

The coverage error, sampling error, and response error are combined to form the non-observation error because the errors are due to not observing the elements in the target population. Measurement error and processing error are combined to form the error of observation because the error occurs even if we observe all elements of the population.

In general, in order to reduce the coverage error among various non-sampling errors,
a good sampling frame should be used. For example, when a yellow book is used to select samples, people who do not have a landline phone will not be sampled, so you will have to get another list to supplement it. Multi-frame survey, using several sampling frames, can be considered in this case, but then we need to worry about duplication problem. 

To reduce nonresponse error, we may need to train the interviewer, publicize the survey, manage the survey target, and give higher incentives to survey participation.  The measurement error was determined by the survey method, the sincerity and training level of the interviewer, and the clarity of the survey questions. 
Sufficient motivation and training is also required from the interviewees.

In general, as the sample size increases, the sampling error decreases, but the non-sampling error becomes more difficult to manage. For example, if the sample size is very large, it becomes difficult to manage interviewers, which increases the risk of non-response errors or measurement errors. 
Therefore, in addition to cost reduction, conducting a sampling of an appropriate size is very important in that non-sampling error can be reduced.

%% file: chapters/chapter2.tex

\setcounter{chapter}{1} 
\chapter{Horvitz-Thompson estimation}

\section{Introduction}

We will study some theory for unbiased estimation under probability sampling designs.  Let $U=\{ 1, \cdots, N\}$ be the set of indexes of the target population. A probability sample is simply a subset of $U$, denoted by $A \subset U$, selected by a probability rule, called the sampling design. Let $\mathcal{A}=\{ A; A \subset U\}$ be the set of all possible samples.  We have the following formal definition of the sampling distribution. 

\begin{definition} 
For a given $\mathcal{A}$ and $U$, sampling
distribution \index{sampling distribution} refers to the  probability mass function defined on $\mathcal{A}$. That is, a sampling distribution $P( \cdot)$ satisfies the following properties: 
\begin{enumerate}
\item $P\left( A \right)  \in \left[ 0, 1\right], \ \ \ \ \ \forall A \in
\mathcal{A} $
\item $\sum_{A \in \mathcal{A}}P\left( A \right)  = 1. $
\end{enumerate}
\end{definition}
If  the sampling distribution satisfies $ P(A)
< 1 , \ \ \ \ \  \mbox{ for all } A \in \mathcal{A}$, then it is called random sampling. \index{random sampling}

If the parameter of interest is a population quantity that can be written as $\theta_N = \theta\left( y_i
; \  i \in U  \right)$, a statistic \index{statistic} is written as
$\hat{\theta}=\hat\theta \left( y_i ;\ i \in A \right)$, which means that it is a function of $y_i$ in the sample. If the statistic is used to estimate $\theta_N$, then it becomes an estimator.  \index{estimator} The sampling distribution of an estimator is obtained from the sample distribution. That is, as discussed in Section 1.2, the probability mass function $P(A)$ applied to obtain $A$ is then used to represent $P\{ \hat{\theta}=\hat\theta \left( y_i ;\ i \in A \right)\}$, the probability distribution or the sampling distribution of $\hat{\theta}$. Using the sampling distribution of an  estimator, we can also compute the expectation and variance of the estimator. 

\begin{definition}  
For parameter  $\theta_N$, let  $ \hat{\theta} \left( A \right)=
\hat\theta \left( y_i ;\ i \in A \right)$ be an estimator of $\theta_N$. 
\begin{itemize}
\item[1.] Expectation : $ E (\thetahat )=\sum_{A\in
\mathcal{A}} P\left(A \right)\hat{\theta} \left( A \right)  $
\item[2.]  Variance: $ V( \thetahat  )= \sum_{A\in
\mathcal{A}} P\left(A\right) \left\{ \hat{\theta} \left( A \right)
-E\left(\hat{\theta}\right) \right\}^{2}
        $
\item[3.] Mean squared error : $ {\rm MSE}( \thetahat ) =\sum_{A\in
\mathcal{A}} P(A) \left\{ \hat{\theta} \left( A \right) -\theta_N
\right\}^{2}
        $
\end{itemize}
\end{definition}
Here, the expectation is taken with respect to the sampling design induced by the probability rule for $A$, treating $ \left\{ y_1, y_2, \cdots, y_N \right\}$  as fixed. As discussed in Chapter 1, the difference between $E( \hat{\theta}) $ and $\theta_N$ is called bias \index{bias} and an estimator is called an unbiased estimator \index{unbiased estimator} when its bias is zero. When an estimator has high precision, it means that its variance is small, but it does not necessarily mean that its accuracy is high. The accuracy of an estimator is related to the small mean squared error of the estimator. 

\section{Horvitz-Thompson estimation}

Does an unbiased estimator always exist for all probability sampling designs? To answer this question, we need the following definition of inclusion probabilities. 

\begin{definition} $ \phantom{} $
\begin{enumerate}
\item First-order inclusion probability:
$$
\pi_i = Pr \left( i \in A \right) = \sum_{ A;\  i \in A} P \left(
A \right)$$ \item Second-order inclusion
probability, or joint inclusion probability:
$$
\pi_{ij} = Pr \left( i, j  \in A \right) = \sum_{ A;\  i,j \in A} P
\left( A \right)
$$
\item Probability sampling design: $\pi_i > 0 , \ \
\ \forall i \in U $ \item Measurable sampling design 
: $\pi_{ij} > 0  \ \ \ \forall i , j \in U $.
\end{enumerate}
\end{definition}
\index{first order inclusion probability}
\index{joint inclusion probability}

That is, the first-order inclusion probability $\pi_i$ refers to the probability that the unit $i$ is included in the sample. 
Furthermore, the second-order inclusion probability $\pi_{ij}$ refers to the probability that both units, unit $i$ and unit $j$, are included in the sample. Note that $\pi_{ii} = \pi_i$ by definition. Probability sampling design is a sampling design in which all first-order inclusion probabilities are strictly greater than zero.  Probability sampling design is a sufficient condition for the existence of a design-unbiased estimator of the population total. Measurable sampling design is a sampling design in which all second-order inclusion probabilities are strictly greater than zero. Measurable sampling design is a sufficient condition for the existence of a design unbiased estimator of sampling variance of an unbiased estimator.

The following lemma presents some algebraic properties of the inclusion probabilities.  
\begin{lemma}
The first order inclusion probabilities satisfy 
\begin{equation}
\sum_{i=1}^N \pi_i = E\left( n \right) \label{2.1}
\end{equation}
where $n$ is the sample size. If the sampling design is a fixed-size
sampling design \index{fixed-size sampling
design} such that $V \left(n \right) = 0 $, then $\sum_{i=1}^N \pi_i=n$ and  
\begin{equation}
\sum_{i=1}^N \pi_{i j} = n \pi_j \label{2.2}
\end{equation}
\end{lemma}
\begin{proof}
Given the sample index set $A$, define the following indicator function
\begin{equation}
I_i = \left\{ \begin{array}{ll} 1 & \mbox{ if } i \in A \\
0 & \mbox{ if } i \notin A.
\end{array}
\right. \label{2.3}
\end{equation}
In this case, $I_i$ is a random variable with $E \left( I_i \right) = \pi_i $ and $ E\left( I_i
I_j \right) = \pi_{ij}$.  Furthermore, by the definition of sample size $n$, 
\begin{equation}
 \sum_{i=1}^N I_i = n.
 \label{2.4}
\end{equation}
Thus, taking expectations of both sides of  (\ref{2.4}), we can obtain  (\ref{2.1}). Also, multiplying both sides of (\ref{2.4}) and taking the expectations again, we obtain (\ref{2.2}).
\end{proof}

When the sample is obtained from a probability sampling design, an unbiased estimator for the total $Y=\sum_{i=1}^N y_i$ is given by 
\begin{equation}
\hat{Y}_{\rm HT}=\sum_{ i \in A} \frac{ y_i}{ \pi_i }.
\label{2.5}
\end{equation}
This is often called  Horvitz-Thompson (HT) estimator,  \index{Horvitz-Thompson estimator} which is originally discussed by \cite{horvitz1952} and also by \cite{narain1951}.  

The following theorem presents the basic statistical properties of the HT estimator. 
\begin{theorem}
The  Horvitz-Thompson estimator, given by (\ref{2.5}), satisfies the following properties: 
\begin{equation}
E\left( \hat{Y}_{\rm HT}  \right) = Y \label{2.6}
\end{equation}
\begin{equation}
V\left( \hat{Y}_{\rm HT}  \right) = \sum_{i=1}^N \sum_{j =1}^N \left(
\pi_{ij} - \pi_i \pi_j \right) \frac{y_i}{\pi_i } \frac{y_j}{\pi_j
} \label{2.7}
\end{equation}
Furthermore, for a fixed-size sampling design (i.e., $V \left(n
\right) = 0 $), we have 
\begin{equation}
V\left( \hat{Y}_{\rm HT}  \right) = -\frac{1}{2} \sum_{i=1}^N \sum_{j
=1}^N \left( \pi_{ij} - \pi_i \pi_j \right)\left( \frac{y_i}{\pi_i
} - \frac{y_j}{\pi_j } \right)^2 . \label{2.8}
\end{equation}
\end{theorem}
\begin{proof}
Using the sample indicator function $I_i$ defined in (\ref{2.3}), the HT estimator can be written as  
\begin{equation*}
\hat{Y}_{\rm HT} = \sum_{i=1}^N \frac{y_i}{\pi_i}  I_i.
\end{equation*}
Treating $\left\{ y_1, y_2, \cdots, y_N \right\}$ as fixed and taking expectations with respect to $I_i$, we have 
\begin{eqnarray*}
E\left(\hat{Y}_{\rm HT} \right) &=&\sum_{i=1}^N
\frac{y_i}{\pi_i}  E \left( I_i \right) = \sum_{i=1}^N \frac{y_i}{\pi_i} \pi_i = Y
\end{eqnarray*}
which shows (\ref{2.6}). Similarly, we have 
\begin{eqnarray*}
V\left(\hat{Y}_{\rm HT}  \right) &=&\sum_{i=1}^N \sum_{j=1}^N
\frac{y_i}{\pi_i}  \frac{y_j}{\pi_j}Cov \left( I_i, I_j  \right) \\
&=&\sum_{i=1}^N \sum_{j=1}^N \frac{y_i}{\pi_i}  \frac{y_j}{\pi_j}
\left( \pi_{ij} - \pi_i \pi_j \right)
\end{eqnarray*}
and (\ref{2.7}) is proved. To show (\ref{2.8}), define $\Delta_{ij} = \pi_{ij} - \pi_i \pi_j$ and express 
 (\ref{2.8}) as 
\begin{equation}
-\frac{1}{2} \sum_{i=1}^N \sum_{j =1}^N \Delta_{ij} \left(
\frac{y_i}{\pi_i } - \frac{y_j}{\pi_j } \right)^2 = -\sum_{i=1}^N
\sum_{j =1}^N \Delta_{ij} \left( \frac{y_i}{\pi_i } \right)^2
+\sum_{i=1}^N \sum_{j =1}^N \Delta_{ij} \frac{y_i}{\pi_i }
\frac{y_j}{\pi_j }.
\label{2.9}
\end{equation}
Now, using (\ref{2.2}) and (\ref{2.1}), we have 
$$\sum_{j =1}^N \Delta_{ij}= \sum_{i=1}^N \pi_{ij} - \sum_{i=1}^N \pi_i \pi_j = n \pi_i - n\pi_i =0 .$$
Thus, the first term on the right side of the equality in (\ref{2.9})  becomes zero and (\ref{2.8}) is established. 
\end{proof}

\begin{example}
 Let $U=\left\{ 1,2 ,3 \right\}$ be the target population and consider the following sampling design.
 $$ P\left( A \right) = \left\{ \begin{array}{ll}
0.5 & \mbox{ if } A=\left\{ 1,2 \right\} \\
0.25 & \mbox{ if }  A=\left\{ 1,3 \right\} \\
0.25 & \mbox{ if }  A=\left\{ 2,3 \right\} \\
\end{array}
\right. $$ 
In this case, we have
$\pi_1 = 0.5 + 0.25 = 0.75$, $\pi_2 = 0.5+ 0.25 =
0.75$, and $\pi_3 = 0.25 + 0.25 = 0.5$.  Thus, the HT estimator for the total is then 
$$ \hat{Y}_{HT}
= \left\{ \begin{array}{ll}
 y_1/0.75 + y_2/0.75&\mbox{ if } A=\left\{ 1,2 \right\} \\
y_1/0.75+y_3/0.5 & \mbox{ if }  A=\left\{ 1,3 \right\} \\
y_2/0.75 + y_3/0.5 &  \mbox{ if }  A=\left\{ 2,3 \right\}\\
\end{array} \right.
$$
Therefore, 
\begin{eqnarray*}
 E\left(  \hat{Y}_{\rm HT} \right)&=& 0.5 \left( y_1/0.75 +
y_2/0.75 \right) +0.25 \left(y_1/0.75+y_3/0.5 \right) \\
&+&0.25 \left( y_2/0.75 + y_3/0.5\right)= y_1 + y_2 + y_3
\end{eqnarray*}
and the HT estimator is unbiased for the population total. 
\end{example}

The HT estimator provides an unbiased estimate under probability sampling. If $\pi_i>0$ does not hold for some elements of the population, the HT estimator cannot be used. Also, the HT estimator is not location-scale invariant. That is, for any constants $a$ and $b$, 
$$ \frac{1}{N} \sum_{ i \in A} \frac{ a+ b y_i}{ \pi_i }
\ne a  + b \frac{1}{N}\sum_{ i \in A} \frac{ y_i}{ \pi_i }.$$

The variance formula (\ref{2.8}) was independently discovered by \cite{sen1953} and \cite{yates1953}.  Thus, it is often called the Sen-Yates-Grundy(SYG) variance formula. The variance will be minimized when $\pi_i \propto y_i$. That is, if the first-order inclusion probability is proportional to $y_i$, the resulting HT estimator under this sampling design will have zero variance. However, in practice, we cannot construct such a design because we do not know the value of $y_i$ in the design stage. If there is an auxiliary variable $x_i$ available in the sample frame and $x_i$   is believed to be closely related to $y_i$, then a sampling design with $\pi_i \propto x_i$ can lead to a very efficient sampling design. 

Now, we discuss an unbiased estimate of the variance of the HT estimator. The variance formula in (\ref{2.7}) or (\ref{2.8}) is a population quantity and must be estimated from the sample. Generally speaking, the variance formula is a quadratic function of $y_i$'s in the population. Thus, to estimate the variance, we need to assume a measurable sampling design that satisfies $\pi_{ij}>0$ for all $i$ and $j$. That is, if the parameter of interest is of the form 
$$ Q= \sum_{i=1}^N \sum_{j=1}^N q(y_i, y_j) $$
 then, under measurable sampling design, an unbiased estimator of $Q$ is 
 \begin{equation} \hat{Q}
= \sum_{i \in A} \sum_{j \in A} \frac{
q\left( y_i,
y_j \right)}{\pi_{ij}} . \label{2.9b}
\end{equation}
Thus, an unbiased estimator of variance in (\ref{2.7}) is 
 \begin{equation}
\hat{V}\left( \hat{Y}_{\rm HT} \right) = \sum_{i \in A} \sum_{j \in A}
\frac{\pi_{ij} - \pi_i \pi_j }{\pi_{ij}} \frac{y_i}{\pi_i}
\frac{y_j}{\pi_j}. \label{2.10}
\end{equation}
Also, for fixed-sized designs, an unbiased estimator of the SYG variance formula is 
\begin{equation}
 \hat{V}\left( \hat{Y}_{\rm HT} \right) = -\frac{1}{2}
\sum_{i \in A} \sum_{j \in A} \frac{\pi_{ij} - \pi_i \pi_j
}{\pi_{ij} }\left( \frac{y_i}{\pi_i } - \frac{y_j}{\pi_j } \right)^2.
\label{eq2.12}
\end{equation}

Another statistical property of the HT estimator is consistency and asymptotic normality, which are established for sufficiently large sample sizes. \index{consistency} \index{asymptotic
normality} For an infinite population, the consistency of the sample mean means that the sample mean converges to the population mean in probability. That is, the probability that the absolute difference between the sample mean and the population mean is greater than a given threshold ($\epsilon$) goes to zero as the sample size increases. That is, for any $\epsilon>0$, 
$$ Pr \left( \left| \bar{y}_n-\bar{Y}_N\right|> \epsilon \right) \rightarrow 0  \ \ \ \mbox{as } n \rightarrow \infty.$$
In the finite population setup, the finite population is conceptualized to be a sequence of finite population with  size $N$  also increasing.  The HT estimator of the finite population mean is given by 
$$\bar{Y}_{\rm HT}=\frac{ \hat{Y}_{\rm HT}}{N}  $$
and, using Chebyshev inequality, 
$$ Pr \left( \left| \bar{Y}_{\rm HT}-\bar{Y}_N\right|> \epsilon
\right)\le \epsilon^{-2} V\left( \bar{Y}_{\rm HT} \right). $$ 
Thus, as long as 
\begin{equation} 
\lim_{n \rightarrow \infty} V\left(
\bar{Y}_{\rm HT} \right) = 0,   
\label{2.13}
\end{equation} 
the consistency of HT estimator follows. For example, under the simple random sampling that we will study in Chapter 3, we have
$$V\left(
\bar{Y}_{\rm HT} \right) = \frac{1}{n} \left( 1- \frac{n}{N} \right) S^2.
$$
Thus, as long as $S^2$ is bounded above in probability, condition (\ref{2.13}) holds and the consistency of the HT estimator follows.

Asymptotic normality is also an important property for obtaining confidence intervals or performing hypothesis testing from the sample. Under some regularity conditions, we can establish 
$$\frac{\hat{Y}_{\rm HT} - Y }{\sqrt{\hat{V}\left( \hat{Y}_{\rm HT} \right)
}} \stackrel{\mathcal{L}}{\longrightarrow} N\left(0,1 \right)$$
for most sampling designs, where $\stackrel{\mathcal{L}}{\longrightarrow}$ denotes convergence in probability.  Thus, in this case, the 95\% confidence interval for the population total is computed by 
$$\left( \hat{Y}_{\rm HT} - 1.96 \sqrt{\hat{V}\left( \hat{Y}_{\rm HT} \right)
}, \ \ \ \hat{Y}_{\rm HT} + 1.96 \sqrt{\hat{V}\left( \hat{Y}_{\rm HT} \right) }
\right).$$
A more in-depth discussion of the asymptotic normality of the HT estimator can be found in Chapter 1 of \cite{fuller2009}. 

\section{Other parameters}

In the previous section, we have studied the estimation of the total of the finite population, $Y=\sum_{i=1}^N y_i$. In many practical situations, we estimate other parameters such as population quantiles or domain totals. For example, one may be interested in estimating certain characteristics (such as household income) for a certain age group (for example, over 60) of the householder. At the time of sampling, we do not know the age of the householders. In this case, one can select a sample from the household population and obtain information about the age and income in the sampled households. Let $z_i=1$ if $i$ belongs to the particular age group of interest and $z_i=0$ otherwise. Also, let $y_i$ be the value of the study variable (income in this example) for the element $i$. The domain mean of $y$ can be written as 
\begin{equation}
 \theta_d = \frac{ \sum_{i=1}^N z_i y_i }{ \sum_{i=1}^N z_i }.
\label{2.14}
\end{equation}
The indicator variable $z$ is used to identify the domain inclusion in the population. If $z_i=1$, then the unit $i$ is eligible for domain estimation.

From (\ref{2.14}), we can express the domain mean as a ratio of two totals. That is,  $\theta_d= Y_d/D$ where  $Y_d=\sum_{i=1}^N  z_i y_i$, $D=\sum_{i=1}^N z_i$. A natural estimator of $\theta_d$ is 
\begin{equation*}
 \hat{\theta}_d = \frac{ \sum_{i \in A} z_i y_i /\pi_i}{ \sum_{ i\in A}z_i /\pi_i}, 
\end{equation*}
which is the ratio of two corresponding HT estimators. That is, $\hat{\theta}_d=\hat{Y}_{d, \rm HT}/ \hat{D}_{\rm HT}$ is a special case of the ratio estimator we will study in Chapter 7. Generally speaking, the ratio estimator is not exactly unbiased, but its relative bias can be asymptotically negligible.

In addition to domain estimation, we are often interested in estimating the population distribution. Suppose, for example, that we are interested in estimating the proportion of the population below a certain poverty level.  In this case, $\theta_p=N^{-1} \sum_{i=1}^N I(y_i <c)$  is the parameter of interest, where $c$ is the threshold for the poverty of the household. We can use the theory of HT estimation to estimate $\theta_p$ by $\hat{\theta}_{p, \rm HT} = N^{-1} \sum_{i\in A} \pi_i^{-1} I(y_i < c)$. If $N$ is unknown, we use its HT estimator $\hat{N}_{\rm HT}  = \sum_{i\in A} \pi_i^{-1}$.  Using this idea, we can estimate the entire cumulative distribution function of the population, given by 
$$ F(y)= \frac{1}{N} \sum_{i=1}^N I(y_i \le y). $$
That is, we can use 
$$ \hat{F}(y)= \frac{1}{ \sum_{i\in A} \pi_i^{-1}  }\sum_{i\in  A} \pi_i^{-1}  I(y_i \le y). $$
Using  $\hat{F} (y)$, we can also perform a quantile estimation. We can define the $q\%$ quantile $(0<q<1)$ of the population as $\theta_q = \inf\{ y;  F(y) \ge q\}$. Its HT estimator can be written as $\hat{\theta}_q= \inf \{ y ; \hat{F}(y) \ge q\}$. 
\index{quantile}

Many population parameters can be written as the solution to the population estimating equation 
\begin{equation} 
\sum_{i=1}^N U(\theta; y_i)=0
\label{2.15a}
\end{equation} 
for some estimating function $U(\theta; y_i)$. For example, the $100\times q\%$ quantile can be defined as $U(\theta; y) = I(y <\theta)-q$. To estimate the parameter defined as the solution to 
$$\sum_{i=1}^N U(\theta; y_i)=0, $$ 
one can use the HT estimation idea to get 
\begin{equation} 
\sum_{i \in A} \pi_i^{-1} U(\theta; y_i)=0
\label{2.16a}
\end{equation} 
as the estimating equation for $\theta_N$ defined through (\ref{2.15a}). \cite{Binder1983} investigated the asymptotic properties of the estimator of the form in (\ref{2.16a}). 

\section{Discussion}

In probability sampling, the HT estimator is regarded a gold standard estimator because of its design-unbiasedness. One may wonder whether there is another method to consider. To investigate this, let us consider the following weighted estimator. 
\begin{equation} 
\hat{Y}_w = \sum_{i \in A} w_i y_i
\label{2.17} 
\end{equation} 
where $w_i$ is the \emph{fixed} weight assigned to unit $i$. Here, ``fixed'' means that it is not random in the sense that it does not depend on the sampling indictor function $I_i$ defined in (\ref{2.3}).

The design-expectation of  $\hat{Y}_w$ is 
\begin{equation} 
E( \hat{Y}_w) = \sum_{i=1}^N \pi_i w_i y_i 
\end{equation} 
where $\pi_i$ is the first-order inclusion probability of unit $i$. To obtain design-unbiasedness, we require $\sum_{i =1}^N \pi_i w_i y_i=\sum_{i=1}^N y_i$ for all $y_i$ in the population, which leads to $w_i=\pi_i^{-1}$.

Now, suppose that we relax $w_i=\pi_i^{-1}$ and only require $\sum_{i=1}^N \pi_i w_i=N$. In this case, 
the bias of $\hat{Y}_w$ is 
\begin{eqnarray*} 
Bias( \hat{Y}_w ) &=& \sum_{i=1}^N \pi_i w_i y_i - \sum_{i=1}^N y_i \\
&=& \sum_{i=1}^N \pi_i w_i y_i - N^{-1} \left(  \sum_{i=1}^N \pi_i w_i \right)\left( \sum_{i=1}^N y_i\right) \\
&=& N \times Cov( q_i, y_i)  ,
\end{eqnarray*} 
where $q_i=\pi_i w_i$ and 
$$ Cov( q_i, y_i) = \frac{1}{N} \sum_{i=1}^N \left( q_i - \bar{q}_N \right) \left( y_i - \bar{y}_N \right)  $$
is the population covariance between $q_i$ and $y_i$, using an operator notation. 

Thus, $\bar{q}_N=N^{-1} \sum_{i=1}^N q_i = 1$, we can express 
\begin{eqnarray*} 
\frac{ Bias( \hat{Y}_w)}{Y}  &=&   Cov( q_i, y_i) / \bar{y}_N \\
&=&  Corr( q_i, y_i) \times  CV( q_i) \times CV(y_i).
\end{eqnarray*} 
Note the HT estimator uses $q_i=1$ which leads to $Corr( q_i, y_i)=0$ and $CV( q_i)=0$. Other estimators can also achieve zero bias if $Corr( q_i, y_i)=0$. For example, if $w_i=N/n$ and $y_i$ is completely independent of $\pi_i$, then the bias of $\hat{Y}_w$ is also zero. 

Another class of unbiased estimator is the difference estimator of the form
\begin{equation} 
\hat{Y}_{{\rm diff}}  = \sum_{i=1}^N x_i + \sum_{i \in A} \frac{y_i-x_i}{\pi_i }  .
\label{eq:diff}
\end{equation} 
\index{difference estimator}
 The difference estimator in (\ref{eq:diff}) is popular in accounting, where $y_i$ is the audit value and $x_i$ is the book value for the reporting unit $i$.  More discussion of the difference estimator will be given in Chapter 8. 

%% file: chapters/chapter3.tex

\setcounter{chapter}{2} 
\chapter{Simple and systematic sampling designs}

\section{Introduction}

Before selecting the sample, the population must be divided into parts that are called \emph{sampling units}, or \emph{units}. These units must cover the whole population, and they must not overlap in the sense that every element in the population belongs to one and only one unit. Sometimes, the appropriate unit is obvious, as in a population of light bulb, in which the unit is the single bulb. Sometimes there is a choice of unit. When sampling people in a town, the unit might be an individual person, the members of a family, or all persons living in the same city block. In sampling an agricultural crop, the unit might be a field, a farm, or an area of land whose shape and dimensions are at our disposal.

The construction of this list of sampling units, called a sampling \emph{frame}, is often one of the main practical problems. We use the term direct element sampling to denote sample selection from a frame that directly identifies the individual elements of the population of interest. That is, in element sampling, the sampling unit is equal to the reporting unit.  A selection of elements can take place directly from the frame. In this chapter, we first consider a simple type of sampling design in which the first-order inclusion probabilities are equal for every element in the population. 


\section{Simple random sampling}

Consider the problem of selecting $n$ units from a finite population of size $N$. There are $N \choose n$  possible samples in this case and the simplest way to select a sample is to select one randomly, that is, to select one realization at random with equal probability. This sampling design is called simple random sampling (SRS) without replacement \index{simple random sampling
without replacement; SRS}, or simple random sampling. The sample distribution of the SRS of size $n$  is given by 
\begin{equation}
  P(A)=
\begin{cases} {N \choose n}^{-1}
& \mbox{ if } \quad \left|A\right|=n \\ 0 & \mbox{ otherwise}.
\label{2.12}
\end{cases}
\end{equation}
From (\ref{2.12}), we can obtain the first-order inclusion probability as 
\begin{eqnarray*}
\pi_i &=& P ( i \in A) \\
&=& \sum_{A \in \mathcal{A}} I( i \in A) P(A) \\
&=& \frac{ {1 \choose 1}{ N-1 \choose n-1} }{ {N \choose n} } \\
&=& \frac{n}{N}.
\end{eqnarray*}
Similarly, we can obtain 
\begin{eqnarray}
\pi_{ij} &=& \begin{cases}
 N^{-1} n & \mbox{ if } i =j \\
N^{-1}\left( N-1 \right)^{-1} n \left( n - 1\right) & \mbox{ if }  i
\neq j.
\end{cases}
\label{2.12b}
\end{eqnarray}
Thus, the 
  Horvitz-Thompson  estimator of the population total $Y=\sum_{i=1}^N y_i$ can be written 
  \begin{equation}
\hat{Y}_{\rm HT} =\frac{N}{n} \sum_{i \in A} y_{i} = N \bar{y}.
\label{3.3}
\end{equation}
and the HT estimator satisfies design-unbiasedness under the SRS design. That is, under SRS, the sample mean $\left( \bar{y} \right)$ is unbiased for the population mean $\bar{Y} = N^{-1} \sum_{i=1}^N y_i$. The sampling variance of the HT estimator is, by (\ref{2.8}), 
\begin{eqnarray*}
V\left( \hat{Y}_{\rm HT} \right) &=& -\frac{1}{2} \sum_{i=1}^N
\sum_{\stackrel{j \neq i}{j =1}}^N \left( \pi_{ij} - \pi_i \pi_j
\right)\left(  \frac{y_i}{\pi_i } - \frac{y_j}{\pi_j }
\right)^2 \\
&=&  \frac{1}{2} \frac{N}{n} \frac{N-n}{N\left( N-1
\right)}\sum_{i=1}^N \sum_{{j =1}}^N \left(  y_i - y_j \right)^2.
\end{eqnarray*}
Since 
\begin{eqnarray}
 \sum_{i=1}^N \sum_{j=1}^N \left( y_i - y_j \right)^2
&=& \sum_{i=1}^N \sum_{j=1}^N \left( y_i - \bar{Y}+ \bar{Y}- y_j \right)^2 \notag \\
&=& 2 N \sum_{i=1}^N \left( y_i - \bar{Y} \right)^2, 
\label{3.3b}
 \end{eqnarray}
we can obtain 
\begin{eqnarray*}
V\left( \hat{Y}_{\rm HT} \right) &=& \frac{N^2}{n} \frac{N-n}{N} S^2
\end{eqnarray*}
where 
\begin{eqnarray*}
S^2 & = & \frac{1}{N-1}\sum_{i=1}^N \left(  y_i - \bar{Y}
\right)^2\notag \\
& = & \frac{1}{2} \frac{1}{N \left( N-1 \right)} \sum_{i=1}^N
\sum_{j=1}^N
 \left( y_i - y_j  \right)^2. 
\end{eqnarray*}
Thus, we can establish 
\begin{equation}
V\left( \bar{y}_n \right) = \frac{1}{n} \frac{N-n}{N} S^2 . 
\label{eq3.5}
\end{equation}
For the special case of $n=1$,  (\ref{eq3.5}) becomes 
$$ \frac{1}{N}\sum_{i=1}^N \left(  y_i - \bar{Y} \right)^2, $$
which is often called population variance, denoted by $\sigma_y^2$.  That is, $\sigma_y^2$ can be understood as the sampling variance of the single sample observation under SRS with size $n=1$. Using the population variance, the
variance formula in (\ref{2.14}) can be written as 
\begin{equation}
V\left( \bar{y}_n  \right) = \frac{1}{n}\left( 1- \frac{n-1}{N-1}
\right)\sigma_y^2, \label{2.15}
\end{equation}
where  $1- \left( N-1 \right)^{-1} \left( n-1 \right)$ is the variance reduction factor due to without-replacement sampling and it is often called  
 FPC (Finite population
correction) term. \index{FPC} The FPC term disappears under simple random sampling with replacement.

Now, consider variance estimation of the HT estimator under SRS. Since SRS is a fixed-sized sampling design, we can use SYG variance estimation formula in (\ref{eq2.12}) to get 
\begin{eqnarray*}
 \hat{V}\left( \hat{Y}_{\rm HT} \right) &=&
-\frac{1}{2} \sum_{i \in A} \sum_{{j \in A}} \frac{ \left(
\pi_{ij} - \pi_i \pi_j \right)}{\pi_{ij}} \left( \frac{y_i}{\pi_i
} - \frac{y_j}{\pi_j }
\right)^2 \\
&=&  \frac{1}{2} \frac{N}{n} \frac{N-n}{n\left( n-1
\right)}\sum_{i\in A} \sum_{{j \in A}} \left(  y_i - y_j \right)^2. 
\end{eqnarray*}
Similarly to (\ref{3.3b}),  we can show 
\begin{equation}
 \sum_{i \in A}  \sum_{j \in A}  \left( y_i - y_j \right)^2
= 2 n \sum_{i \in A} \left( y_i - \bar{y} \right)^2 . \label{2.15b}
\end{equation}
Thus, 
we can obtain 
\begin{eqnarray}
\hat{V}\left( \hat{Y}_{\rm HT} \right) &=& \frac{N^2}{n} \frac{N-n}{N}
s^2
\end{eqnarray}
where 
$$ s^2 =
 \frac{1}{n-1}\sum_{i \in A} \left(  y_i - \bar{y} \right)^2. $$
Thus, under SRS, we have 
 \begin{equation}
  E \left( s^2  \right)
 = S^2.
 \label{eq3.9}
 \end{equation}

If $y$ is dichotomous, taking either 1 or 0, the population mean of $y$ is equal to the proportion of $y=1$ in the population, namely $P=Pr\left( y=1
\right)$. In this case, we can obtain 
$\sigma_y^2=P\left( 1-P \right)$ and the variance of the HT estimator  $\hat{P} = \bar{y}$ of $P$ is then 
equal to 
\begin{equation*}
V\left( \hat{P}\right) =\frac{1}{n} \left( 1-
 \frac{n}{N}\right) \frac{N}{N-1} P \left( 1-P \right)
\end{equation*}
and its unbiased estimator is 
\begin{equation*}
\hat{V}\left( \hat{P}  \right) =\frac{1}{n-1} \left( 1-
 \frac{n}{N}\right) \hat{P} \left( 1-\hat{P} \right).
\end{equation*}

We now discuss the determination of the sample size $n$ under simple random sampling. Under the significant level $\alpha$, the margin of error, denoted by $d$, is defined to satisfy 
$$ Pr \left\{  \left| \hat{\theta} - \theta \right| \le d  \right\} = 1- \alpha .$$
That is, $d$ is half of the length in the confidence interval for $\theta$, Thus, solving 
$$ z_{\alpha /2}  \sqrt{ \frac{1}{n} \left( 1- \frac{n}{N}\right) S^2 }
\le d $$
with respect to $n$ to get 
\begin{equation}
n \ge \frac{S^2 }{\left( d/z_{\alpha/2}\right)^2 + S^2/N,
}
\label{2sw}
\end{equation}
which provide the lower bound of the desired sample size for given $d$. However, since we usually do not know $S^2$ before the sample observation, we need an estimate for $S^2$, from a pilot survey or a similar survey in the same population. In many public opinion surveys, $y$ is a dichotomous variable and the maximum value of $S^2$ in this case is 1/4 and (\ref{2sw}) becomes 
\begin{equation}
n \ge 0.25 \left( \frac{ z_{\alpha/2}  }{ {d} } \right)^2. \label{nsize}
\end{equation}
For $\alpha=0.05$,  $z_{\alpha/2} \doteq 2$ and the above inequality reduces to  $n
\ge d^{-2}$.

\section{Implementation of simple random sampling}

To implement the SRS method in practice, one may consider a draw-by-draw method as follows: In the first draw, select one  element at random from the entire population $U$ with probability $1/N$. Let $k_1$ be the index of the element selected from the first draw. In the second draw, select one element at random from  $U - \{ k_1\}$ with probability $1/(N-1)$. Let $k_2$ be the index of the element selected from the second draw.  We can continue the process until the $n$-th draw. In  the $n$-th draw, select one element at random from $U - \{ k_1, \cdots, k_{n-1} \} $ with probability $(N-n+1)^{-1}$.

In this draw-by-draw procedure,  the probability of selecting $n$ individuals $k_1, \cdots, k_n$ in order is $(N-n)!/N!$, and there are $n!$ ways to order $\{k_1, \cdots, k_n\}$. Thus, the probability of selecting a sample $A$ of $n$ units is $n!$ times $(N-n)!/N!$, which is equal to ${N \choose n}^{-1}$. Such a draw-by-draw method may be quite cumbersome if $N$ is very large, as it would require numbering the elements in the population in advance and then repeating the process $n$ times.

Another method of drawing a sample for SRS is the so-called random sorting method. The idea is to sort the population in a random order and then take the first $n$ units from the ordered population as the final sample. \cite{sunter1977} showed that this random sorting method indeed implements simple random sampling. To see this, 
let $u_i$ be the random number for unit $i$, generated from the $U(0,1)$ distribution. To select a particular sample $A=\{1, \cdots, n\}$ under the descending order sorting, the largest number among $\{u_{n+1}, \cdots, u_N\}$ should be less than the smallest number among $\{u_1, \cdots, u_n\}$.

Let $X$ be the largest number among $\{u_{n+1}, \cdots, u_N\}$. Note that the CDF of $X$ is 
\begin{eqnarray*}
 F(x)  &=& P( u_{n+1} \le x, \cdots, u_N \le x) \\
 &=& \prod_{i=n+1}^N P( u_i \le x) \\
 &=& x^{N-n} 
\end{eqnarray*} 
for $0 < x < 1$. 
The probability of selecting $A=\{1, \cdots, n\}$ is equal to 
\begin{eqnarray*} 
 P(A) &=& \int_0^1 \left\{ \prod_{i=1}^n P( u_i > x)  \right\} d F(x) \\
 &=& \int_0^1 (1-x)^n (N-n) x^{N-n-1} d x \\
 &=& \frac{ n! (N-n)!}{ N!} = \frac{1}{{N \choose n}  }. 
\end{eqnarray*} 
Thus, the random sorting method indeed implements the simple random sampling. The sorting  algorithm can be computationally costly if $N$ is large.

We now consider a different sampling method that does not require reading the population list in advance before sampling. 
The selection-rejection method allows the selection of SRS in a single pass through the population list, but we need to know the population size $N$ and the sample size $n$. The selection-rejection method can be described as follows:

  \begin{description}
        \item{[Step 0]} Start with $k=1$. 
        \item{[Step 1]} Generate $u_k \sim U(0,1)$ independently. 
        \item{[Step 2]} Check if 
          $$ u_k < \frac{ n-\sum_{j=1}^{k-1} I_j}{ N+1 - k}.$$
          If yes,  select $k+1$ into sample and set $I_k=1$. Otherwise, set $I_k=0$. 
        \item{[Step 3]} Set $k=k+1$.  If $\sum_{j=1}^{k} I_j < n$ then goto [Step 1]. Otherwise Stop. 
    \end{description}

Note that 
$$ P( k \in A \mid I_1, \cdots, I_{k-1} ) = \frac{n-\sum_{j=1}^{k-1} I_j }{N+1-k} 
$$
for $k=1, \ldots, N$. Note that $n- \sum_{j=1}^{k-1} I_j$ is the remaining sample size and $N+1-k$ is the remaining population size at the $k$-th pass. 
Thus, writing $n_{k-1}= \sum_{j=1}^{k-1} I_j$, 
\begin{eqnarray*}
P( A) &=& \left[ \prod_{k \in A} \frac{n-n_{k-1}}{N-(k-1)}  \right] \times \left[ \prod_{k \in A^c} \left\{ 1-  \frac{n-n_{k-1}}{N-(k-1)} \right\} \right] \\
&=& \left\{ \prod_{k=1}^{N} ( N-k+1) \right\}^{-1}  \left[ \prod_{k \in A} (n-n_{k-1}) \right] \left[ \prod_{ k \in A^c} (N-k+1-n+n_{k-1}) \right] \\
&=& \left( N!\right)^{-1} n!  (N-n)! = \frac{1}{{N \choose n}  }. 
\end{eqnarray*}
The selection-rejection method implements simple random sampling, but the population size $N$ is needed to compute the selection probability.

\cite{mcleod1983} proposed a novel method for implementing the SRS in a single pass  through the list of records, and does not require a known population size $N$.  The method is later named the reservoir sampling method \citep{vitter1985} in the sense that $n$ sample elements are selected in a reservoir and then replaced if the next element in the population is selected. In the proposed reservoir method, the first $n$ elements of the population are stored in the reservoir.  The $k$-th element ($k=n+1, \cdots, N$) is selected in the reservoir with probability $n/k$ and then one of the $n$ elements is removed from the reservoir with equal probability. The elements in the reservoir that remain after the final selection are the elements in the final sample.  Note that the population size is not necessarily known. You can stop any
time point of the process, then you will obtain a simple random sample
from the finite population considered up to that time point.

To explain the reservoir sampling, let $U_k=\{1, 2, \cdots, k\}$ be the finite population up to element $k$. Let $A_k$ be the index set of the sample elements selected from the reservoir sampling. The probability of selecting element $j(<k)$ in the sample can be written as 
\begin{eqnarray}
P( j \in A_k \mid U_k ) &=& P( j \in A_{k-1} \mid U_{k-1} ) P(j \in A_{k} \mbox{ and } k \in A_{k} \mid U_{k}, A_{k-1} )  \notag \\
&+& P( j \in A_{k-1} \mid U_{k-1} ) P( k \notin A_{k} \mid U_{k}, A_{k-1} ) . \label{3-12} 
\end{eqnarray} 
Now, by the reservoir sampling mechanism, we obtain 
\begin{eqnarray*}
 P(j \in A_{k} \mbox{ and } k \in A_{k} \mid U_{k}, A_{k-1} )  &=& P(j \in A_{k}  \mid U_{k}, A_{k-1} ) P(k \in A_{k} \mid U_{k}, A_{k-1} ) \\
 &=& \frac{ n-1}{n} \times \frac{ n}{k} 
 \end{eqnarray*} 
and 
$$ P( k \notin A_{k} \mid U_{k}, A_{k-1} ) = 1- \frac{n}{k}. $$
 Thus, (\ref{3-12}) reduces to 
 \begin{eqnarray}
 P( j \in A_k \mid U_k ) &=& P( j \in A_{k-1} \mid U_{k-1} ) 
 \times \left(   \frac{ k-1}{k}\right) . 
 \label{3-14a} 
 \end{eqnarray} 
 For $k=n$, $P( j \in A_k \mid U_k ) =1$. Thus, by (\ref{3-14a}), we  obtain 
 \begin{equation} 
 P( j \in A_k \mid U_k ) = \frac{ n}{k} ,
 \label{3-13}
 \end{equation} 
 which is equal to the first-inclusion probability of the SRS of size $n$ from the finite population of size $k$.

\section{Simple random sampling with replacement}

We now consider the sampling design where the sample of size $n$ is selected with equal probability with replacement. The realized sample size can be smaller than $n$ because the sample is selected with replacement, and thus allows for duplication. In the $k$-th sample draw, where $k=1, \cdots, n$, the $i$-th element in the population is selected with probability $1/N$. Let $z_1$ be the realized value of $y_i$ in the first draw. The probability distribution of $z_1$ is given by  
\begin{equation*}
z_1 = \left\{ \begin{array}{ll} y_1 & \mbox{ with probability }
1/N \\
y_2 & \mbox{ with probability }
1/N \\
\vdots & \\
y_N & \mbox{ with probability } 1/N.
\end{array}
\right.
\end{equation*}
Similarly, let $z_k$ be the realized value of $y_i$ in the $k$-th draw. The with-replacement sampling makes $z_1, \cdots, z_n$ be independently and identically distributed (IID). The mean and variance of $z_1$ is 
\begin{eqnarray}
E\left( z_1 \right)&=&N^{-1} \sum_{i=1}^N y_i=\bar{Y}
\label{2.18} \\
V\left( z_1  \right)&=&N^{-1} \sum_{i=1}^N \left( y_i-\bar{y}_N
\right)^2 = \sigma_y^2 .\label{2.19}
\end{eqnarray}
Now, since $z_1, \cdots, z_n$ are IID with mean $\bar{Y}$, we can consider the following class of estimators: 
\begin{equation}
 \hat{\bar{Y}}_w= \sum_{i=1}^n w_i z_i
 \label{3.16b}
 \end{equation}
 where $\sum_{i=1}^n w_i=1$. Note that $\hat{\bar{Y}}_w$ is unbiased regardless of the choice of $w_i$. To find the best choice of $w_i$ in the sense of minimizing its variance, we introduce the following lemma. 
 \begin{lemma}
 Let $X_1, \cdots, X_n$ be $n$ IID realization with $E(X_i) = \mu$ and $V(X_i) = \sigma_i^2$. Let $w_1, \cdots, w_n$ be the fixed constants such that $\sum_{i=1}^n w_i=1$. The  weighted estimator $\hat{\mu}_w=\sum_{i=1}^n w_i X_i$ achieves the minimum at 
     $$ w_i^* = \frac{ \sigma_i^{-2} }{ \sum_{k=1}^n \sigma_k^{-2}}. $$
 \end{lemma}
\begin{proof}
The variance of $\hat{\mu}_w$ is $V(\hat{\mu}_w) = \sum_{i=1}^n w_i^2 \sigma_i^2$. Using Cauchy-Schwarz inequality, we obtain 
$$ \left( \sum_{i=1}^n w_i^2 \sigma_i^2 \right) \left( \sum_{i=1}^n \sigma_i^{-2} \right) \ge \left( \sum_{i=1}^n w_i \right)^2.
$$
Since $\sum_{i=1}^nw_i=1$, we have 
$$ \sum_{i=1}^n w_i^2 \sigma_i^2 \ge \mbox{Constant} $$
and the variance is minimized when $w_i \sigma_i \propto \sigma_i^{-1}$, which is equivalent to $w_i \propto \sigma_i^{-2}$.
\end{proof}

Thus, the best linear unbiased estimator of the population mean $\bar{Y}$ is the sample mean  $\bar{z}=n^{-1} \sum_{i=1}^n z_i$ and its variance is 
\begin{equation}
V\left(\bar{z} \right) = \frac{1}{n} \sigma_y^2 \label{2.20}
\end{equation}
The variance formula in  (\ref{2.15}) under SRS is smaller than the variance formula in (\ref{2.20}) under SRS with replacement. The SRS with replacement is less efficient because the  expected value of the actual sample size is smaller  as it allows for duplication due to with-replacement sampling.

For variance estimation, since $z_1, \cdots, z_n$ are IID, the sample variance 
$$ s_z^{2} = \frac{1}{n-1} \sum_{i=1}^n \left( z_i - \bar{z}
\right)^2 $$ 
can be used to estimate the population variance. Since 
 $ z_1, z_2, \ldots , z_n$ are IID with $(\bar{Y}, \sigma_y^2)$, we have 
\begin{equation}
E\left( s_z^{2} \right) = \sigma_y^2 \label{2.21}
\end{equation}
and the variance estimator of $\bar{z}$ is 
$$\hat{V}\left(\bar{z}\right) = \frac{1}{n}s_z^{2}.
$$
The SRS with replacement is a special case of the PPS sampling that will be covered in Section 5.2. 
\section{Bernoulli sampling}

Bernoulli sampling design is a sampling design based on independent Bernoulli trials for the element in the population. That is, the sample indicator function follows 
\begin{equation}
I_i \iid Bernoulli \left(\pi \right) , \ \ \ i=1,2,\cdots, N,
\label{ber0}
\end{equation}
where $\pi$ is the first order inclusion probability for each unit. We can express $\pi=n_0/N$ where $n_0$ is the expected sample size determined before sample selection. In this Bernoulli sampling, the (realized) sample size follows a binomial distribution $Bin(N, \pi)$ and the fact that the realized sample size, $n= \left| A \right|$,  is a random variable can reduce the efficiency of the resulting HT estimator. 

Under this Bernoulli sampling, the HT estimator of the population mean is 
$${\widehat{\bar{Y}}}_{\rm HT} = \frac{1}{N} \sum_{i \in A} \frac{ y_i}{\pi_i}  . $$ 
Thus, the HT estimator of the mean is not necessarily equal to the simple mean of the sample. Now, the sample mean $\bar{y} = n^{-1} \sum_{i \in A} y_i$
can be expressed as a ratio of two HT estimators: 
$$ \bar{y} = \frac{ \widehat{Y}_{\rm HT} }{\widehat{N}_{\rm HT} } = \frac{ \sum_{i \in A} \pi_i^{-1} y_i}{ \sum_{i \in A } \pi_i^{-1} },  $$
where $\pi_i = n_0/N$. 
Thus, using Taylor linearization, the asymptotic  variance of the sample mean $\bar{y}$ 
is 
\begin{equation}
 V\left( \bar{y} \right) \cong \frac{1}{n_0} \left( 1- \frac{n_0}{N} \right)
S^2 . \label{ber}
\end{equation}
On the other hand, the variance of the HT estimator of the mean is 
\begin{equation}
 V \left( \widehat{\bar{Y}}_{\rm HT} \right) = \frac{1}{n_0} \left( 1- \frac{n_0}{N}
\right) \frac{1}{N} \sum_{i=1}^N y_i^2 \label{ber2} \end{equation}
Thus, the sample mean is generally more efficient than the HT estimator in (\ref{ber2}). 

\begin{example}
Suppose that we are interested in estimating the proportion of students who pass a certain test in a university and there are N=600 of students who took the test in the university.  Using  a
 Bernoulli sampling with $\pi=1/6$, the sample size $n=90$ is realized. Among the 90 sample students, 60 students are found to have passed. In this case, the HT estimator of the mean is  $0.9 \times 2/3$, which is different from the actual passing rate $2/3$ in the sample. In the extreme case, even if all the students pass the exam, the HT estimate is still 0.9. 
\end{example}

If the sampling procedure is such that we repeat the Bernoulli sampling until the realized sample size $n$ is equal to the expected sample size $n_0$, then the resulting sampling procedure is exactly equal to the SRS of size $n_0$. To show this result, note that 
\begin{equation}
Pr\left( I_1, I_2, \ldots, I_N \mid \sum_{i=1}^N I_i = n_0 \right) =
\frac{Pr \left( I_1, I_2, \cdots, I_N , \sum_{i=1}^N I_i = n_0
\right) }{ Pr\left( n=n_0 \right)} .\label{2.23}
\end{equation}
Since 
\begin{eqnarray*}
Pr \left( I_1, I_2, \ldots, I_N , \sum_{i=1}^N I_i = n_0 \right) &=&
\left\{ \begin{array}{ll} \prod_{i=1}^N p^{I_i} \left( 1-p
\right)^{1-I_i} & \mbox{ if }\sum_{i=1}^N I_i = n_0 \\
0 & \mbox{ otherwise}
\end{array}
\right.
\\
&=&  \left\{ \begin{array}{ll} p^{n_0}  \left( 1-p
\right)^{N-n_0} & \mbox{ if }\sum_{i=1}^N I_i = n_0 \\
0 & \mbox{ otherwise}
\end{array}
\right.
\end{eqnarray*}
and $$ Pr\left( n=n_0 \right) =\left( \begin{array}{l} N \\ n_0
\end{array}
 \right) p^{n_0 } \left( 1- p \right)^{N-n_0},
$$
the conditional density in  (\ref{2.23}) is equal to the sampling design 
(\ref{2.12}) under SRS of size $n_0$.

\section{Systematic sampling}

Systematic sampling is an alternative method of selecting an equal probability sample, but it offers several practical advantages, particularly its simplicity of execution. In systematic sampling, a first element is drawn at random with equal probability, among the first $G$ elements in the population list. The positive integer $G$ is fixed in advance and is called the \emph{sampling interval}. \index{sampling interval} \index{systematic sampling}   The rest of the sample is determined by systematically taking every $G$-th element thereafter, until the end of the list. Thus, there are only $G$ possible samples, each having the same probability of being selected. The simplicity of having only one random draw is a great advantage. For example, to select a sample of 200 students from the list of 20,000 students at Iowa State University, we first select one element among the first 100 students. Suppose that the random integer we choose is 73. Then the students numbered 73, 173, 273, $\cdots$,  19,973 would be in the sample.

For a more formal definition of systematic sampling, let $G$ be the fixed sampling interval and let $n$ be the integer part of $N/G$, where $N$ is the population size. Then, 
$$ N=n \cdot G + c$$
where the integer $c$ satisfies $0 \le c < G$. 
In systematic sampling, we first select one integer from $\{1, 2, \ldots, G\}$  with equal probability $1/G$. If $r$ is selected from the selection, the final sample of systematic sampling is
$$ A_r =\left\{ \begin{array}{ll}
\left\{ r, r+G, r+ 2G, \ldots, r + \left( n-1 \right) G \right\} &
\mbox{ if } c < r \le G \\
\\
\left\{ r, r+G, r+ 2G, \ldots, r + n G \right\} &
\mbox{ if } 1 \le r \le c . \\
\end{array}
\right.
$$
The first order inclusion probability for each unit is $\pi_i=1/G$ but the second order inclusion probability is 
$$ \pi_{ij} = \left\{ 
\begin{array}{ll} 
1/G & \mbox{ if } j=i+ k G  \mbox{ for some integer } k \\
0 & \mbox{ otherwise.}
\end{array}
\right.
$$
That is, systematic sampling can be viewed as selecting one cluster at random among the $a$ possible clusters. In this case, the second-order inclusion probability of two units is positive only when the two units belong to the same cluster. Thus, an unbiased estimator of the variance of the HT estimator does not exist. \cite{wolter1984} discuss variance estimation under systematic sampling. 
In addition, the efficiency of systematic sampling depends on the way the list is sorted. Such a concept can be investigated using the intracluster correlation coefficient in cluster sampling, which will be covered in Chapter 6.

In systematic sampling, the finite population $U$ is partitioned into $G$ groups
$$ U = U_1 \cup U_2 \cup \cdots \cup U_G , $$
where $U_i$ are mutually disjoint. The population total is then expressed as
$$ Y = \sum_{i \in U} y_i = \sum_{r=1}^G \sum_{k \in U_r} y_k = \sum_{r=1}^G t_{r} , $$
where $t_{r}= \sum_{k \in U_r} y_k $. Thus, in estimating the total,  the finite population can be treated as a population with $a$ elements with measurements $t_{1}, \cdots, {t}_G$.

The HT estimator can be written
$$ \hat{Y}_{\rm HT} = \frac{ t_r }{1/G} = G \sum_{k \in A} y_k ,$$
if $A=U_r$. Since we select SRS from the population of $G$ elements $\left\{t_1, \cdots, t_G \right\}$, the variance is 
$$ V\left( \hat{Y}_{\rm HT} \right) = \frac{G^2}{1} \left( 1- \frac{1}{G} \right) S_t^2 $$
where
$$S_t^2 = \frac{1}{G-1} \sum_{r=1}^G \left( t_{r} - \bar{t} \right)^2 $$
and $\bar{t} = \sum_{r=1}^G t_{r}/G$.

Now, assuming $N=n \cdot G$
\begin{eqnarray*}
V\left( \hat{Y}_{\rm HT} \right) &=&  G \left( G-1 \right) S_t^2 \\
&=& n^2 G \sum_{r=1}^G \left( \bar{y}_{r} - \bar{y}_u \right)^2
\end{eqnarray*}
where $\bar{y}_{r} = t_{r}/n$ and $\bar{y}_u = \bar{t}/n$. Since 
 $U=\cup_{r=1}^G U_r$, we can use the ANOVA decomposition  to get 
\begin{eqnarray*}
SST &=& \sum_{k \in U} \left( y_k - \bar{y}_u \right)^2  = \sum_{r=1}^G \sum_{k \in U_r}\left( y_k - \bar{y}_u \right)^2  \\
&=& \sum_{r=1}^G \sum_{k \in U_r} \left( y_k - \bar{y}_{r} \right)^2 + n \sum_{r=1}^G \left( \bar{y}_{r} - \bar{y}_u \right)^2 \\
&=& SSW + SSB.
\end{eqnarray*}
Thus, the variance can be written 
$$V\left( \hat{Y}_{\rm HT} \right) = nG \cdot SSB = N \cdot SSB = N\left( SST - SSW \right).$$
 If SSB  is small, then $\bar{y}_{r}$ are more similar and $V( \hat{Y}_{\rm HT} )$ is small.
  If the SSW is small, then $V ( \hat{Y}_{\rm HT} )$ is large.

To compare the systematic sampling and SRS in terms of variance, note that 
\begin{eqnarray*}
V_{\rm SRS} \left( \hat{Y}_{\rm HT} \right)&=& \frac{N^2}{n} \left( 1- \frac{n}{N} \right) \frac{1}{N-1} \sum_{k=1}^N \left( y_k - \bar{Y}_N \right)^2 \\
V_{\rm SYS} \left( \hat{Y}_{\rm HT} \right)&=&  n^2 G \sum_{r=1}^G \left( \bar{y}_{r} - \bar{y}_u \right)^2.
\end{eqnarray*}
We can compare the variance by making additional assumptions about the finite population. \cite{cochran1946}  introduced superpopulation model which the finite population is believed to be generated. The superpopulation model is an assumption about the finite population, and it quantifies the characteristics of the finite population in terms of a smaller number of parameters. 

If the finite population is ordered randomly, then we may use an IID model, denoted by $\zeta$: $\left\{ y_k \right\} \iid \left( \mu, \sigma^2 \right)$. In this case, we can obtain 
\begin{eqnarray*}
 E_{\zeta} \left\{V_{\rm SRS} \left( \hat{Y}_{HT} \right) \right\} &=& \frac{N^2}{n} \left( 1- \frac{n}{N} \right) \sigma^2 \\
  E_{\zeta} \left\{V_{\rm SYS} \left( \hat{Y}_{HT} \right) \right\} &=& \frac{N^2}{n} \left( 1- \frac{n}{N} \right) \sigma^2.
\end{eqnarray*}
Thus, the model expectations for the design variances are the same in the IID model.

\begin{example}

Consider a finite population of size $N$ with linear trend. That is, we assume  the following superpopulation model 
\begin{equation} 
 E_{\zeta} \left( y_k \right) = \beta_0 + \beta_1 k , \ \ \ V_{\zeta} \left( y_k \right) = \sigma^2. 
 \label{eqn:3-25} 
 \end{equation} 
To compute the model expectation of the design variance, we first 
note that,  under the independence model $\zeta$ with $E_\zeta \left( y_i \right) = \mu_i $ and $V_\zeta \left( y_i \right)=\sigma^2$, we can derive
\begin{equation} 
E_\zeta \left( S^2 \right) = \frac{1}{N-1} \sum_{i=1}^N \left( \mu_i - \bar{\mu}_N  \right)^2 + \sigma^2,
\label{eqn:3-25b}
\end{equation} 
where $\bar{\mu}_N = N^{-1} \sum_{i=1}^N \mu_i $. Thus, under (\ref{eqn:3-25}), we obtain 
\begin{eqnarray*}
E_{\zeta} (S^2) &=& \frac{\beta_1^2}{N-1} \sum_{i=1}^N \left( k - \frac{N+1}{2} \right)^2 + \sigma^2 \\
&=& \beta_1^2 \frac{N}{N-1} \frac{N^2-1}{12} + \sigma^2 \\
&=& \frac{ N(N+1) \beta_1^2}{12} + \sigma^2 
\end{eqnarray*}
and so 
\begin{equation} 
 E_{\zeta} \left\{V_{\rm SRS} \left( \widehat{\bar{Y}}_{\rm HT} \right) \right\} = \frac{1}{n} \left( 1- \frac{n}{N} \right) \left\{  \frac{ N(N+1) \beta_1^2}{12} + \sigma^2 \right\} .
\label{eqn:3-26} 
\end{equation} 
Now, to compute the model expectation of 
$$ V_{\rm SYS} ( \widehat{\bar{Y}}_{\rm HT} ) = \frac{1}{G} \sum_{i=r}^G \left( \bar{y}_r- \bar{y}_i  \right)^2 = \frac{G-1}{G} S_z^2, $$
where 
$$ S_{z}^2 = \frac{1}{G-1} \sum_{r=1}^G \left( z_r - \bar{z}_G \right)^2 $$
 $z_r = \bar{y}_r$ and $\bar{z}_G= G^{-1} \sum_{r=1}^G z_r$, we can obtain, similarly to (\ref{eqn:3-25b}),   
$$ E_{\zeta} \left( S_z^2 \right) = \frac{1}{G-1} \sum_{r=1}^G \{ E_\zeta(z_r) - E_\zeta( \bar{z}_G) \}^2 + V_{\zeta} (z) .
$$
Since 
$ z_r = n^{-1} \sum_{k=1}^n y_{r+ (k-1) G} $ for $r=1, \cdots, G$, 
we have 
$$ E( z_r) = \frac{1}{n} \sum_{k=1}^n \{ \beta_0+ \beta_1 r+ \beta_1 (k-1)G \} = \beta_0+ \beta_1 r + \beta_1 G \frac{n(n-1)}{2}  
$$
and $V_{\zeta}(z_r)=\sigma^2/n$. Thus, 
\begin{eqnarray*} 
 E_{\zeta} \left( S_z^2 \right) &=& \frac{\beta_1^2}{G-1} \sum_{r=1}^G \left\{   r - \frac{G+1}{2}  \right\}^2 + \frac{\sigma^2}{n} \\
 &=& \beta_1^2 \frac{G}{G-1} \frac{G^2-1}{12} + \frac{\sigma^2}{n}
\end{eqnarray*} 
and 
\begin{eqnarray} 
 E_{\zeta} \left\{ V_{\rm SYS} ( \widehat{\bar{Y}}_{\rm HT} ) \right\} &=& \frac{G-1}{G} \left\{  \beta_1^2 \frac{G(G+1)}{12} + \frac{\sigma^2}{n}\right\} \notag \\
&=& \frac{1}{n} \left( 1- \frac{n}{N} \right) \left\{  \frac{ N(N+n) \beta_1^2}{12n} + \sigma^2 \right\}. 
\label{eqn:3-27} 
\end{eqnarray} 
Therefore, comparing (\ref{eqn:3-26}) and (\ref{eqn:3-27}), we can establish that 
$$  E_{\zeta} \left\{ V_{\rm SYS} \left( \widehat{\bar{Y}}_{\rm HT} \right) \right\} \le E_{\zeta} \left\{  V_{\rm SRS} \left( \widehat{\bar{Y}}_{\rm HT}  \right) \right\} .$$
\end{example}
\section{Entropy for sampling designs} 

We now introduce a concept which gives a characterization of sampling designs. 

\begin{definition}
For a sample design $p( \cdot)$, the entropy of $p( \cdot)$ is the quantity 
\begin{equation} 
I( p) = - \sum_{A \in \mathcal{A}} p( A) \log \{ p(A) \} 
\label{entropy}
\end{equation} 
\end{definition}

Roughly  speaking, entropy is a measure of the randomness of a probability distribution. The larger the entropy, the more randomly the sample is selected.

The following lemma shows that the Bernoulli sampling design with $\pi=1/2$ is the maximum entropy sampling design among all possible probability sampling designs.  
\begin{lemma} 
The  maximum entropy design in $\mathcal{A} = \{ A; A \subset U\}$ is $P(A)=2^{-N}$, for all $A \subset U$. 
\label{lem:3-2}
\end{lemma} 
\begin{proof}
We are interested in maximizing $I(p)$ in (\ref{entropy}) subject to 
\begin{equation} 
\sum_{A \in \mathcal{A}} p(A) =1.
\label{eqn:3-28}
\end{equation} 
Using Lagrange multiplier method, we have only to find the maximizer of 
$$ Q(p, \lambda)= - \sum_{A \in \mathcal{A}} p( A) \log \{ p(A) \}  + \lambda \left( \sum_{A \in \mathcal{A}} p(A) -1 \right). 
$$
Solving 
$$ \frac{\partial}{ \partial p(A) }  Q(p, \lambda)=0 ,$$
we obtain  
\begin{equation} 
p(A) = \exp ( \lambda-1 ) .
\label{eqn:3-29}
\end{equation} 
Now, inserting (\ref{eqn:3-29}) into (\ref{eqn:3-28}), we get 
$$ \sum_{A \in \mathcal{A}} p(A) = 2^N \exp ( \lambda -1 ) = 1 . $$
Therefore, we obtain $p(A) = 2^{-N}$ for all $A \in \mathcal{A}$. 
\end{proof}

According to Lemma \ref{lem:3-2}, the maximum entropy is the entropy of the Bernoulli sampling design with $\pi=1/2$, which is equal to $$I( p^*) = - \sum_{ A \in \mathcal{A} }p^* (A) \log p^*(A) = N \log 2 . $$
This design is the most random among all sampling designs in the finite population of size $N$. 

The following theorem proves that simple random sampling without replacement is also a maximum entropy sampling design among the class of fixed-sample-size designs. 
\begin{theorem}
The maximum entropy design with fixed sample size $n$ is the simple random sampling design without replacement with a fixed sample size. 
\end{theorem} 
\begin{proof}
Let $\mathcal{A}_n$ be the set of all possible samples of size $n$. Using the Lagrange multiplier method, we only have to find the maximizer of 
$$ Q(p, \lambda)= - \sum_{A \in \mathcal{A}_n} p( A) \log \{ p(A) \}  + \lambda \left( \sum_{A \in \mathcal{A}_n} p(A) -1 \right). 
$$
Solving 
$$ \frac{\partial}{ \partial p(A) }  Q(p, \lambda)=0 ,$$
we obtain (\ref{eqn:3-29}).  
Now, using 
$$ \sum_{A \in \mathcal{A}_n} p(A) = {N \choose n} \exp ( \lambda -1 ) = 1 . $$
 we obtain $p(A) = 1/{N \choose n}$ for all $A \in \mathcal{A}_n$. 
\end{proof}

The entropy of a simple random sampling design is 
$$ I (p_{\rm SRS} ) = - \sum_{ A \in \mathcal{A}_n} \frac{1}{ {N \choose n} } \log {N \choose n}^{-1} = \log {N \choose n} . $$ 
On the other hand, the systematic sampling design has a very small entropy. There are only $G=N/n$ possible samples, and each of these samples has $1/G$ selection probability. Thus, the entropy of systematic sampling is 
$$ I( p_{\rm SYS} ) = - \sum_{ A \in \mathcal{A}} p(A) \log \{ p(A)\} =  G \cdot \frac{1}{G} \cdot \log \left( \frac{1}{G} \right) = \log G,  
$$
which is much smaller than that of simple random sampling of size $n=N/G$. 
In fact, \cite{pea2007} show that systematic sampling designs are minimum entropy designs. 

%% file: chapters/chapter4.tex

\setcounter{chapter}{3} 
\chapter{Stratified Sampling}

\section{Introduction}




Stratified sampling refers to sampling designs in which the finite population is divided into several subpopulations, called strata, and sample draws are made independently across each strata. The formal definition of the stratified sampling can be made as follows.

\begin{definition}{Stratified Sampling satisfies the following two conditions. } $ \phantom{} $
\eopen \item The finite population is stratified into $H$ mutually exclusive and exhaustive subpopulations, called strata. 
$$ U = U_1 \cup \cdots \cup U_H $$
and $U_h \cap U_g = \emptyset$ for $h \neq g$.  

\item Within each population (or stratum), samples are drawn independently across the strata.
$$ Pr \left( i \in A_h , j \in A_g \right) = Pr \left( i \in A_h  \right) Pr \left( j \in A_g \right), \ \ \ \mbox{ for } h \neq g $$
where $A_h$ is the index set of the sample in stratum $h$, $h=1,2, \cdots, H$.
\eclose
\end{definition} 
In stratified sampling, the sample size $n_h$ in stratum $h$ is under  control at the time of sampling design. Thus, we can control the precision of the study domains. Furthermore, by allowing different sampling methods for different strata, we have more flexibility in sample selection and data collection in practice. Generally speaking, stratified sampling improves the efficiency of survey estimates over sample random sampling. For these reasons, stratified sampling is very popular in practice. 

Let $U_h=\{ 1, \cdots, N_h\}$ be the indices in the population elements in stratum $h$ and
let $y_{hi}$ denote the measurement of the study item $y$ for unit $i$ in stratum $h$. The population total $Y=\sum_{h=1}^H Y_h$ is the sum of the stratum totals $Y_h=\sum_{i=1}^{N_h} y_{hi} $. The HT estimator of $Y$ is the sum of the HT estimator of $Y_h$. That is, 
$$ \hat{Y}_{\rm HT} = \sum_{h =1}^H \hat{Y}_{{\rm HT},h}. $$
Note that $\hat{Y}_{HT,h}$ is unbiased for $Y_h$. By the independence assumption, we obtain
$$ V \left( \hat{Y}_{\rm HT}\right) =
\sum_{h =1}^H  V \left( \hat{Y}_{{\rm HT},h} \right) .$$

\begin{example} (Stratified random sampling) 

In stratified random sampling, we have SRS independently for each stratum. In this case, the HT estimator of $Y$ is 
$$ \hat{Y}_{\rm HT} = \sum_{i=1}^H \frac{N_h}{n_h} \sum_{i \in A_h}
y_{hi}= \sum_{i=1}^H N_h \bar{y}_h
$$
where $N_h$ is the size of $U_h$ and $n_h $ is the size of $A_h$. Its variance is 
\begin{eqnarray}
       V(\hat{Y}_{\rm HT} ) &=& \sum_{h=1}^H N_h^2 V( \bar{y}_h ) =
        \sum_{h=1}^H \dfrac{N_h^2}{n_h}(1-\dfrac{n_h}{N_h})S_h^2
        \label{eq4.1}
\end{eqnarray}
where $\bar{Y}_h = N_h^{-1} \sum_{i=1}^{N_h} y_{hi}$ and 
$ S_h^2 =\left( N_h-1 \right)^{-1} \sum_{i=1}^{N_h} (y_{hi} - \bar{Y}_{h})^2.
$

To estimate the variance in (\ref{eq4.1}), we use 
\begin{equation*}
       \hat{V}(\hat{Y}_{\rm HT} ) = \sum_{h=1}^H \hat{V}(\hat{Y}_{HT,h}) =
        \sum_{h=1}^H \dfrac{N_h^2}{n_h}(1-\dfrac{n_h}{N_h})s_h^2
\end{equation*}
where  $ s_h^2 $ is an unbiased estimator of  $S_h^2$ and is given by 
$$ s_h^2 =\dfrac{1}{n_h-1} \sum_{i \in A_h}  (y_{hi} - \bar{y}_{h})^2.
$$
\end{example}

\section{Sample size allocation}

One of the important problems with stratified sampling is that, 
given the total sample size $n$, how to decide the sample size $n_h$ in stratum $h$ such that $\sum_{h=1}^H n_h =n $. Such a problem is called the sample size allocation problem for stratified sampling. 

One simple way is to use the proportional allocation, where the sample size in a stratum is proportional to the population size in the stratum. That is, 
\begin{equation}
n_h = N_h \frac{n}{N} \label{eq4.2}
\end{equation}
In this proportional allocation, the variance in  (\ref{eq4.1}) reduces to 
\begin{equation}
 V(\hat{Y}_{\rm HT} ) = \frac{N^2}{n} \left( 1-
 \frac{n}{N} \right) \sum_{h=1}^H \frac{N_h}{N} S_h^2.
 \label{eq:4.3}
\end{equation}
Assuming that $N_h$ are sufficiently large,  we have 
\begin{eqnarray}
\sum_{h=1}^H \frac{N_h}{N} S_h^2 &=& \frac{1}{N} \sum_{h=1}^H
\frac{N_h}{N_h -1 } \sum_{i=1}^{N_h} \left( y_{hi} - \bar{Y}_h
\right)^2 \notag \\
&\doteq &\frac{1}{N-1} \sum_{h=1}^H \sum_{i=1}^{N_h} \left( y_{hi} -
\bar{Y}_h \right)^2 \notag  \\
&\le &  \frac{1}{N-1} \sum_{h=1}^H \sum_{i=1}^{N_h} \left( y_{hi} -
\bar{Y} \right)^2= S^2. \label{2.29}
\end{eqnarray}
Thus, the variance  (\ref{eq:4.3}) under proportional allocation is no larger than the variance under simple random sampling. That is, stratified sampling under proportional allocation is more efficient than simple random sampling in terms of sampling variance. Here, the inequality in
(\ref{2.29}) is derived from the following equality: 
\begin{eqnarray*}
\sum_{h=1}^H \sum_{i=1}^{N_h} \left( y_{hi} - \bar{Y} \right)^2
&=&\sum_{h=1}^H N_h \left( \bar{Y}_{h} - \bar{Y} \right)^2
+\sum_{h=1}^H \sum_{i=1}^{N_h} \left( y_{hi} - \bar{Y}_h \right)^2,
\end{eqnarray*}
which is often expressed as ${\rm SST}={\rm SSB} + {\rm SSW}$. In 
(\ref{2.29}), the equality holds when ${\rm SSB}=0$ which is the case when $\bar{Y}_h$ are all the same. That is, there is no between-stratum variation. Another extreme case is when ${\rm SSW}=0$ which occurs when $y_{hi} = \bar{Y}_h$, (perfect homogeneity within stratum). In this extreme case, the variance (\ref{eq:4.3}) becomes zero. Hence, when stratum boundaries are formed, its statistical efficiency can be improved if the within-stratum variations are minimized, and the between-stratum variations are maximized. However, stratified sampling does not necessarily have a smaller variance than simple random sampling. In some poor sample size allocation, stratified sampling can have a larger sampling variance than simple random sampling. 

Another advantage of proportional allocation is that the sampling weights are all equal. A sampling design that has equal sampling weights is called a self-weighting design. \index{self-weighting} Self weighting is very convenient and popular in practice. 

To guarantee an unbiased estimate, we need to impose $n_h \ge 1$ for all $h=1, \cdots, H$. The proportional allocation satisfies the following three conditions: 
\begin{enumerate} 
        \item $n_h$ are integer valued with $n_h\ge 1$ for all $h=1, \cdots, H$. 
        \item $\sum_{h=1}^H n_h = n$
        \item $N_h/n_h \cong N/n$ as closely as possible for every $h$.  
        \end{enumerate}

To implement the proportional allocation, the Hintington-Hill method can be used as follows. 
 \begin{enumerate}
      \item  Calculate a set of “priority values” for each stratum $h$, based on the state’s apportionment population. The $s$-th priority value in stratum $h$ is defined as 
     $$ a_{h,s} = \frac{N_h}{\sqrt{s(s+1)}} $$
where $s$ is the number of sample units that the stratum has been allocated so far.
\item Sort these values from largest to smallest.
\item Allocate a seat to a state each time one of its priority values is included in the largest 385 values in the list.
    \end{enumerate}
    
To illustrate the Huntington-Hill method, consider the following toy example of $H=4$ strata, presented in Table \ref{table:4.1}.

\begin{table}[htb]
\begin{center}
\caption{A toy example of $H=4$ strata}
\begin{tabular}{c|r|r|r|r|r}
\hline
   Stratum & $N_h$  & $a_{h,1}$ & $a_{h,2}$ & $a_{h,3}$ & $a_{h,4}$   \\
   \hline 
   1 & 100,000 & 70,710 & 40,825 & 28,868 & 22,361   \\
   2 & 50,000 & 35,355 & 20,412 & 14,434 & 11,180   \\
   3 & 40,000 & 28,284 & 16,330 & 11,547 & 8,944   \\
   4 & 20,000 & 14,142 & 8,165 & 5,774 & 4,472   \\ 
 \hline
\end{tabular}
\end{center}
\label{table:4.1}
\end{table}

Now, suppose that we are interested in selecting a sample of size $n=8$ for proportional allocation. Due to $n_h \ge 1$, we first assign one sample to each stratum. Thus, it remains to select $4$ additional sample elements. Based on the values of $a_{h,s}$, the largest 4 values are $a_{1,1}=70710$, $a_{1,2}=40825$, $a_{2,1}= 35355$, and $a_{1,3}=28868$. Thus, the sample sizes for each stratum are $n_1=4$, $n_2=2$, $n_3=1$ and $n_4=1$. The Huntington-Hill method is currently used to allocate the seats of the House of Representatives in the United States.   \cite{wright2012} provides a statistical interpretation of the Huntington-Hill method for stratified sampling. 

Following the argument of \cite{wright2012}, we can show that 
Huntington-Hill method is essentially minimizing 
    $$ Q = \frac{1}{n} \sum_{h=1}^H n_h \left( \frac{N_h}{n_h} - \frac{N}{n} \right)^2. $$
  It is the average squared distance of the representativeness of individual sample elements from $N/n$.  To see this, note that 
    $$ Q= \frac{1}{n} \sum_{h=1}^H \frac{N_h^2}{n_h}- \frac{N^2}{n^2} .$$
    Thus, for fixed $n$, we only have to minimize the first quantity. 
  Now, using  
\begin{eqnarray*} 
  \sum_{h=1}^H \frac{N_h^2}{n_h} &=&
    \sum_{h=1}^N N_h^2 - \sum_{h=1}^H \sum_{k=1}^{n_h-1} \left( \frac{1}{k} - \frac{1}{k+1} \right) N_h^2\\
&=&  \sum_{h=1}^N N_h^2 - \sum_{h=1}^H \sum_{k=1}^{n_h-1} \frac{N_h^2}{ (k+1)k } 
  \end{eqnarray*} 
  we have only to maximize the second term: 
  \begin{equation} 
   \sum_{h=1}^H \sum_{k=1}^{n_h-1} \frac{N_h^2}{ (k+1)k } =\sum_{h=1}^H \sum_{k=1}^{n_h-1} a_{h,k}^2 
   \label{eqn:4-5}
   \end{equation} 
  Therefore, by selecting the largest $n-H$ elements from $a_{h,k}$' s, we can maximize the above quantity in (\ref{eqn:4-5}).

Now consider another allocation method, which is obtained from an optimization problem. The optimal allocation is obtained by minimizing the variance of the HT estimator under the given constraint. The usual constraint is expressed in terms of the total cost. Total cost is often expressed as $C= c_0 + \sum_{h=1}^H c_h n_h$, where $c_0$ is the initial cost and $c_h$ is the cost of interviewing one unit in stratum $h$. The following theorem is useful in determining the optimal allocation. 
\begin{theorem}
  Assume that the sampling variance of an estimator is 
\begin{equation}
 V\left( \hat{\theta} \right) =  \sum_{h} \frac{Q_h}{n_h}, 
 \label{5-4}
 \end{equation}
 where $Q_h$ is independent of $n_h$, 
and the total cost is 
\begin{equation}
  C= c_0 + \sum_{h=1}^H c_h n_h.
  \label{cost}
\end{equation}
The sample size allocation that minimizes (\ref{5-4}) subject to (\ref{cost}) is 
\begin{equation}
n_h \propto \left( \frac{ {Q_h}} { {n_h} } \right)^{1/2}. \label{neyman}
\end{equation}
\label{thm:4-1} 
 \end{theorem}
\begin{proof}
By the   Cauchy-Schwartz inequality,  we obtain 
$$ \left( \sum_h \frac{Q_h}{n_h} \right)\left( \sum_h c_h n_h \right) \ge \left( \sum_h \sqrt{ Q_h c_h} \right)^2  $$
where the equality holds when 

$$ \frac{c_h n_h }{Q_h/n_h } = \mbox{ constant } $$
which leads to (\ref{neyman}).
\end{proof}

Note that (\ref{neyman}) is the necessary and sufficient condition for the product of $V(\hat{\theta})$ and $C$ to be minimized. Thus, the same conclusion holds either when minimizing the variance subject to the cost constraints or minimizing the cost subject to the constraint on the variance. For the estimation of the population total, using the variance form in (\ref{eq4.1}), we have $Q_h = N_h^2 S_h^2$ and the optimal allocation is 
\begin{equation}
n_h^* \propto N_h S_h / \sqrt{c_h}. \label{neyman2} 
\end{equation}
If $c_h$ are all equal and $S_h$ are the same across the strata, then the optimal allocation reduces to the proportional allocation. The optimal allocation in (\ref{neyman2}) is first proposed by \cite{neyman1934} and is often called the Neyman allocation. \index{Neyman allocation}

 If $c_h$ are all equal, the optimal allocation using 
$$ n_h = n \frac{ N_h S_h}{ \sum_{h=1}^H  N_h S_h}  $$
leads to  
$$ V_{\rm opt}( \hat{Y}_{\rm HT}) = \frac{N^2}{n} \left( \sum_{h=1}^H W_h S_h \right)^2 - N \sum_{h=1}^H W_h S_h^2 ,
$$
where $W_h = N_h/N$.  Note that, under proportional allocation, we have  
$$ V_{\rm prop}( \hat{Y}_{\rm HT}) =  \frac{N^2}{n} \sum_{h=1}^H W_h S_h^2 - N \sum_{h=1}^H W_h S_h^2 .
$$
Using Jensen's inequality, we obtain 
$$ \left( \sum_{h=1}^H W_h S_h \right)^2 \le  \sum_{h=1}^H W_h S_h^2 , 
$$
  which implies  that  
$$  V_{\rm prop}( \hat{Y}_{\rm HT}) \ge  V_{\rm opt}( \hat{Y}_{\rm HT}). $$

Note that the allocation in (\ref{neyman2}) is derived by minimizing the variance of the total estimator. Instead of estimating the population total, if we are more interested in comparing the population means across strata, then the Neyman allocation in (\ref{neyman2}) is not necessarily optimal. For example, if we are interested in comparing the stratum means between stratum 1 and stratum 2, the parameter of interest is $\theta=\bar{Y}_1-\bar{Y}_2$ and its unbiased estimator is 
$\hat{\theta}=\bar{y}_1 - \bar{y}_2$. In this case, given the same cost constraint, the optimal allocation that minimizes the variance of $\hat{\theta}$ is, according to Theorem \ref{thm:4-1}, 
\begin{equation}
n_h^* \propto \frac{S_h }{ \sqrt{c_h}}, \label{2.31}
\end{equation}
which is different from the Neyman allocation in (\ref{neyman2}). Thus, the optimal allocation can be different for different parameters. Assuming that $c_h$ are the same across strata and $S_h^2$ are also the same across the strata, it is better to allocate the sample size proportional to $N_h$ if we are going to estimate the total, while it is better to use equal sample size if we are going to compare different stratum populations. When we want to achieve  the two conflicting goals, we can use an allocation with $n_h \propto N_h^{\alpha} (0 < \alpha < 1)$ as a compromise. This allocation is called a power allocation and $\alpha=1/2$ is a popular choice.  See \cite{bankier1988} for more details about the sample size allocation. 

\begin{example}
Consider the population with $H=3$ strata with the summary data in Table \ref{table:4.2}. We assume that the costs are all equal. 

\begin{table}[htb]
    \begin{center} 
    \caption{A summary of a toy population with $H=3$ strata}
    \begin{tabular}{ccc}
    \hline 
    Stratum & Pop'n Size $(N_h)$ & $S_h$ \\
    \hline 
    1 & 100 & 50 \\
    2 & 110 & 10 \\
    3 & 120 & 5 \\
    \hline 
    
    \end{tabular}
    \end{center}
    \label{table:4.2}
    \end{table}
    
 Suppose that we wish to allocate the sample size $n=140$ to each stratum using optimal allocation. Using (\ref{neyman}) with equal $c_h$, we obtain 
$$ n_1 = 104  \ \ n_2 = 23  \ \ n_3 = 13 $$
 However, $n_1=104$ is greater than $N_1=100$. Thus, we would take $n_1=N_1=100$ and allocate the remaining 40 elements to strata and 3 to obtain
$$ \tiny{n_2 = \frac{ 110 \times 10 }{ 110 \times 10 + 120 \times 5} \cdot 40 \doteq 26} 
$$
and $n_3= 40-26=14$. 
\end{example}

We now briefly discuss the problem of determining the sample size $n$ under stratified random sampling. 
Given the margin of error, we wish to find the sample size $n$ for stratified sampling. That is, we wish to find $n$ such that 
$$ P\left\{ \left| \hat{\theta} - \theta \right| \le e \right\} = 1-
\alpha . $$
 Recall that, under proportional allocation, we have 
$$ V_{\rm prop} ( \hat{Y}_{\rm HT}) = \frac{N^2}{n} \left( 1- \frac{n}{N} \right) S_w^2 
$$
where 
$$ S_w^2 = \sum_{h=1}^H W_h S_h^2 .$$
  Thus, under proportional allocation, you can use $S_w^2$ instead of $S^2$ in the sample size determination formula (\ref{2sw}) under SRS. That is, use
$$ e = z_{\alpha/2} \sqrt{ \frac{S_w^2}{n} \left( 1- \frac{n}{N} \right)}
\Rightarrow n= \frac{z_{\alpha/2}^2 S_w^2 }{ e^2 + z_{\alpha/2}^2
S_w^2/N}
$$
for mean estimation. 

\begin{example}
We illustrate the advantage of the optimal sample-size allocation using the real data example of  the US Agricultural Census data (agpop.dat) which is publicly available. The US agriculatural Census data  can also be obtained in ``SDaA'' package in R. The dataset  consists of 3,078 counties in the US in the year of 1992. From the population of counties, we are interested in using a stratified random sample of size $n=300$ to estimate the population total of variable ``acres92'', which represents  number of acres devoted to farms in 1992. The stratification variable is ``region'' with levels S (south), W (west), NC (north central), NE (northeast).  Table \ref{table4.1} presents some summary statistics for the study variable for each stratum. 

\begin{table}[htb]
\begin{center} 
\caption{Summary statistics for ares devoted to farms in 1992}
\begin{tabular}{lrrr}
\hline 
Region & Population Size & Total Acres  & Standard Dev \\
\hline 
NE & 220 & $0.199 \times 10^8$   & 79,365\\
NC & 1,054 & $3.435 \times 10^8$ & 271,303\\
S & 1,382 & $2.752 \times 10^8$ & 243,956\\
W & 422 & $3.052 \times 10^8$ & 835,638 \\
\hline 
Total & 3,078 & $9.44 \times 10^8$ & 424,686  \\
\hline 
\end{tabular}
\end{center}
\label{table4.1}
\end{table}

Now, the following table presents the variance under proportional allocation and Neyman allocation. 

\begin{table}[htb]
\begin{center} 
\caption{Sample sizes and sampling variances for estimated total acres devoted to farms in 1992}
\begin{tabular}{l|r|rr|rr}
\hline 
Region & Pop. Size & \multicolumn{2}{|c|}{Proportional Allocation} &  \multicolumn{2}{|c}{Optimal Allocation} \\
 & & 
Sample Size & $V ( \hat{Y}_{h, {\rm HT}})$ & Sample Size & $V ( \hat{Y}_{h, {\rm HT}})$  \\
\hline 
NE & 220 & 21  & $0.13 \times 10^{14}$ & 5  & $0.60 \times 10^{14}$ \\
NC & 1,054 & 103 & $7.16 \times 10^{14}$ & 86 & $8.73 \times 10^{14}$\\
S & 1,382 & 135 & $7.60 \times 10^{14}$ &102 & $10.32 \times 10^{14}$ \\
W & 422 &  41 & $27.38 \times 10^{14}$ &107 & $8.67 \times 10^{14}$ \\
\hline 
Total & 3,078 & 300 &$42.28 \times 10^{14}$& 300 &$28.32 \times 10^{14}$ \\
\hline 
\end{tabular}
\end{center}
\label{table4.1}
\end{table}

The efficiency gain due to optimal allocation over the proportional allocation is about 33\%. That is, we can reduce the variance of the HT estimator under proportional allocation by 33\% if we employ the optimal allocation. 

\end{example}
\section{Stratum boundary determination}
\markboth{4. Stratified sampling}{4.3 Stratum boundary determination}

Sample size allocation is a problem of finding the suitable sample size for a given set of stratum boundaries. In some cases, we need to determine the stratum boundaries. The most popular method of finding the stratum boundaries was proposed by \cite{dalenius1959}. 


To explain the idea of the stratification method of Dalenius and Hodges (1959), assume for now that the population values of $y$ are available or at least a proxy values of $y$ are all available. Let $y_0$ and $y_{H}$ be the smallest and largest values of $y$ in the finite population. The problem is to find intermediate stratum boundaries $y_1, \cdots, y_{H-1}$ such that 
$$ V( \hat{\bar{Y}}_{\rm HT} ) = \sum_{h=1}^H W_h^2  \left( \frac{1}{n_h} - \frac{1}{N_h} \right) S_h^2  $$
is a minimum, where $W_h=N_h/N$. 

 Under Neyman allocation, the above variance reduces to 
$$ V( \hat{\bar{Y}}_{\rm HT} ) = \frac{1}{n} \left( \sum_{h=1}^H W_h S_h  \right)^2 - \frac{1}{N} \sum_{h=1}^H W_h S_h^2  .
$$
Thus, if $n_h/N_h$ are ignored, it suffices to minimize $\sum_h W_h S_h$.

Let $f(y)$ be the frequency function of $y$. If the strata are numerous and narrow, $f(y)$ should be approximately constant (rectangular) within a given stratum. Hence, 
\begin{eqnarray*}
W_h &=& \int_{y_{h-1}}^{y_h} f( t) dt  \doteq f_h (y_h - y_{h-1}) \\
S_h & \doteq & (y_h - y_{h-1})  /\sqrt{12}
\end{eqnarray*}
where $f_h$ is the constant value of $f(y)$ in stratum $h$. 
 Thus, writing $Z(y) = \int_{y_0}^y \sqrt{ f(t)} d t ,$  we have 
$$ \sum_{h=1}^H W_h S_h \propto \sum_{h=1}^H f_h (y_{h} - y_{h-1} )^2 \doteq \sum_{h=1}^H (Z_h - Z_{h-1})^2 , $$
where $Z_h=Z(y_h)$ and 
 $$Z(y) = \int_{y_0}^y \sqrt{ f(t)} d t . $$

 Since $(Z_H- Z_0)$ is fixed,  $\sum_{h=1}^H (Z_h - Z_{h-1})^2$ is minimized by making $(Z_{h} - Z_{h-1})$ constant.  To achieve this goal, the rule is to form the cumulative of $\sqrt{f(y)}$ and choose the $y_h$ so that they create equal intervals on the cumulative $\sqrt{f (y)} $ scale. 
\begin{enumerate}
\item Partition the population into $L(> 2H)$ intervals with equal length. 
\item For each interval $l$, compute $\sqrt{f_l}$, the squared root of the frequency, and its cumulative sum. 
\item Choose the boundaries of the stratum so that the sum of the $\sqrt{f_l}$ are about the same in each stratum. 
\end{enumerate}

\section{Number of strata} 
\markboth{4. Stratified sampling}{4.4 Number of strata}

In this section, we explore the relationship between the variance of $\hat{\bar{Y}}_{\rm HT} = \sum_{h=1}^H W_h \bar{y}_h$ and  $H$, the number of strata,  under stratified sampling. To do this, we first consider the case when the stratum boundaries are determined by the values of the study variable $Y$. To simplify the discussion, suppose that the realized population values  $y_1, \cdots, y_N$  are generated from a distribution with density $f(y)$ with support in $[a,b]$. In this case, we can determined the stratum boundaries $a_h$ in such a way that 
$$ \int_{a_{h-1}}^{a_h} f(y) dy = \frac{1}{H} $$
holds. That is, if $y_i \in [a_{h-1}, a_h]$ then unit $i$ belongs to stratum $h$. In this case, $S_h^2$ obtained from $(a_{h-1}, a_h)$ satisfies $S_h^2 \le (a_h - a_{h-1})^2$. Thus, under proportional allocation, by (\ref{eq:4.3}), 
\begin{eqnarray}
V(\hat{\bar{Y}}_{HT} ) &=&  \frac{1}{n} \left( 1- \frac{n}{N} \right) \sum_{h=1}^H W_h S_h^2\label{stvar} \\
&\le &   \frac{1}{n H}  \sum_{h=1}^H (a_h - a_{h-1})^2 \notag
\end{eqnarray}
and, since $\sum_{h=1}^H (a_h - a_{h-1})^2 \le (b-a)^2$ is bounded, we obtain 
\begin{equation}
V(\hat{\bar{Y}}_{HT} ) =  O\left( \frac{1}{nH} \right)
\label{eq:4.12}
\end{equation}
Here, $b_n=O(a_n)$ denotes that $b_n/a_n$ is bounded. 

Therefore, from the result in (\ref{eq:4.12}), we can conclude that the relative efficiency of simple probability sampling increases
proportionally to H. So,  we can conclude that the higher the H, the better the efficiency.
However, in practice, the stratum is not determined by observing the $y$-values of the population in advance, but by observing the values in the frame of the population (which we call x). From the sampling frame, 
 we can only stratify based on the $x$-values in the frame.  
In this case, increasing H does not proportionally improve the efficiency. 
To illustrate, consider the following  linear model 
\begin{equation}
y_i = \alpha + \beta x_i + e_i ,  \ \ \ e_i \sim (0, \sigma_e^2) .
\label{4.15}
\end{equation}
That is, the values of the finite population are generated from model  (\ref{4.15}). Here,  $\alpha, \beta, \sigma_e^2$ are unknown parameters. 

In this case, the model expectation of  $V( \hat{\bar{Y}}_{\rm HT})$ in (\ref{stvar}) is 
\begin{eqnarray*}
 E_\zeta \left\{ V( \hat{\bar{Y}}_{\rm HT}) \right\} &=&  \frac{1}{n}\sum_{h=1}^H W_h E_\zeta( S_h^2 ) \\
  &=&  \frac{1}{n} \sum_{h=1}^H W_h \left( \beta^2 S_{xh}^2 + \sigma_e^2 \right)  \\
  &=&  \frac{1}{n}  \sum_{h=1}^H W_h \left( \beta^2 S_{xh}^2\right) + \frac{1}{n} \sigma_e^2 .
\end{eqnarray*}
So, if we stratify based on x, the first term will be $O(n^{-1}H^{-1})$ as in (\ref{eq:4.12}), but the second term will be $O(n^{-1})$, which is the part that is independent of $H$. In other words, the higher the correlation coefficient between x and y, the smaller the second term.
If the correlation coefficient between x and y is high, the second term will be small, so we can see a lot of efficiency gain due to stratification.
Otherwise, the second term becomes large, so the effect of stratification on reducing variance becomes smaller.

\section{Domain Estimation} 
\markboth{4. Stratified sampling}{4.5 Domain  estimation}

As discussed in Section 2.3, we generally want to make inferences about subpopulations as well as  the whole population.   Let   $U_d \subset U$ be the index set of the subpopulation for domain $d$. Let $N_d = \left| U_d \right|$  be the number of elements in $U_d$, $Y_d=\sum_{i \in U_d} y_i $ be the domain total of $y$ in domain $d$, and 
$\bar{Y}_{d}=Y_d/N_d$ be the domain mean of $y$ in domain $d$. 

To estimate domain parameters,  define
\begin{equation}
z_{kd} = \left\{ \begin{array}{ll}
1 & \mbox{ if } k \in U_d \\
0 & \mbox{ if } k \notin U_d ,
\end{array}
\right. 
\label{id_domain}
\end{equation}
for $k \in U$.  Note that $z_{id}$ is a characteristic of the population. Using $z_{kd}$, we can exexpress $\sum_{k \in U} z_{kd} = N_d$. The HT estimation of $N_d$ is 
$$ \hat{N}_d = \sum_{k \in U} \frac{ z_{kd} I_k}{\pi_k }. $$
Also, HT estimation of $Y_d=\sum_{k \in U_d} y_k = \sum_{k \in U} y_k z_{kd}$ is 
$$ \hat{Y}_d = \sum_{k \in U} \frac{ y_k z_{kd} I_k }{ \pi_k } = \sum_{k \in A} \frac{ y_k z_{kd}  }{ \pi_k }.$$

To estimate $\bar{Y}_{d}=Y_d/N_d$, we may use 
$$ \bar{y}_{d} = \frac{\hat{Y}_d  }{ \hat{N}_d} ,$$
which is a ratio of two HT estimators. Ratio of two unbiased estimator is not exactly unbiased as 
$$ E( \bar{y}_d) = E\left(  \frac{\hat{Y}_d  }{ \hat{N}_d} \right) \neq \frac{ E( \hat{Y}_d )}{ E( \hat{N}_d ) } = \bar{Y}_d , $$
but it is asymptotically unbiased as long as $E( \hat{N}_d) \neq 0$. 

The statistical properties of  $\bar{y}_d$ can be derived from the following approximation:
\begin{eqnarray*}
 \bar{y}_{d} &=& \frac{\hat{Y}_d  }{ \hat{N}_d} = f \left( \hat{N}_d, \hat{Y}_d \right) \\
 &\doteq &  f \left( {N}_d, {Y}_d \right) + \left\{\frac{\partial }{\partial Y_d } f \left({N}_d, {Y}_d \right) \right\} \left( \hat{Y}_d - Y_d \right)\\
  &&
  + \left\{ \frac{\partial }{\partial N_d } f \left({N}_d, {Y}_d \right) \right\} \left( \hat{N}_d - N_d \right) \\
 &=& \frac{Y_d}{N_d}  + \left( \frac{1}{N_d} \right) \left( \hat{Y}_d - Y_d \right) + \left( -  \frac{Y_d}{N_d^2}\right) \left( \hat{N}_d - N_d \right)
 \end{eqnarray*}
 Thus,
 \begin{equation} 
 V\left(  \bar{y}_{d} \right) \doteq V\left\{ \frac{1}{N_d} \left(\hat{Y}_d - \bar{Y}_d \hat{N}_d \right) \right\} .
 \label{eqn:4-11}
 \end{equation} 
 Under SRS, result (\ref{eqn:4-11}) can be written as 
 \begin{eqnarray*} 
 V\left(  \bar{y}_{d} \right) &\doteq & \frac{1}{N_d^2} V \left( \frac{N}{n} \sum_{i\in A} z_{id} (y_i - \bar{Y}_d )   \right)  \\
 &=& \frac{N^2}{N_d^2} \left( \frac{1}{n} - \frac{1}{N} \right) \frac{ 1}{ N-1} \sum_{i=1}^N z_{id} (y_i - \bar{Y}_d )^2 \\
 &\doteq & \left( \frac{1}{E(n_d)} - \frac{1}{N_d} \right) \frac{1}{N_d-1} \sum_{i \in U_d } \left( y_i - \bar{Y}_d \right)^2,
 \end{eqnarray*}
 where $E(n_d) = N_d (n/N)$. 
 Thus, the variance formula suggests that the variance of the domain mean estimator is asymptotically equivalent to the variance of the SRS  mean of size $E(n_d)$ from the subpopulation $U_d$. 

The finite population is decomposed into $G$ groups, as in $U=U_1 \cup U_2 \cup \cdots \cup U_G$. Suppose  that group information was not available at the time of sample selection. The population size of the group $g$, denoted by $N_g$, is known.
If the group information is available from the sampling frame, then we can use stratified sampling. However, if group information is not available, then we can incorporate population information $N_g$ after sample selection. This is called post-stratification. 
\index{post-stratification}

The  post-stratification estimator can be defined as 
\begin{equation} 
 \hat{Y}_{\rm post} = \sum_{g=1}^G N_g  \frac{ \hat{Y}_g}{ \hat{N}_g} .
 \label{eqn:4-12}
 \end{equation} 
Note that, unlike stratified sampling, the denominator terms ($\hat{N}_g$) are random. The post-stratification estimator is a special case of the regression estimator to be introduced in Chapter 8. Using the Taylor expansion again, we obtain 
\begin{eqnarray*} 
\hat{Y}_{\rm post} &=& \sum_{g=1}^G N_g  \frac{ \hat{Y}_g}{ \hat{N}_g}  \\
&\doteq & \sum_{g=1}^G N_g \left\{\frac{Y_d}{N_d} + \frac{1}{N_d} \left( \hat{Y}_d - \bar{Y}_d \hat{N}_d \right)  \right\} \\
&=& Y + \sum_{g=1}^G \frac{1}{\pi_i}  z_{ig} ( y_i - \bar{Y}_g ), 
\end{eqnarray*}
where $z_{ig}$ is the domain indicator of $U_g$, as defined in (\ref{id_domain}). 
Thus, the asymptotic variance can be written as  
\begin{equation}
 V\left( \hat{Y}_{\rm post} \right)  \doteq  V\left\{  \sum_{g=1}^G \sum_{i \in A} \frac{1}{\pi_i} z_{ig} ( y_i - \bar{Y}_g ) \right\}  . 
 \label{eqn:4-13} 
 \end{equation}
 Under SRS, the asymptotic variance (\ref{eqn:4-13}) reduces to   
\begin{eqnarray*}
 V\left( \hat{Y}_{\rm post} \right)  
 &\doteq& \frac{N^2}{n} \left( 1 - \frac{n}{N} \right) \frac{1}{N-1} \sum_{g=1}^G \sum_{i \in U_g} \left( y_i - \bar{Y}_g  \right)^2 .
 \end{eqnarray*}
 Thus, it is essentially equal to the variance under stratified sampling with proportional allocation.

%% file: chapters/chapter5.tex

\chapter{Sampling with Unequal probabilities}

\section{Introduction}


We now consider unequal probability sampling designs, which is very popular in practice. In unequal probability sampling, we can improve the efficiency of the resulting point estimator by properly incorporating the auxiliary information available in the sampling frame into the sampling design.

\begin{example}

Consider the following artificial example of a finite population of business companies presented in Table \ref{table:5.1}.  The total number of employees is available in the whole population, but the annual total income is available only for the sample. 

\begin{table}[htb]
\begin{center}
\caption{An artificial population of business companies}
\begin{tabular}{c|c|c}
\hline  Company & Size (Number of employees)  & y (Income)
  \\ \hline
  A & 100 & 11 \\
  B & 200 & 20 \\
  C & 300 & 24 \\
  D & 1,000 & 245 \\
 \hline
\end{tabular}
\end{center}
\label{table:5.1}
\end{table}

If we wish to select only one company, we can consider the following two approaches.

\begin{enumerate}
\item Equal probability selection
 \item  Unequal probability selection with the selection probability proportional to size. 
\end{enumerate}

The sampling distribution of the total income estimator is then given by 
$$ \hat{Y} = \left\{ \begin{array}{ll}
11 \times 4 & \mbox{ if A is sampled } \\
20 \times 4 & \mbox{ if B is sampled } \\
24 \times 4 & \mbox{ if C is sampled } \\
245 \times 4 & \mbox{ if D is sampled } \\
\end{array}
\right.
$$
and so 
\begin{eqnarray*}
 E ( \hat{Y} ) &=&  11 + 20 + 24 + 245 = 300  \\
 V ( \hat{Y} ) &=& 154,488 .
 \end{eqnarray*}
 On the other hand, for unequal probability sampling,  
$$ \hat{Y} = \left\{ \begin{array}{ll}
11\times 16 & \mbox{ if A is sampled } \\
20\times (16/2) & \mbox{ if B is sampled } \\
24 \times (16/3) & \mbox{ if C is sampled } \\
245\times (16/10) & \mbox{ if D is sampled } \\
\end{array}
\right.
$$
and so
\begin{eqnarray*}
 E ( \hat{Y} ) &=&  11 + 20 + 24 + 245 = 300  \\
 V ( \hat{Y}) &=& 14,248.
 \end{eqnarray*}
 Thus, not surprisingly, unequal probability selection using the number of employees as the size is more efficient than the equal probability sampling mechanism.  A company with many employees is likely to have more income than a company with a smaller number of employees. 
 \label{example:5-1}
 \end{example}

In general, when the sampling frame contains information about the size of the sampling unit, the size information is often considered in the sampling design stage such that the selection probability of a unit is proportional to the size. 
Unequal probability sampling is also popular in cluster sampling, which we plan to cover in Chapter 7. If the cluster sizes are unequal, it is better to use this information in the cluster sampling.  In some cases, unequal probability sampling is implicit. The following example illustrates this point.

\begin{example} 
Suppose that we are interested in estimating the percent of families owing home, by obtaining a sample of school children. Table \ref{table:5.2} presents the  distribution of an artificial  population of school children.  

\begin{table}
\begin{center}
\caption{Artificial population of school children}
\begin{tabular}{c|r|r|r|r|r|r}
\hline 
Number of & \multicolumn{4}{|c}{Families} & \multicolumn{2}{|c}{School children} \\
\cline{2-7}
school children & Number & \%  &  \multicolumn{2}{|c|}{Owing Home} & Number & \% \\
\cline{4-5}
in family   &   & & Number & \% &  & \\
\hline 
4 & 5,000 & 10 & 500 & 10  & 20,000 & 20 \\
3 & 10,000 & 20 & 3,000 & 30 & 30,000 & 30 \\
2 & 15,000 & 30 & 6,000 & 40 & 30,000 & 30 \\
1 & 20,000 & 40 & 12,000 & 60 & 20,000 & 20 \\
\hline 
Total & 50,000 & 100 & 21,500 & 43 & 100,000 & 100 \\
 \hline
\end{tabular}
\end{center}
\label{table:5.2}
\end{table}
In this example, the reporting unit is school children but the analysis unit is family. Thus, if we are selecting a simple random sample of school children then the large-sized families will be over-sampled. Thus, a simple mean of the sample observations will lead to biased estimation. In this example, about 43\% of the families have their home. However, in the school-children level, among the 100,000 school children, the total number of children with home-owning families is
$$ 20,000 \times 0.1 + 30,000 \times 0.3 + 30,000 \times 0.4 + 20,000 \times 0.6 =35,000.$$
Thus, only 35\% of the children belong to families with their home. 

\end{example} 

\section{PPS sampling}

As seen in Example \ref{example:5-1}, a sampling design proportional to the number of employs is quite efficient for estimating the total income of the companies in the population. In this case, the total number of employees serves the role of measure of size (MOS), which is an auxiliary variable that reflects the magnitude of $y_i$ in the population. A sampling that selects elements with probability proportional to MOS with replacement is called probability proportional to size (PPS) sampling. \index{PPS sampling} Since it is easy to select a sample of size one with a selection probability proportional to MOS, we can repeat the selection $n$ times independently with replacement to achieve the PPS sampling of size $n$. PPS sampling is easy to implement, as it is an with-replacement sampling, but it may have duplicated sample elements.

Let $M_i$ be the value of MOS associated with element $i$ in the population. In this case,
  $$p_i = \frac{ M_i }{ \sum_{i=1}^N M_i }$$
is the probability of selecting element $i$ for a single draw of sample selection. Let $a_k$ be the index of population element in the $k$-th draw of the PPS sampling, In this case, $a_k$ is a random variable with probability 
$ P( a_k = i) = p_i$, for $i \in U$. The observed value of $y_i$ in the $k$-th draw is 
$y_{a_k} =  \sum_{i=1}^N I( a_k = i) y_i $. Note that $E(y_{a_k} ) = \sum_{i=1}^N p_i y_i$ which is not necessarily equal to the population mean.

If we define 
$$ z_k = \frac{ y_{a_k}}{p_{a_k}}= \sum_{i=1}^N I( a_k = i) \frac{y_i}{ p_i} , \ \ \  k=1,2, \cdots, n, $$
then   $z_1, z_2, \cdots, z_n$ are independently and identically distributed with the distribution  
\begin{equation*}
z_k = \left\{ \begin{array}{ll} y_1/p_1 & \mbox{ with probability }
p_1 \\
y_2/p_2 & \mbox{ with probability }
p_2 \\
\vdots & \\
y_N/p_N & \mbox{ with probability } p_N.
\end{array}
\right.
\end{equation*}
Thus, for each $k=1, \cdots, n$, we have 
\begin{eqnarray*}
E\left( z_k \right) &=& \sum_{i=1}^N y_i \\
V\left( z_k \right) &=& \sum_{i=1}^N p_i \left( \frac{ y_i}{p_i} -
Y \right)^2.
\end{eqnarray*}
Thus, we can use 
\begin{equation}
 \hat{Y}_{{\rm PPS}} = \frac{1}{n}
\sum_{k=1}^n z_k = \frac{1}{n} \sum_{k=1}^n \frac{y_{a_k}}{ p_{a_k} }
\label{pps}
\end{equation}
as an estimator for $Y=\sum_{i=1}^N y_i$. The estimator in (\ref{pps}) is sometimes called Hansen-Hurwitz (HH) estimator, as it was first proposed by \cite{hansen1943}.  \index{Hansen-Hurwitz estimator}

Alternatively,  we can express 
$$ \hat{Y}_{{\rm PPS} } = \frac{1}{n} \sum_{k=1}^n z_k =  \sum_{i \in A} w_i y_i $$
where 
$ w_i = Q_i/(n p_i)  $ and $Q_i= \sum_{k=1}^n I(a_k=i)$ is the number of times that unit $i$ is selected in the sample. The HH estimator in (\ref{pps}) is a weighted sum of the sample observations.

To discuss variance estimation of HH estimator,  we first prove the following Lemma. 

\begin{lemma}
Let  $X_1, X_2, \cdots, X_n$ be independent random variables with $E(X_i)=\mu$ and $V(X_i) = \sigma_i^2$. An unbiased estimator for the variance of $\bar{X}_n =n^{-1} \sum_{i=1}^n X_i$ is given by 
\begin{equation}
\hat{V} \left(\bar{X}_n \right) = \frac{1}{n} \frac{1}{n-1}
\sum_{i=1}^n \left( X_i - \bar{X}_n \right)^2. 
\label{2.38}
\end{equation} \label{thm2.3}
\end{lemma}
\begin{proof}
Since $X_i$'s are independent, 
$$V \left( \bar{X}_n \right) = \frac{1}{n^2} \sum_{i=1}^n \sigma_i^2. $$
Also, as we can express 
$$\hat{V} \left(\bar{X}_n \right) = \frac{1}{n}
\frac{1}{n-1}\left\{  \sum_{i=1}^n \left( X_i - \mu\right)^2- n
\left( \bar{X}_n - \mu\right)^2 \right\}, 
$$
by taking the expectation on both sides of the above term, we obtain 
$$E \left\{ \hat{V} \left(\bar{X}_n \right)\right\}
= \frac{1}{n} \frac{1}{n-1}\left\{  \sum_{i=1}^n \sigma_i^2 - n
 \frac{1}{n^2} \sum_{i=1}^n \sigma_i^2 \right\}=\frac{1}{n^2} \sum_{i=1}^n \sigma_i^2,
$$
which proves the unbiasedness of $\hat{V}( \bar{X}_n) $ in (\ref{2.38}). 
\end{proof}

By Lemma  \ref{thm2.3}, an unbiased variance estimator of HH estimator is given by 
\begin{equation}
 \hat{V}_{{\rm PPS}} = \frac{1}{n} S_z^2 = \frac{1}{n} \frac{1}{n-1}
\sum_{k=1}^n \left( \frac{y_{a_k}}{p_{a_k} } - \hat{Y}_{{\rm PPS}} \right)^2.
\label{hh_var}
\end{equation}
 Using $Q_i$ notation, we can express the above variance estimation formula as 
$$ \hat{V}_{\rm PPS} = \frac{1}{n} \cdot \frac{1}{n-1} \sum_{i \in A} Q_i \left( \frac{ y_i}{ p_i} - \hat{Y}_{{\rm PPS}}\right)^2  .$$
  If $n/N$ is very small, then $Q_i$ are either 0 or 1. In this case, 
  \begin{equation} 
  \hat{V}_{\rm PPS} = \frac{n}{n-1} \sum_{i \in A}  \left( w_i y_i - \frac{1}{n} \hat{Y}_{\rm PPS}\right)^2  . 
  \label{hh_var2}
  \end{equation}


In some situation, $y_i$ has the meaning of a total value in unit $i$. For example, $y_i$ is the total crop yield in the farm $i$ and $M_i$ is the total size of the crop acres in form $i$. In this case, $\bar{y}_i$ is the average crop yield per acre. In this case, we can express 
$$ \hat{Y}_{\PPS} = \left( \sum_{i=1}^N M_i \right) \times \frac{1}{n}
\sum_{k=1}^n \bar{y}_{a_k} $$
and 
$$ \hat{V} \left( \hat{Y}_{\PPS} \right) = \left( \sum_{i=1}^N M_i \right)^2 \times\frac{1}{n} \frac{1}{n-1}
\sum_{k=1}^n \left( \bar{y}_{a_k} - \hat{\bar{Y}}_{\PPS}  \right)^2
$$
where $\hat{\bar{Y}}_{\PPS}  = n^{-1} 
\sum_{k=1}^n \bar{y}_{a_k}$. If the parameter of interest is the mean 
$$ \bar{Y} = \frac{ \sum_{i=1}^N y_i }{ \sum_{i=1}^N M_i} = \frac{ \sum_{i=1}^N M_i \bar{y}_i }{ \sum_{i=1}^N M_i}  ,$$
then the HH estimator is 
$$ \hat{\bar{Y}}_{\PPS} =  \frac{1}{n}
\sum_{k=1}^n \bar{y}_{a_k} $$
and its variance estimator is 
$$ \hat{V}\left( \hat{\bar{Y}}_{\PPS} \right)  =\frac{1}{n} \frac{1}{n-1}
\sum_{k=1}^n \left( \bar{y}_{a_k} -  \hat{\bar{Y}}_{\PPS}   \right)^2.
$$
That is, in the mean estimation under PPS sampling, we can safely treat
$\bar{y}_{a_1},\bar{y}_{a_2}, \cdots, \bar{y}_{a_n} $ as an IID sample with $E(\bar{y}_{a_k})=\bar{Y}$  and apply Lemma  \ref{thm2.3}.

\section{Poisson sampling}

 Poisson sampling \index{Poisson sampling} is a generalization of Bernoulli sampling by allowing unequal probability selection. In Poisson sampling, the sample selection indicator function $I_i$ follows 
 \begin{equation*}
I_i \stackrel{indep}{\sim}  Bernoulli \left(\pi_i \right) , \ \ \ i=1,2,\cdots, N.
\end{equation*}
Thus, the sampling design is 
\begin{equation} 
P(A) = \left( \prod_{k \in A} \pi_k \right) \times
\left[ \prod_{k \in A^c} (1- \pi_k) \right] 
\end{equation} 
for all $A\subset U$. 
Here,  $\pi_i$ is the first-order inclusion probability.  The following result, also presented in \cite{tille2020},  gives a useful interpretation of Poisson sampling. 
\begin{lemma}
\label{lem:4.1} 
The sampling design that maximizes entropy 
\begin{equation} 
 I( p) = - \sum_{A \in \mathcal{A} } P( A) \log \{ P(A) \} , 
 \label{entropy}
 \end{equation} 
on $\mathcal{A} = \{ A ; A \subset U\}$ with fixed inclusion probabilities $\pi_k, k \in U$, is the Poisson sampling design. 
\end{lemma}
\begin{proof}
We wish to find the maximizer of $I(p)$ in (\ref{entropy}) subject to 
\begin{equation} 
 \sum_{A\in \mathcal{A}} I(k \in A) P(A) = \pi_k, \ \ k \in U 
 \label{eq4.3}
 \end{equation} 
and 
\begin{equation} 
 \sum_{A \in \mathcal{A}} P(A) = 1. 
 \label{eq4.4}
 \end{equation} 
Using the Lagrange multiplier method, we can construct  
\begin{eqnarray*} 
 Q &=& - \sum_{A \in \mathcal{A} } P( A) \log \{ P(A) \} + \sum_{k \in U} \theta_k \left\{ \sum_{A\in \mathcal{A}} I (k \in A) P(A) - \pi_k \right\} \\
 && + \lambda \left\{ \sum_{A \in \mathcal{A}} P(A) - 1\right\} .
\end{eqnarray*} 
Solving 
$$ \frac{ \partial }{ \partial P(A)  } Q = -1 - \log\{ P(A) \} + \sum_{k \in U} \theta_k I( k \in A) + \lambda =0,  $$
we obtain 
\begin{equation} 
P(A) = \exp \left( \sum_{k \in U} \theta_k I_k  -\mu \right) , 
\label{eq4.5}
\end{equation} 
where $I_k = I( k \in A)$ and $\mu=1-\lambda$. 
Using (\ref{eq4.4}), we obtain 
\begin{eqnarray} 
 \exp ( \mu ) &=& \sum_{ A \subset U} \prod_{k \in A} \exp ( \theta_k)   \label{eq:4.6a} 
 \\
 &=& \prod_{k \in U} \{ 1+ \exp ( \theta_k) \} ,
 \label{eq:4.6b} 
 \end{eqnarray} 
 where the second equality follows from 
 $$ \prod_{k \in U} (a_k + b_k) =\sum_{A \subset U}  \left\{ \prod_{i \in A} a_i \cdot  \prod_{j \in U/A} b_j \right\}. 
 $$
Also, using (\ref{eq4.3}), we obtain   
\begin{eqnarray*} 
 \pi_k &=& \frac{ \sum_{ A \in \mathcal{A}} I(k \in A) \prod_{i \in A} \exp ( \theta_i) }{ \exp ( \mu )  }\\
 &=& \frac{ \exp ( \theta_k) \sum_{ A \in U/\{k\} }   \prod_{i \in A} \exp ( \theta_i) }{ \exp ( \mu )  }\\
 &=&  \frac{ \exp ( \theta_k) \prod_{ i \in U/\{k\} }   \{ 1+ \exp ( \theta_i) \} }{ \exp ( \mu )  }\\
 &=& \frac{ \exp ( \theta_k) }{ 1+ \exp( \theta_k) } \cdot  \frac{  \prod_{ i \in U }   \{ 1+ \exp ( \theta_i) \} }{ \exp ( \mu )  } \\
 &=& \frac{ \exp ( \theta_k) }{ 1+ \exp( \theta_k) } ,
 \end{eqnarray*}
 where the third equality follows from (\ref{eq:4.6b}) and the fifth equality follows from (\ref{eq:4.6a}). 
 Thus, we have 
 $$ \exp ( \theta_k) = \frac{ \pi_k}{ 1- \pi_k} $$
 and, by (\ref{eq4.5}),  
\begin{eqnarray} 
P(A) &=& \frac{ \exp\left( \sum_{k \in A} \theta_k  \right)  }{ \sum_{A \subset U} \exp\left( \sum_{k \in A} \theta_k  \right)  }
\label{eq:4.11}
\\
&=& 
\frac{ \prod_{k \in A} \frac{ \pi_k}{1-\pi_k }}{\prod_{k \in U} \frac{ 1}{1-\pi_k } } \notag \\
&=& \left( \prod_{k \in A} \pi_k \right) \times \left\{ \prod_{k \in A^c} (1-\pi_k)\right\} , \notag 
\end{eqnarray} 
which is the sampling design for Poisson sampling. 
\end{proof}
The entropy function in (\ref{entropy}) is a measure of randomnesss of a probability distribution. Thus, 
the Poisson sampling design is the most random design possible which respects a priori fixed first-order inclusion probabilities \citep{tille2020}.  

Poisson sampling is rarely used in practice, but it is useful in understanding the basic nature of unequal probability sampling. Under Poisson sampling, the variance of HT estimator is expressed as 
\begin{equation}
 V\left( \hat{Y}_{\rm HT} \right) = \sum_{i=1}^N \left(
\frac{1}{\pi_i} - 1\right) y_i^2. \label{3.25}
\end{equation}

The following theorem provides a result for the optimal choice of $\pi_i$ under Poisson sampling. 
\begin{theorem}
Consider a Poisson sampling with first-order inclusion probability $\pi_i$. Given the same (expected) sample size, the variance of HT estimator under Poisson sampling is minimized when 
 \begin{equation}
 \pi_i \propto y_i.
 \label{25}
 \end{equation}
\end{theorem}
\begin{proof}
Using the Cauchy-Schwarz inequality, we have 
 $$ \left( \sum_{i=1}^N \frac{y_i^2}{\pi_i} \right) \left( \sum_{i=1}^N \pi_i \right)
 \ge \left( \sum_{i=1}^N y_i \right)^2 $$
 and using the fact that  $\sum_{i=1}^N \pi_i=n$ is a fixed constant, we obtain (\ref{3.25})  is minimized when (\ref{25}) holds. \end{proof}

Thus, under Poisson sampling, we have only to make $\pi_i$ proportional to $y_i$. However, since we never observed $y_i$ at the time of sampling design, we cannot use $y_i$ but instead use $x_i$ in the population that is believed to be proportional to $y_i$.

Poisson sampling, as with Bernoulli sampling, has the disadvantage that the sample size $n$ is random. In the extreme case, we may have $n$ equal to zero. Thus, Poisson sampling has limited usage in practice, but is useful in developing theory. 

If Poisson sampling is used, we can consider the following estimator which is first proposed by \cite{hajek1971}: 
\begin{equation}
\hat{Y}_{\pi}  = N \frac{ \sum_{i \in A} \pi_i^{-1} y_i}{ \sum_{i \in A} \pi_i^{-1} } .
\label{hajek}
\end{equation}
\index{H{\'a}jek estimator} H{\'a}jek estimator is a special case of the ratio estimator that we will study in Chapter 8. The asymptotic variance can be obtained, by applying Taylor method, 
\begin{eqnarray*}
\hat{Y}_{\pi} 
 &=& N\frac{\hat{Y}_{\rm HT} }{ \hat{N}_{\rm HT} }  \\
&\cong &  N \frac{E( \hat{Y}_{\rm HT}) }{ E(\hat{N}_{\rm HT}) }
+ N \frac{1}{E(\hat{N}_{\rm HT}) } \left\{\hat{Y}_{\rm HT} -E( \hat{Y}_{\rm HT})   \right\} \\
&& -  N \frac{E( \hat{Y}_{\rm HT}) }{ \{E(\hat{N}_{\rm HT})\}^2 }  \left\{\hat{N}_{\rm HT} -E( \hat{N}_{\rm HT})   \right\}\\
&\cong & Y + \left(\hat{Y}_{\rm HT} - \bar{Y} \hat{N}_{\rm HT} \right). 
\end{eqnarray*}
Thus, we obtain
\begin{equation}
 V( 
 \hat{Y}_{\pi}  ) \cong V \left\{ \sum_{i \in A} \frac{1}{\pi_i} (y_i - \bar{Y} ) \right\} = \sum_{i=1}^N \left( \frac{1}{\pi_i} -1 \right) \left( y_i - \bar{Y} \right)^2.
\label{vhajek}
\end{equation}
Comparing (\ref{vhajek}) with (\ref{3.25}), the variance of the HT estimator, the only difference is whether we use $y_i^2$ or $(y_i-\bar{Y})^2$ in the variance formula. In many cases, we may say that the H{\'a}jek estimator is more efficient than the HT estimator under Poisson sampling asymptotically. Furthermore, the H{\' a}jek estimator is location invariant after a location transformation of the form $y_i \rightarrow y_i+c$.

\section{Maximum Entropy  Sampling} 

The Poisson sampling introduced in the previous section has the drawback of a random sample size. Using the fact that the Poisson sampling is a particular sampling design maximizing the entropy in (\ref{entropy}) with the constraints on the first-order inclusion probabilities, we can impose an additional constraint on the sample size in the optimization problem. The resulting sampling design is called the maximum entropy sampling design, which was formally studied by \cite{chen1994} and \cite{tille2006}. Maximum entropy sampling is closely related to the rejective Poisson sampling of \cite{hajek1981} and \cite{fuller2009}. In fact, as we can see below, maximum entropy sampling can be implemented using rejective Poisson sampling. 

Let $\mathcal{A}_n=\{A; A \subset U, |A|=n\}$ be the set of all possible samples of size $n$. 
The maximum entropy sampling can be defined as maximizing the entropy in (\ref{entropy}) subject to 
\begin{equation} 
 \sum_{A\in \mathcal{A}_n} I(k \in A) P(A) = \pi_k, \ \ k \in U 
 \label{eq:4.15}
 \end{equation} 
and 
\begin{equation} 
 \sum_{A \in \mathcal{A}_n} P(A) = 1. 
 \label{eq:4.16} 
 \end{equation} 
 Using the same argument for proving Lemma \ref{lem:4.1}, the solution to this optimization problem is 
 \begin{equation} 
 P( A) =
 \left\{ \begin{array}{ll} 
 \frac{ \exp \left( \sum_{k \in A} \theta_k \right) }{ \sum_{A \in \mathcal{A}_n } \exp  \left( \sum_{k \in A} \theta_k \right) } & \mbox{ if } |A| = n \\
 0 & \mbox{ otherwise, } 
 \end{array} 
 \right. 
 \label{maxent} 
 \end{equation} 
 where $\theta_1, \cdots, \theta_N$ are the parameters of the maximum entropy design and are determined by (\ref{eq:4.15}).   
 Thus, (\ref{maxent}) is the sampling design for the maximum entropy sampling. Equation (\ref{eq:4.15}) is a system of $N$ equations to solve $\theta_k$. \cite{tille2006} discussed a fast algorithm for computing $\theta_k$ satisfying (\ref{eq:4.15}).

Now, to compare the maximum entropy sampling with Poisson sampling, define 
\begin{equation} 
 \tilde{\pi}_k = \frac{ \exp ( \theta_k + C) }{ 1+ \exp ( \theta_k + C) }
\label{eq:4.19} 
\end{equation} 
where $C$ is chosen to satisfy 
$$ \sum_{k \in U} \tilde{\pi}_k = n . $$
We can consider the rejective Poisson sampling as follows:
\begin{description} 
\item{[Step 1]} Use $\tilde{\pi}_k$ in (\ref{eq:4.19}) to select a Poisson sample. 
\item{[Step 2]} Check if the realized sample size of the Poisson sample is equal to $n$, the target sample size. If yes, it is the final sample. Otherwise, goto [Step 1].
\end{description}
Because the sampling design of the Poisson sampling in [Step 1] is, by (\ref{eq:4.11}), 
$$ \tilde{P}( A) = \frac{ \exp\left( \sum_{k \in A} \theta_k+C  \right)  }{ \sum_{A \subset U} \exp\left( \sum_{k \in A} \theta_k +C \right)  }, $$
the maximum entropy sampling design in (\ref{maxent}) can be written as
$$ P(A) = \frac{ \tilde{P} (A) }{  \sum_{A \in \mathcal{A}_n } \tilde{P} (A) } ,  \ \ A \in \mathcal{A}_n . 
$$
Thus,  $P(A)$ in (\ref{maxent}) is the conditional Poisson sampling conditional on the fixed sample size $n$.

 In order to construct an algorithm to implement the maximum entropy design, it is necessary to be able to calculate the vector $\tilde{\pi} = ( \tilde{\pi}_1, \cdots, \tilde{\pi}_N)'$ from the vector $\pi = (\pi_1, \cdots, \pi_N)' $. This algorithm is implemented in the function UPmaxentropy of `sampling' package R developed by \cite{tille2016}. 

\section{$\pi$ps  sampling}

The PPS sampling introduced in the previous section has many advantages: it is very easy to implement, the estimation formula is simple. However, since it is a with-replacement sampling, it is inefficient in the sense that it allows for duplicated sample elements.  Let $x_i$ be the size measure that we want to make $\pi_i \propto x_i$ as close as possible. The 
$\pi$ps($\pi$ proportional to
size) sampling refers to a set of sampling designs that satisfies the following  conditions: 
\begin{enumerate}
\item The sampling design is a fixed-size sampling design that does not allow for duplication. 
 \item The first-order inclusion probability $\pi_i$ satisfies 
  $\pi_i \propto x_i$. \item The second order inclusion probability 
  satisfies  $\pi_{ij}>0$ and $\pi_{ij} < \pi_i \pi_j $ ($i \neq
j$).
\end{enumerate}
The third condition guarantees that the SYG variance estimator is always nonnegative. 

For a fixed-size design, $\pi_k \propto x_k$ and $\sum_{i=1}^N \pi_i = n$ leads to 
$$ \pi_k = \frac{n x_k}{ \sum_{i=1}^N x_i } .$$
If some $x_k$ satisfies 
$ x_k > (N/n) \bar{X}_N $, then we have $\pi_i>1$. Thus, the exact proportionality $\pi_i \propto x_i$ is not always possible. In practice, the units with $\pi_k >1$ are automatically selected. The inclusion probabilities are recalculated in the same way excluding the element that is automatically selected. That is, if $k$ satisfies $\pi_k>1$, then we set $\pi_k=1$ and recompute 
$$ \pi_i = (n-1) \frac{x_i}{ \sum_{i \neq k } x_i } . $$
This operation is repeated until $\pi_i \le 1$ for all element in the population. 

For $n=1$, the $\pi$ps sampling is the same as the PPS sampling.  There are two approaches of implementing a PPS sampling of size $n=1$. One is the cumulative total method, and the other is the Lahiri's method. The cumulative total method is described as follows: 
\begin{description}
\item{[Step 1]} Set $T_0=0$ and compute $T_k = T_{k-1} + x_k $, $k=1,2, \cdots, N$.
    \item{[Step 2]} Draw $\epsilon \sim \mbox{Unif}\left( 0, 1 \right)$. If $\epsilon \in (T_{k-1}/T_N, T_k/T_N ) $, element $k$ is selected.
\end{description}
The cumulative total method is very popular because it is easy to understand.  It needs a list of all $x_k$ in the population.

The other method, developed by \cite{lahiri1951}, can be described as follows:
\begin{description}
\item{[Step 0]} Choose $M > \left\{ x_1, x_2, \cdots, x_N \right\}$. Set $r=1$.
\item{[Step 1]} Draw $k_r$ by SRS from $\left\{ 1,2, \cdots, N \right\}$.
\item{[Step 2]} Draw $\epsilon_r \sim \mbox{Unif} \left( 0,1 \right) $.
\item{[Step 3]} If $\epsilon_r \le x_{k_r}/M$, then select element $k_r$ and stop. Otherwise, reject $k_r$ and goto Step 1 with $r = r+1$.
\end{description}
Lahiri's method does not need a list of all $x_k$ in the population, but requires the knowledge of the upper bound of $x_k$, denoted by $M$. 

 A formal justification for Lahiri's method can be described as follows:
  \begin{eqnarray*}
  \pi_k &=& Pr (k \in A) \\
  &=& \sum_{r=1}^\infty Pr ( k \in A, R=r) \\
  &=& \sum_{r=1}^{\infty} Pr \left\{ K_r =k , \epsilon_r < \frac{x_{k_r}}{ M } , \bigcap_{j=1}^{r-1} ( \epsilon_j > \frac{ x_{k_j}}{M} ) \right\} \\
  &=& \sum_{r=1}^{\infty} \frac{1}{N} \frac{x_k}{M} \times \prod_{j=1}^{r-1} Pr \left\{ \epsilon_j > \frac{x_{k_j}}{M} \right\} .
  \end{eqnarray*}  
  Since 
  $$ Pr \left( \epsilon_j > \frac{ x_{k_j}}{M} \right) = \sum_k Pr \left( \epsilon_j > \frac{ x_{k_j}}{ M} \mid k_j = k \right) 
  Pr (K_j = k) = \sum_k \left( 1- \frac{x_k}{M} \right) \frac{1}{N} = 1- \frac{ \bar{x}_U}{ M} , 
  $$
  where $\bar{x}_U = N^{-1} \sum_{i=1}^N x_k $, 
we can obtain  
\begin{eqnarray*}
\pi_k &=& \sum_{r=1}^{\infty} \frac{1}{N} \frac{x_k}{M} \left( 1- \frac{\bar{x}_U}{M} \right)^{r-1} \\
&=& 
\frac{x_k}{ \sum_{i=1}^N x_i } .
\end{eqnarray*}

We now discuss $\pi$ps sampling for $n=2$. Most of the existing schemes for fixed-size $\pi$ps sampling with $n>2$ are quite complicated.  The interested reader is referred to \cite{tille2006}.  To discuss $\pi$ps sampling of size $n=2$, let $\theta_i$ be the probability of selecting unit $i$ in the first draw of the $\pi$ps sampling and let $\theta_{j \mid i}$ be the conditional probability of selecting unit $j$ in the second draw given that unit $i$ is selected in the first draw. Thus, writing $p_i = x_i/( \sum_{j=1}^N x_j )$, 
the problem at hand is to find a set of $\theta_i$ and $\theta_{j \mid i}$ satisfying 
\begin{equation}
 \pi_i = 2 p_i 
 \label{2-31}
 \end{equation}
and 
$$ \sum_{i} \theta_i = \sum_j \theta_{j \mid i} = 1.$$
Since 
 \begin{equation}
\pi_{ij} = \theta_i \theta_{ j \mid i} + \theta_j \theta_{ i \mid
j} \label{2.32}
 \end{equation}
and, as it is a fixed-size sampling design, we can use (\ref{2.2}) to get 
$\sum_{ j \neq i}
\pi_{ij} = \pi_i$,  which implies
\begin{equation*}
\pi_{i} = \theta_i +\sum_{j \neq i} \theta_j \theta_{ i \mid j}.
 \end{equation*}
Thus, constraint (\ref{2-31}) reduces to 
\begin{equation}
 \theta_i +\sum_{j \neq i} \theta_j \theta_{ i \mid j} = 2 p_i.
 \label{2.33}
\end{equation}
Since there are $N^2$ unknowns with $N$ equations, we have many solutions to (\ref{2.33}). 

 \cite{brewer1963} proposed using 
\begin{equation} 
 \theta_i \propto \frac{ p_i \left( 1- p_i \right) }{ 1-2p_i } 
 \label{brewer}
 \end{equation} 
and 
$$ \theta_{j \mid i} \propto p_j $$
to obtain (\ref{2.33}), while \cite{durbin1967} proposed using 
$$ \theta_i = p_i $$
and 
\begin{equation} 
 \theta_{j \mid i} \propto p_j \left( \frac{1}{1-2p_i} + \frac{1}{1-2p_j} \right)  
 \label{durbin}
 \end{equation} 
to achieve the same goal.

In Brewer's method,  we first choose $\theta_{j \mid i} = p_j /(1-p_i)$ for $i \neq j$ and use (\ref{2.33}) to get 
\begin{eqnarray*}
2 p_i &=& \theta_i + \sum_{ j \neq i} \theta_j \frac{p_i}{1-p_j} \\
&=& \theta_i + p_i \left( B - \frac{\theta_i}{1-p_i} \right), 
\end{eqnarray*}
where $B=\sum_{j=1}^N \theta_j/(1-p_j)$. 
Thus, we get 
$$ \theta_i = \frac{ 2p_i-p_iB}{ 1- \frac{p_i}{1-p_i} }= (2-B) \frac{p_i(1-p_i)}{1-2p_i}, $$
which proves (\ref{brewer}).

Using (\ref{2.32}),  Brewer's method leads to 
\begin{equation} 
\pi_{ij} = C \cdot  p_i p_j \left( \frac{1}{1-2p_i} + \frac{1}{1-2p_j} \right) 
\end{equation} 
for some $C$. Now, 
\begin{eqnarray*} 
\sum_{j \neq i} \pi_{ij} &=& C \cdot  p_i \sum_{j \neq i}  p_j \left( \frac{1}{1-2p_i} + \frac{1}{1-2p_j} \right) 
\\ 
&=& C \cdot p_i \left( \sum_{j=1}^N \frac{p_j}{1-2p_j}  -\frac{p_i}{1-2p_i} + \frac{1-p_i}{1-2p_i}  \right) = C \cdot p_i \left( \sum_{j=1}^N \frac{p_j}{1-2p_j}   +1   \right) . 
\end{eqnarray*} 
Because 
\begin{equation}
\pi_i = \sum_{j \neq i} \pi_{ij} = 2p_i, \label{2.36}
\end{equation}
we get
\begin{equation}
\pi_{ij} = \frac{2p_i p_j}{1+K}  \left(\frac{1}{1-2p_i} +
\frac{1}{1-2p_j} \right) \label{2.35}
\end{equation}
where 
 $K=\sum_{i=1}^N \left(1-2p_i\right)^{-1} p_i$.

In \cite{durbin1967}, we set $\theta_i=p_i=\pi_i/2$ and $\theta_{j \mid i} =\pi_{ij}/\pi_i$ with $\pi_{ij}$ in (\ref{2.35}). Thus, we obtain (\ref{durbin}). 
 
\section{Chao's method}
We now introduce one $\pi$ps method proposed by \cite{chao1982}. Chao's sampling is a special case of reservoir sampling with unequal probability of selection. The sampling plan can be described by induction. Let $U_k=\{1, \cdots, k\}$ be the set of elements in the finite population up to the element $k$. Let $A_k$ be the set of sample elements selected from $U_k$ with the first-order inclusion probability proportional to $x$. Let $\pi(k; i)=P( i \in A_k \mid U_k )$ be the corresponding first-order inclusion probability of the unit $i$ from $U_k$ satisfying 
\begin{equation}
 \pi(k ; i) = n \cdot \frac{ x_i}{ \sum_{j=1}^k x_j } . 
 \label{eq:4-23}
 \end{equation} 
Now, for $U_{k+1} = U_k \cup \{ k+1 \}$, we update the sample as follows: 
\begin{enumerate}
    \item Select element $k+1$  with the selection probability $p_{k+1}=n W_{k+1}$, where 
    $$ W_{k+1} =  \frac{x_{k+1}}{ \sum_{j=1}^{k+1} x_j } . $$
    \item  If $k+1$ is selected from Step 1,  remove one element (say $j$) from $A_k$ with probability $1/n$ and set $A_{k+1} = A_k \cup \{k+1\} / \{ j\}$. Otherwise, set $A_{k+1} = A_k$. 
\end{enumerate}
To explain Chao's sampling, note that we can write 
\begin{eqnarray}
P( j \in A_{k+1} \mid U_{k+1} ) &=& P( j \in A_{k} \mid U_{k} ) P(j \in A_{k+1} \mbox{ and } k+1 \in A_{k+1} \mid U_{k+1}, A_{k} )  \notag \\
&+& P( j \in A_{k} \mid U_{k} ) P( k+1 \notin A_{k+1} \mid U_{k+1}, A_{k} ) . \label{4-25} 
\end{eqnarray} 
Now, by the Chao's sampling mechanism, we obtain 
\begin{eqnarray*}
 && P(j \in A_{k+1} \mbox{ and } k+1 \in A_{k+1} \mid U_{k+1}, A_{k} ) \\
 &=& P(j \in A_{k+1}  \mid U_{k+1}, A_{k} ) P(k+1 \in A_{k+1} \mid U_{k+1}, A_{k} ) \\
 &=& \frac{ n-1}{n} \times   p_{k+1} 
 \end{eqnarray*} 
and 
$$ P( k+1 \notin A_{k+1} \mid U_{k+1}, A_{k} ) = 1- p_{k+1}. $$
 Thus, (\ref{4-25}) reduces to 
 \begin{eqnarray}
 P( j \in A_{k+1} \mid U_{k+1} ) &=& P( j \in A_{k} \mid U_{k} ) 
 \times \left( 1- W_{k+1} \right) . 
 \label{3-14} 
 \end{eqnarray} 
 Now, by (\ref{eq:4-23}), we  obtain 
 \begin{equation*} 
 P( j \in A_{k+1} \mid U_{k+1} ) =  n \cdot \frac{x_j}{ \sum_{j=1}^{k+1} x_j } . 
 \end{equation*} 
 Thus, as long as (\ref{eq:4-23}) holds,  the above procedure leads to 
 \begin{equation}
 \pi(k+1 ; i) = n \cdot \frac{ x_i}{ \sum_{j=1}^{k+1}  x_j } 
 \label{eq:4-23b}
 \end{equation} 
 which is a $\pi$ps sampling from population $U_{k+1}$. 
 
 \begin{remark}
 In Chao's sampling,   the initial  sample $S_n=\{1, \cdots, n\}$ for $k=n$ does not satisfy (\ref{eq:4-23}). In fact,  by (\ref{3-14}), 
 $$ \pi( k+1; j) = P( j \in S_j \mid U_j) \prod_{i=j}^{k} \left( 1- W_{i+1} \right) = P( j \in S_j \mid U_j) \frac{ \sum_{i=1}^j x_i}{ \sum_{i=1}^{k+1} x_i} .  $$
 Now, since 
 $$ P( j \in S_j \mid U_j) = \left\{ 
 \begin{array}{ll} 
 1 & \mbox{ if } j \le n \\
 n W_{j} & \mbox{ if } j > n ,\\
 \end{array} 
 \right. 
 $$
 result (\ref{eq:4-23b}) holds only for $i > n$. The first-order inclusion probability of unit $j \le n$ in the $(k+1)$-th population is then 
 $$  \pi( k+1; j) =  \frac{ \sum_{i=1}^n x_i}{ \sum_{i=1}^{k+1} x_i} = n \frac{\bar{x}_n}{\sum_{i=1}^{k+1} x_i } = \frac{1}{n} \sum_{i=1}^n \tilde{\pi} (k+1; i),  $$
 where $\tilde{\pi} (k+1; i) = nx_i/ (\sum_{j=1}^{k+1} x_j )$. 
 \end{remark}

\section{Systematic $\pi$ps sampling}

Systematic $\pi$ps sampling is similar to systematic sampling but allows for unequal probability of sample selection. 
 Let $a=\sum_{i=1}^N x_i / n$ be the sampling interval for systematic sampling. 
  Assume $x_k < a$ for all $k \in U$. (If some of the $x_k$'s are greater than $a$, then such elements are selected in advance and then the systematic sampling is applied in the reduced finite population.) 
 Systematic $\pi$ps sampling can be described as follows. 
 \begin{enumerate}
  \item Choose $R \sim Unif \left( 0,a \right]$
\item   Unit $k$ is selected iff
$$ 
L_k  < R+ l \cdot  a \le  U_k $$
for some $l=0,1, \cdots, n-1$, where $L_k = \sum_{j=1}^{k-1} x_j$ with $L_0=0$ and $U_k = L_k + x_k $. 
 \end{enumerate}
 
 \begin{example}
For example, consider the following artificial finite population of size $N=4$. 

\begin{center}
\begin{tabular}{c|c|c|c}
\hline
 ID & MOS ($x_i$) & L & U \\
\hline 1 & 10 & 0 & 10 \\
2 & 20  & 10 & 30  \\
3 & 30 & 30 & 60  \\
4 & 40 & 60 & 100 \\
\hline
\end{tabular}
\end{center}


To obtain a systematic sample of size $n=2$ with the first-order inclusion probability proportional to $x_i$, note that $a=100/2=50$. Thus, we first generate $R$ from a uniform distribution $(0,50]$. If $R$ belongs to $(0,10]$, we select $A=\{1,3\}$. If $R$ belongs to $(10,30]$, we select $A=\{ 2,4\}$. If $R$ belongs to $(30,50]$, we select $\{3,4\}$.  The sampling distribution of the resulting sample will be 
$$
P \left( A \right) = \left\{ \begin{array}{ll}
0.2, & \mbox{ if } A=\left\{ 1,3 \right\} \\
0.4, & \mbox{ if }  A=\left\{ 2,4 \right\} \\
0.4, & \mbox{ if }  A=\left\{ 3,4 \right\} \\
\end{array}
\right. $$ 
\end{example}

To compute the first-order inclusion probability of unit $k$, let $l$ be the integer satisfying
$  l \cdot a \le L_k < U_k \le (l+1) a $. 
 \begin{eqnarray*}
 Pr \left( k \in A \right) &=& Pr \left\{ L_k < R+ l \cdot a \le U_k\right\} \\
&=& \int_{L_k - l \cdot a}^{U_k - l \cdot a }  \frac{1}{a}  d t = \frac{x_k}{a } = \frac{ n x_k}{\sum_{k \in U} x_k} .
 \end{eqnarray*}
The systematic $\pi$ps sampling is easy to implement but it does not allow for design-unbiased variance estimator, as is the case with the classical systematic sampling.

%% file: chapters/chapter6.tex

\setcounter{chapter}{5} 
\chapter{Cluster Sampling: Single stage cluster sampling}

\section{Introduction}


Element sampling designs are not always feasible when there is no sampling frame for the survey units. Instead, the sampling frame is often available in the form of clusters, such as schools or districts. 
In this case, 
cluster sampling that samples clusters is commonly used. Also, cluster sampling is often cost-effective in terms of reducing the travel cost and also controlling the field work of the interviewers.

In cluster sampling, the finite population is grouped into clusters. A probability sample of clusters is selected, and every element in the  selected clusters is surveyed.  Clusters are also called primary sampling units (PSUs). If a probability sample of PSUs is drawn and every element in the selected PSUs is measured, the sampling design is called single-stage cluster sampling. On the other hand, if another probability sampling is used to draw sample elements in the selected clusters, the sampling design is called a two-stage sampling design. Multi-stage sampling consists of three or more stages of sampling. There is a hierarchy of sampling units: primary sampling units, secondary sampling units within the PUS, and so on. The sampling units in the last-stage sampling are called the ultimate sampling units. In this chapter, we will focus on single-stage sampling. 

Let $U_I =\{ 1, \cdots, N_I\}$ be the index set of clusters in the population. Let $U_i$ be the set of elements in the $i$-th cluster of size $M_i$.  Let $y_{ij}$ be the measurement associated with element $i$ in cluster $i$. The population total of $y$ is denoted by $Y = \sum_{ i =1}^{N_I} \sum_{j \in U_i} y_{ij} = \sum_{i =1}^{N_I} Y_i = \sum_{i=1}^{N_I} M_i \bar{Y}_i $, where $\bar{Y}_i = Y_i/M_i$ is the population mean of $y$ in cluster $i$. 

In the single-stage sampling, we select $A_I$ from $U_I$ by probability sampling. The index set $A$ of the sample elements can be written as $A=\cup_{i \in A_I} U_i$ in single-stage sampling. 
Let $n_I= \left|A_I \right| $ be the number of sampled clusters. Let $n_A=\left| A \right|$ be the number of sampled elements. Under single-stage cluster sampling, we have $n_A= \sum_{i \in A_I} M_i$.

\section{Single-stage cluster sampling: Equal Size Case}

\markboth{6. Single-stage Cluster Sampling  }{6.2 Equal size cluster sampling }

When the cluster size $M_i$ is equal to $M$, the following sampling design can be considered. 
\begin{enumerate}
\item Simple random sampling of $n_I$ clusters from $N_I$ clusters. 
\item Observe all the elements in the selected clusters. 
\end{enumerate}
Such a sampling design is called simple random cluster sampling. In this case, the HT estimator of $Y=\sum_{i =1}^{N_I} Y_i$ is
$$ \hat{Y}_{\rm HT} = \sum_{i \in A_I} \frac{Y_i}{\pi_{Ii} } = \frac{N_I}{n_I} \sum_{i \in A_I} Y_i $$
where $\pi_{Ii}=P( i \in A_I)$ is the cluster-level first-order  inclusion probability and is equal to $\pi_{Ii}=n_I/N_I$ under simple random cluster sampling. Thus, the HT estimator of $\bar{Y}_U=Y/(N_I M) $ is 
$$ \hat{\bar{Y}}_{U} = \frac{\hat{Y}_{HT}}{N_I M}  = \frac{1}{n_I} \frac{1}{M} \sum_{i \in A_I} Y_i  =  \frac{1}{n_I} \sum_{i \in A_I} \bar{Y}_i .$$
Note that $\hat{\bar{Y}}_{U}$ is the sample mean of $\bar{Y}_i$. Thus, 
the sampling variance of $\hat{\bar{Y}}_{U}$ is 
\begin{equation}
 V ( \hat{\bar{Y}}_{U}  ) = \frac{1}{n_I} \left( 1 - \frac{n_I}{N_I} \right)\frac{1}{N_I - 1} \sum_{i=1}^{N_I} \left( \bar{Y}_i - \bar{Y}_{U} \right)^2  
 \label{7-1}
 \end{equation}
where $\bar{Y}_i = \sum_{j=1}^M y_{ij}/M$ and $\bar{Y}_{U} = Y/(N_I M ) = \sum_{i=1}^{N_I} \bar{Y}_i/N_I$.

To discuss the variance form in (\ref{7-1}),  the following ANOVA table is considered. 

\begin{table}[htb]
\begin{center} 
\caption{ANOVA table for the population of clusters with equal size}

\begin{tabular}{c|c|c|c}
\hline
 Source & d.f    & Sum of Squares  & Mean SS \\
   \cline{1-4}
Between cluster &$N_I-1$&  SSB &  $S_b^2$ \\
Within cluster & $N_I \left(M-1\right)$ & SSW & $S_w^2$ \\
\hline
 Total & $N_I M-1$ & SST & $S^2$\\
 \hline
\end{tabular}
\end{center}
\label{table:6.1}
\end{table}

Here, the sum of squares are defined as below and the mean sum of squares are computed by dividing the sum of squares by their degrees of freedom 
\begin{eqnarray*}
SST &=& \sum_{i=1}^{N_I} \sum_{j=1}^M (y_{ij}-\bar{Y})^2 \\
 SSB &=& \sum_{i=1}^{N_I} \sum_{j=1}^M  (\bar{Y}_{i}-\bar{Y})^2 \\
 SSW &=& \sum_{i=1}^{N_I}  \sum_{j=1}^M (y_{ij}-\bar{Y}_i )^2 .
\end{eqnarray*}
Note that, since $SST=SSB + SSW$,  $S^2$ is a weighted average of two mean sum of squares terms: 
\begin{eqnarray*}
 S^2 &=& \frac{SST}{ N_I M-1} \\
 &=& \frac{\left( N_I-1 \right) S_b^2 + N_I \left( M-1 \right) S_w^2 }{N_I M -1 } \\
 &\cong  & \frac{S_b^2 + \left( M-1 \right) S_w^2}{M }. 
 \end{eqnarray*}
Thus, since $SSB=M \cdot \sum_{i=1}^{N_I} ( \bar{Y}_i - \bar{Y} )^2 $, we have
$$\frac{1}{N_I - 1} \sum_{i=1}^{N_I} \left( \bar{Y}_i - \bar{Y}_{U} \right)^2  = \frac{ SSB}{ (N_I-1) M } = \frac{S_b^2}{M} 
$$
and 
the variance expression in (\ref{7-1}) can be written as 
\begin{equation}
V\left( \hat{\bar{Y}}_U  \right) = \frac{1}{n_I M} \left( 1 - \frac{n_I}{N_I} \right) S_b^2. 
\label{7-2}
\end{equation}

Now, let's compare the variance in (\ref{7-2}) with the variance of another HT estimator under simple random sample of the same size, which is 
\begin{equation}
V_{\rm SRS} \left( \hat{\bar{Y}}_U   \right) = \frac{1}{n_I M} \left( 1 - \frac{n_I M}{N_I M} \right) S^2, 
\label{7-3}
\end{equation}
as $n=n_I M$ is the size of the sampled elements. To compare (\ref{7-3}) with (\ref{7-2}), we first introduce the concept of intracluster correlation coefficient. The intracluster correlation coefficient is the population correlation coefficient  between two units in the same cluster. \index{intraclass correlation coefficient}

\begin{definition}
Intraclass correlation coefficient: 
\begin{align*}
\rho & = \dfrac{\cov[y_{ij},y_{ik} \mid j \ne
k]}{\sqrt{\var(y_{ij})}\sqrt{\var(y_{ik})} } \\
& = \dfrac{\sum_{i=1}^{N_I} \sum_{j \ne k} \sum (y_{ij} - \Ybar) (y_{ik} -
\Ybar)  / \sum_{i =1}^{N_I}  M(M-1)}{\sum_{i=1}^{N_I}  \sumjM  (y_{ij} - \bar{Y})^2/\sum_{i=1}^{N_I} 
M}\\
&= \frac{1}{M-1} \dfrac{\sum_{i=1}^{N_I} \sum_{j \ne k} \sum (y_{ij} - \Ybar) (y_{ik} -
\Ybar)  }{\sum_{i=1}^{N_I}  \sumjM (y_{ij} - \bar{Y})^2}
\end{align*}
\end{definition}

Using the following identity 
 \begin{eqnarray*}
 \sum_i  \sum_{j \neq k}\sum_k (y_{ij} - \Ybar) (y_{ik} -
\Ybar) &=& \sum_i \left\{ \sum_j(y_{ij} - \Ybar) \right\}^2 - \sum_i \sum_j(y_{ij} -
\Ybar)^2\\
&=& M \cdot SSB - SST ,
\end{eqnarray*}
we can express 
\begin{equation}
 \rho = \frac{M \cdot SSB - SST}{(M-1) SST}= 1-\dfrac{M}{M-1} \dfrac{SSW}{SST}
\label{eq:6.4}
\end{equation}
and so 
$$ -\dfrac{1}{M-1} \le \rho \le 1. $$
The minimum value of $\rho$ is achieved when $SSB=0$ which occurs when $\bar{Y}_i$ are all equal. The maximum value of $\rho$ is achieved when $SSW=0$ which occurs when all elements are homogeneous within clusters (i.e. $y_{ij} = \bar{Y}_i$). Note that we can also express 
\begin{equation}
\rho = 1- \frac{SSW/\{N_I(M-1)\}}{SST/N_I M}\doteq   1- \frac{S_w^2}{S^2}.
\label{icc}
\end{equation}

The following lemma gives alternative expressions for the mean sum of square terms of the ANOVA table in Table \ref{table:6.1} in terms of the intracluster correlation coefficient $\rho$. 
\begin{lemma}
Using $\rho$, we can express  
\begin{equation}
    S_b^2 \doteq S^2 \{ 1+ (M-1) \rho \} .
    \label{ssb} 
\end{equation}
\end{lemma}
\begin{proof} 
From   (\ref{eq:6.4}), we can use  $SST=SSB+SSW$ to get 
$$ SSB = \frac{1}{M} \left[ 1 + \rho \left( M-1 \right) \right] SST. $$
Now, using $S_b^2=SSB/(N_I-1)\doteq SSB/N_I$ and $S^2 \doteq SST/(N_IM)$, we get (\ref{ssb}). 
\end{proof}

By (\ref{ssb}), combining (\ref{7-2}) and (\ref{7-3}), we can establish the following theorem, which
gives an alternative expression of the sampling variance of the HT estimator in a single-stage cluster sampling design.  

\begin{theorem}
Under simple random cluster sampling, for sufficiently large $N$, the variance of HT estimator can be written 
 \begin{equation}
  \var\left(\hat{Y}_{\rm HT} \right) \doteq
\var_{\rm SRS} \left(\hat{Y}_{\rm HT} \right) \cdot \left\{1+(M-1)\rho\right\}
\label{3.6}
\end{equation}
where $\rho$ is the intracluster correlation coefficient  and 
 $\var_{\rm SRS} \left(\hat{Y}_{\rm HT} \right)$ is the variance of HT estimator under SRS of equal size. 
\end{theorem}

The above theorem shows that the sampling variance of the HT estimator under simple random cluster sampling is $\left\{ 1+ \left(M-1
\right)\rho \right\}$ times larger than that of HT estimator under SRS of equal size. Thus, $\left\{1+ \left(M-1 \right)\rho \right\}$ is the ratio of two variances and expresses the inverse of the relative efficiency of the simple random cluster sampling over the simple random sampling. 
\cite{kish1965} first introduced the concept of the design effect as follows. 

\begin{definition} Design effect (deff) of the sampling design $p( \cdot)$ with the sample size $n=n_p$: 
\begin{equation*}
{\rm deff } \left(p, \hat{Y}_{HT} \right) = \frac{ V_p \left( \hat{Y}_{\rm HT} \mid n_p \right) }{ V_{\rm SRS} \left( \hat{Y}_{\rm HT} \mid n_p \right) }. 
\end{equation*}
\end{definition}
In simple random cluster sampling, we have 
\begin{equation}
\mbox{deff}= 1+(M-1) \rho .
\label{deff}
\end{equation}
For example, if $\rho=0.1$ and $M=11$, deff = $2$. Thus, even if $\rho$ is low, the design effect can be large if $M$ is large.

There are two usages of the design effect. One is to compare designs. For example, if deff $>1$, then $p\left( \cdot \right) $ is less efficient than SRS. If 
 deff $<1 $, then $p\left( \cdot \right) $ is more efficient than SRS.

The other usage is to determine the sample size. When the design effect of a design $p( \cdot) $ is known, the sample size under the design is determined as follows. 
\begin{enumerate}
\item Have some desired variance $V^*$
\item Under SRS, you can easily find required sample size $n^*$
\item Choose $n_p^* = \mbox{ deff} \cdot n^* $.
\end{enumerate}

Then, ignoring the finite population correction term, 
\begin{eqnarray*}
V_p \left( \hat{Y}_{\rm HT} \mid n_p^* \right) &=& V_{\rm SRS} \left( \hat{Y}_{\rm HT} \mid n_p^* \right) \cdot  \mbox{ deff} \\
&=& \frac{N^2}{n_p^*} S^2 \cdot  \mbox{deff} \\
&=& \frac{N^2}{n^*} S^2 \\
&=& V^* .
\end{eqnarray*}
The sample size  $n^*$ is often called the \emph{effective sample size}. \index{effective sample size} It is the sample size required for the given $V^*$ if the sample design is SRS. Given the current sample size $n$, the effective sample size $n^*$ is calculated by $$ n^* = \frac{n}{\mbox{ deff } }. $$

\begin{example}

Suppose that we are interested in estimating the population proportion of certain characteristics by simple random cluster sampling with $M=200$ elements in the cluster. We want to have a margin of error equal to 2\% at the significant level $\alpha=0.05$. How many sampling clusters are needed for this design? Assume that the intracluster correlation coefficient $\rho=0.05$.

To answer this question, first calculate the effective sample size. The effective sample size is equal to $n^*=(0.02)^{-2}=2,500$. The design effect is $1+ 199 \times 0.05 \doteq 11$. Thus, the number of sample clusters is $2,500 \times 11/200=137.5$. \end{example}

\begin{example}

Assume that a simple random sample of clusters is selected from a population of clusters of equal size. Suppose that there are $N_I$ clusters in the population and that there are $M$ elements in each cluster.
Within the sampled cluster, we used simple random sampling to obtain the observed values of the sample elements. The sample observations are summarized in the following ANOVA table.

\begin{table}[htb]
\begin{center}
\caption{Sample ANOVA table (equal cluster size $M$)}
\begin{tabular}{c|c|c|c}
\hline
Source  & D.F.    & Sum of Squares    & Mean S.S. \\
   \cline{1-4}
Between  & $n_I-1$ &  Sample SSB (=SSSB) &  $s_b^2=SSSB/(n_I-1)$ \\
Within  & $n_I (M-1)$ & Sample SSW (=SSSW) &  $s_w^2=SSSW/\{ n_I(M-1)\} $\\
\hline
Total & $n_I M-1$ & Sample SST  (=SSST) &  $s^2 = SSST/(n_IM-1)$ \\
\hline 
\end{tabular}
\end{center}
\label{table3-1b}
\end{table}

In this case, we have 
\begin{eqnarray*}
\mbox{Sample SSB} &=& \sum_{ i=1}^{n_I} M\left\{ \bar{Y}_i  -    ( n^{-1} \sum_{k=1}^n \bar{Y}_k) \right\}^2  \\
 \mbox{Sample SSW} &=&\sum_{ i=1}^{n_I} \sum_{j=1}^M (y_{ij} - \bar{Y}_i)^2 
\end{eqnarray*}
and Sample SST = Sample SSB + Sample SSW. Since the sampling design is the simple random cluster sampling, 
$$E(s_b^2) = \frac{1}{N_I-1} \sum_{i=1}^{N_I} \left( \bar{Y}_i - \bar{Y} \right)^2  = S_b^2$$
and 
$$ E(s_w^2) = \frac{1}{N_I (M-1)} \sum_{i=1}^{N_I}  \left( y_{ij}- \bar{Y}_i \right)^2 = S_w^2 $$
but we do not have $E(s^2)=S^2$. In this case, we can use 
$$ \hat{\rho} = 1- \frac{ s_w^2}{ \hat{S}_y^2 } $$
where 
\begin{eqnarray*}
 \hat{S}_y^2 &=& \frac{ (N-1) s_b^2 +  N(M-1) s_w^2 }{ NM} \\
&\doteq & \frac{1}{M} s_b^2 + \left( 1- \frac{1}{M} \right) s_w^2 .
\end{eqnarray*}
In other words, we have 
$$ \hat{\rho} =  \frac{ s_b^2 - s_w^2 }{ s_b^2 + (M-1) s_w^2} $$
as an estimator of the intracluster correlation. Once $\hat{\rho}$ is obtained, we can use (\ref{diff}) to compute the design effect of the single-stage cluster sampling.

\end{example}

\section{Single-stage cluster sampling: Unequal Size Case}

\markboth{6. Cluster Sampling 1 }{6.3 Unequal size cluster sampling }

We now consider the case when the cluster sizes $M_i$ are unequal and we are interested in estimating the population total or population mean. If the cluster-level first-order inclusion probability is given by $\pi_{Ii} = \Pr\left( i \in A_I \right)$, the HT estimator of the population total is 
$$  \hat{Y}_{\rm HT} = \sum_{i \in A_I } \frac{Y_i}{\pi_{Ii}}$$
and its variance is, under fixed-size cluster
sampling, 
$$ V\left(\hat{Y}_{\rm HT} \right)
= - \frac{1}{2}\sum_{i=1}^N \sum_{j=1}^N  \left( \pi_{Iij} -
\pi_{Ii} \pi_{Ij} \right) \left(\frac{Y_i}{\pi_{Ii} } - \frac{Y_j
}{\pi_{Ij}} \right)^2.
$$
The variance is minimized when $\pi_{Ii} \propto
Y_i$. Since we do not know $Y_i$ in practice and   $Y_i = M_i
\bar{Y}_i$, we may use 
$\pi_{Ii} \propto M_i$ if $\bar{Y}_{Ii}$ is believed to be independent of $M_i$. 

To formally discuss the problem of optimal choice of the first-order inclusion probability, 
assume the following superpopulation model 
\begin{equation}
y_{ij} = \mu + a_i + e_{ij} 
\label{6-7}
\end{equation}
where $\mu$ is an unknown parameter, $a_i \sim (0, \sigma_a^2)$, and $e_{ij} \sim (0, \sigma_e^2)$. 
The  total variance of $\hat{Y}_{HT}$ under the model (\ref{6-7}) is 
\begin{eqnarray*}
V(\hat{Y}_{\rm HT}) &=& V\left( \sum_{i \in A_I} \pi_{Ii}^{-1}M_i \mu \right) + E\left(  \sum_{i \in A_I} \pi_{Ii}^{-2} \gamma_i \right) \\
&=& V\left( \sum_{i \in A_I} \pi_{Ii}^{-1} M_i \mu \right) + E\left(  \sum_{i \in U_I} \pi_{Ii}^{-1} \gamma_i \right) ,  \end{eqnarray*}
where $\gamma_i = V(Y_i) = M_i^2 \sigma_a^2 + M_i \sigma_e^2$. 
The second term of the total variance is minimized when 
\begin{equation}
\pi_{Ii} \propto \gamma_i^{1/2} = M_i \sigma_a \left( 1+ \frac{\sigma_e^2}{\sigma_a^2} \cdot \frac{1}{ M_i}   \right)^{1/2}. 
\label{6-8}
\end{equation}
The solution (\ref{6-8}) is obtained by applying the following Cauchy-Schwartz inequality 
$$ \left( \sum_{i \in U_I} \pi_{Ii}^{-1} \gamma_i \right) \left( \sum_{i \in U_I} \pi_{Ii} \right) \ge \left( \sum_{i \in U_I} \sqrt{\gamma}_i \right)^2 
$$
and using that $ \sum_{i \in U_I} \pi_{Ii}$ is fixed. 
Thus, if $\sigma_e^2/\sigma_a^2 $ is small, then $\pi_{Ii} \propto M_i$ lead to an optimal sampling design.
On the other hand, if 
$\sigma_e^2/\sigma_a^2 $ is large, then $\pi_{Ii} \propto M_i^{1/2}$ can be a better design. 

We now consider the situation when the simple random cluster sampling is used in spite of the fact that $M_i$ are unequal. The HT estimator of the total is 
$$ \hat{Y}_{\rm HT} = \frac{N_I}{n_I} \sum_{i \in A_I} Y_i$$
where  $Y_i = M_i \bar{Y}_i$. The variance is 
\begin{eqnarray}
 V_{\rm SRC}( \hat{Y}_{\rm HT} ) &=& \frac{N_I^2}{n_I} \left( 1- \frac{n_I}{N_I} \right) S_I^2
 \label{5.10b}
\end{eqnarray}
 where 
$$ S_I^2 =   \frac{1}{N_I-1} \sum_{i \in U_I} \left( Y_i - \bar{Y}_I  \right)^2 =
\frac{1}{N_I-1} \sum_{i \in U_I} \left( M_i \bar{Y}_i - \bar{M} \bar{Y}_U   \right)^2  ,$$
 $\bar{M} = N_I^{-1} \sum_{ i \in U_I} M_i $, $\bar{Y}_I = N_I^{-1} \sum_{i \in U_I} Y_i $, and  $\bar{Y}_U = \hat{Y}_I/ \bar{M} $.

Since $M_i$ can be different, we  extend the definition   of the intracluster correlation coefficient to define the cluster homogeneity coefficient as follows: 
 \index{cluster homogeneity coefficient}
\begin{equation}
 \delta = 1- \frac{\sum_{i \in U_I} \sum_{j \in U_i} \left( y_{ij} - \bar{Y}_i \right)^2/ \left( N - N_I \right) }{
\sum_{i \in U_I} \sum_{j \in U_i} \left( y_{ij} - \bar{Y}_U \right)^2/ \left( N - 1 \right) }
= 1- \frac{SSW/(N-N_I)}{SST/(N-1)}  .
\label{chcoff}
\end{equation}
If $M_i$ are all equal, then $\delta$ is equal to the intracluster correlation in  (\ref{eq:6.4}).

 Using $\delta$ in (\ref{chcoff}), we can express the variance in (\ref{5.10b}) as 
\begin{equation}
 V_{\rm SRC} \left( \hat{Y}_{\rm HT} \right)
 = \left( 1+ \frac{ N - N_I}{N_I - 1 } \delta \right) \bar{M} S_y^2 K_I + C^* \cdot K_I
 \label{5.result}
 \end{equation}
 where $K_I= \frac{N_I^2}{n_I} \left( 1- \frac{n_I}{N_I} \right) $ and 
$$ C^* = \frac{1}{N_I-1} \sum_{i=1}^{N_I} \left( M_i - \bar{M} \right) M_i \bar{Y}_i^2 $$
is the population covariance between  $M_i$ and   $M_i \bar{Y}_i^2$. If  $\bar{Y}_i^2$ are all equal, then we have 
 $C^* =Cov \left( M_i, M_j \right) $. Roughly speaking, the second term in (\ref{5.result}) is the additional variance due to the unequal cluster sizes.

Now, consider simple random sampling  using the same sample size  $n=n_I \bar{M} $. In this case, the variance under SRS is 
\begin{eqnarray*}
V_{\rm SRS} \left( \hat{Y}_{\rm HT} \right) &=& \frac{ N^2}{n_I \bar{M} } \left( 1 - \frac{ n_I \bar{M} }{N} \right) S_y^2 \\
&=& \frac{N}{N_I} S_y^2 \frac{N_I^2}{n_I} \left( 1- \frac{n_I}{N_I} \right) = \bar{M} S_y^2 K_I. 
\end{eqnarray*}
Thus, the design effect of the simple random cluster sampling design is \index{design effect}
$$ \mbox{deff} = \frac{ V_{\rm SRC} \left( \hat{Y}_{\rm HT} \right) }{ V_{\rm SRS} \left( \hat{Y}_{\rm HT} \right)}
= \left( 1+ \frac{ N-N_I }{N_I-1} \delta \right) + \frac{ C^* }{ \bar{M} S_y^2 }.  $$
Therefore, there are two source of having  deff $>$ 1 in this case.
\begin{enumerate}
\item 
$\delta >0$: The homogeneity of the $y$ values within the cluster reduces efficiency. 
\item $C^* >0$: The variability of cluster size reduces efficiency. 
\end{enumerate}
If  $M_i=M$, then $C^*=0$ and the design effect is equal to  $1+ (M-1) \delta$, which  is consistent with the result in  (\ref{deff}).

We now consider mean estimation under cluster sampling. 
If the parameter of interest is population mean, we may have two different concepts:
 \begin{eqnarray*}
 \mu_1 &=& \frac{1}{N_I} \sum_{i=1}^{N_I} \frac{Y_i}{M_i} = \frac{1}{N_I} \sum_{i=1}^{N_I} \bar{Y}_i \\
\mu_2 &=& \frac{1}{N} \sum_{i =1}^{N_I} Y_i = \frac{ \sum_{i=1}^{N_I}  \sum_{j=1}^{M_i} y_{ij} }{  \sum_{i=1}^{N_I} M_i }.  
\end{eqnarray*}
 If $M_i=M$, then $\mu_1=\mu_2$. Otherwise, the two parameters have different meanings. 
  The first parameter $\mu_1$ is the cluster-level mean while the second parameter $\mu_2$ is the element-level mean.
 
 Suppose that we are interested in estimating $\bar{Y}_U=\mu_2$.  From each sampled cluster $i$, we observe $\left( M_i, \bar{Y}_i \right)$. We can estimate the mean by a ratio of the estimated total of $y$ to the estimated total size: 
$$ \hat{\bar{Y}}_{R} =  \frac{\sum_{i \in A_I} Y_i/\pi_{Ii} }{\sum_{i \in A_I} M_i/\pi_{Ii}  } :=  \frac{ \hat{Y}_{\rm HT}}{ 
\hat{N}_{\rm HT}} .$$
Since the ratio is a nonlinear function of two HT estimators, we use a Taylor linearization 
to get the following approximation
$$\frac{ \hat{Y}_{\rm HT}}{ 
\hat{N}_{\rm HT}}  \cong \frac{Y}{N} + \frac{1}{N} \left( \hat{Y}_{\rm HT}-  \frac{Y}{N} \hat{N}_{\rm HT} \right). 
$$
Thus, the approximate variance is
$$ V \left( \hat{\bar{Y}}_{R}  \right) 
\doteq 
 \frac{1}{N^2} 
 \left\{  V ( \hat{Y}_{\rm HT} ) - 2 \mu  Cov(\hat{Y}_{\rm HT}, \hat{N}_{\rm HT}) + \mu^2 V( \hat{N}_{\rm HT} ) \right\}.
 $$

Under simple random cluster sampling, for example, the variance is 
\begin{eqnarray*}
 V\left( \hat{\bar{Y}}_{R} \right) &\doteq & \frac{N_I^2}{n_I N^2} \left( 1- \frac{n_I}{N_I} \right) \frac{1}{N_I-1}\sum_{i=1}^{N_I} \left( Y_i - M_i \bar{Y}_U \right)^2 \\
 &=& \frac{1}{n_I \bar{M}^2 } \left( 1- \frac{n_I}{N_I} \right) \frac{1}{N_I-1}\sum_{i=1}^{N_I}M_i^2  \left( \bar{Y}_i - \bar{Y}_U \right)^2 
 \end{eqnarray*}
 with $\bar{M}=N_I^{-1} \sum_{i =1}^{N_I} M_i=N/N_I$. For variance estimation, we can use 
 \begin{eqnarray*}
 \hat{V}\left( \hat{\bar{Y}}_{R} \right) 
 &=& \frac{1}{n_I \bar{M}^2 } \left( 1- \frac{n_I}{N_I} \right) \frac{1}{n_I-1}\sum_{i \in A_I} M_i^2  \left( \bar{Y}_i - \bar{Y}_U \right)^2 .
 \end{eqnarray*}

%% file: chapters/chapter7.tex
\setcounter{chapter}{6} 
\chapter{Cluster Sampling: Two-stage cluster sampling}

\section{Introduction}





In this chapter, we consider two-stage cluster sampling where the sample clusters are selected in the first stage, and the sample elements are selected in the second stage sampling. To formally state this, two-stage sampling can be described as follows: 
\begin{enumerate}
\item Stage 1: Draw $A_I \subset U_I$ via $p_I \left( \cdot \right)$
\item Stage 2: For every $i \in A_I$, draw $A_i \subset U_i$ via $p_i \left( \cdot \mid A_I \right)$
\end{enumerate}
The set of sample  elements is now given by $A=\cup_{i \in A_I} A_i $.  

In two-stage sampling, we have two simplifying assumptions about the second stage sampling design $p_i( \cdot \mid A_I)$: 
\begin{enumerate}
\item \emph{Invariance} of the second-stage design $p_i \left( \cdot \mid A_I \right) = p_i \left( \cdot \right)$ for every $i \in U_I$ and for every $A_I$ such that $i \in A_I$
\item \emph{Independence} of the second-stage design
$$ P \left( \bigcup_{i \in A_I} A_i \mid A_I  \right) = \prod_{i \in A_I} P \left( A_i \mid A_I \right) $$
\end{enumerate}
Under independence, we have 
$$ P ( k \in A_i, l \in A_j) = P( k \in A_i ) P ( l \in A_j) , \   \ i \neq j.$$
If the invariance assumption does not hold, the sampling design is called two-phase sampling design. Under two-phase sampling, the second-phase sampling design depends on the outcome of the first-phase sample. 
The two-phase sampling design will be covered in Chapter 11. 

In two-stage sampling, we use $n_I$ to denote the cluster sample size in the first-stage sampling and use $m_i=\left| A_i \right| $ to denote the sample size in cluster $i$. The number of sampled elements is equal to  $\sum_{i \in A_I} m_{i}= \left| A \right|$. The first-order inclusion probability of element $k$ in cluster $i$ is a product of the cluster-level inclusion probability and the conditional inclusion probability given the cluster:
$$ \pi_{ik} = P \left\{ (ik) \in A \right\} = P \left( k \in A_i \mid i \in A_I \right) P \left( i \in A_I \right) = \pi_{k \mid i} \pi_{Ii} ,$$
where $\pi_{Ii}$ is the cluster-level inclusion probability and $\pi_{k \mid i} = P \left[ k \in A_i \mid i \in A_I \right] $ is the element-level conditional inclusion probability. In general, $\pi_{k \mid i}$ is a random variable (in the sense that it is a function of $A_I$). Under invariance, it is fixed.

The second-order inclusion probability between two elements can be expressed as 
$$ \pi_{ik,jl} = \left\{
\begin{array}{ll}
\pi_{Ii} \pi_{k \mid i} & \mbox{ if } i=j \mbox{ and }  k=l  \\
\pi_{Ii} \pi_{kl \mid i} & \mbox{ if } i=j \mbox{ and }  k \neq l\\
\pi_{Iij} \pi_{k \mid i} \pi_{l \mid j} & \mbox{ if }  i \neq j \end{array}
\right.
$$
where $\pi_{Iij}$ is the cluster level joint inclusion probability and $
\pi_{kl \mid i}= P \left( k,l \in A_i \mid i \in A_I \right)$.

\section{Estimation}

In two-stage cluster sampling, we do not observe $Y_i$. Instead, we obtain $\hat{Y}_i$ from the second stage sampling such that $E( \hat{Y}_i \mid A_I) = Y_i$, where the conditional expectation is taken with respect to the second-stage sampling. For simplicity, we use $E_2( \hat{Y}_i ) = E( \hat{Y}_i \mid A_I)$.

The HT estimation for $Y=\sum_{i \in U_I} Y_i = \sum_{i \in U_I} \sum_{k \in U_i} y_{ik} $ is given by 
\begin{equation}
 \hat{Y}_{\rm HT} = \sum_{i \in A_I} \frac{\hat{Y}_i }{\pi_{Ii}} = \sum_{i \in A_I} \sum_{k \in A_i} \frac{ y_{ik} }{ \pi_{k \mid i} \pi_{Ii} } .
 \label{3.8}
 \end{equation}
 The HT estimator in (\ref{3.8}) is unbiased 
 and its variance can be computed by 
 \begin{equation}
V\left(\hat{Y}_{\rm HT} \right) = V\left\{ E\left( \hat{Y}_{\rm HT}\mid
A_I \right)  \right\} +E\left\{ V\left( \hat{Y}_{\rm HT}\mid A_I
\right) \right\} . 
\end{equation}
The first term is the variance due to the first-stage sampling (sampling of PSUs) and the second term is the variance due to the second-stage sampling (sampling of SSUs). Thus, we can write  
 \begin{equation}
  V\left( \hat{Y}_{\rm HT} \right)= V_{\rm PSU} + V_{\rm SSU} 
  \label{3.10}
  \end{equation}
where
\begin{eqnarray*}
V_{\rm PSU} &=&  V \left\{ \sum_{i \in A_I} \frac{{Y}_i }{\pi_{Ii}} \right\} = 
\sum_{i \in U_I} \sum_{j \in U_I} (\pi_{Iij} - \pi_{Ii} \pi_{Ij}) \frac{Y_i}{\pi_{Ii}} \frac{ Y_j}{\pi_{Ij}} \\
V_{\rm SSU} &=& E\left\{ \sum_{i \in A_I} \frac{1}{\pi_{Ii}^2} V_i \right\} = 
 \sum_{i \in U_I} \frac{ V_i}{\pi_{Ii}},  \   \ \end{eqnarray*}
and
$$ V_i = V\left( \hat{Y}_i \mid A_i \right) = \sum_{k \in U_i} \sum_{l \in U_i} ( \pi_{kl\mid i} - \pi_{k \mid i} \pi_{l \mid i} ) \frac{y_{ik}}{\pi_{k \mid i}} \frac{y_{il} }{\pi_{l \mid i}}. $$

\begin{example}
Consider the following two-stage sampling design. 
\begin{enumerate}
\item Stage One: Select $n_I$ sample clusters from $N_I$ population clusters by simple random cluster sampling. 
 \item Stage Two: Within sampled cluster $i$, select $m_i$ sample elements from $M_i$ population elements independently. 
\end{enumerate}

Under this two-stage sampling, we have 
\begin{align*}
\hat{Y}_{\rm HT} & = \dfrac{N_I}{n_I} \sum_{i \in A_I} \hat{Y}_i =
 \sum_{i \in A_I} \sum_{j \in A_i} \dfrac{N_I}{n_I}\dfrac{M_i}{m_i}
            y_{ij}
\end{align*}
and its variance is 
\begin{equation}
V \left( \hat{Y}_{\rm HT} \right)= \frac{N_I^2}{n_I} \left(1-\dfrac{n_I}{N_I}\right)S_I^2 
+
                 \left( \dfrac{N_I}{n_I}\right)  \sum_{i=1}^{N_I} \frac{ M_i^2}{m_i} (1-\dfrac{m_i}{M_i})
                 S_i^2 , 
                 \label{3.13}
\end{equation}
where 
\begin{eqnarray*}
 S_I^2 &=& \left( N_I-1 \right)^{-1} \sum_{i=1}^{N_I} \left( Y_i -
 \bar{Y}_N \right)^2 \\
 S_i^2&=& \left(M_i-1\right)^{-1} \sum_{j=1}^{M_i}
                 (y_{ij} - \bar{Y}_{i})^2.
 \end{eqnarray*}
 
Now, consider the estimation of the population mean  $\bar{Y}=N^{-1} \sum_{i=1}^{N_I} \sum_{j=1}^{M_i} y_{ij} $,  where  $N=\sum_{i=1}^{N_I} M_i$ is assumed to be known. In this case,  
$$ \hat{\bar{Y}}_{\rm HT} = \frac{\hat{Y}_{\rm HT}}{ N} =
\frac{N_I}{n_I N } \sum_{i \in A_I} \hat{Y}_i  = \frac{1}{n_I } \sum_{i \in A_I} \frac{\hat{Y}_i}{\bar{M}}  ,  $$
where $\bar{M} = N_I^{-1} \sum_{i=1}^{N_I} M_i$. Its variance is, using (\ref{3.13}),  
\begin{eqnarray*}
 V\left(\hat{\bar{Y}}_{\rm HT} \right)  &=& \frac{1}{n_I } \left( 1 - \frac{n_I}{N_I} \right) S_{q1}^2  + \frac{1}{n_I N_I \bar{M}^2 }  \sum_{i=1}^{N_I}  \frac{M_i^2}{m_i} \left( 1- \frac{m_i}{M_i} \right)S_{2i}^2, 
  \end{eqnarray*} 
where $S_{q1}^2=(N_I-1)^{-1} \sum_{i=1}^{N_I} ( q_i - \bar{q}_{1} )^2 $ with $q_i=Y_i/\bar{M}$, $\bar{q}_1 = N_I^{-1} \sum_{i=1}^{N_I} q_i$,  and $S_{2i}^2 = (M_i-1)^{-1}  \sum_{k \in U_i} (y_{ik}- \bar{Y}_i)^2$.

 If the sampling rate for the second stage sampling is constant such that $m_i/M_i=f_2$, then we can write \begin{eqnarray*}
 V\left( \hat{\bar{Y}}_{\rm HT} \right)  &=& \frac{ 1}{n_I} (1-f_1)S_{q1}^2  + \frac{1 }{n_I \bar{m} }  (1- f_2 ) \frac{1}{N} \sum_{i=1}^{N_I} M_i S_{2i}^2\\
 & := & \frac{1}{n_I} (1-f_1)B^2  + \frac{1 }{n_I \bar{m} } (1- f_2 )   W^2 
  \end{eqnarray*} 
where $f_1= n_I/N_I$ and $\bar{m} = N_I^{-1} \sum_{i=1}^{N_I} m_i $. 
 
\label{example3-2}
\end{example}

\begin{example}
We now consider a special case of  Example  \ref{example3-2} where $ M_i
= M $ and $m_i=m $. In this case,  (\ref{3.13}) is further simplified 
\begin{eqnarray}
V \left( \hat{Y}_{\rm HT} \right)&=&
\frac{N_I^2}{n_I}\left(1-\dfrac{n_I}{N_I}\right) \frac{M \times SSB}{N_I-1} +
                 \left( \dfrac{N_I}{n_I}\right)  \left(1-\dfrac{m}{M}\right)
                 \dfrac{M^2}{m \left( M-1 \right)} SSW \notag  \\
 & =&\frac{N_I^2}{n}\left(1-\dfrac{n_I}{N_I}\right) M S_b^2 + \left( \dfrac{N_I^2}{n_I}\right)  \left(1-\dfrac{m}{M}\right)
                 \frac{M^2}{m} S_w^2
                 \label{3.13b}
\end{eqnarray}
For the case of mean estimation, we can simply divide 
(\ref{3.13b}) by $N^2= N_I^2 M^2$ to get 
\begin{equation}
 V ( \hat{\bar{Y}}_{\rm HT}  ) =\left(1-\dfrac{n_I}{N_I}\right)
 \dfrac{S_b^2}{n_I M}+\left(1-\dfrac{m}{M}\right)\dfrac{S_w^2}{n_Im}
\label{3.15a}
\end{equation}
Note that the sample size associated with the first term ($V_{PSU}$ term) is $n_I M$ while the sample size associated with the second term ($V_{SSU}$ term) is $n_I m $. \label{example3-3}
\end{example}

Now, we can express the variance term in (\ref{3.15a}) in terms of the intracluster correlation coefficient. Using (\ref{icc}) and (\ref{ssb}), 
 the variance term in   (\ref{3.15a}) reduces to, ignoring 
$n_I/N_I$ term, 
\begin{eqnarray}
 V \left( \hat{\bar{Y}}_{\rm HT}  \right) &\doteq &
\frac{1}{n_IM} \left\{ 1 + \left( M-1 \right) \rho \right\} S^2 + \left(
1 - \frac{m}{M} \right) \frac{1}{n_I m} \left( 1- \rho \right) S^2
\notag
\\
&=& \frac{S^2}{n_I m}
 \left\{ 1+ \left(m-1 \right)\rho \right\}.
\label{3.15b}
\end{eqnarray}
Thus, the design effect becomes $1+ \left( m-1 \right)
\rho$.

In this case of  $ M_i = M $ and $m_i=m $, the problem of finding the optimal choice of  $ m $ given the cost function 
$ C=c_0 + c_1 n_I + c_2 n_Im$ can be formulated as minimizing 
\begin{eqnarray*}
 V \left( \hat{\bar{Y}}_{\rm HT}  \right)&=& 
 \dfrac{S_b^2}{n_I M}+(1-\dfrac{m}{M})\dfrac{S_w^2}{n_Im}\\
 &=& \frac{1}{n_I} \left\{ \frac{1}{M} \left( S_b^2 - S_w^2 \right) + \frac{1}{m} S_w^2  \right\} 
 \end{eqnarray*}
 subject to 
 $$ C=c_0 + c_1 n_I + c_2 n_I m.
$$
When the total cost $C$ is fixed, we have 
$$  n =\dfrac{C-c_0}{c_1+c_2 m} $$
and the optimal choice is given by 
\begin{equation}
          m^* =\sqrt{\dfrac{c_1}{c_2}\dfrac{M \times S_w^2}{S_b^2-S_w^2}}.
\label{3.15e}
\end{equation}
The optimal solution (\ref{3.15e}) is obtained by applying 
$$ \frac{a}{m} + b m \ge 2\sqrt{ab} $$
with equality if and only if $m=\sqrt{a/b}$. That is, since 
\begin{eqnarray*}
V ( \hat{\bar{Y}}_{\rm HT} ) \cdot (C-c_0)  &=& \left\{ \frac{1}{M} \left( S_b^2 - S_w^2 \right) + \frac{1}{m} S_w^2  \right\} \left( c_1 + c_2 m \right) 
\\
&=& \mbox{const.} +  \frac{ c_1}{m} S_w^2 +  \frac{c_2 }{M} \left( S_b^2 - S_w^2 \right) m ,
\end{eqnarray*}
the lower bound is achieved when 
$$ m = \left\{  c_1 S_w^2 \right\}^{1/2} \left\{ \frac{c_2 }{M} \left( S_b^2 - S_w^2 \right)  \right\}^{-1/2} $$
which equals to (\ref{3.15e}). 
 For sufficiently large $M$, the optimal solution becomes  
$$ m^*\cong \sqrt{\dfrac{c_1}{c_2}\left(\dfrac{1}{\rho}-1\right)}.$$
More generally, the objective function can be written as 
 $$V ( \hat{\bar{Y}}_{\rm HT} )=  \frac{1}{n_I} B^2  + \frac{1 }{n_I \bar{m} } \left(1- f_2 \right)   W^2. $$
 In this case, the optimal solution becomes 
 \begin{equation}
          m^* =\sqrt{\dfrac{c_1}{c_2}\dfrac{ W^2}{B^2-  W^2/M}}.
\label{3.15g}
\end{equation} 

We now discuss variance estimation under two-stage cluster sampling. 
\begin{theorem}
An unbiased estimator for the variance of HT estimator in  (\ref{3.8}) under two-stage sampling design is 
\begin{equation}
\hat{V} \left( \hat{Y}_{\rm HT} \right) =\sum_{i \in A_I } \sum_{j \in
A_I } \frac{ \pi_{Iij} - \pi_{Ii} \pi_{Ij} }{  \pi_{Iij}}
\frac{\hat{Y}_i}{\pi_{Ii}} \frac{\hat{Y}_j}{\pi_{Ij}}+ \sum_{i \in
A_I} \frac{1}{\pi_{Ii}} \hat{V}_i \label{3.15}
\end{equation}
where $\hat{V}_i $ satisfies  $E(  \hat{V}_i \mid i \in A_I ) = V(
\hat{Y}_i \mid i \in A_I )$.
\end{theorem}
\begin{proof}
By (\ref{3.10}), 
\begin{equation}
V\left(\hat{Y}_{\rm HT}  \right)  = \sum_{i=1}^{N_I} \sum_{j=1}^{N_I} \left(
\pi_{Iij} - \pi_{Ii} \pi_{Ij} \right) \frac{Y_i}{\pi_{Ii}}
\frac{Y_j}{\pi_{Ij}} +  \sum_{i=1}^{N_I} \frac{V_i}{\pi_{Ii}}.
\label{3.16}
\end{equation}
By the independence assumption, 
$$ E_2 \left( \hat{Y}_i \hat{Y}_j \right)
= \left\{ \begin{array}{ll} Y_i^2 + V_i & \mbox{ if } i=j \\
Y_i Y_j & \mbox{ if } i \neq j \\
\end{array}
\right.
$$
where $E_2 ( \cdot)$ denotes the expectation with respect to the second-stage sampling. 
Thus, 
\begin{eqnarray*}
\sum_{i \in A_I } \sum_{j \in A_I } \frac{ \pi_{Iij} - \pi_{Ii}
\pi_{Ij} }{  \pi_{Iij}} \frac{E_{2} \left( \hat{Y}_i\hat{Y}_j
\right)}{\pi_{Ii}\pi_{Ij}}
 &=& \sum_{i \in A_I } \sum_{j \in A_I } \frac{
\pi_{Iij} - \pi_{Ii} \pi_{Ij} }{  \pi_{Iij}} \frac{{Y}_i Y_j
}{\pi_{Ii} \pi_{Ij}}  \\
&+& \sum_{i \in A_I } \frac{\pi_{Ii} - \pi_{Ii}^2 }{\pi_{Ii}}
\frac{V_i}{\pi_{Ii}^2}
\end{eqnarray*}
and, since $E_{2} \left( \hat{V}_i \right) = V_i$, we have 
$$ E_2 \left\{
 \hat{V} \left( \hat{Y}_{\rm HT} \right)\right\} =\sum_{i \in A_I } \sum_{j \in A_I } \frac{
\pi_{Iij} - \pi_{Ii} \pi_{Ij} }{  \pi_{Iij}} \frac{{Y}_i Y_j
}{\pi_{Ii} \pi_{Ij}} + \sum_{i \in A_I} \frac{V_i}{\pi_{Ii}^2}.$$
Taking the expectation of the above term with respect to the first-stage sampling design, it equal to the variance term in  (\ref{3.16}).
\end{proof}

The variance estimation formula in  (\ref{3.15}) is the sum of two terms. The first term is the variance estimation formula for the first-stage sampling applied to $\hat{Y}_i$ and the second term is the point estimator for the first-stage sampling applied to $\hat{V}_i$. The validity of the variance estimation formula (\ref{3.15}) also holds even when $\hat{Y}_i$ and $\hat{V}_i$ are obtained from multi-stage sampling. That is, as long as $E ( \hat{Y}_i \mid A_I) = Y_i$ and $E ( \hat{V}_i \mid A_I) = V( \hat{Y}_i \mid A_I ) $ hold, the variance estimation formula in (\ref{3.15}) remains unbiased. Such a phenomenon was first discovered by \cite{raj1966}. 

If we use only the first term of   (\ref{3.15}) 
$$\hat{V}_1 = \sum_{i \in
A_I } \sum_{j \in A_I } \frac{ \pi_{Iij} - \pi_{Ii} \pi_{Ij} }{
\pi_{Iij}} \frac{\hat{Y}_i}{\pi_{Ii}} \frac{\hat{Y}_j}{\pi_{Ij}},$$
to estimate  the total variance, the bias can be written  
\begin{equation}
 Bias\left( \hat{V}_1 \right) = - \sum_{i=1}^{N_I} V_i.
\label{3.17}
 \end{equation}
 and the bias term is of order $O(N_I)$. 
Since $\var \left(
\hat{Y}_{HT} \right)$ is of the order $O\left(n_I^{-1} N_I^{2}  \right)$, the bias term is negligible when $n_I/N_I=o\left( 1\right)$.

 Under the setup of Example \ref{example3-2} where  $ M_i = M $, $m_i=m $, the variance estimation formula in (\ref{3.15}) reduces to 
 \begin{equation}
\hat{V} \left( \hat{Y}_{\rm HT} \right) = \frac{N_I^2}{n_I} \left( 1- \frac{n_I}{N_I} \right) \frac{1}{n_I-1} \sum_{i \in A_I} \left( \hat{Y}_i - \frac{1}{n_I} \sum_{j \in A_I} \hat{Y}_i \right)^2 + \frac{N_I}{n_I} \sum_{i \in A_I} \hat{V}_i  
 \label{3.15c}
\end{equation}
where $\hat{Y}_i = M \bar{y}_i$ and 
$$\hat{V}_i = \frac{M^2}{m} \left( 1- \frac{m}{M} \right) \frac{1}{m-1} \sum_{j  \in A_i} ( y_{ij} - \bar{y}_i )^2 . $$
 In the case of mean estimation, we can divide (\ref{3.15}) by $N^2=N_I^2 M^2$ to get 
\begin{equation}
 \hat{V} \left( \hat{\bar{Y}}_{\rm HT} \right)   =\left(1-\dfrac{n_I}{N_I}\right)
 \dfrac{s_b^2}{n_I}+ \dfrac{n_I}{N_I} \left(1-\dfrac{m}{M}\right)\dfrac{s_w^2}{n_I m}
\label{3.15c}
\end{equation}
where 
\begin{eqnarray*}
s_b^2 &=& \left( n_I-1 \right)^{-1} \sum_{i \in A_I } \left( \bar{y}_i -\hat{\bar{Y}} \right)^2   \\
s_w^2 &=& n_I^{-1} \left( m-1 \right)^{-1} \sum_{i \in A_I } \sum_{j
\in A_i } \left( y_{ij} - \bar{y}_i \right)^2.
\end{eqnarray*}
If  $f_1=n_I/N_I $ is negligible, then we can use $$ \hat{V} \left( \hat{\bar{Y}}_{\rm HT} \right)   =
 \dfrac{s_b^2}{n_I}$$
as a variance estimator for the mean estimator under simple random sampling.

When the cluster sizes are unequal, simple random sampling in the first stage sampling is not preferable. 
The following example is a very popular method of two-stage sampling in the case of unequal cluster sizes. 
\begin{example}
\label{example7.3}
Consider the following two-stage sampling design. 
\begin{enumerate}
\item Stage One: Select clusters (of size $n_I$) by PPS sampling with size measure $M_i$. 
\item Stage Two: Select elements by SRS of size $m$ from $M_i$ elements in the sample cluster $i$. 
\end{enumerate}

We first consider estimation of population total $Y=\sum_{i =1}^{N_I} \sum_{j=1}^{M_i} y_{ij}$. Under  single-stage cluster sampling, we would have observed $Y_i=\sum_{j=1}^{M_i} y_{ij}$. In this case, an unbiased estimator of $Y$
is given by 
\begin{equation}
\hat{Y}_{\rm PPS} = \frac{N}{n_I} \sum_{k=1}^{n_I}  \frac{Y_{a_k}}{M_{a_k}}
\label{3.18}
\end{equation}
where $a_k$ is the index of population cluster in the $k$-th draw of the PPS sampling. In the two-stage sampling, we do not observe $Y_i$ but we obtain $\hat{Y}_i = M_i \bar{y}_i$, where $\bar{y}_i$ is the sample mean of elements in the cluster $i$. Thus, we can use 
\begin{equation}
\hat{Y}_{\rm PPS} = \frac{N}{n_I} \sum_{k=1}^{n_I}  \frac{\hat{Y}_{a_k}}{M_{a_k}}= \frac{N}{n_I} \sum_{k=1}^{n_I}  \bar{y}_{a_k} \label{3.19}
\end{equation}
to estimate the total $Y$. Assuming that there is no duplication of the selected clusters, the sampling weights are all equal to $N/(n_I m)$, which implies that every element in the population has the same probability of selection. 
 The sampling design that leads to equal sampling weights is called a self-weighting design. 

For estimation of the population mean $\bar{Y}=Y/N$, 
we have
\begin{equation}
\hat{\bar{Y}}_{\rm PPS} =  \frac{1}{n_I} \sum_{k=1}^{n_I}  \bar{y}_{a_k}   \label{3.20}
\end{equation}
which takes the sample mean of the cluster means. 

To discuss variance estimation, note that the point estimator (\ref{3.19})  can be written as the sample mean of $z_1, \cdots, z_{n_I}$ where $z_k$ are independently and identically distributed with the following discrete distribution: 
$$ z_1 = \hat{Y}_{i}/ p_{i}  \   \   \   \  \mbox{ with probability } p_i = M_i/N, \ i=1, \cdots, N_I. $$
Note that $E(z_1) = \sum_{i=1}^{N_I} \hat{Y}_i$, which is unbiased for $Y=\sum_{i=1}^{N_I} Y_i$ as $E_2( \hat{Y}_i) = Y_i$. For variance estimation, since $\hat{Y}_{\rm PPS}$ in (\ref{3.19}) can be written as the  mean of $n_I$ independent sample of $z_k$s, we have 
\begin{equation}
 \hat{V}_{\rm PPS}\left( \hat{{Y}}_{\rm PPS} \right)  = \frac{1}{n_I} S_z^2 = \frac{1}{n_I} \frac{1}{n_I-1}
\sum_{k=1}^{n_I} \left(z_k - \bar{z} \right)^2 .\label{3.22}
\end{equation}
Variance estimation of the mean estimator (\ref{3.20}) can be similarly constructed. Specifically, we can use  
\begin{equation}
 \hat{V}_{\rm PPS}\left( \hat{\bar{Y}}_{\rm PPS} \right)  =  \frac{1}{n_I} \frac{1}{n_I-1}
\sum_{k=1}^{n_I} \left(\bar{y}_{a_k} - \hat{\bar{Y}}_{\rm PPS}   \right)^2 \label{3.22b}
\end{equation}
as an unbiased estimator for the variance of the mean estimator (\ref{3.20}). 
\end{example}

To illustrate the use of two-stage sampling in Example \ref{example7.3},  consider a finite population of households in a city. The city consists of $N_I$ clusters of houses, and the cluster $i$ consists of $M_i$ houses. We used the following two-stage cluster sampling. 
\begin{description}
\item{[Stage 1]} Select $n_I=3$ sample clusters by the PPS sampling with the measure of size equal to $M_i$. 
\item{[Stage 2]} Within each selected cluster $i$, select $m_i=4$ sample houses by simple random sampling. 
\end{description}
Once the sample households are selected, we obtain two information.  One is the number of household members in the house ($t_{ij}$)  and the other is the number of household members under 6 years of age ($y_{ij}$). We are interested in estimating the proportion of the population with age under 6 in the city. That is, the parameter of interest is 
$$ P= \frac{ \sum_{i=1}^{N_I} \sum_{j=1}^{M_i} y_{ij} }{ \sum_{i=1}^{N_i} \sum_{i=1}^{M_i} t_{ij} }:= \frac{ Y}{ T}  .$$

The following table gives the summary of the realized sample household from the above two-stage sampling.

\begin{table}[htb]
\begin{center} 
\caption{An illustrative example of two-stage cluster sample}
\centering 
\begin{tabular}{c | c | c | c } 
\hline \
 Sample Cluster ID & Sample household ID & $t_{ij}$  & $y_{ij}$   \\
\hline 
1 &  1 & 8 & 2 \\
1 &  2 & 7 & 2 \\
1 &  3 & 7 & 1 \\
1 &  4 & 6 & 1 \\
\hline
2 & 1  & 8  & 0 \\
2 & 2  & 12 & 1 \\
2 & 3  & 10 & 3 \\
2 & 4  & 11 & 1 \\
\hline
3 & 1 & 4 & 2 \\
3 & 2 & 5 & 3 \\
3 & 3 & 5 & 2 \\
3 & 4 & 6 & 1 \\
\hline
\end{tabular}
\end{center} 
\end{table}

The proportion of the population with age under 6 in the city is estimated by 
$$ \hat{P} = \frac{ \hat{Y} }{ \hat{T} }
 = \frac{N n_I^{-1} \sum_{k=1}^{n_I} \bar{y}_{k} }{ 
N n_I^{-1} \sum_{k=1}^{n_I} \bar{t}_{k} } = \frac{6/4+5/4+8/4 }{ 28/4+ 41/4+20/4 } \doteq 0.213
$$
where the second equality follows from (\ref{3.19}). To estimate the variance of $\hat{P}$, we use 
$$\hat{V}(\hat{P})=\frac{1}{n_{I}}\frac{1}{n_{I}-1} \left(\frac{1}{n_{I}}\sum_{i=1}^{n_{I}}\bar{t}_{i} \right)^{-2}\sum_{i=1}^{n_{I}}(\bar{y}_{i}-\hat{\theta}\bar{t}_{i})^2=0.005302. $$
The design effect can be computed by the ratio of $\hat{V}(\hat{P})$ under the current sampling design to the 
variance of $\hat{P}$ under simple random sampling, which is computed by  
$$\hat{V}_{\rm SRS}(\hat{p})=\frac{1}{n}\hat{p}(1-\hat{p})=\left(\sum_{i=1}^{n_I}\sum_{j=1}^{m_i}t_{ij}\right)^{-1}\hat{p}(1-\hat{p})=0.001887.$$
Thus, the estimated design effect is $0.005302/0.001887=2.8105$.
\vspace{.5in}

%% file: chapters/chapter8.tex

\setcounter{chapter}{7} 
\chapter{Estimation: Part 1 }

\section{Introduction}




So far, we have considered HT estimators for various sampling designs. HT estimator is design-unbiased but does not necessarily achieve small variance. To improve the efficiency of the resulting estimator, auxiliary information is often incorporated into the estimation if the population total of the auxiliary variable is known from external sources. An estimator is called a linear estimator if it is a linear function of the sample observations of $y$. \index{linear estimator} That is, 
\begin{equation}
 \hat{Y}= \sum_{i  \in A} w_i y_i 
 \label{8-1}
 \end{equation} 
for some $w_i\ge 0 $ that does not depend on $y$. If $z_i = x_i + y_i$, then the linear estimator applied to $z$ satisfies $\hat{Z} = \hat{X} + \hat{Y}$. This property is called internal consistency. \index{internal consistency} HT estimator is the only linear estimator that is design unbiased. 

If auxiliary variable $\mathbf{x}_i$ is observed in the sample and the population total of $\mathbf{x}_i$ is known from an external source such as census or administrative data, we may want to achieve 
\begin{equation}
 \sum_{i \in A} w_i \mathbf{x}_i = \sum_{i=1}^N \mathbf{x}_i.
\label{8-2}
\end{equation}
Property (\ref{8-2}) is sometimes called external consistency. We will consider two examples of estimators satisfying (\ref{8-2}). One is the ratio estimator, and the other is the regression estimator.

\section{Ratio estimation}

Suppose that a scalar auxiliary variable $x_i$ is available whose population total $X=\sum_{i=1}^N x_i$ is known from an external source. In this case, the HT estimator of $X$, $\hat{X}_{\rm HT}=\sum_{i \in A} \pi_i^{-1} x_i$ is not necessarily equal to $X$. We can use $X$ information to modify the HT estimator to get 
\begin{equation}
 \hat{Y}_{\rm R} = X  \frac{\hat{Y}_{\rm HT} }{\hat{X}_{\rm HT} }.
 \label{8-3}
\end{equation}
Thus, ratio estimator is computed by multiplying $\hat{X}_{\rm HT}^{-1} X$ to HT estimator of $Y$ and $\hat{X}_{\rm HT}^{-1} X$ is sometimes called ratio adjustment factor. \index{ratio adjustment factor} The ratio adjustment factor is close to one,  but not necessarily equal to one. If the realized sample satisfies  $\hat{X}_{\rm HT} < X$, then the ratio adjustment factor is greater than one and $\hat{Y}_{\rm R} > \hat{Y}_{\rm HT}$. In this case, the HT estimator is inflated by the ratio adjustment factor.

Ratio estimator is a linear estimator of the form in (\ref{8-1}) and the weight is 
\begin{equation}
w_i = \frac{1}{\pi_i} \times \frac{X}{ \hat{X}_{\rm HT}} .
\label{8-4}
\end{equation}
Thus, the final weight is computed by a product of the design weight $\pi_i^{-1}$ and the ratio adjustment factor. 

We now consider the statistical properties of the ratio estimator. Note that the ratio estimator is a nonlinear function of $\hat{X}_{\rm HT}$ and $\hat{Y}_{\rm HT}$. Thus, 
$$E\left(\frac{\hat{Y}_{\rm HT} }{\hat{X}_{\rm HT}
}  \right) \neq  \frac{E \left( \hat{Y}_{\rm HT} \right) }{E \left(
\hat{X}_{\rm HT} \right) }
$$
and the ratio estimator is not design unbiased for $Y$. To discuss the bias of the ratio estimator, define 
$$\hat{R}= \frac{\hat{Y}_{\rm HT}}{\hat{X}_{\rm HT}}  \mbox{ and } R =  \frac{ E \left( \hat{Y}_{\rm HT} \right) }{ E \left( \hat{X}_{\rm HT} \right) }.$$
The bias of $\hat{R}$ as an estimator of $R$ is equal to  $E(\hat{R})-R$ and it is often called ratio bias. \index{ratio bias} 
To discuss ratio bias, note that 
\begin{eqnarray*}
Cov \left( \hat{R}, \hat{X}_{\rm HT} \right) &=& E \left( \hat{R}
\hat{X}_{\rm HT}
\right) -E \left( \hat{R}\right) E \left( \hat{X}_{\rm HT} \right) \\
&=&E \left( \hat{Y}_{\rm HT} \right) -E \left( \hat{R}\right) E \left(
\hat{X}_{\rm HT} \right).
\end{eqnarray*}
Dividing both sides of the above equation by $-E\left( \hat{X}_{\rm HT} \right) $, we get 
\begin{equation}
Bias \left( \hat{R} \right) = - \frac{Cov \left( \hat{R},
\hat{X}_{\rm HT} \right) }{ E\left( \hat{X}_{\rm HT} \right) } 
\label{8-5}
\end{equation}
and so 
\begin{equation*}
Bias \left( \hat{Y}_R \right) = - Cov \left( \hat{R}, \hat{X}_{\rm HT}
\right).
\end{equation*}
Even if the ratio estimator is biased, but if the bias is smaller order than its standard error, the bias can be negligible. Generally speaking, we can express 
\begin{equation}
\frac{\hat{\theta} - \theta }{ \sqrt{V( \hat{\theta} )
} }  =\frac{\hat{\theta} - E \left( \hat{\theta} \right) }{
\sqrt{V( \hat{\theta} ) } } +R.B.(\hat{\theta})
\label{8-6}
\end{equation}
where $$ R.B.(\hat{\theta})=\dfrac{Bias(\hat{\theta})}{\sqrt{V(\hat{\theta})}} $$
is the relative bias of $\hat{\theta}$. The first term on the right hand side of (\ref{8-6}) converges to the standard normal distribution by the central limit theorem. Thus, when the second term converges to zero, the bias term can be safely ignored. We formalize the concept in the following definition.

\begin{definition}
$ \mbox{Bias}( \hat{\theta}) \mbox{  is negligible} $ 
$$ \iff {\rm R.B.} (\hat{\theta})=\dfrac{Bias(\hat{\theta})}{\sqrt{V(\hat{\theta})}}
\rightarrow 0 \quad as \quad n \rightarrow \infty .$$
\end{definition}
Note that, since
$$ \frac{ {\rm MSE} \left( \hat{\theta} \right) }{V(\hat{\theta})}
= 1+ \left[ R.B.(\hat{\theta}) \right]^2, 
$$
the MSE of  $\hat{\theta}$ is approximately equal to $V( \hat{\theta})$ if the bias of $\hat{\theta}$ is negligible. 

Now, to discuss the relative bias of the ratio estimator, first note that the relative bias of the ratio estimator is equal to the relative bias of $\hat{R}$. Thus, by (\ref{8-5}), we have 
\begin{equation}
\left|  R.B.(\hat{Y}_{R}) \right|=\left| R.B.(\hat{R})
\right|=\left|Corr \left( \hat{R}, \hat{X}_{\rm HT} \right)
\frac{\sqrt{V(\hat{X}_{HT} )}}{E\left( \hat{X}_{HT}
\right)}\right|\le CV\left(\hat{X}_{HT} \right),  \label{4.11a}
\end{equation}
where $CV( \hat{\theta}) = \sqrt{V( \hat{\theta}) }/ E( \hat{\theta})$ is the coefficient of variation. 
For example, under simple random sampling, 
$$ CV\left( \hat{X}_{\rm HT} \right) = \sqrt{ \frac{1}{n} \left( 1-
\frac{n}{N} \right) } \frac{ S_x}{\bar{X}}. $$ 
Thus, the bias of the ratio estimator of negligible if one of the following conditions holds: 
\begin{enumerate}
\item $n$ is large
\item $n/N$ is large (close to one) 
\item $CV\left( x\right)= S_x/\bar{X} $ is small. 
\end{enumerate}
For general sampling designs other than SRS, $CV( \hat{X}_{\rm HT})$ will converge to zero as $n$ increases and the bias of the ratio estimator is negligible.

To further discuss relative bias of the ratio estimator, we use the second order Taylor expansion to get 
\begin{eqnarray}
\widehat{\bar{Y}}_R &\cong &\bar{Y} + \left( \widehat{\bar{Y}}_{\rm HT} - \bar{Y} \right) - R
\left( \widehat{\bar{X}}_{\rm HT}- \bar{X}  \right) \label{8-8} \\
&& - \bar{X}^{-1} \left[ \left( \widehat{\bar{X}}_{\rm HT}-\bar{X} \right) \left(
\widehat{\bar{Y}}_{\rm HT}-\bar{Y}  \right)-R\,  \left( \widehat{\bar{X}}_{\rm HT}-\bar{X}
\right)^2\right] , \notag
\end{eqnarray}
where  $\widehat{\bar{Y}}_R = N^{-1} \hat{Y}_R$, $\widehat{\bar{X}}_{\rm HT} =
N^{-1} \hat{X}_{\rm HT}$, and $\widehat{\bar{Y}}_{\rm HT} = N^{-1} \hat{Y}_{\rm HT}$.  Thus, 
\begin{equation}
Bias\left(\widehat{\bar{Y}}_R \right) \doteq  - \bar{X}^{-1} \left[  Cov \left(
\widehat{\bar{X}}_{\rm HT}, \widehat{\bar{Y}}_{\rm HT} \right)-R \cdot  V\left(
\widehat{\bar{X}}_{\rm HT}\right)\right] . 
\label{8-9}
\end{equation}
Note that the leading term of the bias of the ratio estimator is of order $n^{-1}$.

Next, we consider the variance of the ratio estimator. By (\ref{8-8}), we obtain 
\begin{eqnarray}
\widehat{\bar{Y}}_R &=& \bar{Y} + \left( \widehat{\bar{Y}}_{\rm HT} - R
 \widehat{\bar{X}}_{\rm HT}   \right)+ O_p \left( n^{-1} \right),
\label{4.14}
\end{eqnarray}
where $\hat{\theta}_n= O_p( 1) $ denotes that $\hat{\theta}_n$ is bounded in probability. Ignoring $O_p(n^{-1})$ terms, we have 
    \begin{equation}
        \var (\widehat{\bar{Y}}_R - \bar{Y} )  \cong \var \left(\widehat{\bar{Y}}_{\rm HT} -R \widehat{\bar{X}}_{\rm HT} \right).
        \label{rvar}
        \end{equation}
        Therefore, the ratio estimator is better than the HT estimator if
$$ -2 R \cdot  Cov \left( \widehat{\bar{Y}}_{\rm HT}, \widehat{\bar{X}}_{\rm HT} \right) + R^2
\var \left(  \widehat{\bar{X}}_{\rm HT} \right) < 0 ,$$ 
which reduces to, under the simple random sampling, 
\begin{equation}
 \frac{1}{2} {\frac{CV\left( x\right) }{CV\left( y\right) }} <
Corr\left( x, y \right).
\label{8-12}
\end{equation}
That is, under (\ref{8-12}), the ratio estimator is more efficient than the HT estimator under SRS. 

For variance estimation, using (\ref{rvar}), we may use 
$$ \sum_{i \in A} \sum_{j \in A} \frac{\pi_{ij} - \pi_i \pi_j
 }{\pi_{ij}} \frac{y_i - R x_i}{\pi_i}\frac{y_j - R x_j}{\pi_j}.
$$
However, since we do not know $R$, we use $\hat{R}$ to obtain 
 \begin{equation}
 \hat{V} \left( \hat{Y}_R \right)=
 \sum_{i \in A} \sum_{j \in A} \frac{\pi_{ij} - \pi_i \pi_j
 }{\pi_{ij}} \frac{y_i - \hat{R} x_i}{\pi_i}\frac{y_j - \hat{R} x_j}{\pi_j}
\label{rvar2}
\end{equation}
as a variance estimator.

We now introduce Haj{\'e}k estimator of the population mean as a special case of the ratio estimator. Haj{\'e}k estimator uses $x_i=1$ in the ratio estimator. 
\index{Haj{\'e}k estimator} 
That is, 
\begin{equation}
\widehat{\bar{Y}}_{\pi} = \frac{\sum_{i \in A} \pi_i^{-1} y_i}{\sum_{i \in A}
\pi_i^{-1}} \label{4.16}
\end{equation}
is the  Haj{\'e}k estimator of the population mean and is often more efficient than the HT estimator 
 $\widehat{\bar{Y}}_{\rm HT} = N^{-1} \sum_{i \in A} \pi_i^{-1} y_i$. 
 Using (\ref{4.14}), we can obtain 
 $$ \widehat{\bar{Y}}_{\pi} = \bar{Y} + N^{-1} \sum_{i \in A} \pi_i^{-1} \left( y_i -  \bar{Y} \right) + O_p
\left( n^{-1} \right).$$
Thus, 
the variance of the Haj{\'e}k estimator is obtained using $y_i-\bar{Y}$ instead of $y_i$ in the variance formula for the HT estimator.

In some cases, ratio itself is a parameter of interest. Domain estimation is an example of using a ratio for parameter estimation. Suppose that we are interested in the mean of $y$ in a particular domain $D$. That is, the parameter of interest is 
$$\bar{Y}_D = \frac{\sum_{i =1}^N z_i y_i}{\sum_{i=1}^N
z_i}$$
where 
$$ z_i = \left\{ \begin{array}{ll}
1 & \mbox{ if } i \in D \\
0 & \mbox{ if } i \notin D . \\
\end{array} \right. $$
To estimate $\bar{Y}_D$, we use 
$$\widehat{\bar{Y}}_{D,\pi} =\frac{\sum_{i \in A} \pi_i^{-1} z_i
y_i}{\sum_{i \in A} \pi_i^{-1} z_i}.$$
By the Taylor linearization, we have  
$$ \bar{Y}_{D, \pi} = \bar{Y}_D + N_D^{-1} \sum_{i \in A} \pi_i^{-1} z_i
\left( y_i -  \bar{Y}_D \right) + O_p \left( n^{-1} \right)$$ 
where $N_D = \sum_{i=1}^N z_i$.  Under SRS, 
\begin{eqnarray*}
\var(\widehat{\bar{Y}}_{D, \pi})
&\doteq & \dfrac{1}{n} (1-f) \left(\dfrac{N}{N_d}\right)^2 \dfrac{1}{N-1}
 \sumiN z_i (y_i - \bar{Y}_D )^2 \\
& =&\dfrac{1}{n} (1-f) \left(\dfrac{N}{N_d}\right)^2 \dfrac{N_d -1}{N-1} S_D^2, 
\end{eqnarray*}
where $ S_D^2= \left( N_D -1 \right)^{-1} \sum_{i \in D} (y_i
- \bar{Y}_D )^2 $. If the sample size is large such that
$$ \frac{n_D}{n} \cong \dfrac{N_D}{N}
\cong \dfrac{N_D-1}{N-1} $$
holds, we have 
$$\var(\widehat{\bar{Y}}_{D, \pi})
\doteq
 \dfrac{1}{n_D} (1-f) S_D^2 $$
 where 
  $S_D^2$ is estimated by $s_D^2 = \left( n_D-1
\right)^{-1} \sum_{i \in A} z_i \left( y_i -\bar{y}_D
\right)^2$.

\section{Regression estimation}

Ratio estimator is useful when there is strong positive correlation between $x$ and $y$. If the correlation is negative, the ratio estimator is actually worse than using the HT estimator. Also, ratio estimator is not directly applicable for the vector $x$ cases. To overcome such limitation, we consider regression estimation in this section.  
 
Suppose that the auxiliary information for unit $i$ is denoted by a $p$-dimensional column vector  $\mathbf{x}_i$  and its population total $\mathbf{X} = \sum_{i=1}^N \bx_i$ is available. In the sample, we observe $\bx_i$ and $y_i$. In this case, regression estimator of $Y=\sum_{i=1}^N y_i$ is defined as follows: 
  \begin{equation}
\hat{Y}_{\rm reg} = \sum_{i=1}^N \hat{y}_i \label{eq:reg}
 \end{equation}
where $\hat{y}_i = \bx_i^{\top} \hat{\mathbf{B}} $ and 
 $$ \hat{\mathbf{B}} = \left( \sum_{i \in A} \pi_i^{-1} \mathbf{x}_i
 \mathbf{x}_i^{\top} \right)^{-1}\sum_{i \in A} \pi_i^{-1} \mathbf{x}_i y_i. $$
If $\sum_{i \in A} \pi_i^{-1} \mathbf{x}_i
 \mathbf{x}_i^{\top}  $ is singular, then its inverse does not exist. In such case, one may use the generalized inverse and still define regression estimator accordingly.  For estimation of the population mean $\bar{Y} = N^{-1} \sum_{i=1}^N y_i$, the regression estimator of $\bar{Y}$ is defined by 
 \begin{equation}
\widehat{\bar{Y}}_{\rm reg} =  N^{-1} \sum_{i=1}^N \bx_i^{\top} \hat{\mathbf{B}} = \bar{\mathbf{X}}^{\top} \hat{\mathbf{B}} , 
\label{eq:reg2}
 \end{equation}
where  $\bar{\mathbf{X}}=N^{-1}\sum_{i=1}^N \mathbf{x}_i$. 

In many cases, $\bx_i$ includes an intercept term. If we write 
$\mathbf{x}_i^\top
 = \left( 1, \mathbf{x}_{i1}^\top \right)$,  the regression estimator in 
   (\ref{eq:reg2}) reduces to 
\begin{equation}
\widehat{\bar{Y}}_{\rm reg} = \hat{B}_0 + \bar{\mathbf{X}}_1^\top \hat{\mathbf{B}}_1 
 \label{4.19}
 \end{equation}
 where  $\bar{\mathbf{X}}_1 = N^{-1} \sum_{i=1}^N \mathbf{x}_{i1} $, 
  $$ 
\hat{B}_0 = \left( \sum_{i \in A} \pi_i^{-1}  \right)^{-1} \sum_{i \in A} \pi_i^{-1} \left( y_i - \bx_{i1}^\top \hat{\mathbf{B}}_1 \right) ,
$$  
 $$ \hat{\mathbf{B}}_1 = \left[ \sum_{i \in A} \pi_i^{-1} \left(
 \mathbf{x}_{i1}
 -\widehat{\bar{\mathbf{X}}}_{\pi 1} \right)\left(
 \mathbf{x}_{i1}
 -\widehat{\bar{\mathbf{X}}}_{\pi 1} \right)^\top \right]^{-1}\sum_{i \in A} \pi_i^{-1} \left(
 \mathbf{x}_{i1}
 -\widehat{\bar{\mathbf{X}}}_{\pi 1} \right) y_i, $$
and $$\widehat{\bar{\mathbf{X}}}_{\pi 1} = \dfrac{\sum_{i \in A} \pi_i^{-1} \bx_{i1} }{ \sum_{i \in A} \pi_i^{-1} } . 
$$
Thus, the regression estimator in (\ref{4.19}) can be written as 
\begin{equation}
\widehat{\bar{Y}}_{\rm reg} = \frac{1}{N} \sum_{i =1}^N \bx_{i1}^\top \hat{\mathbf{B}}_1 + \frac{1}{ \hat{N}} \sum_{i \in A} \frac{y_i - \bx_{i1}^{\top} \hat{\mathbf{B}}_1 }{\pi_i} , 
\label{eq:8-18}
\end{equation}
where $\hat{N}= \sum_{i \in A} \pi_i^{-1}$. Note that the second term of (\ref{eq:8-18}) is equal to $\hat{B}_0$, which plays the role of the bias-correction in the prediction using $\hat{y}_i = \bx_{i1}^{\top} \hat{\mathbf{B}}_1$. In other words, including an intercept term in the regression model makes the resulting regression estimator design consistent. This is closely related to the concept of internally bias calibration condition that will be covered in Chapter 9.

Before we discuss statistical properties of the regression estimator, we briefly discuss some algebraic properties of the regression estimator. First, regression estimator is a linear estimator. That is, we can express the regression estimator in (\ref{eq:reg}) as the linear form in (\ref{8-1}) with 
\begin{equation}
 w_i =\bar{\mathbf{X}}^{\top} \left( \sum_{i \in A} \pi_i^{-1} \mathbf{x}_i
 \mathbf{x}_i^\top \right)^{-1} \pi_i^{-1} \mathbf{x}_i.
\label{4.20}
\end{equation}
 Note that the weights satisfy 
 \begin{equation}
 \sum_{i \in A} w_i \mathbf{x}_i =\bar{\mathbf{X}}.
\label{4.21} \end{equation}
Thus, regression estimator satisfies the external consistency discussed in (\ref{8-2}). Property (\ref{4.21}) is also called \emph{calibration property}  or \emph{benchmarking property} in the  survey literature \citep{deville1992}. \index{calibration} 
\index{benchmarking} 
If an intercept term is included in $\bx_i$, then $\sum_{i \in A} w_i =1 $ holds and the regression estimator is location-scale invariant. That is, 
$$ \sum_{i \in A} w_i \left( a + b y_i \right) = a + b \sum_{i \in
A} w_i y_i.$$


We now discuss statistical properties of the regression estimator. First assume that the sampling design and the HT estimators are such that 
\begin{equation}
N^{-1} \left( \sum_{i \in A}\pi_i^{-1}  \mathbf{z}_i \mathbf{z}_i^\top -
\sum_{i=1}^N  \mathbf{z}_i \mathbf{z}_i^\top\right) = O_p \left(
n^{-1/2} \right) \label{4.23}
\end{equation}
where $\mathbf{z}_i^\top = \left( \mathbf{x}_{i}^\top, y_i \right)$. Condition (\ref{4.23}) means that the HT estimator of the population means converge in probability to their population mean and the order of convergence of the same as that of the sample mean in the IID case.

The following lemma dervies Taylor linearization of the regression estimator. 
\begin{lemma}
Under  (\ref{4.23}), the regression estimator in (\ref{eq:reg2}) satisfes 
\begin{equation}
\widehat{\bar{Y}}_{\rm reg} = \bar{\mathbf{X}}^\top B + \bar{\mathbf{X}}^\top \left( \sum_{i=1}^N
\mathbf{x}_i \mathbf{x}_i^\top \right)^{-1} \sum_{i \in A} \pi_i^{-1}
\mathbf{x}_i \left( y_i - \mathbf{x}_i^\top \mathbf{B}  \right) + O_p \left(n^{-1}
\right), 
 \label{4.24}
\end{equation}
where 
$\mathbf{B}= \left(\sum_{i=1}^N  \mathbf{x}_{i} \mathbf{x}_i^\top 
\right)^{-1}\sum_{i=1}^N  \mathbf{x}_i y_i $.
\label{lem:8-1}
\end{lemma}
\begin{proof}
Let 
\begin{eqnarray*}
\left( \hat{\mathbf{C}},  \hat{\mathbf{d}}\right) &=& \sum_{k \in A} \pi_k^{-1} \left( \mathbf{x}_k \mathbf{x}_k^\top ,  \mathbf{x}_k y_k \right)
 \end{eqnarray*}
and 
\begin{eqnarray*}
\left( \mathbf{C}, {\mathbf{d}}\right)  &=& \sum_{k =1}^N \left( \mathbf{x}_k \mathbf{x}_k^\top, \mathbf{x}_k y_k \right).
 \end{eqnarray*}
 The estimated regression coefficient 
 $\hat{\mathbf{B}}$ can be written as 
$$ \hat{\mathbf{B}}= f\left( \hat{\mathbf{C}}, \hat{\mathbf{d}}
\right) = \hat{\mathbf{C}}^{-1} \hat{\mathbf{d}} . $$
To compute Taylor linearization of vector $\hat{\mathbf{B}}$, let $\hat{c}_{ij}$ be the $(i,j)$-th element of $\hat{\mathbf{C}}$ and let $\hat{d}_i$ be the $i$-th element of 
 $\hat{d}_i$.  Taylor linearization of $\hat{\mathbf{B}}$ can be expressed as 
\begin{eqnarray}
f\left( \hat{\mathbf{C}}, \hat{\mathbf{d} }\right) &\cong & f \left(
\mathbf{C}, \mathbf{d} \right) + \sum_i \left( \frac{\partial f
}{\partial {d}_i } \right) \left( \hat{d}_i - d_i \right) + \sum_i
\sum_j \left( \frac{ \partial f }{ \partial c_{ij} } \right) \left(
\hat{c}_{ij} - c_{ij} \right). \notag \\
\label{proof1}
\end{eqnarray}
Here, we have 
$$ \frac{\partial f }{ \partial d_i } = \mathbf{C}^{-1} \mathbf{e}_i,$$
where   $\mathbf{e}_i$ is a vector where the $i$-th element is 1 and the remaining elements are 0. Also, 
$$\frac{\partial f }{ \partial c_{ij} } = -\mathbf{C}^{-1} E_{ij} \mathbf{C}^{-1}
\mathbf{d} $$
where  $E_{ij}=\mathbf{e}_i \mathbf{e}_j^\top$ is a matrix where the $(i,j)$th element is 1 and the remaining elements are zero. 
Thus,  (\ref{proof1}) reduces to 
\begin{eqnarray*}
f \left( \hat{\mathbf{C}}, \hat{\mathbf{d} }\right) &\cong & f \left(
\mathbf{C}, \mathbf{d} \right) + \mathbf{C}^{-1} \left(
\hat{\mathbf{d}}- \mathbf{d} \right) - \mathbf{C}^{-1}
\left( \hat{\mathbf{C}}- \mathbf{C} \right) \mathbf{C}^{-1}\mathbf{d}
 \\
&=& \mathbf{B} + \mathbf{C}^{-1} \left(\hat{\mathbf{d}}-
\hat{\mathbf{C}} \mathbf{B} \right),
\end{eqnarray*}
which proves (\ref{4.24}).
\hfill 
\end{proof}

If $\bx_i$ includes 1, we can write  $\mathbf{a}^\top  \mathbf{x}_i =
1$ for some $\mathbf{a}$ which implies $X \mathbf{a} = \mathbf{1}_N$ (where $\mathbf{1}_N$ is an $N$-dimensional vector of ones). Thus, 
\begin{eqnarray*}
\bar{\mathbf{X}}^\top \left( \sum_{i=1}^N \mathbf{x}_i \mathbf{x}_i^\top
\right)^{-1} &=& N^{-1} \left( \mathbf{1}_N^\top  X \right) \left( X^\top X
\right)^{-1} \\
&=& N^{-1} \mathbf{a}^\top  X^\top X  \left( X^\top X \right)^{-1} =
N^{-1} \mathbf{a}^\top,
\end{eqnarray*} 
and 
$$\bar{\mathbf{X}}^\top \mathbf{B} = \bar{Y}. $$
Thus, 
(\ref{4.24}) reduces to 
$$  \widehat{\bar{Y}}_{\rm reg} - 
\bar{Y}= 
 N^{-1}   \sum_{i \in A} \pi_i^{-1}   \left(y_i - \bx_i^\top \mathbf{B}\right)+ o_p(n^{-1/2}).$$
The leading term is unbiased to zero. 
The asymptotic variance of the regression estimator is then 
\begin{equation}
V\left( \widehat{\bar{Y}}_{\rm reg} \right) \doteq  V\left\{ N^{-1} \sum_{i \in A}
\pi_i^{-1} e_i \right\} =N^{-2}  \sum_{i \in U} \sum_{j \in U} \left( \pi_{ij} - \pi_i \pi_j \right) \frac{e_i}{\pi_i} \frac{e_j}{\pi_j} ,
 \label{4.26}
\end{equation}
where $e_i= y_i - \bx_i^\top  \mathbf{B}$. 

Under SRS, the asymptotic variance can be written 
\begin{eqnarray}
V\left(\widehat{\bar{Y}}_{\rm reg} \right) &\doteq & \frac{1}{n} \left( 1 -
\frac{n}{N} \right) \frac{1}{N-1} \sum_{i=1}^N \left( y_i -
\mathbf{x}_i^\top  \mathbf{B} \right)^2,
\end{eqnarray}
which leads to 
$$
V\left( \widehat{\bar{Y}}_{\rm reg} \right)
 \doteq V \left( \bar{y} \right) \times
\left( 1 -R^2 \right)
$$
where $R^2$ is the coefficient of determination for the regression of $y$ on $\bx$ in the population level. Thus, the regression estimator is efficient if $R^2$ is large, i.e., when there is a strong linear relationship between $y$ and $\mathbf{x}$. If the regression relation does not hold, the regression estimator is still asymptotically unbiased. Thus,  the regression estimator is model-assisted, not model-dependent \citep{sarndal1992model}.

To estimate the asymptotic variance in (\ref{4.26}), we use 
 \begin{equation*}
 \hat{V} \left( \hat{Y}_{\rm reg} \right)=
 \sum_{i \in A} \sum_{j \in A} \frac{\pi_{ij} - \pi_i \pi_j
 }{\pi_{ij}} \frac{y_i - \bx_i^\top \hat{\mathbf{B}}}{\pi_i}\frac{y_j - \bx_i^\top \hat{\mathbf{B}} }{\pi_j}
, \end{equation*}
where $\hat{\mathbf{B}}$ is defined in (\ref{eq:reg}). 

Next, we consider post-stratification which is a special case of regression estimation. Suppose that the finite population is partitioned into $G$ mutually exclusive and exhaustive groups.
Assume that $N_g$, the population size of group $g$, are known for all $g=1,\cdots, G$. In this case, the auxiliary information can be written as  $\mathbf{x}_i = \left( x_{i1}, x_{i2}, \cdots,
x_{iG}\right)'$ where $x_{ig}$ is the indicator function for group $g$. 
Since 
\begin{eqnarray*}
\sum_{i=1}^N \mathbf{x}_i' &=&  \left( N_1, N_2, \cdots,
N_G \right)'\\
 \sum_{i \in A}\pi_i^{-1}
\mathbf{x}_i \mathbf{x}_i' &=& diag \left(\hat{N}_1, \hat{N}_2,
\cdots, \hat{N}_G \right)
\end{eqnarray*}
where $\hat{N}_g = \sum_{i \in A} \pi_i^{-1} x_{ig} $, 
the regression estimator in (\ref{eq:reg}) reduces to 
\begin{equation}
\hat{Y}_{\rm post}= \sum_{g=1}^G \sum_{i \in A_g} \pi_i^{-1}
\frac{N_g}{\hat{N}_g}
 y_i
 \label{post}
\end{equation}
where $A_g$ is the set of sample indices in group $g$. Thus, the final weights in post-stratified estimator are obtained by
 multiplying the design weight by the adjustment factor $\hat{N}_g^{-1} N_g$. Since $\mathbf{x}_i$ includes one, we can use 
(\ref{4.26}) to obtain 
\begin{equation}
V\left( \hat{Y}_{\rm post} \right) = V\left\{ \sum_{g=1}^G  \sum_{i
\in A_g} \pi_i^{-1}  \left( y_i -\bar{Y}_g \right) \right\}  + o
\left(n^{-1} \right).
 \label{4.30}
\end{equation}
Under SRS, the variance is 
\begin{eqnarray}
V\left(\hat{Y}_{\rm post} \right) &\doteq & \frac{N^2}{n} \left( 1 -
\frac{n}{N} \right) \frac{1}{N-1} \sum_{g=1}^G \sum_{i \in U_g}
\left( y_i - \bar{Y}_g  \right)^2
\end{eqnarray}
which is asymptotically equal to the variance of stratified sampling under proportional allocation.

\begin{example}
(Raking ratio estimation) 

Suppose that we have $I \times J$ groups or cells. Cell counts $N_{ij}$ are not known. Marginal counts $N_{i\cdot} = \sum_{j=1}^J N_{ij} $ and $N_{ \cdot j} = \sum_{i=1}^I N_{ij} $ are known. In this case, we may consider the following two-way additive model 
  \begin{eqnarray*}
E_\zeta \left( Y_k \right) &=& \alpha_i + \beta_j \\
V_\zeta \left( Y_k \right) &=&  \sigma^2 
\end{eqnarray*}
where $\alpha_i$, $\beta_j$, and $\sigma^2$ are unknown parameters.   
Define 
$$z_{ijk}  = \left\{ \begin{array}{ll}
    1 & \mbox{ if } k \in U_{ij} \\
    0 & \mbox{ otherwise .} 
    \end{array}
    \right. $$
    Unfortunately, we do not observe $\delta_{ijk}$ in the population. Let 
    $$ \mathbf{x}_k = \left( z_{1 \cdot,k}, z_{2 \cdot k}, \cdots, z_{I \cdot k} , z_{ \cdot 1 k}, z_{\cdot 2 k }, \cdots, z_{\cdot J k} \right) $$
and we know    $ \sum_{k=1}^N \mathbf{x}_k$.

The  regression  estimator in this case can be written as 
    $$ \hat{Y}_{\rm reg} = \sum_{i \in A} \frac{1}{\pi_i} g_i \left( A \right) y_i $$
    where 
    $$g_i \left( A \right) =  \left( \sum_{k=1}^N \mathbf{x}_k  \right)^\top \left( \sum_{k \in A}  \frac{1}{\pi_k } \mathbf{x}_k \mathbf{x}_k^\top  \right)^{-1}   \mathbf{x}_i .$$
    Unfortunately, we cannot compute the inverse of $\sum_{k \in A} \pi_k^{-1} \mathbf{x}_k \mathbf{x}_k^\top$ because its rank is $I+J-1$, which is not full rank. Thus, there is no unique solution $\hat{B}$ to 
    $$ \sum_{i \in A} \pi_i^{-1} \bx_i \bx_i^\top  \hat{\mathbf{B}} = \sum_{i \in A} \pi_i^{-1} \bx_i y_i .$$
    The goal is to  find $g_{kA}=g_{k} \left( A \right)$ such that 
\begin{eqnarray}
\sum_{k \in A} \frac{g_{kA} }{\pi_k} 
z_{i \cdot,k}&=& \sum_{k=1}^N z_{i \cdot k} , \ \ i=1,2,\cdots,I \label{rake-1}\\
\sum_{k \in A} \frac{g_{k  A }}{\pi_k} z_{ \cdot j k} &=& \sum_{k=1}^N z_{ \cdot j k}, \ \ j=1,2,\cdots,J. \label{rake-2}
\end{eqnarray}
One way to obtain the solution to (\ref{rake-1}) and (\ref{rake-2}) is to solve the equations iteratively  as follows: \begin{enumerate}
\item Start with $g_{kA}^{(0)}  = 1$. 
\item For $z_{i \cdot k}=1$, 
$$ g_{kA}^{(t+1)} = g_{kA}^{(t)}\frac{ \sum_{k=1}^N z_{i \cdot k}}{ \sum_{k \in A}g_{kA}^{(t)}  z_{i \cdot k}/\pi_k }.$$
It satisfies (\ref{rake-1}), but not necessarily satisfy (\ref{rake-2}).  
\item For $\delta_{ \cdot j k}=1$,
$$ g_{kA}^{(t+2)} = g_{kA}^{(t+1)}\frac{ \sum_{k=1}^N z_{\cdot j k}}{ \sum_{k \in A}g_{kA}^{(t+1)} z_{\cdot j k}/\pi_k }.$$
It satisfies (\ref{rake-2}), but not necessarily satisfy (\ref{rake-1}).  
\item Set $t \leftarrow t+2 $ and go to Step 2. Continue until convergence. 
\end{enumerate}
This computation method is called raking ratio estimation and was first considered by \cite{deming1940} in the Census application. See also \cite{deville1993}. 
\end{example}

%% file: chapters/chapter9.tex

\setcounter{chapter}{8} 
\chapter{Estimation: Part 2 }

\section{GREG estimation}




In Chapter 8, we have seen that the regression estimator is an efficient estimator when there is a linear relationship between $y$ and $\mathbf{x}$. In this chapter, we generalize the concept of regression estimation to a more general class of models for developing model-assisted estimation. 

To motivate the generalized regression  estimator, we first introduce difference estimator. Suppose that a proxy value of $y_i$, denoted by $y_i^{(0)}$, throughout the population. The difference  estimator of $Y=\sum_{i=1}^N y_i$ using $y_1^{(0)}, \ldots, y_N^{(0)}$ is defined as \index{difference estimator} 
\begin{equation}
\hat{Y}_{\rm diff } = \sum_{i=1}^N y_i^{(0)} + \sum_{i \in A} \frac{1}{\pi_i} \left( y_i - y_i^{(0)} \right) .
\label{diff} 
\end{equation} 
The difference estimator is unbiased regardless of the choice of $y_i^{(0)}$. The variance of the difference estimator is 
$$ V\left( \hat{Y}_{\rm diff} \right) = V\left( \sum_{i \in A} \frac{y_i - y_i^{(0)} }{\pi_i}  \right). $$
Thus, the difference estimator is unbiased regardless of $y_i^{(0)}$ but its variances are different for different choice of $y_i^{(0)}$. If $y_i^{(0)}$ is close to the true value of $y_i$, then the variance of the difference estimator  will be smaller than that of the HT estimator. 

In practice, we do not know $y_i^{(0)}$. Instead, we use auxiliary variable $\mathbf{x}_i$ to construct a model for $y_i$ and develop $y_i^{(0)}$ from the model. That is, we assume that the finite population is a random sample of a superpopulation model that has generated the current finite population. One of the commonly used superpopulation models is \begin{eqnarray}
E_\zeta \left( y_i \mid \bx_i \right) &=& \mathbf{x}_i^\top  \bm \beta \label{4.33} \\
Cov_\zeta \left( y_i, y_j \mid \bx  \right) &=& \left\{ \begin{array}{ll}
v_i \sigma^2 & \mbox{ if } i = j \\
0 & \mbox{ if } i \neq j
\end{array}\right.
\label{4.34}
\end{eqnarray}
where  $v_i=v\left(
\bx_i \right)$ is a known function of $\bx_i$,  and $\bm \beta$ and $ \sigma^2$ are unknown parameters. Model (\ref{4.33})-(\ref{4.34}) is often called the generalized regression (GREG) model.

Under the GREG model, the (model-based) \emph{projection  estimator} is defined to be the sum of the predicted values in the GREG model.
\index{projection estimator}
That is, we define 
\begin{equation}
\widehat{Y}_{\rm proj} = \sum_{i=1}^N \hat{y}_i , 
\label{greg}
\end{equation}
 where  $\hat{y}_i= \mathbf{x}_i^\top  \hat{\bm \beta}_{\rm c}$ and 
\begin{equation}
\hat{\bm \beta}_{\rm c} = \left( \sum_{i \in A} \frac{1}{c_i } \bx_i
\bx_i^\top \right)^{-1} \sum_{i \in A} \frac{1}{c_i} \bx_i y_i, 
\label{4.35}
\end{equation}
where $c_i$ is a function of $v_i$ and $\pi_i$. The choice of $c_i$ is somewhat arbitrary.  The choice of $c_i=v_i$ leads to the generalized least square (GLS) estimator of $\bbeta$, assuming that the regression model also holds in the sample.  Since 
the model-based projection estimator in (\ref{greg}) is developed under the GREG  model in (\ref{4.34}), it is also called the generalized regression estimator (GREG). \index{generalized regression estimator} Note that the probability limit of $\hat{\bm \beta}_c$ in (\ref{4.35}) is 
\begin{equation}
 \mathbf{B}_c = \left( \sum_{i=1}^N  \frac{\pi_i}{c_i } \bx_i
\bx_i^\top \right)^{-1} \sum_{i=1}^N  \frac{\pi_i}{c_i} \bx_i y_i. 
\label{eq:9-5b}
\end{equation}
Thus, the choice of $c_i = \pi_i v_i$ leads to the population-level GLS estimator of $\bm \beta$. 


To ensure unbiasedness in the face of model misspecification, it is  crucial that the resulting estimator is also  asymptotically design unbiased (ADU). A key condition for achieving ADU is:
\begin{equation}
\sum_{i \in A} \frac{1}{\pi_i} \left( y_i - \hat{y}_i \right)=0. 
\label{ibc}
\end{equation}
To understand why condition (\ref{ibc}) leads to ADU, 
note that (\ref{ibc}) implies that 
\begin{eqnarray} 
\hat{Y}_{\rm proj} &=&  \sum_{i=1}^N \bx_i^\top  \hat{\bm \beta}_c + \sum_{i \in
A} \frac{1}{\pi_i} \left( y_i -\bx_i^\top  \hat{\bm \beta}_c \right)  \label{greg2} \\
&=& \hat{Y}_{\rm HT} + \left( \mathbf{X} - \hat{\mathbf{X}}_{\rm HT}
\right)^\top  \hat{\bm \beta}_c. \notag 
\end{eqnarray} 
The second term of  (\ref{greg2}) is a bias-correction term of the mode-based projection estimator in (\ref{greg}). Note that (\ref{greg}) is equivalent to (\ref{greg2}) under condition (\ref{ibc}). Condition (\ref{ibc}) essentially makes the bias correction term identically equal to zero, which implies that the projection estimator in (\ref{greg}) is design consistent. In this sense, the condition (\ref{ibc}) can be  called the \emph{internal bias calibration}  (IBC) condition, which is termed by \cite{firth1998}. \index{Internal Bias Calibration} \index{IBC}

The following lemma presents a sufficient condition for the IBC condition in (\ref{ibc}).  
\begin{lemma}
If $c_i$ satisfies 
\begin{equation} 
\frac{c_i}{ \pi_i} = \bm \lambda^\top  \mathbf{x}_i  
\label{ibc1}
\end{equation} 
for some $\bm \lambda$,  then we have 
\begin{equation}
\sum_{i=1}^N \left(y_i - \bx_i^\top  \mathbf{B}_c \right) = 0 
\label{eq:9-6}
\end{equation}
and 
\begin{equation} 
\sum_{i \in A} \frac{1}{ \pi_i} \left( y_i - \bx_i^\top  \hat{\bm \beta}_c \right) = 0 .
\label{eq:9-7} 
\end{equation} 
\label{lemma9.1}
\end{lemma}
\begin{proof}
First, by the definition of $\mathbf{B}_c$, we obtain 
$$
\sum_{i=1}^N \left( y_i - \bx_i^\top  \mathbf{B}_c \right) \bx_i \frac{\pi_i}{c_i} = \mathbf{0}.  
$$
Thus, pre-multiplying both sides of the above equation by $\bm \lambda^\top $, we get 
\begin{eqnarray*}
0 &=& \sum_{i=1}^N \left( y_i - \bx_i^\top \mathbf{B}_c \right) \bm \lambda^\top  \bx_i \frac{\pi_i}{c_i}  = \sum_{i=1}^N \left( y_i - \bx_i^\top  \mathbf{B}_c \right)\end{eqnarray*}
where the second equality follows from  (\ref{ibc1}). Thus, (\ref{eq:9-6}) is proved.

Now, to show (\ref{eq:9-7}),  
by the definition of $\hat{}$, we have 
$$ \sum_{i \in A}  \left( y_i - \bx_i^\top  \hat{\bm \beta}_c \right) \bx_i^\top / c_i =  \mathbf{0} , 
$$
which implies 
$$ \sum_{i \in A} \frac{1}{\pi_i}  \left( y_i - \bx_i^\top  \hat{\bm \beta}_c  \right) \bx_i^\top \bm \lambda/ c_i =  0 . 
$$
Thus, by  (\ref{ibc1}), we have (\ref{eq:9-7}). 
\end{proof}

If (\ref{ibc1}) holds, then the projection estimator  satisfies ADU. If condition (\ref{ibc1}) 
does not hold, the projection estimator does  not necessarily satisfy ADU.  To construct a design-consistent projection estimator, we add a bias-correction term to get
\begin{equation} 
\widehat{Y}_{\rm GREG} = \sum_{i=1}^N \bx_i^\top \hat{\bm \beta}_c +  \sum_{i \in A} \frac{ y_i - \bx_i^\top \hat{\bm \beta}_c}{\pi_i }
 .
\label{eq:greg} 
\end{equation}
The above estimator is the sum of the two terms: projection estimator and its bias-correction term. 
Thus, we can call (\ref{eq:greg}) a bias-corrected GREG  estimator or \emph{debiased GREG  estimator}, to emphasize its ADU property.   \index{debiased generalized regression estimator}
 By Lemma 9.1, the second term, the bias-correction term of the projection estimator,  is algebraically equal to zero if (\ref{ibc1}) is satisfied.

Since we can express (\ref{eq:greg}) as 
\begin{equation}
\widehat{Y}_{\rm GREG} = 
\widehat{Y}_{\rm HT} + \left( \mathbf{X} - \widehat{\mathbf{X}}_{\rm HT}
\right)^\top  \hat{\bm \beta}_c , 
\label{eq:greg2}
\end{equation}
 we can obtain 
\begin{eqnarray}
\widehat{Y}_{\rm GREG} &=& \widehat{Y}_{\rm HT} + \left( \mathbf{X} - \widehat{\mathbf{X}}_{\rm HT}
\right)^\top  \mathbf{B}_c +  \left( \mathbf{X} - \widehat{\mathbf{X}}_{\rm HT}
\right)^\top  \left( \hat{\bm \beta}_c - \mathbf{B}_c \right)\notag \\
&=& \widehat{Y}_{\rm HT} + \left( \mathbf{X} - \widehat{\mathbf{X}}_{\rm HT}
\right)^\top  \mathbf{B}_c + O_p(n^{-1}N ). 
\label{9-5}
\end{eqnarray} 
Thus,  regardless of whether (\ref{ibc1}) holds or not,  we obtain 
$$
\widehat{Y}_{\rm GREG}  = \sum_{i=1}^N \bx_i^\top \mathbf{B}_c + 
\sum_{i \in A} \pi_i^{-1} \left( y_i - \bx_i^\top  \mathbf{B}_c \right) + O_p(n^{-1}N) ,
$$
which justifies the asymptotic unbiasedness of $\widehat{Y}_{\rm GREG}$ in (\ref{eq:greg2}). Since $\widehat{Y}_{\rm GREG}$ is motivated from the GREG model and satisfies the design consistency, it is also called the GREG estimator.

The  variance of the GREG estimator  is approximated by 
\begin{equation}
V\left( \widehat{Y}_{\rm GREG} \right) \cong  V\left\{  \sum_{i \in A}
\pi_i^{-1} \left( y_i - \mathbf{x}_i^\top \mathbf{B}_c \right) \right\} . \label{4.37}
\end{equation}
If $e_i=y_i- \bx_i^\top \mathbf{B}_c$ were observed,  the variance (\ref{4.37}) would be estimated  by
$$ V\left( \widehat{Y}_{\rm GREG} \right) \cong \sum_{i \in A} \sum_{j \in A} \frac{\pi_{ij} - \pi_i \pi_j
}{\pi_{ij}} \frac{e_i}{\pi_i} \frac{e_j }{\pi_j}. $$ 
If we use $\hat{e}_i = y_i - \bx_i^\top  \hat{\bm \beta}_c$ instead of $e_i$, 
 a consistent variance estimator is computed by 
\begin{equation}
\widehat{V}_{\rm GREG} = \sum_{i \in A} \sum_{j \in A} \frac{\pi_{ij} -
\pi_i \pi_j }{\pi_{ij}} \frac{\hat{e}_i}{\pi_i} \frac{\hat{e}_j }{\pi_j}.
\label{4.38}
\end{equation}

 Now, let's discuss some algebraic properties of the GREG estimator. Using (\ref{eq:greg2}), we can express 
\begin{equation} 
 \widehat{Y}_{\rm GREG}= \sum_{i \in A} \hat{\omega}_i   y_i ,
\label{greg3}
\end{equation} 
where
\begin{equation}
 \hat{\omega}_i  = d_i   + \left( \mathbf{X} - \widehat{\mathbf{X}}_{\rm HT}
\right)^\top  \left( \sum_{k\in A} \frac{1}{ c_k} \bx_k
\bx_k^\top  \right)^{-1} \frac{1}{c_i} \bx_i 
\label{eq:9-16}
\end{equation}
and  $d_i = \pi_i^{-1}$ is the design weight of unit $i$.  
The second term in (\ref{eq:9-16})  is the weight adjustment term incorporating the population auxiliary information $\mathbf{X}$. The final weights $\{ \hat{\omega}_i; i \in A \}$ satisfy  the calibration property 
$$ \sum_{i \in A} \hat{\omega}_i \mathbf{x}_i
= \sum_{i=1}^N \mathbf{x}_i. $$ 

Since we can express 
\begin{equation} 
 \hat{\omega}_i = d_i + {\bm \lambda}^{\top} \bx_i/ c_i 
 \label{gfactor}
 \end{equation}
 for some ${\bm \lambda}$, 
    the final weight $\hat{\omega}_i$ is linear in $\bx_i$.  Thus, if some individual values of $\bx_i$ are extreme (too large or too small), then the final weights can be too large or too small, even take negative values. Also, if $c_i$ is large, then the effect of $\bx_i$ on the final weight is reduced.

The GREG weight in (\ref{eq:9-16}) can be viewed as the solution to an optimization problem that minimizes 
\begin{equation} 
Q ( \bm \omega)= \sum_{i \in A}   \left( \omega_i - d_i \right)^2  c_i  
\label{objective}
\end{equation} 
subject to 
\begin{equation} 
 \sum_{i \in A} \omega_i
\mathbf{x}_i = \sum_{i=1}^N \mathbf{x}_i.
\label{calib3}
\end{equation} 
Constraint (\ref{calib3}) is attractive as it incorporates the external auxiliary information $\mathbf{X} = \sum_{i=1}^N \bx_i$. The calibration condition (\ref{calib3}) will be satisfied with $w_i=d_i$ approximately for sufficiently large $n$. Thus,  the solution $\hat{w}_i$ to the above optimization problem will converge in probability to $d_i$. 

By the Lagrange multiplier method, it is equivalent to  minimizing 
    \begin{equation}
     Q( \bm \omega, \bm \blambda ) =  \frac{1}{2} \sum_{i \in A}  \left( \omega_i - d_i \right)^2  c_i + \bm \lambda^\top  \left( \mathbf{X} -   \sum_{i \in A} \omega_i
\bx_i \right) 
\label{eq:9-17}
\end{equation}
with respect to $\bm \omega$ and $\bm \lambda$. We can solve the optimization problem by computing the partial derivatives: 
\begin{eqnarray*}
\frac{ \partial}{ \partial \omega_i } Q &=&   (\omega_i-d_i) c_i- \bm \lambda^\top  \bx_i = 0 
\end{eqnarray*} 
to obtain the form 
\begin{equation} 
\omega_i^\star = d_i +  \bm \lambda^\top  \bx_i/ c_i   . 
\label{eq:9-17} 
\end{equation}
To compute $\bm \lambda$, we insert (\ref{eq:9-17}) to (\ref{calib3}) and obtain 
$$ \hat{\bm \lambda}^\top  = \left( \mathbf{X} - \widehat{\mathbf{X}}_{\rm HT} \right)^\top  \left(\sum_{i \in A}  \bx_i \bx_i' / c_i  \right)^{-1} . $$ 
 Therefore,  (\ref{greg3}) is obtained. Note that $\hat{\bm \lambda}$ will converge to zero in probability as $\widehat{\mathbf{X}}_{\rm HT}$ will converge to $
 \mathbf{X}$ in probability. 
 
 Constraint (\ref{calib3}) is called the calibration constraint, and the estimator obtained from an optimization problem of finding the final weights subject to the calibration constraint is called the \emph{calibration estimator}. \index{calibration estimation} 

More generally, instead of using the squared error  distance  in (\ref{objective}), we may consider minimizing 
\begin{equation}
    Q(\bm \omega) = \sum_{i \in A} d_i   G\left( \frac{\omega_i }{d_i} \right)  v_i  \label{DS}
\end{equation}
subject to the calibration constraints in (\ref{calib3}), where $d_i= \pi_i^{-1}$ and   $G(\cdot): \mathcal V \to \mathbb R$ is a nonnegative function that is strictly convex, differentiable, and $G'(1) = 0$. The domain $\mathcal V$ of $G(\cdot)$ is an interval in $\mathbb R$.  For example, $G(\omega) = \omega \log(\omega) - \omega + 1$, with possible domain $\mathcal V \subseteq (0, \infty)$, corresponds to the Kullback-Leibler divergence, while $G(\omega) = (\omega - 1)^2$, with possible domain $\mathcal V \subseteq  (-\infty, \infty)$, corresponds to the Chi-squared distance from 1.

Let $\hat{\omega}_i$ be the solution to the above optimization problem using (\ref{DS}) 
and $\hat{Y}_{\rm cal} = \sum_{i \in A} \hat{\omega}_i y_i$ be the resulting calibration estimator. 
Under some conditions on $G(\cdot)$, 
$\hat{Y}_{\rm cal} $
is asymptotically equivalent to the GREG  estimator in (\ref{eq:greg2}). The result is summarized in the following theorem. 
\begin{theorem}
Under some regularity conditions, the calibration estimator $\hat{Y}_{\rm cal} $ satisfies 
\begin{equation}
  \widehat{Y}_{\rm cal}  =  \widehat{Y}_{\rm HT} + \left( \mathbf{X} - \widehat{\mathbf{X}}_{\rm HT} \right)^\top \mathbf{B}_v + o_p(n^{-1/2} N )
 \label{thm:9-1}
 \end{equation}
 where 
 $$ \mathbf{B}_v  = \left( \sum_{i=1}^N \bx_i \bx_i^\top / v_i \right)^{-1} \sum_{i=1}^N \bx_i y_i / v_i . $$  
 \label{thm:9.1}
 \end{theorem}
\begin{proof}
Using the Lagrange multiplier method, the constrained optimization problem can be expressed as maximizing 
$$ 
Q_1\left( \bm \omega, \bm \lambda \right) = - \sum_{i \in A}  d_i G\left( \frac{\omega_i }{d_i} \right) v_i + \bm \lambda^\top  \left( \sum_{i \in A} \omega_i \bx_i - \sum_{i=1}^N \bx_i  \right) .$$
Since 
$$ \frac{ \partial }{ \partial \omega_i} Q_1 = - g \left( \omega_i / d_i \right) v_i   + \bm \lambda^\top  \bx_i $$
where $g( \omega ) = d G(\omega)/ d \omega$, the maximizer can be expressed as 
$$ 
\omega_i^\star ( \bm \lambda) = d_i g^{-1} \left( \bm \lambda^\top \bx_i / v_i \right). $$

Now,  we can express 
$$ \hat{Y}_{\rm cal} = \hat{Y}_{\rm cal} ( \hat{\bm \lambda} ) = \sum_{i \in A}   {\omega}_i^{\star} ( \hat{\bm \lambda} ) y_i $$
where $\hat{\bm \lambda}$ satisfies (\ref{calib3}), we can express 
\begin{eqnarray*} 
\widehat{Y}_{\rm cal} &=& \sum_{i \in A}  \omega_i^\star (  \hat{\bm \lambda} ) y_i+ \underbrace{ \left( \sum_{i =1}^N \bx_i - \sum_{i \in A}  \omega_i^\star (  \hat{\bm \lambda} ) \bx_i\right)^\top  }_{=\mathbf{0}^\top }{{\bm \gamma}} \\
&:=& \hat{Y}_{\ell} ( \hat{\bm \lambda}, {{\bm \gamma}} ) . 
\end{eqnarray*} 
That is, $\hat{Y}_{\ell } ( \hat{\bm \lambda}, {{\bm \gamma}} )= \hat{Y}_{\rm cal} (\hat{\bm \lambda})$ for all ${{\bm \gamma}}$.

     Let $\bm \lambda^*$ be the probability limit of $\hat{\bm \lambda}$. 
     Since $\hat{\bm \lambda}$ satisfies (\ref{calib3}),  $\bm \lambda^* $ should satisfy     
     $$ \underbrace{E \left\{ \sum_{i=1}^N I_i \omega_i^\star \left(   {\bm \lambda^*}  \right) \bx_i \mid \mathcal{F}_N \right\}}_{=\sum_{i=1}^N \pi_i \omega_i^\star \left(  {\bm \lambda^*}  \right) \bx_i} = \sum_{i=1}^N \bx_i , 
    $$
    which implies that 
    $$ \omega_i ^\star ( \bm \lambda^*) =\pi_i^{-1} = d_i $$
    or $g^{-1}\left(\bx_i^\top  \bm \lambda^*/ v_i \right)=1$. Since $g(1)=0$ and $g( \cdot)$ is one-to-one, we get $\bm \lambda^*=\mathbf{0}$.  
     
     Now, if we can find $\bm \gamma^*$ such that  
    \begin{equation} 
    E \left\{ \frac{\partial}{\partial \bm \lambda} \hat{Y}_{\ell} ( {\bm \lambda}^*, {\bm \gamma} )  \mid \mathcal{F} \right\} = \mathbf{0}  
    \label{eq:9-23} 
    \end{equation}
    at ${\bm \gamma} = \bm \gamma^*$,  then the sampling error of $\hat{\bm \lambda}$ can be safely ignored \citep{randles1982aysmptotic}
    and we can obtain 
\begin{equation}
\hat{Y}_{\rm cal} = \hat{Y}_{\ell } (\bm \lambda^*, \bm \gamma^*) + o_p(n^{-1/2} N). 
\label{eq:9-24} \end{equation}    
Now, since  
    \begin{eqnarray*}
   \frac{\partial}{\partial \bm \lambda} \hat{Y}_{\ell} ( {\bm \lambda}^*, {\bm \gamma}  )  
&=& \sum_{i=1}^N I_i d_i \frac{1}{g'(\bx_i^\top  \bm \lambda^*/v_i)} \left( y_i - \bx_i^\top {\bm \gamma}\right) \bx_i/ v_i ,   \end{eqnarray*}
and $g'(0)>0$ is constant,  by (\ref{eq:9-23}),    we obtain 
    $$
    \bm \gamma^* = \left( \sum_{i=1}^N \bx_i \bx_i^\top  / v_i \right)^{-1}  \sum_{i=1}^N \bx_i y_i  / v_i  = \mathbf{B}_v . $$ 
Therefore, by (\ref{eq:9-24}),  
we obtain  (\ref{thm:9-1}).  
\end{proof}

See \cite{deville1992} and \cite{fuller2002} for further discussion of the calibration weighting method.

\section{Optimal estimation }

So far, we have discussed a class of model-assisted estimators that improve the efficiency of the HT estimator by incorporating auxiliary information. 
In this section, we explore the optimality of the Generalized Regression (GREG) estimator within a certain class of estimators. We initially demonstrate the nonexistence of the Uniformly Minimum Variance Unbiased Estimator (UMVUE) in a strictly design-based context, which was originally discovered by  \cite{godambe1965} and then also proved by \cite{basu1971} with a simpler proof. 

\begin{theorem}
 Let any noncensus design with $\pi_k>0$ $(k=1,2,\ldots, N)$ be given. Then
no uniformly minimum variance estimator exists in the class of all unbiased estimators of $Y=\sum_{i=1}^N y_i$.
\end{theorem}
\begin{proof}
For a given value $\mathbf{y}^* = \left(y_1^* , y_2^*, \ldots, y_N^* \right)$, consider the following difference estimator:
\begin{equation}
\hat{Y}_{\text{diff}} = \sum_{i=1}^N y_i^*  + \sum_{i \in A} \frac{1}{\pi_i} \left( y_i - y_i^* \right)
\end{equation}
This estimator is unbiased regardless of $\mathbf{y}^* = \left(y_1^*, y_2^*, \ldots, y_N^*\right)$, and its variance is zero when $\mathbf{y}=\mathbf{y}^*$.

For an unbiased estimator $\hat{Y}$ to be considered a UMVUE, it must satisfy:
$$ V\left( \hat{Y} \right) \le V\left( \hat{Y}_{\rm diff} \right) , \ \ \
\forall \, \by . $$ Given that $V( \hat{Y}_{\text{diff}})=0$ for $\mathbf{y}=\mathbf{y}^*$, it implies that $V( \hat{Y} ) =0$ for $\mathbf{y}=\mathbf{y}^*$. Since any arbitrary $\mathbf{y}^*$ can be chosen, it follows that $V( \hat{Y} ) =0$ for all $\mathbf{y}$, a condition that only holds for a census.
\end{proof}
The theorem discussed above demonstrates that it is not feasible to identify the optimal estimator within the class of unbiased estimators in terms of minimizing the design variance. To address this challenge, we adjust our approach for evaluating the efficiency of estimators by incorporating the superpopulation model into our considerations. More precisely, we will examine the expected value of the design variance under the superpopulation model. This type of variance is referred to as the \emph{anticipated variance} and is formally defined as follows. 
\index{anticipated
variance} 

\begin{definition}
Anticipated variance of  $\hat{\theta}$ is defined by 
$$ AV\left( \hat{\theta} \right) = E_{\zeta} \left\{  V\left( \hat{\theta} \mid \mathcal{F}_N \right) \right\} ,$$
where subscript $\zeta$ denotes the distribution with respect to the superpopulation model and the conditional expectation given $\mathcal{F}= \{y_1, \ldots, y_N\}$ denotes the design-based expectation  under  the sampling mechanism. \end{definition}

\begin{lemma}
Let $\hat{\theta}$ be design-unbiased for $\theta_N$. The anticipated variance of $\hat{\theta}$ can be written as 
\begin{equation}
AV\left( \hat{\theta} \right) =
E_p \left\{ V_\zeta \left( \hat{\theta} \right) \right\} + V_p \left\{ E_\zeta \left( \hat{\theta} \right) \right\}- V_\zeta \left( \theta_N \right) .
\label{av3}
\end{equation}
\end{lemma}
\begin{proof}
 Since $\hat{\theta}$ is design-unbiased for $\theta_N$, we can write 
$$ AV\left( \hat{\theta} \right) = E_\zeta V_p \left( \hat{\theta} \right) =  E_\zeta E_p \left( \hat{\theta}- \theta_N \right)^2. $$
Thus, 
\begin{eqnarray*}
 AV\left( \hat{\theta} \right) &=& E_p E_\zeta \left( \hat{\theta}- \theta_N \right)^2\\
 &=&  E_p E_\zeta \left\{ \hat{\theta}-E_\zeta( \hat{\theta} ) + E_\zeta ( \hat{\theta} )- E_\zeta \left( {\theta}_N \right)+ E_\zeta \left( {\theta}_N \right) -   \theta_N \right\}^2
\end{eqnarray*}
and 
\begin{eqnarray*}
 AV\left( \hat{\theta} \right) &=& E_p E_\zeta \left\{ \hat{\theta}-E_\zeta( \hat{\theta} ) \right\}^2
 + E_p \left\{ E_\zeta ( \hat{\theta} )- E_\zeta \left( {\theta}_N \right) \right\}^2+ E_p\left\{  E_\zeta \left( {\theta}_N \right) -   \theta_N \right\}^2 \\
 && + 2 E_p\left\{ \left( \hat{\theta}-E_\zeta( \hat{\theta} ) \right)\left(  E_\zeta \left( {\theta}_N \right) -   \theta_N\right)  \right\}
\end{eqnarray*}
and the remaining cross product terms are zero. Since
$$E_p\left\{ \left( \hat{\theta}-E_\zeta( \hat{\theta} ) \right) \right\}=- \left(  E_\zeta \left( {\theta}_N \right) -   \theta_N\right) $$
and 
\begin{eqnarray*}
2 E_p\left\{ \left( \hat{\theta}-E_\zeta( \hat{\theta} ) \right)\left(  E_\zeta \left( {\theta}_N \right) -   \theta_N\right)  \right\}&=& 2\left(  E_\zeta \left( {\theta}_N \right) -   \theta_N\right) E_p\left\{ \left( \hat{\theta}-E_\zeta( \hat{\theta} ) \right)  \right\}\\
&&= -2\left(  E_\zeta \left( {\theta}_N \right) -   \theta_N\right)^2,
\end{eqnarray*}
we obtain 
\begin{eqnarray*}
 AV\left( \hat{\theta} \right) &=& E_p E_\zeta \left\{ \hat{\theta}-E_\zeta( \hat{\theta} ) \right\}^2
 + E_p \left\{ E_\zeta ( \hat{\theta} )- E_\zeta \left( {\theta}_N \right) \right\}^2- \left\{  E_\zeta \left( {\theta}_N \right) -   \theta_N \right\}^2
\end{eqnarray*}
which proves (\ref{av3}).
\end{proof}

The following theorem presents the lower bound of the anticipated variance for a design unbiased estimator. 
\begin{theorem}
Let $y_i$ be independently distributed in the superpopulation model. The anticipate variance of any design-unbiased estimator $\hat{Y}$ of $Y=\sum_{i=1}^N y_i$ satisfies 
\begin{equation}
E_\zeta \left\{  V\left( \hat{Y} \mid \mathcal{F}_N   \right)\right\} \ge \sum_{i =1}^N \left(
\frac{1}{\pi_i} -1 \right) V_{\zeta} \left( y_i \right). 
\label{4.41}
\end{equation}
\end{theorem}
\begin{proof}
Write 
$\hat{Y}$ as  $\hat{Y} = \hat{Y}_{\rm HT} + R $ where  $\hat{Y}_{\rm HT}$ is the HT estimator of $Y$. Since $\hat{Y}$ is design unbiased, we have 
 $E\left( R \right) = 0$ and, for fixed $j \in U$,
 \begin{eqnarray*}
0 &=& E\left( R \right) \\
&=& \sum_{A \in \mathcal{A}} p \left( A \right) R\left( A \right) \\
&=& \sum_{A \in \mathcal{A}; j \in A}p \left( A \right) R\left( A
\right)  +\sum_{A \in \mathcal{A}; j \notin A}p \left( A \right)
R\left( A \right).
\end{eqnarray*}
Now, since 
$$ V_\zeta \left( \hat{Y} \right) = V_\zeta \left( \hat{Y}_{\rm HT} \right)
+ V_\zeta \left( R \right) + 2 Cov_{\zeta} \left( \hat{Y}_{\rm HT}, R
\right), $$
we obtain 
\begin{eqnarray*}
E_p \left\{ Cov_\zeta \left( \hat{Y}_{\rm HT}, R \right)  \right\} &=&
E_p\left[ E_\zeta \left\{ \left( \hat{Y}_{HT} - E_\zeta\left(
\hat{Y}_{HT} \right) \right) R \right\} \right]\\
&=& E_p \left[ \sum_{j \in U} E_\zeta \left\{ \frac{\left( y_j - E_\zeta \left(y_j  \right) \right)I_j }{ \pi_j} R \right\}  \right] \\
&=&\sum_{j \in U} E_\zeta \left\{ \frac{\left( y_j - E_\zeta
\left(y_j  \right) \right) }{ \pi_j} E\left\{ I_j \left( A \right)
R\left( A \right) \right\} \right\}\\
&=& \sum_{j \in U} E_\zeta \left\{ \frac{\left( y_j - E_\zeta
\left(y_j  \right) \right) }{ \pi_j}  \sum_{A \in \mathcal{A} ; j \in A} R\left( A \right) p \left( A \right) \right\}\\
&=& -\sum_{j \in U} E_\zeta \left\{ \frac{\left( y_j - E_\zeta
\left(y_j  \right) \right) }{ \pi_j}  \sum_{A \in \mathcal{A} ; j \notin A} R\left( A \right) p \left( A \right) \right\}\\
&=& 0 ,
\end{eqnarray*}
where the last equality follows because 
 $\sum_{A \in \mathcal{A} ; j
\notin A} R\left( A \right) p \left( A \right)$ does not depend on $y_j$. Thus, 
\begin{eqnarray*}
E_p\left\{  V_\zeta \left( \hat{Y} \right) \right\} &=& E_p\left\{
V_\zeta \left( \hat{Y}_{HT} \right) \right\}+ E_p\left\{  V_\zeta
\left( R \right) \right\} \\
&\ge &E_p\left\{ V_\zeta \left( \hat{Y}_{HT} \right) \right\}\\
&=& E_p \left\{  V_\zeta \left( \sum_{i =1}^N \frac{ y_i I_i}{
\pi_i}
\right)\right\}\\
&=& E_p\left\{ \sum_{i=1}^N \frac{ \sigma_i^2 I_i}{ \pi_i^2}
\right\}= \sum_{i=1}^N \frac{\sigma_i^2}{\pi_i}
\end{eqnarray*}
and we have 
\begin{eqnarray*}
AV \left( \hat{Y} \right) &=& E_p V_\zeta \left( \hat{Y} \right) +
V_p E_\zeta \left( \hat{Y} \right) - V_\zeta \left( Y \right) \\
&\ge & E_p V_\zeta \left( \hat{Y} \right) - V_\zeta \left( Y \right)
\\
&\ge & \sum_{i=1}^N \frac{\sigma_i^2}{\pi_i} - \sum_{i \in U}
\sigma_i^2.
\end{eqnarray*}
\end{proof}

The lower bound in (\ref{4.41}) is the lower bound of the anticipated variance of any unbiased estimator. The lower bound was first discovered by Godambe and Joshi (1965) and is often called Godambe-Joshi lower bound. For a fix-ed size probability sampling design, the Godambe-Joshi lower bound is minimized when 
\begin{equation}
 \pi_i \propto \{  V_{\zeta} (y_i) \}^{1/2}. 
 \label{9-16}
 \end{equation}
To show this, we minimize $\sum_{i=1}^N V_{\zeta} (y_i)/ \pi_i$ subject to $\sum_{i=1}^N \pi_i=n$. The solution can be obtained by applying Cauchy -Schwarz inequality to get 
$$ \left\{ \sum_{i=1}^N \sigma_i^2/ \pi_i  \right\} \left( \sum_{i=1}^N \pi_i \right) \ge \left\{ \sum_{i=1}^N \sigma_i  \right\}^2 $$
with equality when (\ref{9-16}) holds. 

The following theorem, which was first proved by \cite{isaki1982}, shows that the GREG estimator achieves the Godambe-Joshi lower bound  asymptotically.

\begin{theorem}
Suppose that  $\zeta$ is a superpopulation model
with $y_i$'s independent and $E_\zeta \left( y_i \mid \bx_i \right) =
\mathbf{x}_i^\top  \bm \beta$ and $V_\zeta \left( y_i \mid \bx_i \right) = v_i
\sigma^2$. Then, the anticipated variance of the GREG estimator in (\ref{eq:greg2}) asymptotically attains the Godambe-Joshi lower bound.
\label{thm9.3} 
\end{theorem}

\begin{proof} 
By (\ref{9-5}),  
 the GREG estimator  is asymptotically equivalent to the difference estimator  in (\ref{9-5}). 
Thus, 
\begin{eqnarray*}
 E_\zeta \left\{ V\left( \hat{Y}_{\rm GREG} \right) \right\}
&\doteq & E_\zeta \left\{ \sum_{i=1}^N \sum_{j =1}^N \left( \pi_{ij} -
\pi_i \pi_j \right) \frac{y_i - \bx_i^\top  \mathbf{B}_c }{\pi_i}  \frac{y_j -
\bx_j^\top  \mathbf{B}_c }{\pi_j}\right\}\\
&\doteq& E_\zeta \left\{ \sum_{i=1}^N \sum_{j =1}^N \left( \pi_{ij}
- \pi_i \pi_j \right) \frac{y_i - \bx_i^\top  \bm \beta }{\pi_i}  \frac{y_j
- \bx_j^\top  \bm \beta}{\pi_j}\right\}\\ &=& \sum_{i=1}^N \left(
\frac{1}{\pi_i} -1 \right) v_i \sigma^2
\end{eqnarray*}
which is equal to the Gobambe-Joshi lower bound under the superpopulation model. 
\end{proof} 

\section{Model-assisted calibration} 

We now turn our attention to a model-assisted approach for calibration estimation. To illustrate the concept of calibration estimation within the framework of the superpopulation model specified in equations  (\ref{4.33})-(\ref{4.34}), let's consider a linear estimator defined as 
$$ \hat{Y}_\omega = \sum_{i \in A} \omega_i y_i $$
for some $\omega_i$. Now, note that 
\begin{equation}
\hat{Y}_\omega - Y = \left( \sum_{i \in A} \omega_i \bx_i  - \sum_{i=1}^N  \bx_i  \right)^\top \bm \beta + \left\{ \sum_{i \in A} \omega_i e_i - \sum_{i=1}^N e_i \right\}:= C + D .
\label{eq:9-22} 
\end{equation} 
The first term, $C$, can be eliminated if the weights $\omega_i$ satisfy the  calibration constraint (\ref{calib3}). Consequently, our goal shifts to minimizing the model variance of the term $D$. 
 Note that 
\begin{eqnarray*} 
  E_{\zeta} ( D^2  \mid I_1, \cdots, I_N)
 &=& \sum_{i \in A} \omega_i^2 v_i \sigma^2 - 2\sum_{i \in A} \omega_i v_i \sigma^2 + \sum_{i =1}^N v_i \sigma^2  
\end{eqnarray*} 
If the condition 
\begin{equation}
v_i = \bm \lambda^\top  \mathbf{x}_i
\label{eq:ibc1}
\end{equation}
is met for some $\bm \lambda$, then  the calibration constraint in (\ref{calib3}) implies that 
\begin{equation}
\sum_{i \in A} \omega_i v_i  = \sum_{i=1}^N v_i  .
\label{normalizing}
\end{equation}
If $v_i=1$, then (\ref{eq:ibc1}) means that $\bx_i$ includes an intercept term and constraint (\ref{normalizing}) is a normalization constraint for $\omega_i$. 

Therefore, under (\ref{eq:ibc1}), minimizing the model variance of $D$ is equivalent to minimizing 
\begin{equation}
Q(w) = \sum_{i \in A} \omega_i^2 v_i  
\label{eq:9-30}
\end{equation}
subject to the calibration constraint (\ref{calib3}). 
The optimal calibration estimator is then given by 
\begin{equation}
\hat{Y}_{\rm opt} = \sum_{i \in A} \hat{\omega}_i y_i = \sum_{i=1}^N \bx_i^\top  \hat{\bm \beta}_{v} 
\label{res3}
\end{equation}
where 
$
\hat{\bm \beta}_{v} = \left( \sum_{i \in A}  v_i^{-1} \bx_i \bx_i^\top    \right)^{-1}\sum_{i \in A}  v_i^{-1} \bx_i y_i 
$. This formulation reveals that the final calibration estimator can be interpreted as a projection estimator in Section 9.1. 

 The optimal calibration estimator in (\ref{res3}) is not necessarily design consistent. One way to achieve the design consistency is to impose the IBC condition in (\ref{ibc}) in the regression model. 
 For the regression projection estimator with $\hat{y}_i= \bx_i^\top  \hat{\bm \beta}_{v}$, by Lemma 9.1,  
 the IBC condition in (\ref{ibc}) can be met by augmenting $\mathbf{x}_i$ to include $ \pi_i^{-1} v_i$. 
In other words, 
by 
defining $\mathbf{z}_i^\top  = (\mathbf{x}_i^\top , \pi_i^{-1} v_i )$ and calculating $\hat{\boldsymbol{\gamma}}_v = \left( \sum_{i \in A} v_i^{-1} \mathbf{z}_i \mathbf{z}_i^\top \right)^{-1}\sum_{i \in A} v_i^{-1} \mathbf{z}_i y_i $, the property of the residuals ensures that $\hat{y}_i= \mathbf{z}_i^\top  \hat{\boldsymbol{\gamma}}_v$ satisfies  
$$ \sum_{i \in A}  \left( y_i - \hat{y}_i \right) \bz_i v_i^{-1} = \bm{0} ,
$$thereby implying the IBC condition in (\ref{ibc}). The augmented regression approach effectively integrates additional calibration into the estimation process, achieving the ADU of the resulting regression projection estimator. 
\index{Asymptotic Design Unbiasedness} \index{ADU} 

To minimize the model variance $\hat{\theta}_{\omega}$ while also adhering to the Asymptotically Design Unbiased (ADU) condition, our objective narrows down to minimizing 
\begin{equation}
\sum_{i \in A} \omega_i^2 v_i
\end{equation}
subject to 
\begin{equation}
\sum_{i \in A} \omega_i  \bz_i   = \sum_{i=1}^N \bz_i .
\label{eq:debiased}
\end{equation}

Let $\hat{\omega}_i$ be the solution to the optimization problem described above. Using the Lagrangian multiplier method, using the same argument for obtaining (\ref{res3}),  the resulting estimator   can be written as 
\begin{equation}
        \widehat Y_{\rm cal} = \sum_{i \in A} \hat{\omega}_i y_i = \sum_{i=1}^N \bz_i^{\top} \hat{\bm \gamma}_v ,  \label{opt3}
\end{equation}
    where 
    $$ \hat{\omega}_i =\left( \sum_{i=1}^N \mathbf{z}_i\right)^\top \left( \sum_{i \in A}  v_i^{-1} \bz_i \bz_i^\top   \right)^{-1} \bz_i/v_i $$ 
and 
$$\hat{\bm \gamma}_v = \left( \sum_{i \in A}  v_i^{-1} \bz_i \bz_i'  \right)^{-1}\sum_{i \in A}  v_i^{-1} \bz_i y_i. $$

Since the IBC condition (\ref{ibc})  holds for $\hat{y}_i= \bz_i^\top \hat{\bm \gamma}_v$, we can express (\ref{opt3}) as 
   $$  \widehat Y_{\rm cal} =  \sum_{i=1}^N \bz_i^{\top} \hat{\bm \gamma}_v  + \sum_{i \in A} \frac{1}{\pi_i} \left( y_i - \bz_i^{\top} \hat{\bm \gamma}_v \right) .  $$
Let $\bm \gamma^*$ be the probability limit of $\hat{\bm \gamma}_v$. Using the standard argument similar to (\ref{9-5}), we can obtain 
$$ 
N^{-1} \widehat{Y}_{\rm cal} = N^{-1} \widehat{Y}_{\rm diff} + o_p (n^{-1/2}) , 
$$
where 
$$ 
\widehat{Y}_{\rm diff} = 
\sum_{i \in U} \bz_i^\top  {\bm \gamma}^* + \sum_{i \in A} \frac{1}{\pi_i} \left( y_i - \bz_i^\top   {\bm \gamma}^* \right).$$
Now, note that 
\begin{eqnarray*} 
\hat{Y}_{\rm diff} - Y &=& \sum_{i=1}^N \left( \frac{I_i}{\pi_i} - 1 \right) \left( y_i - \bz_i^\top  {\bm \gamma}^* \right)\end{eqnarray*}
and 
\begin{eqnarray*} 
E \left\{ \left(\hat{Y}_{\rm diff} - Y \right)^2 \right\} &=&  E \left\{ \sum_{i=1}^N \left( \frac{I_i}{\pi_i} - 1 \right)^2 v_i^\star \right\} \\
&=& \sum_{i=1}^N  \left( \frac{1}{\pi_i} - 1 \right) v_i^\star 
\end{eqnarray*}
where $v_i^\star = E_\zeta \left\{ \left( y_i - \bz_i^\top  \bm \gamma^* \right)^2 \mid \bz_i \right\}   $ is the conditional variance under the augmented regression model.  

If the Generalized Regression (GREG) model specified in equations  (\ref{4.33})-(\ref{4.34})
is correctly specified, then we have $\bz_i^\top \bm \gamma^* = \bx_i^{\top} \bm \beta$ and 
the variance component for each unit, $v_i^\star$, is equal to $v_i \sigma^2$. However, if the GREG model does not accurately represent the underlying data structure, it is possible to observe that $v_i \sigma^2 > v_i^\star$. This discrepancy arises because the additional covariate $\pi_i^{-1} v_i$ improves the prediction of $y_i$, thus contributing to a more precise estimation.
Therefore,  the model-assisted calibration estimator, which employs the augmented covariate $\mathbf{z}_i$ for calibration, exhibits greater efficiency compared to the GREG estimator, especially when the superpopulation model delineated in equations (\ref{4.33})-(\ref{4.34})  does not perfectly align with the actual data.

For general models, \cite{wu2001model} introduced the concept of model calibration, which integrates the working regression model directly into the calibration constraint. This approach involves using $\hat{m}_i = m(\mathbf{x}_i; \hat{\boldsymbol{\beta}})$ as an estimated predictor for $y_i$, derived from the working regression model where $E(Y \mid \mathbf{x}) = m(\mathbf{x}; \boldsymbol{\beta})$.
If it is possible to calculate $\hat{m}_i$ for every unit  of the population, then the following equation can be employed as the calibration constraint for the weighting problem: 
\begin{equation} 
\sum_{i \in A} w_i ( 1, \hat{m}_i) = \sum_{i=1}^N (1, \hat{m}_i) \label{mcalib} 
\end{equation} 
 The uncertainty of $\hat{\beta}$ in $\hat{m}_i$ can be safely ignored in the final inference.

Instead of using model calibration,  an alternative approach involves calibrating for basis functions. This method is applicable when the expected value of $Y$ on $\mathbf{x}$ belongs to the span of a set of basis functions $b_1( \bx), \cdots, b_L ( \bx)$.  
In such scenarios, the calibration estimation can utilize the following constraint:
\begin{equation}
\sum_{ i \in A} w_i\left[b_1( \mathbf{x}_i), \cdots, b_L ( \mathbf{x}_i) \right] = \sum_{i=1}^N \left[ b_1( \mathbf{x}_i), \cdots, b_L ( \mathbf{x}_i) \right]
\end{equation}
This condition ensures that the $C$ term in (\ref{eq:9-22}) is nullified. As the sample size increases, the dimension $L$ of the basis functions may also need to be increased. In such instances, regularization methods can be employed to select a suitable $L$. 

For example, \cite{montanari2005nonparametric} explored the use of neural network models, while \cite{breidt05}
 applied penalized spline models for nonparametric calibration estimation. \cite{breidt17} gave a comprehensive review of the nonparametric calibration methods. 
These methodologies offer a flexible framework for calibration estimation, allowing for the accommodation of complex relationships between the response variable. 

\section{Generalized entropy calibration}

Instead of the squared error loss in (\ref{eq:9-30}), we can consider the  we now consider maximizing the generalized entropy that does not depend on the design weights, which was proposed by \cite{kwon2024}. Let $G: \mathcal V \to \mathbb R$ be a prespecified function that is strictly convex and twice-continuously differentiable. The domain of $G$ is an open interval $\mathcal V = (\nu_1, \nu_2)$ in $\mathbb R$, where $\nu_1$ and $\nu_2$ are allowed to be $-\infty$ and $\infty$ respectively. 
Using (\ref{4.33})-(\ref{4.34}) 
as a working model for model-assisted estimation, the generalized entropy method can be formulated as minimizing 
\begin{equation}
\sum_{i \in A} v_i G( \omega_i) 
\label{wel}
\end{equation}
subject to (\ref{calib3}) and 
\begin{equation}
    \sum_{i \in A}\omega_i g(d_i) v_i = \sum_{i=1}^N  g(d_i) v_i, 
    \label{dcc2}
\end{equation}
where $g( \omega ) = d G(\omega)/ d \omega$ and $d_i = \pi_i^{-1}$.  
Constraint (\ref{dcc2}) plays the role of achieving the ADU condition of the resulting calibration estimator and it is also called  the debiasing constraint under model heterogeneity \citep{kwon2024}. 

Examples of generalized entropies and their debiasing calibration constraints can be found in Table \ref{tab1}. 

\begin{table}[!t]
\centering

\begin{tabular}{ccc}
\hline\hline
Entropy                                             & $G(\omega)$     & $g_i = g (\pi_i^{-1})$     \\ \hline
Squared Loss                                        & $\omega^2 / 2$                  & $\pi_i^{-1}$               \\
Empirical likelihood                                & $-\log \omega$              & $- \pi_i$              \\
Exponential tilting                                 & $\omega\log (\omega) - \omega$ & $-\log \pi_i$            \\
Cross entropy                                       & $(\omega - 1)\log(\omega - 1) - \omega \log(\omega)$    & $\log(1 - \pi_i)$    \\
Hellinger distance                                  & $-4\sqrt{\omega}$            & $-2\pi_i^{1/2}$         \\
Pseudo-Huber loss & $M^2 \{1 + (\omega / M)^2\}^{1/2}$ & $\pi_i^{-1} \{1 + (M\pi_i)^{-2}\}^{-1/2}$ \\
Inverse  & $1 / (2\omega)$   & $-\pi_i^2 / 2$   \\
R\'enyi entropy & $\alpha^{-1}(\alpha + 1)^{-1}\omega^{\alpha + 1}$   & $\alpha^{-1} \pi_i^{-\alpha}$  \\ \hline\hline
\end{tabular}

\caption{Examples of generalized entropies with the corresponding $G(\omega)$ and  the calibration covariates $g_i = g(\pi_i^{-1})$. R\'enyi entropy requires $\alpha \neq 0, -1$.} 
\label{tab1}
\end{table}

The following theorem presents the $\sqrt{n}$-consistency of the generalized entropy calibration estimator.


\begin{theorem}
Let $\hat{\omega}_i$ be obtained by minimizing (\ref{wel}) subject to (\ref{calib3}) and (\ref{dcc2}). The resulting calibration estimator $\hat{Y}_{\rm gec} = \sum_{i \in A} \hat{\omega}_i y_i$ satisfies 
$$ \widehat{Y}_{\rm gec} = \widehat{Y}_{\rm gec,  \ell} + o_p(n^{-1/2} N ), $$
where 
 \begin{equation}
    \widehat Y_{\rm  gec, \ell}  =\sum_{i=1}^N  \bz_i^\top  {\bm \gamma}^*   +  \sum_{i \in A} d_i  \left(   y_i - \bz_i^\top   {\bm \gamma}^* \right)  , 
    \label{eq:9-41}
    \end{equation}
 $\bz_i^\top  =\left(\bx_i^\top , g(d_i) v_i\right) $ and ${\bm \gamma}^*$ is the probability limit of $\hat{\bm \gamma}$ given by      \begin{equation} 
\hat{\bm \gamma} 
= \left( 
\sum_{i \in A}   \frac{1}{g'(d_i) v_i }  \bz_i \bz_i^\top  \right)^{-1} \sum_{i \in A} \frac{1}{g'(d_i) v_i }\bz_i y_i . \label{eq:9-42} 
\end{equation}
\label{thm9.5}
 \end{theorem} 
\begin{proof}
The proof is very similar to that of Theorem \ref{thm:9.1}. 
Using Lagrange multiplier method, the optimization problem can be expressed as maximizing 
$$ 
Q_2\left( \bm \omega, \bm \lambda \right) = - \sum_{i \in A}  G(\omega_i ) v_i + \bm \lambda^\top  \left( \sum_{i \in A} \omega_i \bz_i - \sum_{i=1}^N \bz_i  \right) .$$
Since 
$$ \frac{ \partial }{ \partial \omega_i} Q_2 = - g \left( \omega_i  \right) v_i  + \bm \lambda^\top  \bz_i,  $$
 the maximizer can be expressed as 
$$ 
\omega_i^\star ( \bm \lambda) =  g^{-1} \left( \bm \lambda_1^\top  \bx_i/ v_i + \lambda_2 g_i \right) $$
where $g_i=g(d_i)$.

Now,  we can express 
$$ \widehat{Y}_{\rm gec} = \widehat{Y}_{\rm gec} ( \hat{\bm \lambda} ) = \sum_{i=1}^N I_i  {\omega}_i^{\star} ( \hat{\bm \lambda} ) y_i $$
where $\hat{\bm \lambda}$ satisfies (\ref{calib3}), we can express 
\begin{eqnarray*} 
\widehat{Y}_{\rm gec} &=& \sum_{i=1}^N I_i \omega_i^\star (  \hat{\bm \lambda} ) y_i+ \underbrace{ \left( \sum_{i =1}^N \bz_i - \sum_{i=1}^N I_i \omega_i^\star (  \hat{\bm \lambda} ) \bz_i\right)^\top }_{=\mathbf{0}^{\top} }{{\bm \gamma}} \\
&:=& \hat{Y}_{\rm gec, \ell} ( \hat{\bm \lambda}, {{\bm \gamma}} ) . 
\end{eqnarray*}
   Let $\bm \lambda^*$ be the probability limit of $\hat{\bm \lambda}$.  
     Since $\hat{\bm \lambda}$ satisfies (\ref{calib3}),  $\bm \lambda^* $ should satisfy     
     $$ \underbrace{E \left\{ \sum_{i=1}^N I_i \omega_i^\star \left(   {\bm \lambda^*}  \right) \bz_i \mid \mathcal{F}_N \right\}}_{=\sum_{i=1}^N \pi_i \omega_i^\star \left(  {\bm \lambda^*}  \right) \bz_i} = \sum_{i=1}^N \bz_i , 
    $$
    which implies that 
    $$ \omega_i ^\star ( \bm \lambda^*) =\pi_i^{-1} = d_i $$
    or $$g^{-1}\left(\bx_i^\top  \bm \lambda_1^*/ v_i + g (d_i) \lambda_2^*  \right)= d_i.  $$ Since  $g( \cdot)$ is one-to-one, we get $\bm \lambda_1^*=\mathbf{0}$ and $\lambda_2^*=1$.  
     
     Now, we wish to find $\bm \gamma^*$ such that 
\begin{equation}
E \left\{  \frac{\partial}{\partial \bm \lambda} \hat{Y}_{\rm gec, \ell} ( {\bm \lambda}^*, {\bm \gamma}^*  )   \right\} = \mathbf{0} , 
\label{eq:randles2}
\end{equation}
which can be used to establish 
$$ \widehat{Y}_{\rm gec} = \widehat{Y}_{\rm gec, \ell} \left( \bm \lambda^*, \bm \gamma^* \right) + o_p (n^{-1/2} N ) . 
$$
To find $\bm \gamma^*$ satisfying (\ref{eq:randles2}), since  
    \begin{eqnarray*}
   \frac{\partial}{\partial \bm \lambda} \hat{Y}_{\rm gec, \ell} ( {\bm \lambda}^*, {\bm \gamma}  )  
&=& \sum_{i=1}^N I_i  \frac{1}{g' \{ g^{-1} (\bz_i^\top  \bm \lambda^*/v_i)\}} \left( y_i - \bz_i^\top {\bm \gamma}\right) \bz_i/ v_i ,   \end{eqnarray*}
    we obtain 
    $$
    \bm \gamma^* = \left( \sum_{i=1}^N \frac{\pi_i}{g' ( d_i ) v_i  } \bz_i \bz_i'  \right)^{-1}  \sum_{i=1}^N\frac{\pi_i}{g' (d_i) v_i  } \bz_i y_i      . $$ 
Therefore,  we obtain  (\ref{eq:9-42}). 
\end{proof}
Note that Theorem \ref{thm9.5} does not use the superpopulation model in (\ref{4.33})-(\ref{4.34}) as an assumption. If the GREG superpopulation model  is  indeed correct, then we obtain $\bm \gamma^*=(\bm \beta', 0)'$  and 
$$
 \hat Y_{\rm gec, \ell}  =\sum_{i=1}^N  \bm x_i' {\bm \beta}   +  \sum_{i \in A} d_i \left( y_i - \bm x_i' {\bm \beta} \right) + o_p\left(n^{-1/2} N\right), $$
 which achieves the Godambe-Joshi lower bound of the anticipate variance  \citep{godambe1965}. 

 If the superpopulation model  is incorrect, Theorem \ref{thm9.5} is still applicable and the asymptotic variance of $\hat{Y}_{\rm gcal}$ is equal to 
\begin{equation}
V \left( \hat{Y}_{\rm greg, \ell} \right)   = V \left( \sum_{i=1}^N y_i \right) +  E \left\{ \sum_{i=1}^N \left( \pi_i^{-1} - 1 \right) \left( y_i - {\bm z}_i^\top   {\bm \gamma}^* \right)^2 \right\} . 
\label{var1}
\end{equation}
On the other hand, by  Theorem \ref{thm:9.1}, the classical calibration estimator $\hat{\theta}_{\rm DS}$ of \cite{deville1992}, ignoring the smaller order terms,  satisfies  
\begin{equation}
V \left( \hat{Y}_{\rm DS} \right)   = V \left( \sum_{i=1}^N y_i \right) +  E \left\{ \sum_{i=1}^N \left( \pi_i^{-1} - 1 \right) \left( y_i - {\bm x}_i^\top   \bm \beta^*   \right)^2 \right\} , 
\label{var2}
\end{equation}
where 
 $$
 \bm \beta^* 
  = \left( \sum_{i=1}^N \bm x_i \bm x_i^\top/ v_i \right)^{-1}\sum_{i=1}^N \bm x_i y_i /v_i.
 $$
 Comparing (\ref{var1}) with (\ref{var2}), the additional covariate $g(d_i) v_i$ in $\bz_i$ can improve the prediction power for $y_i$. Thus, the proposed calibration estimator is more efficient than the classical calibration estimator when the superpopulation model is incorrect. 

%% file: chapters/chapter10.tex
\setcounter{chapter}{9} 
\chapter{Variance Estimation}

\section{Introduction}

Variance estimation is an important practical problem in survey sampling. Variance estimates are used for two purposes. One is an analytic purpose, such as constructing confidence intervals or performing hypothesis tests. The other is descriptive purposed to evaluate the efficiency of survey designs or estimates for planning surveys. 

Desirable variance estimates should satisfy the following properties:
\begin{itemize}
\item It should be unbiased, or approximately unbiased. 
\item The variance estimator should be small. That is, the variance estimator is stable. 
\item It should not take negative values. 
\item The computation should be simple. 
\end{itemize}

The HT variance estimator is unbiased, but it can take negative values. In addition, computing the joint inclusion probabilities $\pi_{ij}$ can be cumbersome when the sample size is large.

\begin{example}
Consider a finite population of size
$N=3$ with $y_1=16, y_2=21$ and $y_3=18$  and consider the following sampling design. 

\begin{table}[htb]
\caption{A sampling design for Example 10.1 }
\begin{center}
\begin{tabular}{c|c|c|c}
\hline
   Sample (A)  & $P\left( A \right) $ & HT estimator & HT variance estimator    \\
   \hline
   $A_1=\left\{ 1,2 \right\}$ & 0.4 & 50 & 206 \\
$A_2=\left\{ 1,3 \right\}$ & 0.3 & 50 & 200 \\
$A_3=\left\{ 2, 3 \right\}$ & 0.2 & 60 & -90 \\
$A_4=\left\{ 1,2,3 \right\}$ & 0.1 & 80 & -394 \\
 \hline
\end{tabular}
\end{center}
\label{table5.1}
\end{table}

The sampling variance of the HT estimator is 85. Note that the HT variance estimator has expectation
$$206\times 0.4 + 200 \times 0.3 + (-90)\times 0.2 + (-394)\times
0.1 = 85$$ 
but it can take negative values in some samples. 
\end{example}

The variance estimator under PPS sampling is always nonnegative and can be computed without computing the joint inclusion probability. In practice, the PPS sampling variance estimator is often applied as an alternative variance estimator even for non-replacement sampling. The resulting variance estimator can be written 
\begin{equation}
\hat{V}_0 = \frac{1}{n\left( n -1 \right)} \sum_{i \in A} \left(
\frac{y_i}{p_i } - \hat{Y}_{\rm HT} \right)^2 =  \frac{n}{\left( n -1 \right)} \sum_{i \in A} \left(
\frac{y_i}{\pi_i } - \frac{1}{n} \hat{Y}_{\rm HT} \right)^2, \label{5.1}
\end{equation}
where $p_i = \pi_i/n$. The following theorem expresses the bias of the simplified variance estimator in (\ref{5.1}) as an estimator of the variance of the HT estimator. 

\begin{theorem}
The simplified variance estimator in   (\ref{5.1}) satisfies 
\begin{equation}
  E\left( \hat{V}_0 \right) - Var\left( \hat{Y}_{\rm HT} \right)
=\frac{n}{n-1} \left\{ Var \left( \hat{Y}_{\rm PPS}  \right) - Var\left(
\hat{Y}_{\rm HT} \right) \right\} \label{5.2}
\end{equation}
 where   $$Var \left( \hat{Y}_{\rm PPS}  \right) = \frac{1}{n} \sum_{i=1}^N p_i
 \left(
 \frac{y_i}{p_i} - Y \right)^2 $$
and 
 $p_i=\pi_i/n$.
 \end{theorem}
\begin{proof}
Note that 
$\hat{V}_0$ satisfies 
\begin{eqnarray*}
&&
 E\left\{ \sum_{k\in A} \left( \frac{y_k }{p_k}  -Y + Y-
 \hat{Y}_{\rm HT} \right)^2 \right\} \\
 &=&
 E\left\{ \sum_{k\in A} \left( \frac{y_k }{p_k} - Y  \right)^2
 \right\}
 +2 E\left\{ \sum_{k\in A} \left( \frac{y_k }{p_k} - Y  \right) \left(
 Y - \hat{Y}_{\rm HT} \right)
 \right\} +E\left\{ \sum_{k\in A}  \left(
 Y - \hat{Y}_{\rm HT} \right)^2
 \right\}. 
\end{eqnarray*}
The first term is 
\begin{eqnarray*}
E \left\{\sum_{k\in A} \left( \frac{y_k }{p_k}  -Y \right)^2
\right\} &=& \sum_{k=1}^N \left( \frac{y_k}{p_k} - Y \right)^2
\pi_k
\\ &=& n^2 Var\left( \hat{Y}_{\rm PPS} \right)
\end{eqnarray*}
and, using  
$$ \sum_{k \in A} \left( \frac{y_k}{p_k} - Y \right) = n \left(
\sum_{k \in A} \frac{y_k}{\pi_k} - Y \right)= n \left( \hat{Y}_{\rm HT}
- Y \right), $$
the second term equals to  $ - 2n Var\left(
\hat{Y}_{\rm HT} \right)$ and the last term is equal to  $n Var\left( \hat{Y}_{\rm HT}
\right)$, which proves the result.
\end{proof}

In many cases, the bias term in (\ref{5.2}) is positive and the simplified variance estimator conservatively estimates the variance. Under SRS, the relative bias of the simplified variance estimator (\ref{5.1}) is 
 \begin{equation}
  \frac{\hat{V}_0 - Var\left( \hat{Y}_{\rm HT} \right)}{Var\left( \hat{Y}_{\rm HT}
 \right)} = \frac{n}{N-n}\label{5.3}
 \end{equation}
 and the relative bias is negligible when $n/N$ is negligible.

 The simplified variance estimator can be directly applicable to multistage sampling designs. Under multistage sampling design, the HT estimator of the total can be written 
$$ \hat{Y}_{\rm HT} = \sum_{i \in A_I}\frac{ \hat{Y}_i}{\pi_{Ii}} $$
where $\hat{Y}_i $ is the estimated total for the PSU $i$. The simplified variance estimator is then given by 
\begin{equation*}
\hat{V}_0 = \frac{n}{\left( n -1 \right)} \sum_{i \in A_I} \left(
\frac{\hat{Y}_i}{\pi_{Ii} } - \frac{1}{n}\hat{Y}_{\rm HT} \right)^2. 
\end{equation*}
Under stratified multistage cluster sampling, the simplified variance estimator can be written
\begin{equation}
\hat{V}_0 =\sum_{h=1}^H  \frac{n_h}{ n_h -1 } \sum_{i =1}^{n_h}
\left( w_{hi} \hat{Y}_{hi}  - \frac{1}{n_h}\sum_{j=1}^{n_h} w_{hj}
\hat{Y}_{hj} \right)^2 \label{simplev}
\end{equation}
where  $w_{hi}$ is the sampling weight for cluster $i$ in stratum $h$. 

\section{Linearization approach to  variance estimation}

When the point estimator is a nonlinear estimator, such as a ratio estimator or regression estimator, exact variance estimation for such estimates is very difficult. Instead, we often rely on linearization methods to estimate the sampling variance of such estimators. 

Roughly speaking, if $\by$ is a $p$-dimensional vector and when $ \bar{\mathbf{y}}_n =
\bar{\mathbf{Y}}_N + O_p \left( n^{-1/2} \right)$ holds, the Taylor linearization of $ g\left( \bar{\mathbf{y}}_n  \right) $ 
can be written as 
$$ g\left( \bar{\mathbf{y}}_n  \right) = g\left( \bar{\mathbf{Y}} \right) + \sum_{j=1}^{p}
\frac{\partial g \left( \bar{\mathbf{Y}} \right)}{\partial y_j}
\left( \bar{y}_{j}- \bar{Y}_j \right) + O_p \left( n^{-1} \right).
$$
Thus, the variance of the linearized term of  $g\left( \bar{\mathbf{y}}_n
\right)$ can be written 
$$ V\left\{ g\left( \bar{\mathbf{y}}_n  \right) \right\} \doteq \sum_{i=1}^p \sum_{j=1}^{p}
\frac{\partial g \left( \bar{\mathbf{Y}} \right)}{\partial y_i}
\frac{\partial g \left( \bar{\mathbf{Y}} \right)}{\partial y_j}
Cov\left\{  \bar{y}_{i},  \bar{y}_{j} \right\}
$$
and is estimated by 
\begin{equation}
 \hat{V}\left\{ g\left( \bar{\mathbf{y}}_n  \right) \right\} \doteq \sum_{i=1}^p \sum_{j=1}^{p}
\frac{\partial g \left( \bar{\mathbf{y}}_n \right)}{\partial y_i}
\frac{\partial g \left( \bar{\mathbf{y}}_n \right)}{\partial y_j}
\hat{C}\left\{  \bar{y}_{i},  \bar{y}_{j} \right\}. \label{5.5}
\end{equation}
In summary, the linearization method for variance estimation can be described as follows: 
\begin{enumerate}
\item Apply Taylor linearization and ignore the remainder terms. 
\item Calculate the variance and covariance terms for each component of $\bar{y}_n$ using the standard variance estimation formula. \item Estimate the partial derivative terms 
($\partial g/ \partial y $)  from the sample observation. 
\end{enumerate}

\begin{example}
If the parameter of interest is $R= \bar{Y}/ \bar{X}$ and we use 
$$\hat{R} = \frac{ \bar{Y}_{\rm HT}}{\bar{X}_{\rm HT} }    $$
to estimate $R$, we can apply Taylor linearization to get
$$ \hat{R} = R + \bar{X}^{-1} \left( \bar{Y}_{\rm HT} -
R \bar{X}_{\rm HT} \right) + O_p \left( n^{-1} \right) $$ 
and the variance estimation formula in (\ref{5.5}) reduces to 
\begin{equation}
 \hat{V} \left( \hat{R} \right) \doteq  \bar{X}_{\rm HT}^{-2} \hat{V} \left( \bar{Y}_{\rm HT} \right) +
\bar{X}_{\rm HT}^{-2} \hat{R}^2 \hat{V} \left( \bar{X}_{\rm HT} \right) - 2
\bar{X}_{\rm HT}^{-2} \hat{R} \widehat{Cov} \left( \bar{X}_{\rm HT}, \bar{Y}_{\rm HT}
\right). \label{5.6} \end{equation}
If the variances and covariances of $\bar{X}_{\rm HT}$ and $\bar{Y}_{\rm HT}$ are estimated by HT variance estimation formula, (\ref{5.6}) can be estimated by  
$$\hat{V} \left( \hat{R} \right) \doteq  \frac{1}{\hat{X}_{\rm HT}^2} \sum_{
i \in A} \sum_{j \in A} \frac{\pi_{ij} - \pi_i \pi_j }{\pi_{ij} }
\frac{{e}_i }{\pi_i } \frac{ {e}_j}{\pi_j} $$ where   $e_i= y_i - \hat{R}
x_i$.
\label{example10.2}
\end{example}

Using the result in Example \ref{example10.2}, the variance of the ratio estimator  $\hat{Y}_R = X
\hat{R}$ is estimated by 
\begin{equation}
\hat{V} \left( \hat{Y}_{R} \right) \doteq
\left(\frac{X}{\hat{X}_{\rm HT}} \right)^2 \sum_{ i \in A} \sum_{j \in
A} \frac{\pi_{ij} - \pi_i \pi_j }{\pi_{ij} } \frac{e_i }{\pi_i }
\frac{e_j }{\pi_j} , \label{rvar3}
\end{equation}
which is obtained by multiplying $\hat{X}_{\rm HT}^{-2} X^2$ to the variance formula in 
 (\ref{rvar2}).  The variance estimator in (\ref{rvar3}) is asymptotically equivalent to the linearization variance estimator in (\ref{rvar2}), but it gives a more adequate measure of the conditional variance of the ratio estimator, as advocated by  \cite{royall1981}. 
 More generally,  
\cite{sarndal1989} 
 proposed using 
 \begin{equation}
\hat{V}\left( \hat{Y}_{\rm GREG} \right) = \sum_{i \in A} \sum_{j \in A}
\frac{\pi_{ij} - \pi_i \pi_j }{\pi_{ij}} \frac{g_i e_i}{\pi_i}
\frac{g_j e_j }{\pi_j} \label{vgreg}
\end{equation}
as the conditional variance estimator of the GREG estimator of the form 
$ \hat{Y}_{\rm GREG} = \sum_{i \in
A}\pi_i^{-1} g_i y_i$, where 
\begin{equation}
 g_i = \mathbf{X}'\left( \sum_{i \in A} \frac{1}{ \pi_i c_i } \bx_i \bx_i'\right)^{-1}
 \frac{1}{ c_i} \bx_i
\label{5.10}
\end{equation}
  and 
  $$e_i=y_i - \bx_i'\left( \sum_{i \in
A} \frac{1}{ \pi_i c_i } \bx_i \bx_i'\right)^{-1}\sum_{i \in A}
\frac{1}{ \pi_i c_i } \bx_i y_i .$$
The  $g_i$ in 
 (\ref{5.10}) is the factor to applied to the design weight to satisfy the calibration constraint.

\begin{example}
We now consider the estimation of the variance of the post-stratification estimator in (\ref{post}). To estimate the variance, the unconditional variance estimator is given by 
\begin{equation}
 \hat{V}_u = \frac{N^2}{n} \left( 1- \frac{n}{N} \right)
\sum_{g=1}^G \frac{n_g-1}{n-1} s_g^2 \label{uncondv}
\end{equation}
where 
$ s_g^2 =
\left( n_g - 1\right)^{-1} \sum_{i \in A_g} \left( y_i - \bar{y}_g
\right)^2 $. On the other hand, the conditional variance estimator in 
 (\ref{vgreg}) is given by 
\begin{equation}
\hat{V}_c =  \left( 1- \frac{n}{N} \right) \frac{n}{n-1}
\sum_{g=1}^G \frac{N_g^2}{n_g}  \frac{n_g-1}{n_g} s_g^2. 
\label{condv}
\end{equation}
Note that the conditional variance formula in (\ref{condv}) is very similar to the variance estimation formula under stratified sampling.  \label{expost}
\end{example}

\section{Replication approach to variance estimation}

We now consider an alternative approach to variance estimation that is based on several replicates of the original sample estimator. Such a replication approach does not use Taylor linearization, but instead generates several resamples and computes a replicate to each resample. Variability between replicates is used to estimate the sampling variance of the point estimator. Such a replication approach often uses repeated computation of the replicates using a computer. Replication methods include the random group method, jackknife, balanced repeated replication, and the bootstrap method. More details of the replication methods can be found in \cite{wolter2007}.

\subsection{Random group method}\index{Random group method}

The random group method was first used in jute acreage surveys in Bengal (Mahalanobis; 1939). The random group method is useful in understanding the basic idea for the replication approach to variance estimation. In the random group method, $G$ random groups are constructed from the sample, and the point estimate is calculated for each random group sample and then combined to obtain the final point estimate. Once the final point estimate is constructed, its variance estimate is calculated using the variability of the $G$ random group estimates. There are two versions of the random group method. One is independent random group method and the other is non-independent random group method. We first consider the independent random group method. 

The independent random group method can be described as follows. 
\begin{description}
\item{[Step 1]} Using the given sampling design, select the first random group sample, denoted by $A_{(1)}$, and compute $\hat{\theta}_{(1)}$, an unbiased estimator of $\theta$ from the first random group sample. 
 \item{[Step 2]}  Using the same sampling design, select the second random group sample, independently from the first random group sample, and compute $\hat{\theta}_{(2)}$ from the second random group sample. 
\item{[Step 3]} Using the same procedure, obtain $G$ independent estimate $\hat{\theta}_{(1)}, \cdots, \hat{\theta}_{(G)}$ from the $G$ random group sample.  
\item{[Step 4]} The final estimator of $\theta$ is 
\begin{equation}
 \hat{\theta}_{\rm RG} = \frac{1}{G} \sum_{k=1}^G \hat{\theta}_{(k)}
\label{rgest}
\end{equation}
and its unbiased variance estimator is 
\begin{equation}
 \hat{V} \left(
\hat{\theta}_{\rm RG} \right) = \frac{1}{G} \frac{1}{G-1} \sum_{k=1}^G
\left( \hat{\theta}_{(k)}  - \hat{\theta}_{RG} \right)^2. \label{vrg}
\end{equation}
\end{description}

Since $\hat{\theta}_{(1)}, \cdots, \hat{\theta}_{(G)}$ are independently and identically distributed, the variance estimator in (\ref{vrg}) is unbiased for the variance of $\hat{\theta}_{RG}$ in (\ref{rgest}). Such independent random group method is very easy to understand and is applicable for complicated sampling designs for selecting $A_{(g)}$, but the sample allows for duplication of sample elements and the variance estimator may be unstable when $G$ is small. 

We now consider non-independent random group method, which does not allow for duplication of the sample elements. In the non-independent random group method, the sample is partitioned into $G$ groups, exhaustive and mutually exclusive, denoted by 
$A=\cup_{g=1}^G A_{(g)}$ and then apply the sample estimation method for independent random group method, by treating the non-independent random group samples as if they  are independent. The following theorem expresses the bias of the variance estimator for this case. 

\begin{theorem}
Let $\hat{\theta}_{(g)}$ be an unbiased estimator of $\theta$, calculated from the $g$-th random group sample $A_{(g)}$ for $g=1, \cdots, G$.  Then the random group variance estimator (\ref{vrg}) satisfies 
\begin{equation}
 E\left\{\hat{V} \left(
\hat{\theta}_{\rm RG} \right) \right\} -V\left(\hat{\theta}_{\rm RG} \right)
= - \frac{1}{G\left( G-1 \right)} {\sum\sum}_{ i \neq j} Cov \left(
\hat{\theta}_{(i)}, \hat{\theta}_{(j)} \right) \label{biasrg}
\end{equation}
\end{theorem}
\begin{proof}
By (\ref{rgest}), we have 
\begin{equation}
 V\left( \hat{\theta}_{\rm RG} \right) = \frac{1}{G^2}\left\{  \sum_{i=1}^G
V\left( \hat{\theta}_{(i)} \right)  +  {\sum\sum}_{i \neq j} Cov
\left( \hat{\theta}_{(i)}, \hat{\theta}_{(j)}\right)\right\}
\label{vrg2}
\end{equation}
and, since 
$$ \sum_{i=1}^G \left( \hat{\theta}_{(i)} - \hat{\theta}_{\rm RG}
\right)^2 =\sum_{i=1}^G \left( \hat{\theta}_{(i)} - {\theta}
\right)^2  - G \left(  \hat{\theta}_{\rm RG} - \theta \right)^2
$$
and using $E\left( \hat{\theta}_{(i)} \right)=\theta$, 
\begin{eqnarray*}
&& E \left\{\sum_{i=1}^G \left( \hat{\theta}_{(i)} - \hat{\theta}_{RG}
\right)^2 \right\} \\ &=&\sum_{i=1}^G V\left( \hat{\theta}_{(i)}
\right)- G  \times V\left( \hat{\theta}_{RG} \right) \\
&=&\left( 1 - G^{-1} \right) \sum_{i=1}^G V\left( \hat{\theta}_{(i)}
\right)- G^{-1}{\sum\sum}_{i \neq j} Cov \left( \hat{\theta}_{(i)},
\hat{\theta}_{(j)}\right)
\end{eqnarray*}
which implies 
\begin{equation*}
 E\left\{\hat{V} \left(
\hat{\theta}_{\rm RG} \right) \right\} =G^{-2} \sum_{i=1}^G V\left(
\hat{\theta}_{(i)} \right)- G^{-2}\left( G-1 \right)^{-1}
{\sum\sum}_{i \neq j} Cov \left( \hat{\theta}_{(i)},
\hat{\theta}_{(j)}\right).
\end{equation*}
Thus, using (\ref{vrg2}), we have (\ref{biasrg}).
\end{proof}

The right side of (\ref{biasrg}) is the bias of $\hat{V} \left( \hat{\theta}_{\rm RG} \right)
$ as an estimator for the variance of $\hat{\theta}_{\rm RG}$. Such bias becomes zero if the sampling for random groups is a with-replacement sampling. For without-replacement sampling, the covariance between the two different replicates is negative, and so the bias term becomes positive. The relative amount of bias is often negligible. The following example computes the amount of the relative bias.

\begin{example}
Consider a sample of size $n$ obtained from simple random sampling of a finite population of size $N$. 
Let $b=n/G$ be an integer value that is the size of $A_{(g)}$ such that $A=\cup_{g=1}^G A_{(g)}$. The sample mean of $y$ obtained from $A_{(g)}$ is denoted by $\bar{y}_{(g)}$ and the overall mean of $y$ is given by 
$$ \hat{\theta} = \frac{1}{G} \sum_{g=1}^G \bar{y}_{(g)}. $$
In this case, $\bar{y}_{(1)}, \cdots,
\bar{y}_{(G)}$ are not independently distributed but are identically distributed with the same mean. By  (\ref{vrg2}), we have 
$$ Var
\left( \hat{\theta} \right) = \frac{1}{G} V\left(\bar{y}_{(1)}
\right) + \left( 1 - \frac{1}{G} \right) Cov \left( \bar{y}_{(1)},
\bar{y}_{(2)} \right)$$ and since 
\begin{equation*}
 V\left(\bar{y}_{(1)} \right) = \left( \frac{1}{b} - \frac{1}{N} \right) S^2,
 \end{equation*}
we have 
\begin{equation*}
 Cov\left(\bar{y}_{(1)}, \bar{y}_{(2)} \right) = -\frac{ S^2}{N} .
 \end{equation*}
 Thus, the bias in (\ref{biasrg}) reduces to 
\begin{eqnarray*}
Bias\left\{ \hat{V} \left( \hat{\theta}_{RG} \right) \right\}
 &=&- Cov\left(\bar{y}_{(1)}, \bar{y}_{(2)} \right) \\
 & =& \frac{S^2}{N} 
\end{eqnarray*}
which is often negligible. Therefore, the random group variance estimator (\ref{vrg}) can be safely used to estimate the variance of $\bar{y}_n$ in simple random sampling. 
\end{example}

The random group method provides a useful computational tool for estimating the variance of point estimators. However, the random group method is applicable only when the sampling design for $A_{(g)}$ has the same structure as the original sample $A$. The partition $A=\cup_{g=1}^G A_{(g)}$ that leads to the unbiasedness of $\hat{\theta}_{(g)}$ is not always possible for complex sampling designs.

\section{Jackknife method}

Jackknife was first introduced by  \cite{quenouille1956} 
to reduce the bias of the ratio estimator and then was suggested by \cite{tukey1958} to be used for variance estimation. Jackknife is very popular in practice as a tool for variance estimation. 

To introduce the idea of \cite{quenouille1956}, suppose that $n$ independent observations of $(x_i, y_i)$ are available that are generated from a distribution with mean $(\mu_x, \mu_y)$. If the parameter of interest is
$\theta = \mu_y / \mu_x $, then the sample ratio $\hat{\theta} =
\bar{x}^{-1} \bar{y} $ has a bias of order $O\left(n^{-1} \right)$. That is, we have 
$$ E\left( \hat{\theta} \right) = \theta + \frac{C}{n} + O\left(
n^{-2} \right).$$ 
If we delete the $k$-th observation and recompute the ratio
$$\hat{\theta}^{(-k)}= \left( \sum_{i \neq k } x_i
\right)^{-1} \sum_{i \neq k} y_i ,$$ 
we obtain 
$$ E\left( \hat{\theta}^{(-k)} \right) = \theta + \frac{C}{n-1} + O\left(
n^{-2} \right).$$
Thus, the jackknife pseudo value defined by $\hat{\theta}_{(k)} = n
\hat{\theta} - \left( n-1 \right)\hat{\theta}^{(-k)}$ can be used to compute 
\begin{equation}
\hat{\theta}_{(.)} = \frac{1}{n} \sum_{k=1}^n
\hat{\theta}_{(k)}
\label{5.3.2}
 \end{equation}
 which has bias of order $O\left( n^{-2} \right)$. Thus, the jackknife can be used to reduce the bias of nonlinear estimators.

Note that if  $\hat{\theta}=\bar{y}$ then $\hat{\theta}_{(k)}=y_k$.  
Tukey (1958) suggested using $\hat{\theta}_{(k)}$ as an approximate IID  observation to obtain the following jackknife variance estimator. 
\begin{eqnarray*}
 \hat{V}_{\rm JK} \left( \hat{\theta}  \right) &\doteq&
 \frac{1}{n} \frac{1}{n-1} \sum_{k=1}^n \left( \hat{\theta}_{(k)} - \bar{\theta}_{(.)}
 \right)^2\\
&=& \frac{n-1}{n}  \sum_{k=1}^n \left( \hat{\theta}^{(-k)} -
\bar{\theta}^{(.)}
 \right)^2
 \end{eqnarray*}
 For the special case of 
  $\hat{\theta}_n = \bar{y} $, we obtain 
$$ \hat{V}_{\rm JK} = \frac{1}{n} \frac{1}{n-1} \sum_{i=1}^n \left(
y_i - \bar{y} \right)^2 = \frac{s^2}{n}  $$
and the jackknife variance estimator is algebraically equivalent to the usual variance estimator of the sample mean under simple random sampling ignoring the finite population correction term. 

If we are interested in estimating the variance of $\hat{\theta}= f\left( \bar{x} , \bar{y} \right) $, using 
 $\hat{\theta}^{(-k)}= f\left(
\bar{x}^{(-k)} , \bar{y}^{(-k)} \right)$, we can construct 
\begin{equation}
 \hat{V}_{\rm JK} \left( \hat{\theta}  \right)
= \frac{n-1}{n}  \sum_{k=1}^n \left( \hat{\theta}^{(-k)} -
\hat{\theta} \right)^2 \label{jk}
 \end{equation}
 as the jackknife variance estimator of $\hat{\theta}$. The jackknife replicate 
  $\hat{\theta}^{(-k)}$  is computed by using the same formula for 
  $\hat{\theta}$ using the jackknife weight $w_i^{(k)}$ instead of the original weight 
 $w_i=1/n$, where 
 $$ w_i^{(-k)} = \left\{ \begin{array}{ll}
\left(n-1\right)^{-1} & \mbox{ if } i \neq k \\
0 & \mbox{ if } i =k .
\end{array}
\right. $$

To discuss the asymptotic property of the jackknife variance estimator in (\ref{jk}),  we use the following Taylor expansion, which is often called second-type Taylor expansion. 
\begin{lemma}
 Let $\left\{ X_n , W_n \right\}$ be a sequence of random
 variables such that
 $$ X_n = W_n + O_p \left( r_n \right) $$
 where $r_n \rightarrow 0 $ as $n \rightarrow \infty $. If
 $g\left( x\right)$ is a function with $s$-th continuous
 derivatives in the line segment joining $X_n$ and $W_n$ and the
 $s$-th order partial derivatives are bounded, then
\begin{eqnarray*}
 g\left( X_n \right) &=& g\left( W_n \right) + \sum_{k=1}^{s-1}
\frac{1}{k !}  g^{(k)} \left( W_n \right) \left( X_n - W_n
\right)^k + O_p \left( r_n^s \right)
\end{eqnarray*}
where $g^{(k)}\left( a \right)$ is the $k$-th derivative of $g\left(
x \right)$ evaluated at $x=a$.
\end{lemma}

Now,   since  $\bar{y}^{(-k)} - \bar{y}=
\left(n-1 \right)^{-1} \left( \bar{y}-y_k \right)$, we have  
$$ \bar{y}^{(-k)} - \bar{y} = O_p \left( n^{-1}
\right). $$ 
For the case of 
 $\hat{\theta}= f\left(
\bar{x} , \bar{y} \right) $, we can apply the above lemma to get 
\begin{eqnarray*}
\hat{\theta}^{(-k)} - \hat{\theta} &=& \frac{ \partial f }{\partial
x} \left( \bar{x}, \bar{y} \right) \left(
\bar{x}^{(-k)} - \bar{x} \right) \\
&& +\frac{ \partial f }{\partial y} \left( \bar{x}, \bar{y} \right)
\left( \bar{y}^{(-k)} - \bar{y} \right) + o_p \left( n^{-1} \right)
\end{eqnarray*}
so that 
\begin{eqnarray*}
&& \frac{n-1}{n} \sum_{k=1}^n \left( \hat{\theta}^{(-k)} -
\hat{\theta}\right)^2  \\ &=& \left\{ \frac{
\partial f }{\partial x} \left( \bar{x}, \bar{y} \right)\right\}^2 \frac{n-1}{n}
\sum_{k=1}^n  \left( \bar{x}^{(-k)} - \bar{x} \right)^2 +\left\{
\frac{ \partial f }{\partial y} \left( \bar{x}, \bar{y} \right)
\right\}^2 \frac{n-1}{n} \sum_{k=1}^n \left( \bar{y}^{(-k)}
- \bar{y} \right)^2\\
&& + 2  \left\{ \frac{
\partial f }{\partial x} \left( \bar{x}, \bar{y} \right)\right\}
\left\{ \frac{ \partial f }{\partial y} \left( \bar{x}, \bar{y}
\right) \right\} \frac{n-1}{n} \sum_{k=1}^n \left( \bar{x}^{(-k)} -
\bar{x} \right) \left( \bar{y}^{(-k)} - \bar{y} \right)+ o_p \left(
n^{-1} \right).
\end{eqnarray*}
Thus, the jackknife variance estimator is asymptotically equivalent to the linearized variance estimator. 
That is, the second-type Taylor linearization leads to 
\begin{eqnarray*}
\hat{V}_{\rm JK} \left(\hat{\theta}\right) &=& \left\{ \frac{
\partial f }{\partial x} \left( \bar{x}, \bar{y} \right)\right\}^2 \hat{V}\left( \bar{x} \right)
 +\left\{
\frac{ \partial f }{\partial y} \left( \bar{x}, \bar{y} \right)
\right\}^2 \hat{V}\left( \bar{y} \right)\\
&& + 2  \left\{ \frac{
\partial f }{\partial x} \left( \bar{x}, \bar{y} \right)\right\}
\left\{ \frac{ \partial f }{\partial y} \left( \bar{x}, \bar{y}
\right) \right\} \hat{C}\left( \bar{x}, \bar{y} \right)+ o_p \left(
n^{-1} \right).
\end{eqnarray*}

The above jackknife method is constructed under simple random sampling. For multistage stratified sampling, jackknife replicates can be constructed by removing a cluster for each replicate. Let 
$$ \hat{Y}_{\rm HT} = \sum_{h=1}^H \sum_{i=1}^{n_h} w_{hi} \hat{Y}_{hi}
$$
be the HT estimator of $Y$ under stratified multistage cluster sampling. The jackknife weights are constructed by deleting a cluster sequentially as 
$$w_{hi}^{(-gj)} = \left\{
\begin{array}{ll}
0 & \mbox{ if } h=g \mbox{ and } i=j \\
\left(n_h-1\right)^{-1} n_h w_{hi} &\mbox{ if } h=g \mbox{ and } i\neq j \\
w_{hi} & \mbox{otherwise}
\end{array}
\right.
$$
and the jackknife variance estimator is computed by 
\begin{equation}
\hat{V}_{\rm JK} \left( \hat{Y}_{\rm HT} \right) = \sum_{h=1}^H
\frac{n_h-1}{n_h} \sum_{i=1}^{n_h} \left( \hat{Y}_{\rm HT}^{(-hi)} -
\frac{1}{n_h} \sum_{j=1}^{n_h}\hat{Y}_{\rm HT}^{(-hj)} \right)^2
\label{jk2}
\end{equation}
where 
$$\hat{Y}_{\rm HT}^{(-gj)} =\sum_{h=1}^H \sum_{i=1}^{n_h} w_{hi}^{(-gj)} \hat{Y}_{hi}  .$$

The following theorem presents the algebraic property of the jackknife variance estimator in (\ref{jk2}). 

\begin{theorem}
The jackknife variance estimator  in (\ref{jk2}) is algebraically equivalent to the simplified variance estimator in (\ref{simplev}).
\end{theorem}

If  $\hat{\theta}=f\left( \hat{X}_{\rm HT}, \hat{Y}_{\rm HT} \right)$ then the jackknife replicates are constructed as
$\hat{\theta}^{(-gj)}=f\left(
\hat{X}_{\rm HT}^{(-gj)}, \hat{Y}_{\rm HT}^{(-gj)} \right)$. In this case, the jackknife variance estimator is computed as 
\begin{equation}
\hat{V}_{\rm JK} \left( \hat{\theta} \right) = \sum_{h=1}^H
\frac{n_h-1}{n_h} \sum_{i=1}^{n_h} \left( \hat{\theta}^{(-hi)} -
\frac{1}{n_h} \sum_{j=1}^{n_h}\hat{\theta}^{(-hj)} \right)^2
\label{jk3}
\end{equation}
or, more simply, 
\begin{equation}
\hat{V}_{\rm JK} \left( \hat{\theta} \right) = \sum_{h=1}^H
\frac{n_h-1}{n_h} \sum_{i=1}^{n_h} \left( \hat{\theta}^{(-hi)}
-\hat{\theta} \right)^2. \label{jk4}
\end{equation}
 The asymptotic properties of the above jackknife variance estimator under stratified multistage sampling can be established. For references, see \cite{krewski1981} or \cite{shao1995}.

\begin{example}
We now revisit Example 10.3 to estimate the variance of the post-stratification estimator using the jackknife. 
For simplicity, assume that SRS for the sample. The post-stratification estimator is calculated as
$$ \hat{Y}_{\rm post} = \sum_{g=1}^G N_g \bar{y}_g , $$
where
$\bar{y}_g = n_g^{-1} \sum_{i \in A_g} y_i$. Now, the $k$-th replicate of $\hat{Y}_{\rm post}$ is 
$$ \hat{Y}_{\rm post}^{(-k)} = \sum_{g \neq h} N_g \bar{y}_g + N_h \bar{y}_h^{(-k)}$$
where  $\bar{y}_h^{(-k)} = \left(n_h -1 \right)^{-1}
 \left( n_h \bar{y}_h - y_k \right)$. Thus, 
 \begin{eqnarray*}
 \hat{Y}_{\rm post}^{(-k)} -\hat{Y}_{\rm post} &= &
 N_h \left( \bar{y}_h^{(-k)} - \bar{y}_h \right) \\
 &=& N_h \left(n_h -1 \right)^{-1}\left(\bar{y}_h - y_k \right)
 \end{eqnarray*}
and 
 \begin{eqnarray*}
 \hat{V}_{\rm JK} \left(\hat{Y}_{\rm post} \right)
 &=& \frac{n-1}{n} \sum_{k=1}^n  \left( \hat{Y}_{\rm post}^{(-k)}
 -\hat{Y}_{\rm post} \right)^2 \\
 &=&\frac{n-1}{n} \sum_{g=1}^G N_g^2 \left(n_g -1 \right)^{-1} s_g^2, 
\end{eqnarray*}
which, ignoring $n/N$ term, is asymptotically equivalent to the conditional variance estimator in   (\ref{condv}).
\end{example}

%% file: chapters/chapter11.tex
\setcounter{chapter}{10} 
\chapter{Two-phase sampling}

\section{Introduction}

The two-phase sampling design is a sampling design where sample selection is performed in two phases. 
In the first phase,  the auxiliary variable $\bx$ is observed,  and in the second phase, the study variable $y$ is observed. The second phase sample is a subset of the first phase sample. 

Two-phase sampling is particularly attractive when the cost of observing $\bx$ is relatively low compared to  the cost of observing $y$. To formalize the concept, two-phase sampling can be described as follows:
\begin{description}
\item{[Step 1]}  From the finite population, s a first-phase sample $A_1$ of size $n_1$ is selected, and the auxiliary variable  $\mathbf{x}$ is observed. 
\item{[Step 2]} The first-phase sample $A_1$ is treated as a finite population, and a second-phase sample $A_2$ of size $n_2$ is selected from it. Study variable $y$ is observed in the second-phase sample. The probability of selection for the second phase sample is often determined by the values of $\mathbf{x}$ obtained from the first-phase sample. 
\end{description}

Since the second-phase sample selection probability depends on the observed values of the first-phase sample, the sample inclusion probability for the second-phase sample is a random variable that changes as the first-phase sample changes. In this case, the standard Horvitz-Thompson (HT) estimation theory is not directly applicable.

To discuss this further, note that the overall first-order inclusion probability for the two-phase sampling design is
$$ \pi_i = Pr \left( i \in A_2 \right) = \sum_{A_2; i \in A_2 } P\left( A_2 \right) $$
where 
$$ P\left( A_2 \right) = \sum_{A_1 ; A_2 \subset A_1 } P_2\left( A_2 \mid A_1 \right) P_1 \left( A_1 \right) .$$
Here, $P_1 \left( \cdot \right)$ is the sample selection probability for the first-phase sample, and $P_2\left( \cdot \mid A_1 \right)$ is the sample selection probability for the second-phase sample, which is conditional on the first-phase sample. Thus, 
\begin{eqnarray*}
 \pi_i &=&\sum_{ \left\{ A_2; \ i \in A_2 \right\}} \sum_{\left\{ A_1 ; A_2 \subset A_1 \right\} } P_2\left( A_2
\mid A_1 \right) P_1 \left( A_1 \right) \\
&=&\sum_{\left\{ A_1; \ i \in A_1 \right\} } \sum_{\left\{ A_2 ; A_2
\subset A_1 \& i \in A_2 \right\}} P_2\left(
A_2 \mid A_1 \right) P_1 \left( A_1 \right)\\
&=&\sum_{\left\{ A_1; \ i \in A_1 \right\} } \sum_{\left\{ A_2 ;\ i
\in A_2 \right\}} P_2\left(
A_2 \mid A_1 \right) P_1 \left( A_1 \right).
\end{eqnarray*}
If we define the conditional first-order inclusion probability for the second-phase sampling as 
$$ \pi_{i \mid A_1 }^{(2)} = P \left( i \in A_2 \mid A_1 \right) $$
then the first order inclusion probability is 
\begin{eqnarray}
 \pi_i
&=&\sum_{ A_1; \ i \in A_1  } \pi_{i \mid A_1 }^{(2)}  P_1 \left( A_1 \right) \label{7.1}\\
&=& E_1\left( \pi_{i \mid A_1}^{(2)} \right), \notag
\end{eqnarray}
where  $E_1 \left( \cdot \right)$ is the expectation with respect to the first-phase sampling. 
Here, the conditional first-order inclusion probability  $\pi_{i \mid A_1 }^{(2)}$
 is a random variable in the sense that it is a function of $\bx$ in the first phase sample $A_1$. The conditional expectation (\ref{7.1}) cannot be computed because we have a single  realization of  the first-phase sample $A_1$. 

If the sampling design satisfies the invariance condition, as is the case in two-stage sampling, then the conditional first-order inclusion probability $\pi_{i|A_1}^{(2)}$ is equal to the unconditional first-order inclusion probability $\pi_i^{(2)}$. On this case, by (\ref{7.1}), we have 
\begin{eqnarray*}
 \pi_i
&=&\sum_{ A_1; \ i \in A_1  }  P_1 \left( A_1 \right) \pi_{i}^{(2)} \\
&=& \pi_i^{(1)} \pi_i^{(2)}.
\end{eqnarray*}
In this scenario, where the sampling design satisfies the invariance condition, the Horvitz-Thompson (HT) estimator can be directly implemented.

To discuss unbiased estimation for two-phase sampling, let's first consider the following quantity
$$ \hat{Y}_1 = \sum_{i \in A_1} \frac{y_i}{\pi_i^{(1)}} .$$
This is an unbiased estimator for the population total of $y$. The task is now to construct an unbiased estimator of $\hat{Y}_1$ using the  two-phase sample. 

Applying the Horvitz-Thompson (HT) estimation idea conditionally,  we obtain 
\begin{equation}
 \hat{Y}_{\rm DEE} = \sum_{ i \in A_2 }\frac{y_i}{\pi_i^{(1)} \pi_{i\mid A_1}^{(2)}}.
 \label{7.2}
 \end{equation}
If we define $\pi_i^* =\pi_i^{(1)} \pi_{i\mid A_1}^{(2)} $,  then (\ref{7.2}) can be written as $\hat{Y}^* = \sum_{ i \in A_2 } y_i/ \pi_i^*$. This  conditionally unbiased estimator 
 (\ref{7.2}) is sometimes called  the \emph{double expansion estimator} \citep{kott1997}. 

 The double expansion estimator (DEE) is conditional unbiased to $\hat{Y}_1$, which is itself an unbiased estimator of the population total $Y$. Therefore, the  DEE is unbiased unconditionally.

 The variance of the DEE  is given by  
\begin{eqnarray}
V\left( \hat{Y}_{\rm DEE} \right) &=& V \left\{ E\left( \hat{Y}_{\rm DEE} \mid A_1
\right) \right\} + E \left\{ V \left( \hat{Y}_{\rm DEE} \mid A_1 \right)
\right\} \notag \\
&=& V \left\{  \sum_{i \in A_1}
\frac{y_i}{\pi_i^{(1)}}  \right\} + E \left\{ \sum_{i \in A_1}
\sum_{j \in A_1} \left( \pi_{ij  \mid A_1}^{(2)} -  \pi_{i  \mid
A_1}^{(2)} \pi_{j  \mid A_1}^{(2)} \right) \frac{y_i}{\pi_i^*}
\frac{y_j}{\pi_j^*} \right\}. \notag \\ \label{eq:11.3}
\end{eqnarray}
Here, $\pi_{ij|A_1}^{(2)} = Pr\left( i \in A_2, j \in A_2 \mid A_1 \right)$ is the conditional joint inclusion probability. The variance expression in (\ref{eq:11.3}) has two parts: the first part is the variance due to the first-phase sampling, and the second part is the variance due to the second-phase sampling.

\section{Two-phase sampling for stratification} 

Stratified sampling is one of the most popular sampling methods for improving the efficiency of point estimators. To apply stratified sampling, the stratification variables need to be available for the finite population. If that is not the case, the two-phase sampling approach can be used.

Let $\mathbf{x}_i = (x_{i1}, \ldots, x_{iH})$ be the vector of stratification variables, where $x_{ih}$ takes the value 1 if unit $i$ belongs to stratum $h$, and 0 otherwise. In this case, the auxiliary variable $\mathbf{x}$ is not available in the finite population.

The two-phase sampling for stratification can be described as follows: 

\begin{enumerate}
\item  Perform a simple random sample (SRS)  of size $n$ from the finite population and obtain
 $\sum_{i \in A_1}
\mathbf{x}_i = \left( n_{1}, n_2, \ldots, n_H \right)$ where
$n=\sum_{h=1}^H n_{h}$.
\item Among the   $n_h$ elements in each stratum,  select  $r_h$ elements   independently by SRS, where $r_h$ is determined after when the first-phase sample is selected. 
\end{enumerate}
This two-phase sampling approach allows the stratification variables to be obtained, even when they are not directly available in the finite population.

In this two-phase sampling design, the realized sample size $n_h$ in stratum $h$ is a random variable, and it follows an approximate multinomial distribution:
$$ (n_1, n_2, \ldots, n_H) \sim MN \left( n; W_1, W_2, \ldots, W_H \right) $$ 
where $MN \left( n; \mathbf{p} \right)$ denotes the multinomial distribution with parameter $\mathbf{p}$, and $W_h = N_h/N$ is the population proportion of stratum $h$.

This means that the realized sample sizes across the strata follow a multinomial distribution, with the population stratum proportions $W_h$ as the underlying probabilities, and the total sample size $n$ as the number of trials.

In this two-phase sampling, the double expansion estimator of the population mean of $y$ is 
\begin{equation}
\hat{\bar{Y}}_{\rm tp} = \sum_{h=1}^H  w_h  \bar{y}_{h2} \label{7.4}
\end{equation}
where  $w_h={n_h}/{n}$ and $\bar{y}_{h2} = r_h^{-1} \sum_{i \in
A_2} x_{ih} y_i $. Since the expectation of $w_h = n_h/n$ is  $W_h =
N_h/N$, the double expansion estimator in (\ref{7.4}) can be viewed as the stratified sample estimator when the stratum size $W_h$ is unknown. The total variance is, by (\ref{eq:11.3}), obtained as 
\begin{equation}
V\left(\hat{\bar{Y}}_{\rm tp} \right) = \left( \frac{1}{n} - \frac{1}{N}
\right)S^2  + E\left\{ \sum_{h=1}^H \left( \frac{n_h}{n} \right)^2
\left( \frac{1}{r_h} - \frac{1}{n_h} \right)s_{h1}^2 \right\}
\label{7.5}
\end{equation}
where 
$$s_{h1}^2 = \frac{1}{n_h-1} \sum_{i \in A_1} x_{ih} \left( y_i -
\bar{y}_{h1} \right)^2
$$
and  $\bar{y}_{h1}=n_h^{-1} \sum_{i \in A_1} x_{ih} y_i $.

Also, if we define $\bar{y}_1 =n^{-1} \sum_{h=1}^H n_h \bar{y}_{h1}$ and 
\begin{eqnarray*}
s_1^2 &=&   \frac{1}{n-1} \sum_{i \in A_1} \left( y_i - \bar{y}_1
\right)^2
\end{eqnarray*}
then we have  $E\left( s_1^2 \right) = S^2$ and 
\begin{eqnarray*}
s_1^2
 & = &
\sum_{h=1}^{H}\left\{ \frac{n_{h}}{n-1}  \left(
 \bar{y}_{h1} - \bar{y}_1 \right)^2 + \frac{ n_h -1 }{n-1}  s_{h1}^2
 \right\}.
\end{eqnarray*}
Thus,  (\ref{7.5}) is approximately equal to 
\begin{equation}
V\left(\hat{\bar{Y}}_{\rm tp} \right) = E \left\{   n^{-1}
\sum_{h=1}^{H} w_{h} \left(
 \bar{y}_{h1} - \bar{y}_1 \right)^2 +
 \sum_{h=1}^{H} r_{h}^{-1} w_{h}^2 s_{h1}^2 \right\}.
\label{7.6}
\end{equation}
Here, the finite population correction term is ignored. The variance formula in (\ref{7.6}) is expressed as the sum of two terms. One is a function of the between-stratum variance, and the other is a function of the within-stratum variance.

In computing the between-stratum variance, we used the full sample size $n$. However, in computing the within-stratum variance, we only used the smaller sample size $r_h$ within each stratum.

Therefore, if the between-stratum variance is larger than the within-stratum variance, the efficiency of the two-phase sampling estimator increases. This is because the between-stratum variance is estimated more precisely using the larger sample size $n$, compared to the within-stratum variance estimated with the smaller sample size $r_h$.

In addition, the variance formula in (\ref{7.6}) provides a way to estimate the variance. Since (\ref{7.6}) is expressed as an expectation of  quantities that can be computed from the first-phase sample, we can use $\bar{y}_{h2}$ and $s_{h2}^2$ instead of $\bar{y}_{h1}$ and $s_{h1}^2$, respectively, in (\ref{7.6}). Thus, 
\begin{equation}
\hat{V}\left(\hat{\bar{Y}}_{\rm tp} \right) = E \left\{   n^{-1}
\sum_{h=1}^{H} w_{h} \left(
 \bar{y}_{h2} - \hat{\bar{Y}}_{tp} \right)^2 +
 \sum_{h=1}^{H} r_{h}^{-1} w_{h}^2 s_{h2}^2 \right\}
\label{7.6b}
\end{equation}
is an approximately unbiased estimator of the variance in (\ref{7.6}). \cite{rao1973} developed a formal theory for the two-phase sampling for stratification. Alternatively,   
instead of (\ref{7.6b}), we can also use the Jackknife method to estimate the variance of the two-phase sampling estimator. See \cite{kim2006}
 for more details.

To compare the variance (\ref{7.6}) of the two-phase sampling estimator with that of the simple random sampling (SRS) estimator of equal sample size $r=\sum_{h=1}^H r_h$, note that 
\begin{equation*}
V\left(\hat{\bar{Y}}_{\rm SRS} \right) = E \left\{   r^{-1}
\sum_{h=1}^{H} w_{h} \left(
 \bar{y}_{h1} - \bar{y}_1 \right)^2 +r^{-1}
 \sum_{h=1}^{H}  w_{h} s_{h1}^2 \right\}. 
\end{equation*}
Therefore, the difference in variances is 
\begin{equation*}
V\left(\hat{\bar{Y}}_{\rm SRS} \right) - V\left(\hat{\bar{Y}}_{\rm tp}
\right) = E\left\{\left( \frac{1}{r}- \frac{1}{n} \right)
\sum_{h=1}^{H} w_{h} \left(
 \bar{y}_{h1} - \bar{y}_1 \right)^2 +  \sum_{h=1}^{H}
\left( \frac{1}{r} - \frac{w_h}{r_h}\right) w_{h}s_{h1}^2 \right\}.
\end{equation*}
The first term is always positive, as $r < n$. The second term is zero under proportional allocation ($r_h = r w_h$), but it can be made positive for optimal allocation.

This shows that the two-phase sampling estimator can be more efficient than the SRS estimator, depending on the relative magnitudes of the between-stratum and within-stratum variances and by choosing the optimal sample-size allocation for the second-phase sampling.  

To discuss optimal allocation, first consider the cost function. The total cost can be expressed as 
$$ C= c_1 n + \sum_{h=1}^H c_{2h} r_h $$
and, writing $\nu_h = r_h/n_h$, the optimal allocation can be determined by minimizing 
$$ V= \frac{1}{n} \left\{  \left( S^2 - \sum_{h=1}^H W_h S_h^2 \right)
+ \sum_{h=1}^H W_h S_h^2 \frac{1}{\nu_h}  \right\}   $$ 
subject to 
$$ C= n \left( c_1  + \sum_{h=1}^H c_{2h} W_h \nu_h \right) .$$
To find the optimal allocation, we have only to find the set of $\nu_h$'s that minimizes  $V \times C$. The optimal solution is 
\begin{equation}
\nu_h^* = \left( \frac{c_1}{c_{2h} } \times \frac{ S_h^2 }{ S^2 -
\sum_{h=1}^H W_h S_h^2   } \right)^{1/2} \label{7.7}
\end{equation}
which lead to 
\begin{equation}
\frac{r_h^*}{n^*} =  W_h \nu_h^* = W_h \left(
\frac{c_1}{c_{2h}}\right)^{1/2} \left( \frac{ S_h^2 }{ S^2 -
\sum_{h=1}^H W_h S_h^2   } \right)^{1/2}. \label{7.8}
\end{equation}
If $c_{2h}=c_2$ and $S_h^2 = S_w^2$ for all $h$, then 
\begin{equation}
\frac{r^*}{n^*} =     \left( \frac{c_1}{c_{2}}\right)^{1/2} \left(
\frac{ 1}{\phi -  1 } \right)^{1/2}  \label{7.9}
\end{equation}
where 
$$\phi = \frac{S^2}{S_w^2}$$
denotes the relative efficiency due to stratification (under proportional allocation). If $c_2=10 c_1$ and $\phi=2$ then
$r/n=\sqrt{0.1}=0.32$.

The two-phase sampling for stratification is also a useful method for estimating parameters of subpopulations defined by specific eligibility criteria. For example, if the population of interest is not the entire population, but a subpopulation with certain attributes, and information on these attributes is not available in the population frame, the two-phase sampling approach can be employed. In the first phase, the information on eligibility can be obtained from the initial sample. This allows the population to be stratified based on the eligibility criteria. Then, in the second phase, a sample is selected from the stratum of eligible groups, and the variable of interest $y$ is observed.

For instance, consider a survey where the target population is households with members over the age of 60. Since there may not be a household population frame with age group information, the two-phase sampling for stratification can be applied. The first phase would identify the eligible households, and the second phase would sample from this stratum of eligible households to observe the variable of interest.

This approach enables efficient estimation of parameters for subpopulations defined by specific eligibility criteria, even when that information is not directly available in the population frame.

\section{Regression estimator for two-phase sampling}

In the previous section, the auxiliary variable $\mathbf{x}$ obtained from the first-phase sample was used to design the sampling mechanism for the second-phase sampling. In this section, we consider the case where the auxiliary variable is used in the estimation stage, rather than used in the sampling design.

To illustrate the idea, assume the first-phase sample is a simple random sample of size $n$, and the second-phase sample is also a simple random sample of size $r$ from the first-phase sample. Since we observe $\mathbf{x}_i$ in the first-phase sample, we can compute:
$$\bar{\mathbf{x}}_1 = \frac{1}{n} \sum_{i \in A_1} \mathbf{x}_i$$
from the first-phase sample, and
$$\left(\bar{\mathbf{x}}_2, \bar{y}_2\right) = \frac{1}{r} \sum_{i \in A_2} \left(\mathbf{x}_i, y_i\right)$$
from the second-phase sample.

The two-phase regression estimator is then computed as 
\begin{equation}
\bar{y}_{\rm reg, tp} = \bar{y}_2 + \left(\bar{\bx}_1 - \bar{\bx}_2\right)^\top \hat{\mathbf{B}}
\label{7.10}
\end{equation}
where 
$$ \hat{\mathbf{B}} = \left(\sum_{i \in A_2} \bx_i \bx_i^\top \right)^{-1}\sum_{i \in A_2} \bx_i y_i. $$ 
This two-phase regression estimator utilizes the auxiliary information $\mathbf{x}$ obtained from the first-phase sample to improve the estimation of the population mean $\bar{Y}$.

Now, to discuss its statistical properties, note that $\hat{\mathbf{B}}- \mathbf{B} = O_p (r^{-1/2})$, where $\mathbf{B}$ is the population-level regression coefficient for the regression of $y$ on $\bx$.  Thus, we have 
\begin{equation*}
\bar{y}_{\rm reg, tp} = \bar{y}_2 + \left( \bar{\bx}_1 - \bar{\bx}_2
\right)^\top \mathbf{B} + O_p \left( r^{-1} \right)
\end{equation*}
and obtain 
\begin{equation}
V\left(\bar{y}_{\rm reg, tp} \right) \doteq  \mathbf{B}^\top V \left( \bar{\bx}_1
\right) \mathbf{B} + V\left( \bar{e}_2 \right)
\label{eq:11-12}
\end{equation}
where 
$$\bar{e}_2 = \frac{1}{r} \sum_{i \in A_2} \left(  y_i - \mathbf{x}_i^\top  \mathbf{B} \right) .$$
The variance of $\bar{y}_{\rm reg, tp}$ is then determined. 
In (\ref{eq:11-12}), the total variance  consists of two components:
\begin{enumerate}
\item The variance explained by the regression on $\bar{\bx}_1$, given by $B' V(\bar{\bx}_1) B$.
\item The variance of the mean of the residuals from the second-phase sample, $V(\bar{e}_2)$.
\end{enumerate}

Under the SRS under both phases,  the specific forms of these variance terms are provided below: 
\begin{equation}
V\left(\bar{y}_{\rm reg, tp} \right) \doteq  \left( \frac{1}{n} -
\frac{1}{N} \right) \mathbf{B}^\top  S_{xx} \mathbf{B} +\left( \frac{1}{r} - \frac{1}{N}
\right) S_{ee} \label{7.11}
\end{equation}
where 
\begin{eqnarray*}
 S_{xx} &=& \frac{1}{N-1} \sum_{i=1}^N \left( \bx_i - \bar{\bx}_N \right)\left( \bx_i - \bar{\bx}_N
\right)^\top  \\
 S_{ee} &=& \frac{1}{N-1} \sum_{i=1}^N \left( e_i - \bar{e}_N \right)^2.
\end{eqnarray*}
The first term in  (\ref{7.11}) is the variance term explained by $\bar{\bx}_1$ and the second term in (\ref{7.11}) is  
the variance of $\bar{y}_{\rm reg,tp}$ that is not explained by $\bx$.

Note that the efficiency gain due to two-phase sampling can be computed as
\begin{eqnarray*}
V\left( \bar{y}_2 \right) - V \left( \bar{y}_{\rm reg, tp} \right) &=&
\left( \frac{1}{r} - \frac{1}{n} \right)\mathbf{B}^\top  S_{xx} \mathbf{B}.
\end{eqnarray*}
The gain is larger when the regression relationship is strong (large $B' S{xx} B$) and when the second-phase subsample is much smaller than the first-phase sample (small $r/n$).

For variance estimation, we can use (\ref{7.11}) and estimate each component of the sample separately. That is, we may use 
\begin{equation}
\hat{V}\left(\bar{y}_{\rm reg, tp} \right) \doteq  \left( \frac{1}{n} -
\frac{1}{N} \right) \hat{\mathbf{B}}^\top  \hat{S}_{xx,1} \hat{\mathbf{B}}  +\left( \frac{1}{r} - \frac{1}{N}
\right) \hat{S}_{ee,2} \label{7.12}
\end{equation}
where 
\begin{eqnarray*}
 \hat{S}_{xx,1} &=& \frac{1}{n-1} \sum_{i \in A_1}  \left( \bx_i - \bar{\bx}_1 \right)\left( \bx_i - \bar{\bx}_1
\right)^\top  \\
 \hat{S}_{ee,2} &=& \frac{1}{r-1} \sum_{i \in A_2} \left( y_i - \mathbf{x}_i^\top  \hat{\mathbf{B}} \right)^2.
\end{eqnarray*}
If jackknife is used for variance estimation, one should take into account the sampling variability of $\bar{\bx}_1$ in the two-phase regression estimator. See 
 \cite{sitter1997}  and \cite{fuller1998} for more details.

Now, suppose the total cost $C$ is given by $C=c_0 + c_1 n + c_2 r$, where $c_0, c_1$ and $c_2$ are known constants, and the sample sizes $n$ and $r$ are to be determined. We wish to find the sample sizes that minimize the variance expression in (\ref{7.11}), given the total cost $C$.

Defining $\nu = r/n$, we can express the variance times the term $(C-c_0)$ as 
\begin{eqnarray*}
 V \times (C-c_0)  &=& \left( c_1 + \nu c_2 \right) \left( \mathbf{B}^\top  S_{xx} \mathbf{B} + \frac{1}{ \nu } S_{ee} \right) \\
 &=& \mbox{Const.} + c_2 \mathbf{B}^\top  S_{xx} \mathbf{B} \nu + c_1 S_{ee} \frac{1}{\nu}. 
 \end{eqnarray*}
The optimal value of $\nu$ that minimizes this expression is
\begin{equation}
\nu^* = \left( \frac{c_1}{c_2} \times \frac{S_{ee}}{ \mathbf{B}^\top  S_{xx} \mathbf{B} }  \right)^{1/2}.
\label{9.14b}
\end{equation}
This optimal $\nu^*$ can be interpreted as follows:
\begin{enumerate}
\item If the regression model has strong predictive power, then $\nu^*$ can be reduced, as the regression estimation becomes more effective.
\item If the cost of observing $y$ is quite high compared to the cost of observing $\bx$ (large $c_2/c_1$), then $\nu^*$ can be reduced.
\end{enumerate}
Once $\nu^*$ is obtained from (\ref{9.14b}), the optimal value of $n^*$ can be found by solving the total cost equation 
$$ C= c_0 + c_1 n + c_2 n \nu^* $$
with respect to $n$. 

 We now consider the general situation where the first-phase sampling is not necessarily the simple random sampling. 
 For general two-phase sampling designs, the (debiased) two-phase regression estimator can be written as 
\begin{equation}
\hat{Y}_{\rm reg, tp} = \sum_{i \in A_1} w_{1i} \mathbf{x}_i^{\top} \hat{\bm \beta}_2 + \sum_{i \in A_2} w_{1i} \frac{1}{\pi_{2i \mid 1} } \left( y_i - \mathbf{x}_i^\top  \hat{\bm \beta}_2 \right) .
\label{9.15a}
\end{equation}
where $w_{1i} = 1/\pi_i^{(1)}$ are the first-phase sampling weights, $\pi_{2i|1} = \pi_{i|A_1}^{(2)}$ are the conditional second-phase inclusion probabilities, and $\hat{\bm \beta}_2$ is the regression coefficient estimated from the second-phase sample. 
This formulation generalizes the previous expression for the two-phase regression estimator, which assumed simple random sampling in both phases. The key distinction is the incorporation of the general sampling weights and probabilities to account for more complex sampling designs in the first and second phases.

Notably, the asymptotic unbiasedness of the two-phase regression estimator in equation \eqref{9.15a} holds regardless of the specific method used to obtain the estimate $\hat{\bm \beta}_2$ from the second-phase sample. This property enhances the flexibility and applicability of this estimator in practical settings with diverse sampling schemes.

 Now,  suppose that $\hat{\bm \beta}_2$ takes the form of 
 \begin{equation}
 \hat{\bm \beta}_2 = \left( \sum_{i \in A_2} w_{1i} \mathbf{x}_i \mathbf{x}_i^\top  / c_i \right)^{-1} \sum_{i \in A_2} w_{1i} \mathbf{x}_i y_i / c_i . 
 \label{eq:11-18}
 \end{equation}
Note that, by construction, $\hat{\bm \beta}_2$ in (\ref{eq:11-18}) satisfies 
$$ \sum_{i \in A_2} w_{1i} \left( y_i - \bx_i^\top \hat{\bm \beta}_2  \right) \bx_i/ c_i = \mathbf{0}. $$
 Thus,  
 if we assume that 
 the vector $\bx_i$ includes $c_i \pi_{2i|1}^{-1}$ such that $\bx_i^\top  \mathbf{a} = c_i \pi_{2i|1}^{-1}$ for some vector $\mathbf{a}$, then 
\begin{equation}
\sum_{i \in A_2} w_{1i} \frac{1}{\pi_{2i \mid 1} } \left( y_i - \mathbf{x}_i^\top \hat{\bm \beta}_2 \right)=0
\label{eq:9-ibc}
\end{equation}
holds and 
 the two-phase regression estimator in equation \eqref{9.15a} is algebraically equivalent to the following form: 
\begin{equation}
  \hat{Y}_{\rm  reg, tp} = \sum_{i \in A_1} w_{1i} \mathbf{x}_i^\top  \hat{\bm \beta}_2 .\label{9.15b}
  \end{equation}

This alternative form in equation \eqref{9.15b} takes the structure of a projection estimator, as discussed in \cite{kimrao12}. In other words,  
the two-phase regression estimator can be expressed as a projection of the observations onto the space spanned by the covariates $\mathbf{x}_i$, with the projection coefficients given by $\hat{\bm \beta}_2$. This projection estimator formulation in \eqref{9.15b} is asymptotically design-unbiased, as the IBC  condition (\ref{eq:9-ibc}) holds. 

To compute the asymptotic variance of the two-phase regression estimator in equation \eqref{9.15a}, we define $\bm \beta^*$ to be the probability limit of the regression coefficient estimate $\hat{\bm \beta}_2$. In this case, since $\hat{\bm \beta}_2 - \bm \beta^* = O_p(n^{-1/2})$, we can establish the following linearized representation of the two-phase regression estimator:
\begin{equation}
\hat{Y}_{\rm reg, tp} = \hat{Y}_{{\rm reg, tp, \ell}} + O_p(n^{-1} N)
\label{11-lin}
\end{equation}
where the linearized estimator is given by \begin{equation}
\widehat{Y}_{{\rm reg, tp, \ell}}  = \sum_{i \in A_1} w_{1i}  \mathbf{x}_i^\top  {\bm \beta}^* + \sum_{i \in A_2} w_{1i} \frac{1}{\pi_{2i \mid 1} } \left( y_i - \mathbf{x}_i^\top  {\bm \beta}^*\right). 
\label{eq:11-19}
\end{equation}
This linearized representation will facilitate the subsequent derivation of the asymptotic variance of the two-phase regression estimator, which can be obtained by analyzing the properties of the component terms in equation \eqref{eq:11-19}.

Under some regularity conditions, the asymptotic variance of the two-phase regression estimator $\hat{Y}_{\rm reg, tp}$ is equivalent to the variance of the two-phase linearized  estimator (or difference estimator) defined in equation \eqref{eq:11-19}. This variance can be expressed as 
\begin{equation}
V \left( \hat{Y}_{\rm reg, tp, \ell} \right) = V \left( \sum_{ i \in A_1} w_{1i} y_i \right) + E\left[ V \left\{ \sum_{i \in A_2} w_{1i} \frac{1}{\pi_{2i \mid 1} } \left( y_i - \mathbf{x}_i^\top  {\bm \beta}^*\right) \mid A_1 \right\} \right]. 
\label{9.15d}
\end{equation}
If the second-phase sampling follows a Poisson scheme, then the second term in the above expression can be simplified to
$$
V_2=   E\left\{ \sum_{i \in A_1} w_{1i}^2 \left( \frac{1}{ \pi_{2i \mid 1}} -1 \right) e_i^2 \right\}$$
where $e_i = y_i - \bx_i^\top  \bm \beta^*. $ 

The optimal choice of the conditional second-phase inclusion probabilities $\pi_{2i|1}^*$ that minimizes the variance $V_2$ is 
\begin{equation}
\pi_{2i \mid 1}^* \propto  w_{1i}\left[ E\left\{  e_i^2 \mid \bx_i \right\} \right]^{1/2}
. 
\end{equation}
This optimal design of the second-phase sampling probabilities depends on the conditional variance of the residuals $e_i$ given the covariates $\bx_i$, as well as the first-phase sampling weights $w_{1i}$.


If both phase samples are SRS, the above variance formula reduces to 
\begin{eqnarray*}
V \left(  \hat{Y}_{\rm reg, tp, \ell} \right) &=& \frac{N^2}{n} \left( 1- \frac{n}{N} \right) S_y^2 +  \left( \frac{N^2}{n} \right)^2 \frac{n^2}{r} \left( 1- \frac{r}{n} \right) S_e^2 \\
&=& N^2 \left( \frac{1}{n} - \frac{1}{N} \right) \left( S_y^2 - S_e^2 \right) + N^2 \left( \frac{1}{r} - \frac{1}{N} \right) S_e^2
\end{eqnarray*}
which is consistent with the result in (\ref{7.11}).

We now consider variance estimation for two-phase regression estimator. Note that using the linearization formula in (\ref{11-lin}), we can express 
    \begin{equation} 
    \hat{Y}_{\rm reg, tp, \ell} = \sum_{i \in A_1} w_{1i} \eta_i 
    \label{eq:11-23b}
    \end{equation}
    where 
    $$ \eta_i = \bx_i^\top  \bm \beta^* + \frac{ \delta_i}{ \pi_{2i \mid 1}} \left( y_i - \bx_i^\top  \bm \beta^* \right)$$
    and 
    $$ \delta_i =
    \left\{ 
    \begin{array}{ll}
  1 & \mbox{ if } i \in A_2 \mbox{ when 
 is sampled in } A_1  \\
  0 & \mbox{otherwise.}
    \end{array}
    \right. 
    $$
This formulation expresses the linearized two-phase regression estimator as a weighted sum of  $\eta_i$, where the weights are the first-phase sampling weights $w_{1i}$. 
This representation of the linearized two-phase regression estimator will facilitate the subsequent development of variance estimation procedures, as the variance can be expressed in terms of the influence functions $\eta_i$.

Note that the indicators $\delta_i$ are defined across the entire finite population, not just the sampled units in $A_1$. While we only observe the values of $\delta_i$ for $i \in A_1$, we can still conceptualize the $\delta_i$ as being defined for all $i=1,\ldots,N$ in the population.
    
 Given this view of the finite population, we can apply the \textit{reverse framework} of \cite{fay92}, \cite{shao99}, and \cite{kim2006}. 
   In this framework, the finite population consists of two components:   $\mathcal{R}_N=\{\delta_1,\delta_2,\ldots,\delta_N\}$ and $\mathcal{F}_N = \{ (\bx_i, y_i); i=1, \ldots, N \}$. 
   The sample $A_1$ is then selected from this population according to a probability sampling design. We observe $\delta_i$ and $\bx_i$ for $i \in A_1$, and observe $y_i$ for the units with $\delta_i=1$.
   
Using this setup, the total variance of the linearized two-phase regression estimator $\hat{Y}_{\rm reg, tp, \ell}$ can be expressed as

    \begin{eqnarray*} 
   &&  V \left( \hat{Y}_{\rm reg, tp, \ell} \right) \\
   &=& E \left\{ V \left( \sum_{i \in A_1} w_{1i} \eta_i \mid \mathcal{R}_N, \mathcal{F}_N \right) \mid \mathcal{F}_N \right\}  \\
   && + V\left\{ E \left( \sum_{i \in A_1} w_{1i} \eta_i \mid \mathcal{R}_N, \mathcal{F}_N  \right)  \mid \mathcal{F}_N \right\}   \\
   &=& E \left\{ \sum_{i=1}^N \sum_{j=1}^N \left( \pi_{1ij} - \pi_{1i} \pi_{1j} \right) w_i w_j \eta_i \eta_j \mid \mathcal{F}_N \right\} +V\left\{  \sum_{i=1}^N \eta_i\mid \mathcal{F}_N \right\} \\
   &:=& V_1 + V_2 .
   \end{eqnarray*}
The first term $V_1$ represents the sampling variance of the linearized two-phase regression estimator, treating the $\delta_i$ indicators as fixed. The second term $V_2$ reflects the variance due to the randomness in the $\delta_i$ indicators in the influence function $\eta_i$. 

Observe that if the influence functions $\eta_i$ were observed, the first component $V_1$ of the total variance could be easily estimated using the standard variance estimation formula. Therefore, a linearization-based variance estimator for two-phase sampling can be developed as
    \begin{equation}
     \hat{V} = \sum_{i \in A_1} \sum_{j \in A_1} \frac{\pi_{1ij}- \pi_{1i} \pi_{1j}}{ \pi_{1ij} } \frac{ \hat{\eta}_i}{ \pi_{1i}} 
    \frac{ \hat{\eta}_j}{ \pi_{1j}}     \label{vlin}
    \end{equation}
where the estimated influence function $\hat{\eta}_i$ is given by 
      $$ \hat{\eta}_i = \bx_i^\top  \hat{\bm \beta}_2 + \frac{ \delta_i}{ \pi_{2i \mid 1}} \left( y_i - \bx_i^\top  \hat{\bm \beta}_2   \right).$$

      This linearization variance estimator $\hat{V}$ approximates the first component $V_1$ of the total variance, which can be expressed as
 $$ E \left( \hat{V} \mid \mathcal{R}_N, \mathcal{F}_N \right) \cong 
  V \left( \sum_{i \in A_1} w_{1i} \eta_i \mid \mathcal{R}_N, \mathcal{F}_N \right)  .  $$     
And taking the expectation with respect to the first-phase sampling design, we have
 $$ E \left( \hat{V} \mid  \mathcal{F}_N \right) \cong 
 E \left\{  V \left( \sum_{i \in A_1} w_{1i} \eta_i \mid \mathcal{R}_N, \mathcal{F}_N \right)  \right\} = V_1.  $$
 This implies that the bias of the linearization variance estimator $\hat{V}$ in (\ref{vlin})  is given by 
\begin{eqnarray*} 
\mbox{Bias} \left( \hat{V} \right) 
&=& - V\left\{ \sum_{i=1}^N \eta_i \mid \mathcal{F}_N  \right\} \\
&=& - \sum_{i=1}^N \sum_{j=1}^N Cov( \delta_i, \delta_j) \pi_{2i}^{-1} \pi_{2j}^{-1} e_i e_j . 
\end{eqnarray*} 
This bias term is of order $O(N)$, which is smaller than the leading order of $V_1 = O(n^{-1} N^2)$ when the first-phase sampling rate is negligible (i.e., $n/N \doteq 0$).  
Under Poisson sampling in the second phase, we can estimate the bias by 
$$ 
\widehat{\mbox{Bias}} \left( \hat{V} \right) = - \sum_{i\in A_1} w_{1i} \frac{\delta_i}{\pi_{2i}} \left( \frac{1}{ \pi_{2i} }-1 \right) \left( y_i - \bx_i^\top  \hat{\bm \beta} \right)^2 . 
$$

Replication variance estimator can be easily applied in this case. To do this, we use (\ref{eq:9-ibc}) to get 
$$ \widehat{Y}_{\rm reg, tp} = \sum_{i \in A_1} w_i \mathbf{x}_i^\top  \hat{\bm \beta}_2= \widehat{\mathbf{X}}_1^\top  \hat{\bm \beta}_2 $$
which is a product of two random variables. We can easily construct the replicates of $\hat{Y}_{\rm reg, tp}$ by replacing the original weight $w_i$ by its replication weight  $w_i^{(k)}$ to get 
$$ \hat{Y}_{\rm reg, tp}^{(k)} = \left( \hat{\mathbf{X}}_1^{(k)} \right)^\top \hat{\bm \beta}_2^{(k)}, $$
where 
$$ \hat{\mathbf{X}}_1^{(k)}  =  \sum_{i \in A_1} w_i^{(k)} \mathbf{x}_i $$
and 
$$ \hat{\bm \beta}_2^{(k)}  = \left( \sum_{i \in A_2} w_i^{(k)}  \mathbf{x}_i \mathbf{x}_i^\top / c_i \right)^{-1} \sum_{i \in A_2} w_i^{(k)}  \mathbf{x}_i y_i / c_i .$$ 
Note that we did not  construct the replicate of  $\pi_{2i \mid 1}$ in (\ref{9.15a}). Using (\ref{eq:9-ibc}), the sampling variability of $\pi_{2 i \mid 1}$ is safely transferred to $\hat{\bm \beta}_2$. 
Once the replicates are calculated, we can use the replication variance formula to estimate the variance.   See 
\cite{park2019} for more details of replication variance estimation under two-phase sampling. 

\section{Calibration estimation under two-phase sampling}
The two-phase regression estimator can be implemented using a calibration weighting approach. Specifically, we can obtain the second-phase weights $\omega_{2i}$ by minimizing the following quadratic objective function 
\begin{equation}
 Q(\bm \omega_2 ) = \sum_{i \in A_2} w_{1i}   \left(   \omega_{2i} -     \pi_{2i \mid 1}^{-1}  \right)^2  c_i 
 \label{eq:11-25}
 \end{equation}
subject to the calibration constraint 
\begin{equation}
 \sum_{i \in A_2 }  w_{1i} \omega_{2i}  \bx_i= \sum_{i \in A_1} w_{1i}  \bx_i .
\label{11-calib}
\end{equation}
Using the Lagrange multiplier method, 
 the solution to the optimization problem is expressed as  
 \begin{equation} 
\omega_{2i}^\star =  \pi_{2i \mid 1}^{-1}  +  \bm \lambda^\top  \bx_i/ c_i   ,
\label{eq:11-28} 
\end{equation}
where $\hat{\bm \lambda}$ is determined from (\ref{11-calib}). Thus, the final calibration weights can be written as  
$$ \hat{\omega}_{2i} =  \pi_{2i \mid 1}^{-1}  + \left(\sum_{i \in A_1} w_{1i}  \bx_i  -\sum_{i \in A_2} w_{1i}\pi_{2i \mid 1}^{-1} \bx_i  \right)^\top  \left( \sum_{i \in A_2} w_{1i} \bx_i \bx_i^\top  / c_i  \right)^{-1}  \bx_i/ c_i   $$
and we have
\begin{equation}
\sum_{i \in A_2} w_{1i} \hat{\omega}_{2i} y_i = \sum_{i \in A_1} w_{1i} \mathbf{x}_i^{\top} \hat{\bm \beta}_2 + \sum_{i \in A_2} w_{1i} \frac{1}{\pi_{2i \mid 1} } \left( y_i - \mathbf{x}_i^\top  \hat{\bm \beta}_2 \right) 
\end{equation}
where $\hat{\bm \beta}_2$ is defined in \eqref{eq:11-18}. That is, the weighted sum $\sum_{i \in A_2} w_{1i} \hat{\omega}_{2i} y_i$ is algebraically equivalent to the two-phase regression estimator in equation \eqref{9.15a}. 
This calibration weighting approach provides an alternative implementation of the two-phase regression estimator, which can be useful as the regression coefficients are estimated indirectly from the data. The calibration constraint \eqref{11-calib} ensures that the weighted sum of the first-phase covariates in the second-phase sample matches the total from the first-phase sample, thereby achieving the desired  calibration.

More generally, we can minimize 
\begin{equation}
 Q_G ( \bm \omega_2 ) =  \sum_{i \in A_2} w_{1i} G \left( \omega_{2i} \right) c_i 
 \label{eq:11-20}
 \end{equation}
subject to (\ref{11-calib}) and 
\begin{equation}
\sum_{i \in A_2} w_{1i} \omega_{2i} g_i c_i = \sum_{i \in A_1} w_{1i} g_i c_i 
\label{11-calib2}
\end{equation}
where $g_i = g \left( \pi_{2i \mid 1}^{-1} \right)$ and $g( \omega) = d G( \omega)/ d \omega$. The second constraint (\ref{11-calib2}) is the debiasing calibration constraint to achieve the IBC condition. 

Similarly to Theorem \ref{thm9.5}, we can establish the following theorem. 
\begin{theorem}
Let $\hat{\omega}_{2i}$ be obtained by minimizing (\ref{eq:11-20}) subject to (\ref{11-calib}) and (\ref{11-calib2}). The resulting calibration estimator $\hat{Y}_{\rm gec, tp} = \sum_{i \in A_2} w_{1i} \hat{\omega}_{2i} y_i$ satisfies 
$$ \widehat{Y}_{\rm gec, tp } = \widehat{Y}_{\rm gec, tp,   \ell} + o_p(n^{-1/2} N ), $$
where 
 \begin{equation}
    \widehat Y_{\rm  gec, tp,  \ell}  =\sum_{i \in A_1} w_{1i} \left\{  \bz_i^\top  {\bm \gamma}^*   +   \frac{\delta_i}{\pi_{2i \mid 1}}  \left(   y_i - \bz_i^\top   {\bm \gamma}^* \right) \right\} , 
    \label{eq:11-41}
    \end{equation}
 $\bz_i^\top  =\left(\bx_i^\top , g_i c_i\right) $ and ${\bm \gamma}^*$ is the probability limit of $\hat{\bm \gamma}$ given by      \begin{equation*} 
\hat{\bm \gamma} 
= \left( 
\sum_{i \in A_2} w_{1i}   \frac{1}{g'(d_{2i}) c_i }  \bz_i \bz_i^\top  \right)^{-1} \sum_{i \in A_2} w_{1i} \frac{1}{g'(d_{2i}) c_i }\bz_i y_i , 
\end{equation*}
where $d_{2i} = \pi_{2i \mid 1}^{-1}$. 
\label{thm:11-1}
\end{theorem}
The proof of Theorem 11.1 is very similar to that of Theorem \ref{thm9.5} and is skipped here.  The linearization in (\ref{eq:11-41}) gives the asymptotic equivalence with a two-phase regression estimator. Thus, the variance estimation method in Section 11.3 can be used directly. 

When the auxiliary variables $\bx_i$ exhibit a monotone missing pattern, the calibration weighting approach can be implemented sequentially. 
Suppose the vector of auxiliary variables $\bx_i$ can be partitioned as $\bx_i = (\bx_i^{(1)}, \bx_i^{(2)})$, and the population total of $\bx_i^{(1)}$ is known. In this case,  we can apply the following  two-step calibration method:  
\begin{description}
    \item{[Step 1]} Obtain $w_{1i}$ by solving the calibration problem by minimizing 
    $$Q_1( \omega_1 ) = \sum_{i \in A_1} \left( \omega_{1i} - \pi_{1i}^{-1} \right)^2 $$
    subject to 
    $$ \sum_{i \in A_1} \omega_{1i} \bx_i^{(1)} = \sum_{i=1}^N \bx_i^{(1)}. $$
    \item{[Step 2]} 
    Once the first-phase weights    $w_{1i} = \hat{\omega}_{1i}$ are obtained from [Step 1], 
obtain $\omega_{2i}$ by minimizing $Q(\bm \omega_2)$ in (\ref{eq:11-25}), subject to the calibration constraint 
(\ref{11-calib}). 
\end{description}

This two-step calibration approach leverages the known population total of the first-part of the auxiliary variables $\bx_i^{(1)}$ to first determine the appropriate first-phase weights $w_{1i}$. Then, in the second step, the second-phase weights $\omega_{2i}$ are obtained by calibrating to the full set of auxiliary variables $\bx_i$.

This sequential calibration procedure is particularly useful when the auxiliary information exhibits a monotone missing pattern, allowing for efficient utilization of the available data at each phase of the sampling design.

\section{Non-nested two-phase sampling}

In contrast to the classical two-phase sampling framework, non-nested two-phase sampling involves two independent surveys conducted on the same target population. The key distinction is that the two samples, denoted as $A_1$ and $A_2$, are drawn independently rather than sequentially.  Table \ref{table:11-1} presents the data structure for non-nested two-phase sampling. 

In the non-nested two-phase sampling, 
a large probability sample $A_1$ is drawn from a finite population,  collecting  only the $\bx$ variable, and  a  smaller sample $A_2$ is drawn from the same population, providing information  on both the $y$ and $\bx$ variables.
It is assumed that the observed variable $x$ is comparable in the two surveys. \cite{renssen1997} formally addressed this non-nested two-phase sampling problem  and 
\cite{merkouris2004} extended the idea further to develop regression estimation combining information from multiple surveys. 
\cite{kimrao12} considered the non-nested two-phase sampling in the context of mass imputation combining two independent surveys at the population and domain levels.

\begin{table}[htb]
\caption{Data Structure for non-nested two-phase sampling}
\label{table:1}\par
\vskip .2cm
\centerline{\tabcolsep=3truept\begin{tabular}{|c|cc|}
\hline
Sample  & $X$ & $Y$ \\
\hline
$A_1$ & \checkmark &  \\
$A_2$ & \checkmark & \checkmark  \\
\hline
\end{tabular}}
\label{table:11-1}
\end{table}

To illustrate the non-nested two-phase sampling approach, let's consider the data structure shown in Table \ref{table:11-1}. This setup involves two independent samples, $A_1$ and $A_2$, drawn from the same target population.

From these two samples, we can compute two unbiased estimators of the population total $\mathbf{X} = \sum_{i=1}^N \bx_i$ for the auxiliary variable x: $\hat{\mathbf{X}}_1 = \sum_{i \in A_1} \pi_{1i}^{-1} \bx_i$
and 
$\hat{\mathbf{X}}_2 = \sum_{i \in A_2} \pi_{2i}^{-1} \bx_i$. 
Here, $\pi_{1i}$ and $\pi_{2i}$ represent the inclusion probabilities for samples $A_1$ and $A_2$, respectively.

Both $\hat{\mathbf{X}}_1$ and $\hat{\mathbf{X}}_2$ are unbiased estimators of the population total $\mathbf{X}$ under the respective sampling designs. 
The availability of these two unbiased estimators is a key feature of the non-nested two-phase sampling design, as it provides opportunities for developing enhanced estimation procedures combining information from different sources.  

We can  construct a combined estimator of X, denoted as $\widehat{\mathbf{X}}_c$, as follows:
\begin{equation}
\widehat{\mathbf{X}}_c = W \hat{\mathbf{X}}_1 + (I - W) \hat{\mathbf{X}}_2, 
\label{xgls}
\end{equation}
where $W$ is a $p \times p$ symmetric matrix of constants, and $p = \text{dim}(\bx)$ is the dimension of the auxiliary variable x. The optimal choice of the matrix W can be determined using the Generalized Least Squares (GLS) method. However, other choices of W can also be used. The key idea is to leverage the information from these two independent surveys to obtain a more accurate and efficient estimator of the population total X for the auxiliary variable x, compared to using only one of the surveys alone.

Using the combined estimator  $\widehat{\mathbf{X}}_{\rm c} $ in (\ref{xgls}), we can construct the following projection estimator:  
\begin{equation}
 \widehat{Y}_p = \widehat{\mathbf{X}}_{\rm c}^\top  \hat{\bm \beta}_q  
 \label{11-proj}
 \end{equation}
where the regression coefficient estimator $\hat{\bm \beta}_q$ is defined as 
$$ 
\hat{\bm \beta}_q = \left( \sum_{i \in A_2} \bx_i \bx_i^\top  / q_i \right)^{-1} 
\sum_{i \in A_2} \bx_i y_i / q_i. 
$$
The choice of $q_i$ in the regression coefficient estimator is somewhat arbitrary. Two possible choices are:
\begin{enumerate}
\item Using the model variance under a regression superpopulation model.
\item Using $q_i =\left(  \pi_{2i}^{-2} - \pi_{2i}^{-1}\right)^{-1} $ to compute the design-optimal regression estimator \citep{montanari1987} under Poisson sampling.
\end{enumerate}
The key idea is that by using the combined estimator $\widehat{\mathbf{X}}_c$ in the projection estimator $\widehat{Y}_p$, we can leverage the information from both the $A_1$ and $A_2$ samples to obtain a more accurate prediction of the variable of interest Y. The choice of $q_i$ allows for some flexibility in how the regression coefficient is estimated.

To ensure the design-consistency of the projection estimator in (\ref{11-proj}), we can use the following regression estimator under non-nested two-phase sampling:
\begin{equation}
\widehat{Y}_{\rm tp, reg} = \widehat{Y}_2 + \left( \widehat{\mathbf{X}}_{\rm c} - \widehat{\mathbf{X}}_2 \right)^\top  \hat{\bm \beta}_q
\label{eq:11-28}
\end{equation} 
By the definition of $\widehat{\mathbf{X}}_{\rm c}$, we can also express this as: 
\begin{equation}
\widehat{Y}_{\rm tp, reg} = \widehat{Y}_2 + \left( \widehat{\mathbf{X}}_{1} - \widehat{\mathbf{X}}_2 \right)^\top  \hat{\bm \alpha}_q,
\label{eq:11-28}
\end{equation} 
where $\hat{\bm \alpha}_q = W \hat{\bm \beta}_q$.
The key points are:
\begin{enumerate}
\item The design-consistent regression estimator $\widehat{Y}_{\rm tp, reg}$ is constructed by adding a correction term to the projection estimator $\widehat{Y}_p$ from the second sample.
\item The regression estimator improves the efficiency of the  design unbiased estimator  $\hat{Y}_2$ by substracting the projection of $\hat{Y}_2$ onto the augmentation space \citep{tsiatis2006}, 
the linear space generated by 
 the difference between the combined estimator $\widehat{\mathbf{X}}_{\rm c}$ and the estimator $\widehat{\mathbf{X}}_2$ from the second sample.
\item 
Alternatively, the augmentation space  can be expressed using the difference between the estimators $\widehat{\mathbf{X}}_{1}$ and $\widehat{\mathbf{X}}_2$, weighted by $\hat{\bm \alpha}_q$.
\end{enumerate} 
The goal is to leverage the information from both samples to obtain a design-consistent regression estimator for the variable of interest Y.

Using the standard argument, we can obtain 
\begin{eqnarray}
\widehat{Y}_{\rm tp, reg} 
&=& \widehat{Y}_2 +  \left( \widehat{\mathbf{X}}_{1} - \widehat{\mathbf{X}}_2 \right)^\top  {\bm \alpha}_q^* + O_p(n^{-1} N) \label{eq:11-29}
\end{eqnarray}
where $\bm \alpha_q^*$ is the probability limit of $\hat{\bm \alpha}_q = W \hat{\bm \beta}_q$. By (\ref{eq:11-29}), 
 we can obtain 
\begin{equation}
V \left( \widehat{Y}_{\rm tp, reg}  \right) 
= ({\bm \alpha}_q^* )^\top  V \left( \widehat{\mathbf{X}}_{1}  \right) {\bm \alpha}_q^* + V\left( \hat{u}_2 \right) 
\label{eq:11-31}
\end{equation}
where $\hat{u}_2 = \sum_{i \in A_2} \pi_{2i}^{-1} \left( y_i -  \bx_i'\bm \alpha_q^* \right) $. From the formula in (\ref{eq:11-31}), we can construct a linearized variance estimator. 

Now, we can use the calibration weighting to construct the regression estimator under non-nested two-phase sampling. For  given the design weights $d_{2i} = \pi_{2i}^{-1}$, we find the minimizer of 
    $$Q \left( {\bm \omega } \right) = \sum_{i \in A_2} \left(  {\omega_i} - d_{2i}  \right)^2 q_i $$
    subject to 
    $$ \sum_{i \in A_2} {\omega_i} \bx_i = \widehat{\mathbf{X}}_{\rm c} .$$
    The solution is 
    $$ \hat{\omega}_i = d_{2i} + \left( \widehat{\mathbf{X}}_{\rm c} - \widehat{\mathbf{X}}_{\rm 2} \right)^\top  \left( \sum_{i \in A_2} q_i^{-1} \bx_i \bx_i^\top  \right)^{-1} \bx_i q_i^{-1} . $$
     Note that 
    $$ \sum_{i \in A_2} \hat{\omega}_i y_i = \widehat{Y}_{\rm tp, reg} , $$
    where  $\widehat{Y}_{\rm tp, reg}$ is defined in (\ref{eq:11-28}). 
    Thus, the algebraic equivalence between the  regression estimator and the calibration weighting estimator is established under non-nested two-phase sampling.

We now consider an extension combining information from three independent samples. Suppose that we have three independent samples, denoted by $A, B$, and $C$, and the observed data structure is summarized in Table \ref{table:11-2}.

 \begin{table}[htb] 
  
\caption{Data Structure for three independent samples}

\begin{center}
  \begin{tabular}{c|ccc | cc  c }
  \hline 
     & \multicolumn{3}{|c|}{Case 1} & \multicolumn{3}{|c}{Case 2} \\
     \cline{2-7} 
    Sample& $X_1$&$X_2$ & Y & $X_1$& $X_2$ & Y \\
    \hline
    A& \checkmark & \checkmark & \checkmark  &\checkmark & \checkmark &   \\ \hline
    B& \checkmark & & & \checkmark &  & \checkmark \\
    \hline 
    C& &\checkmark &  & & \checkmark & \\ \hline
  \end{tabular}
\end{center}
\label{table:11-2}
\end{table}

We first consider the data structure in Case 1. 
In this case, we can use two-step estimation procedure for combining all available information. In the first step, we use the generalized least squares (GLS) method to combine information and obtain the best linear unbiased estimator of the population total of $\bx^\top  = (\bx_1^\top , \bx_2^\top )$. Let $\hat{\mathbf{X}}_{\rm GLS}$ be the GLS estimator of $\mathbf{X} = \sum_{i=1}^N \bx_i$.

In the second step, we construct the calibration weights in sample $A$. Given the design weight $d_{i}^{(A)}$ for sample A, we find the minimizer of 
    $$Q_A \left( {\bm \omega } \right) = \sum_{i \in A} \left(  {\omega_i} - d_{i}^{(A)}   \right)^2 q_i $$
    subject to 
    $$ \sum_{i \in A} {\omega_i} \bx_i = \widehat{\mathbf{X}}_{\rm GLS} .$$ 
    
The solution is 
    $$ \hat{\omega}_i = d_{i}^{(A)} + \left( \hat{\mathbf{X}}_{\rm GLS} - \hat{\mathbf{X}}_{A} \right)^\top \left( \sum_{i \in A} q_i^{-1} \bx_i \bx_i^\top  \right)^{-1} \bx_i q_i^{-1} . $$
    Note that 
    $$ \sum_{i \in A} \hat{\omega}_i y_i =\hat{Y}_{\rm reg}    $$
where 
$$ \hat{Y}_{\rm reg} =\hat{Y}_A + \left( \hat{\mathbf{X}}_{\rm GLS} - \hat{\mathbf{X}}_A \right)^\top \hat{\bm \beta}_q  $$
and 
$$\hat{\bm \beta}_q   = \left( \sum_{i \in A} q_i^{-1} \bx_i \bx_i^\top  \right)^{-1}\sum_{i \in A} q_i^{-1} \bx_i  y_i. $$
Thus, the above calibration estimator is algebraically equivalent to the bias-corrected regression  estimator using  $\widehat{\mathbf{X}}_{\rm GLS}$ as the estimated  control for $\mathbf{X}$.

Now, in Case 2, we wish to develop calibration weights for sample $B$. The GLS step is the same as Case 1.  Now, to construct the calibration weights for sample B,  given the design weight $d_{i}^{(B)}$ for sample B, we find the minimizer of 
    $$Q_B \left( {\bm \omega }^{(B)} \right) = \sum_{i \in B} \left(  \omega_i^{(B)} - d_{i}^{(B)}   \right)^2 q_i $$
    subject to 
    $$ \sum_{i \in B} {\omega_i^{(B)} } \bx_{1i} = \hat{\mathbf{X}}_{\rm 1, GLS} .$$
The solution is 
    $$ \hat{\omega}_i^{(B)} = d_{i}^{(B)} + \left( \hat{\mathbf{X}}_{1, \rm GLS} - \hat{\mathbf{X}}_{1 B} \right)^\top \left( \sum_{i \in B} q_i^{-1} \bx_{1i} \bx_{1i}^\top  \right)^{-1} \bx_{1i} q_i^{-1}. $$
Note that 
 $$ \sum_{i \in B} \hat{\omega}_i^{(B)}  y_i = \hat{Y}_B + \left( \hat{\mathbf{X}}_{1, \rm GLS} - \hat{\mathbf{X}}_{1, B} \right)^\top  \hat{\bm \beta}_{1q}  
    $$
where $\hat{Y}_B = \sum_{i \in B} d_i^{(B)} y_i$, $\hat{\mathbf{X}}_{1, B} = \sum_{i \in B} d_i^{(B)} \bx_{1i}$, and 
$$\hat{\bm \beta}_{1q}    = \left( \sum_{i \in B} q_i^{-1} \bx_{1i} \bx_{1i}^\top  \right)^{-1}\sum_{i \in B} q_i^{-1} \bx_{1i}  y_i. $$
  
Therefore, the calibration weighting method can be used to combine the information in the multiple surveys effectively.

\section{Repeated surveys} 

Repeated surveys means that the survey measurement is taken for the same population at different times. For example, in the US Current Population Survey, the employment rates are announced monthly. In this case, the sample selection can be repeated at different times. In the repeated surveys, suppose that there are two different years, there are three different parameters of interest. 
\begin{enumerate}
\item $\bar{Y}_1-\bar{Y}_2$: the difference of the population mean over two different years. 
\item $(\bar{Y}_1+\bar{Y}_2)/2$: overall mean over the two different years. 
\item $\bar{Y}_2$: the population mean at the most recent year. 
\end{enumerate}
The optimal sampling design for $\theta_1= \bar{Y}_1-\bar{Y}_2$ can be quite different from the optimal sampling design for $\bar{Y}=(\bar{Y}_1+\bar{Y}_2)/2$. Let $\bar{y}_1$ and $\bar{y}_2$ be an unbiased estimator of $\bar{Y}_1$ and $\bar{Y}_2$, respectively. If we use $\hat{\theta}=\bar{y}_1-\bar{y}_2$ to estimate $\bar{Y}_1-\bar{Y}_2$, the variance of $\hat{\theta}$ is minimized when $Corr( \bar{y}_1, \bar{y}_2) = 1$. The correlation increases when the sample for $t=1$ is the same as the sample for $t=2$. That is, it is the case where the same sample is used to obtain the measurement for $t=1$ and for $t=2$. Such a sample is often called a panel sample. On the other hand, if the parameter of interest is $\bar{Y}=(\bar{Y}_1+\bar{Y}_2)/2$, the design of the panel sample increases the variance. Independent sample selection, leading to $Corr (\bar{y}_1, \bar{y}_2)=0$, is more efficient than panel sample design if we are interested in estimating $\bar{Y}=(\bar{Y}_1+\bar{Y}_2)/2$.

Now, if we are interested in estimating $\bar{Y}_2$, the following partial overlap sampling design is more efficient than the previous two sampling designs. 

\begin{enumerate}
\item At $t=1$, using an SRS of size $n$ to obtain $A_1$. 
\item At $t=2$, first stratify the finite population into two strata. One is $A_1$ and the other is $U -A_1$. From the first stratum $A_1$, select an SRS of size $n_m$ to obtain $A_{2m}$. From the second stratum $U-A_1$, select an SRS of size $n_u=n-n_m$, independently of the first stratum, to obtain $A_{2u}$. The final sample is $A_2 = A_{2m} \cup A_{2u}$.  The first stratum is called ``matched'' stratum and the second stratum is called ``unmatched'' stratum. 
\end{enumerate}

In this case, the final sample in the matched stratum can be viewed as a two-phase sample where the first phase sample is $A_1$ and the second phase sample is $A_{2m}$. Also, the final sample in the unmatched sample is also a two-phase stratified sample, where the first phase sample is $U -A_1$ and the second-phase sample is $A_{2u}$. 
The following table presents a summary of the two estimators in each two-phase sample. 

\begin{table}[htb]
\begin{center}
\caption{A summary of the two estimators in the setup of sampling two time periods}
\begin{tabular}{c|c|c|c}
\hline Stratum & Population Size  & Sample Size &  Estimator of $\bar{Y}$   \\
 \hline
 Matched  & $n$ &  $n_m$ & $\hat{\bar{Y}}_{m}$ \\
 Unmatched & $N-n$ &  $n_u$ &$\hat{\bar{Y}}_{u}$   \\
 \hline
  & $N$ & $n$ & $\alpha\hat{\bar{Y}}_{u}+ \left( 1- \alpha \right)
  \hat{\bar{Y}}_m $ \\
  \hline 
\end{tabular}
\end{center}
\end{table}

Now, consider the following class of estimators indexed by a constant $\alpha$: 
\begin{equation}
\hat{\bar{Y}}_\alpha =\alpha\hat{\bar{Y}}_{u}+ \left( 1- \alpha
\right)
  \hat{\bar{Y}}_m. 
\label{composite}
\end{equation}
Such an estimator is a weighted average of two unbiased estimators, and is often called a composite estimator. The composite estimator is (approximately) unbiased if the two components, $\hat{\bar{Y}}_{u}$ and $\hat{\bar{Y}}_m$, are (approximately) unbiased. The optimal value of $\alpha$ that minimizes the variance of the composite estimator is 
\begin{equation}
\alpha^* = \frac{V\left( \hat{\bar{Y}}_{m} \right) -
Cov\left(\hat{\bar{Y}}_{u}, \hat{\bar{Y}}_{m} \right) }{V\left(
\hat{\bar{Y}}_{u}\right) + V\left( \hat{\bar{Y}}_{m}\right) -2
Cov\left(\hat{\bar{Y}}_{u}, \hat{\bar{Y}}_{m} \right) }. 
\label{alpha}
\end{equation}
 In this case, the optimal composite estimator is 
\begin{equation*}
\hat{\bar{Y}}_\alpha^* =\alpha^* \hat{\bar{Y}}_{u}+ \left( 1-
\alpha^*  \right)
  \hat{\bar{Y}}_m
\end{equation*}
and its variance is 
\begin{equation}
V \left( \hat{\bar{Y}}_\alpha^* \right) = \frac{V\left(
\hat{\bar{Y}}_{m} \right)V\left( \hat{\bar{Y}}_{u} \right) -\left\{
Cov\left(\hat{\bar{Y}}_{u}, \hat{\bar{Y}}_{m} \right)\right\}^2
}{V\left( \hat{\bar{Y}}_{u}\right) + V\left(
\hat{\bar{Y}}_{m}\right) -2 Cov\left(\hat{\bar{Y}}_{u},
\hat{\bar{Y}}_{m} \right) } .\label{vcomposite}
\end{equation}

To discuss the choice of unbiased estimators, we first note that the measurement at $t=1$ can be treated as the auxiliary variable $x$ and the measurement at $t=2$ can be treated as the study variable $y$.  In the unmatched stratum, there is no auxiliary information, so we use 
$$\hat{\bar{Y}}_{u}=\frac{1}{n_u}  \sum_{ i \in A_{2u}} y_i \equiv \bar{y}_u.  $$
On the other hand, in the matched stratum, we can use auxiliary information to get 
$$\hat{\bar{Y}}_{m}=\bar{y}_m + \left( \bar{x}_1 - \bar{x}_m \right) b $$
where 
\begin{eqnarray*}
\left( \bar{x}_m, \bar{y}_m \right) &=& n_m^{-1} \sum_{ i \in A_{2m}
} \left( x_i, y_i \right) \\
b & =& \left\{ \sum_{ i \in A_{2m} } \left( x_i -\bar{x}_m
\right)^2\right\}^{-1}\sum_{ i \in A_{2m} } \left( x_i -\bar{x}_m
\right) y_i.
\end{eqnarray*}

Thus, the following summary can be made to the two estimators. 
\begin{center}
\begin{tabular}{c|c|c|c}
\hline Stratum  &  Estimator of $\bar{Y}$ &  Expectation &  Variance     \\
 \hline
 Matched  &  $\hat{\bar{Y}}_{m}= \bar{y}_m + \left( \bar{x}_1 - \bar{x}_m \right) b $
 & $\bar{Y} + O\left( n^{-1} \right)$ & $n_m^{-1} \left( 1- \rho^2 \right) S^2 + n^{-1} \rho^2 S^2 $  \\
 Unmatched & $\hat{\bar{Y}}_{u}= \bar{y}_u$    & $\bar{Y}$ &
    $n_m^{-1} S^2$\\
 \hline
  &$\alpha^* \hat{\bar{Y}}_{u}+ \left( 1- \alpha^*  \right)
  \hat{\bar{Y}}_m $  & $\bar{Y} + O(n^{-1} )$ &
  $ (2n)^{-1} S^2 \left( {1+\sqrt{1-\rho^2}} \right)  $ \\
  \hline 
\end{tabular}
\end{center}

Also, we can show that 
\begin{equation}
Cov\left( \hat{\bar{Y}}_{m}, \hat{\bar{Y}}_{u} \right)=0.
\label{compcov}
\end{equation}
Thus, the optimal solution 
 (\ref{alpha}) is 
$$ \alpha^* = \frac{n n_u - n_u^2 \rho^2  }{ n^2 - n_u^2 \rho^2 }$$
which is equal to $\alpha^*=n_u/n$ for 
$\rho=0$ and equal to  $
\alpha^*=n_u/\left( n + n_u \right)$ for $\rho =1$. 

The variance of the optimal composite estimator is then, by  
 (\ref{vcomposite}), 
\begin{equation}
V \left( \hat{\bar{Y}}_\alpha^* \right) = \frac{ n  - n_u \rho^2 }{ n^2 - n_u^2 \rho^2 }
S^2 \ge \frac{1}{n} S^2 
\label{vcomposite2}
\end{equation}
with the equality holds when  $n_m=n$ or $n_m=0$ for $\rho\neq 0$. 

The optimal allocation minimizing the variance 
(\ref{vcomposite2}) is 
$$ \frac{n_u}{n} = \frac{1}{1+\sqrt{1-\rho^2}}, \ \ \ \ \   \frac{n_m}{n} =
\frac{\sqrt{1-\rho^2}}{1+\sqrt{1-\rho^2}}.$$ 
If $\rho$ is large then more sample is selected for the matched stratum. Under this optimal allocation, the variance (\ref{vcomposite2}) reduces to 
\begin{equation}
V \left( \hat{\bar{Y}}_\alpha^* \right) = \frac{ S^2 }{ 2n } \left(
1+ \sqrt{1-\rho^2} \right) \label{vcomposite3}
\end{equation}
which takes the value between $S^2/(2n)$ and $S^2/n$. More discussion on this type of repeated surveys can be found in 
\cite{fuller1990}. 

%% file: chapters/chapter12.tex
\chapter{Unit Nonresponse}
\section{Introduction}

\index{unit nonresponse} 

Most surveys will have nonresponse. There are two types of nonresponse: unit nonresponse, where the survey unit itself is nonresponsive, and item nonresponse, where only some items are nonresponsive. These nonresponses can be handled in two ways: in the case of unit nonresponse, a call-back or
follow-up survey or nonresponse weighting adjustment can be used. For item nonresponse, imputation is commonly used. 

To understand the effect of nonresponse, consider the data structure in Table \ref{table12.1}.  

\begin{table}[htb]
\caption{Data structure with nonresponse}
\begin{center}
\begin{tabular}{cccc}
\hline Stratum  & Population Size  & Population Mean & Sample Size \\
 \hline
Respondents  & $N_R$ & $\bar{Y}_{R}$ & $n_R$\\
Non-respondents  & $N_M$ & $\bar{Y}_M $ & $n_M$  \\
 \hline
 Population  & $N$ & $\bar{Y}$ & $n$ \\
 \hline 
\end{tabular}
\end{center}
\label{table12.1} 
\end{table} 

In Table \ref{table12.1}, the entire population consists of a population of respondents and a population of non-respondents. If we were to draw a sample from the entire population using simple random sampling
from the entire population, we would only be able to observe the respondents in the sample. In this case, if the respondent mean $\bar{y}_R$ is used to estimate the population mean, 
\begin{eqnarray*}
 Bias \left(\bar{y}_R \right) &\doteq&
\left( 1 - \frac{N_R}{N} \right) \left( \bar{Y}_R - \bar{Y}_M \right)\\
V\left(\bar{y}_R \right) &\doteq& \frac{1}{n_R} S_R^2
\end{eqnarray*}

Here, two problems arise. One is that the estimate is biased. The only way the bias becomes zero is if the mean of the respondents and the mean of the nonrespondents become equal. The other problem is that the efficiency of the estimation decreases as the sample size decreases ($n_R < n$). Compensating for this bias and trying to regain lost efficiency is the goal of nonresponse adjustment.

\section{Call-back}\index{call-back}
 When an investigator calls or conducts an interview, there are several sources  for nonresponse. One is refusal, which declines to participate the survey. Another one is not-at-home or non-available at the time of survey. 
  In these cases, one may need to resurvey a subset of the non-respondents reinterviewing a subset of non-respondents, which is often called follow-up survey or callback. The callback    
  can be very effective in reducing non-response bias. 

Suppose that we randomly  select $\nu n_M$ units  from the nonrespondents of size $n_M$, where $0 < \nu < 1$, to obtain the final sample, the final data set can be described as in Table 12.2.

\begin{table}[htb] 
\begin{center} 
\caption{Data structure after a callback from nonrespondents}
\begin{tabular}{ccccc}
\hline
  Stratum   & Population    & Original sample & Final sample  & Final sample  \\
 &  size    & mean & size  & mean  \\
 \hline
Respondents  & $N_R$ & $n_R$ & $r_1=n_R$ & $\bar{y}_1$ \\
 Nonrespondents  & $N_M$  & $n_M$ & $r_2 = \nu n_M$ &  $\bar{y}_2$ \\
 \hline
Population  & $N$  & $n$ & $r$ & \\
\hline 
\end{tabular}
\end{center}
\label{table12.2}
\end{table}

If the original sample was selected by SRS, the final estimator of the final sample can be constructed as
\begin{equation}
\hat{\bar{Y}}_{\rm cb} = \frac{n_R}{n} \bar{y}_1 +
\frac{n_M}{n} \bar{y}_2 .
\label{callback}
\end{equation}
Here, we also assume that $r_2= \nu  n_M$ units are selected by SRS for callbacks. In this case, we can apply the theory of two-phase sampling for stratification in Section 11.2 to show that $\hat{\bar{Y}}_{\rm cb}$  in (\ref{callback}) is design unbiased. Also, by (\ref{7.5}), we obtain 
$$ Var\left(\hat{\bar{Y}} \right) = \frac{1}{n} \left( 1 -
\frac{n}{N} \right) S^2 + \frac{W_2 S_2^2}{n} \left( \frac{1}{\nu} -
1\right)
$$
where  $W_2 = N^{-1} N_M $. In many practical situations, the second phase sample (which is the callback sample) has a higher survey cost than the original sample. Thus, the total cost can be written as 
$$ C = c_0 n + c_1 W_1 n + c_2 W_2 \nu n $$
where $c_1$ is the unit-level survey cost for the original sample and $c_2$ is the unit-level survey cost for the callback sample.

Using  (\ref{7.7}), the optimal sampling rate for the callback sample is calculated as 
$$ \nu^* = \sqrt{ \frac{c_0+c_1 W_1}{c_2} \times \frac{S_2^2 }{ S^2- W_2 S_2^2 } }.$$
The optimal sampling rate for the callback sample balances between the statistical efficiency and the cost efficiency.

\section{Nonresponse weighting adjustment}
\index{nonresponse weighting adjustment;
NWA}

We now consider the case of unit nonresponse in survey sampling. Assume that $\mathbf{x}_i$ is  observed throughout the sample,  and $y_i$ is observed only  if $\delta_i=1$. We  assume that the response mechanism does not depend on $y$. Thus, we assume that
    \begin{equation}
        P( \delta=1 \mid  \bx,y ) = P( \delta=1 \mid  \bx )
        =        p(\bx; \bm \phi_0)
        \label{four}
    \end{equation}
for some unknown vector $\bm \phi_0$. The first equality implies that the data are  missing-at-random (MAR) in the sense of \cite{rubin1976} at the population level.

Given the response model (\ref{four}), a consistent estimator of $\bm \phi_0$ can be obtained by maximizing the pseudo log-likelihood function 
\begin{equation} 
\ell_{\rm p} \left( \bm \phi \right) = \sum_{i \in A} w_i  \left[ \delta_i \log  \{ p(\bx_i; \bm \phi) \}+ (1- \delta_i) \log \{ 1-p(\bx_i; \bm \phi)\} \right]
\label{eq:12-3}
\end{equation} 
over the parameter space of $\bm \phi$, where $w_i$ is the sampling weight for element $i$ in the original sample.  Note that $\hat{\bm \phi}$ obtained from (\ref{eq:12-3}) can be expressed as the solution to the following pseudo-score equation 
\begin{equation}
\hat{S}_p (\bm  \phi) \equiv \sum_{i \in A} w_i    \left\{ \frac{\delta_i}{ p(\bx_i; \bm \phi)} -1   \right\}\mathbf{h}(\bx_i; \bm \phi) = \mathbf{0}
\label{eq:12-4}
\end{equation}
where 
$$\mathbf{h}(\bx; \bm \phi)=p (\bx; \bm \phi)  \cdot \frac{\partial}{ \partial \bm \phi} \mbox{logit} \{ p (\bx;\bm  \phi)\} .$$

Note that (\ref{eq:12-4}) is a calibration equation using $\mathbf{h} (\bx_i; \bm \phi)$ as a control variable for calibration.  We can extend it to the more general case when $\bm \phi$ is estimated through a calibration equation. For example, suppose that $\bm \phi_0$ is estimated by solving the following equation 
\begin{equation}
\hat{U}_b ( \bm \phi) \equiv  \sum_{i \in A} w_i    \left\{ \frac{\delta_i}{ p(\bx_i; \bm \phi)} -1   \right\}\mathbf{b}(\bx_i) = \mathbf{0}
\label{eq:12-4b}\end{equation}
for some $\mathbf{b} (\bx)$ such that the solution to (\ref{eq:12-4b}) exists uniquely almost surely. Note that (\ref{eq:12-4}) is a special case of (\ref{eq:12-4b}) with $\mathbf{h} ( \bx_i) = \mathbf{b} ( \bx_i)$. 

Once $\hat{\bm \phi}$ is computed from  (\ref{eq:12-4b}), the propensity score  (PS) estimator of $Y=\sum_{i=1}^N y_i$ is given by  \index{PS estimator! Survey sampling|(}
\begin{equation}
\widehat{Y}_{\rm PS} = \sum_{i \in A_R} w_i \{  p(\bx_i; \hat{\bm \phi} )\}^{-1}y_i,
\label{8-psa}
\end{equation}
where $A_R=\{ i \in A:  \delta_i =1 \}$ is the set of respondents. Roughly speaking, the response mechanism in (\ref{four}) can be treated as the second-phase sampling in the two-phase sampling setup. The parameter in the second-phase sampling is estimated using the pseudo maximum likelihood method.

The following theorem presents the asymptotic properties of the PS estimator in (\ref{8-psa}). 

\begin{theorem}
Assume that the response model  (\ref{four}) is correctly specified. Let $\hat{\bm \phi}$ be the solution to the estimating equation in (\ref{eq:12-4b}). Under some regularity conditions, the PS estimator in (\ref{8-psa}) satisfies 
\begin{equation}
\widehat{Y}_{\rm PS} = \widehat{Y}_{\rm PS, \ell} + o_p (n^{-1/2} N ) ,
\label{eq:12-5}
\end{equation}
where 
\begin{eqnarray*}
 \widehat{Y}_{\rm PS, \ell} 
 &=&\sum_{i \in A} w_i \left\{   \mathbf{b}_i^\top  \mathbf{B}_1^* + \frac{\delta_i}{p_i } \left( y_i    -   \mathbf{b}_i^\top  \mathbf{B}_1^*  \right) \right\}, 
 \end{eqnarray*}
 with 
\begin{equation}
\mathbf{B}_1^*   =  \left\{
\sum_{i=1}^N p_i^{-1} (1-p_i) \mathbf{h}_i \mathbf{b}_i^\top  \right\}^{-1} \sum_{i=1}^N p_i^{-1} (1-p_i) \mathbf{h}_i y_i
\label{eq:12-7} 
\end{equation}
$p_i=p(\bx_i; \bm \phi_0)$, 
and $\mathbf{h}_i = \mathbf{h}(\bx_i; \bm \phi_0)$. 
\end{theorem}
\begin{proof}
Write $\hat{Y}_{\rm PS} (\bm \phi)= \sum_{i \in A_R} w_i \{  p(\bx_i; {\bm \phi} )\}^{-1}y_i$ and   
define   
$$ 
\widehat{Y}_B (\bm \phi) = \widehat{Y}_{\rm PS} (\bm \phi) -  \hat{U}_b (\bm \phi)^\top  \mathbf{B}  
$$
indexed by $\mathbf{B}$, where $\hat{U}_b (\bm \phi)$ is defined in (\ref{eq:12-4b}).   Note that $\widehat{Y}_B (\hat{\bm \phi})= \widehat{Y}_{\rm PS}$ regardless of the choice of $\mathbf{B}$. If we can   find $\mathbf{B}_1^*$ such that 
\begin{equation}
E\left\{ \frac{\partial}{\partial \bm \phi} \hat{Y}_B( \bm \phi_0) \right\}={\bf 0} 
\label{randles}
\end{equation}
holds 
at $\mathbf{B}=\mathbf{B}_1^*$, then, according to \cite{randles1982aysmptotic}, the effect of estimating $\bm \phi_0$ can be safely ignored for the choice of $\mathbf{B}=\mathbf{B}_1^*$.  
Now, using  
\begin{equation} \frac{\partial}{ \partial \bm \phi} 
\left\{ p(\bx; \bm \phi_0) \right\}^{-1} = - \frac{(1-p_i)}{p_i^2}  \mathbf{h} (\bx; \bm \phi_0),
\label{eq:12-9}
\end{equation}
where $p_i=p(\bx_i; \bm \phi_0)$, we can express  
\begin{eqnarray*} 
E\left\{ \frac{\partial}{\partial \bm \phi} \hat{Y}_B( \bm \phi_0) \right\}
&=& - \sum_{i=1}^N p_i^{-1} (1-p_i) \mathbf{h}_i y_i + \sum_{i=1}^N p_i^{-1} (1-p_i) \mathbf{h}_i \mathbf{b}_i^\top   \mathbf{B} .
\end{eqnarray*} 
Thus, Randles' condition in (\ref{randles}) is satisfied at $\mathbf{B}=\mathbf{B}_1^*$ in (\ref{eq:12-7}). 
Therefore, we have shown that 
$$
\widehat{Y}_{\rm PS } = \widehat{Y}_{\rm PS} (\bm \phi_0) - \hat{U}_b (\bm \phi_0)^\top  \mathbf{B}_1^* + o_p(n^{-1/2} N) $$
which proves (\ref{eq:12-5}). 
\end{proof}

Now, using the two-phase sampling theory, we can obtain that 
$$
E \left( \widehat{Y}_{\rm PS, \ell} \right) = Y 
$$
and 
\begin{eqnarray*}
V\left( \widehat{Y}_{\rm PS, \ell}  \right) &=& V\left( \sum_{i \in A} w_i y_i \right) + E\left\{ \sum_{i \in A} w_i^{2}  (p_i^{-1}-1) \left( y_i  - \mathbf{b}_i^\top  \mathbf{B}^*   \right)^2 \right\},
\end{eqnarray*}
where $p_i = p(\bx_i; \bm \phi_0)$. Therefore, if we choose $\mathbf{b}_i$ in the basis function of $E(Y_i \mid \bx_i)$, the second term of the variance will be reduced.

We now discuss the estimation of the variance of the PS estimators of the form (\ref{8-psa}) where
$\hat{p}_i=p_i(\hat{\bm \phi})$ is constructed to
satisfy (\ref{eq:12-4b}). By (\ref{eq:12-5}), we can write
\begin{equation}
    \widehat{Y}_{\rm PS} =  \sum_{i \in A} w_i {\eta}_i(\bm \phi_0) + O_p \left(n^{-1/2} N \right),
    \label{psalin}
\end{equation}
where
\begin{equation}
     {\eta}_i(\bm \phi) = \mathbf{b}_i(\bm \phi)'\mathbf{B}^* + \frac{\delta_i}{p_i(\bm \phi)} \left\{ y_i - \mathbf{b}_i(\bm \phi)' \mathbf{B}^* \right\} .
     \label{8-eta}
\end{equation}
To derive the variance estimator,  we assume that the variance estimator $\hat{V}=\sum_{i \in A}\sum_{j \in A}\Omega_{ij} q_i q_j$ satisfies
$\hat{V}/V(\hat{q}_{\rm HT}|\mathcal{F}_N)=1+o_p(1)$
for some $\Omega_{ij}$ related to the joint inclusion probability, where $\hat{q}_{HT}=\sum_{i \in A} w_i q_i$ for any $q$ with a finite fourth moment.

\index{Reverse framework}
To obtain the total variance, the \textit{reverse framework} of \cite{fay92}, \cite{shao99}, and \cite{kim2006} is considered. In this framework, the finite population is divided into two groups, a population of respondents and a population of nonrespondents, so the response indicator is extended to the entire population as $\mathcal{R}_N=\{\delta_1,\delta_2,\ldots,\delta_N\}$. Given the population, the sample $A$ is selected according to a probability sampling design. Then, we have both respondents and nonrespondents in the sample $A$.
The total variance of $\hat{\eta}_{HT}=\sum_{i \in A} w_i \eta_i$ can be written as
\begin{equation}
    V(\hat{\eta}_{\rm HT})=E\{V(\hat{\eta}_{\rm HT} \mid  \mathcal{R}_N) \}+V\{E(\hat{\eta}_{\rm HT} \mid \mathcal{R}_N) \}:=V_1 + V_2. 
    \label{totalvar}
\end{equation}
The conditional variance term $V(\hat{\eta}_{HT}|  \mathcal{R}_N)$  in (\ref{totalvar})
can be estimated by
\begin{equation}
    \hat{V}_1= \sum_{i \in A} \sum_{j \in A}\Omega_{ij}\hat\eta_i\hat\eta_j,
    \label{varh1}
\end{equation}
where $\hat{\eta}_i =\eta_i (\hat{\bm \phi})$ is defined in (\ref{8-eta}) with $B^*$ replaced by a consistent estimator such as
$$ \hat{\mathbf{B}}^* = \left( \sum_{i \in A_R} w_i \hat{p}_i^{-2} (1- \hat{p}_i)  \hat{\mathbf{h}}_i{\mathbf{b}}_i^\top  \right)^{-1}  \sum_{i \in A_R} w_i \hat{p}_i^{-2} (1- \hat{p}_i) \hat{\mathbf{h}}_i y_i $$
and $\hat{\mathbf{h}}_i=\mathbf{h}(\bx_i;\hat{\bm \phi})$.
  The second term $V_2$ in (\ref{totalvar}) is
\begin{eqnarray*}
    V\{E(\hat{\eta}_{\rm HT} \mid  \mathcal{R}_N)  \}
    &=&V\left(\sum_{i=1}^N\eta_i \right)= \sum_{i=1}^N\frac{1-p_i}{p_i}\left(y_i  - \mathbf{b}_i^\top  \mathbf{B}^* \right)^2.
\end{eqnarray*}
  A consistent estimator of $V_2$ can be derived as
\begin{equation}
    \hat{V}_2=\sum_{i \in A_R} w_i  \frac{1-\hat{p}_i}{\hat{p}_i^2}
    \left(y_i -  \mathbf{b}_i^\top  \hat{\mathbf{B}}^*  \right)^2.
    \label{varh2}
\end{equation}
Therefore,
\begin{equation}
    \hat{V}\left( \widehat{Y}_{\rm PS} \right) = \hat{V}_1+\hat{V}_2 
    \label{varh}
\end{equation}
is consistent for the variance of the PS estimator defined in (\ref{8-psa}) with $\hat{p}_i=p_i(\hat\phi)$ satisfying (\ref{eq:12-4}), where $\hat{V}_1$ is in (\ref{varh1}) and $\hat{V}_2$ is in (\ref{varh2}). When the sampling fraction $nN^{-1}$ is negligible, that is, $nN^{-1}=o(1)$, the second term $V_2$ can be ignored and $\hat{V}_1$ is a consistent estimator of total variance. Otherwise, the second term $V_2$ should be taken into account so that a consistent variance estimator can be constructed as in (\ref{varh}).

\section{Calibration weighting for nonresponse adjustment}

Let $\hat{Y}_{n}=\sum_{i \in A} w_{1i}  y_i$ be any design consistent  estimator of $Y$ where $w_{1i}$ can be either design weight or the calibration weight. Unfortunately, we cannot compute $\hat{Y}_{n}$ due to nonresponse. Our goal is to construct the final weight $w_{1i} \omega_{2i}$ in $A_R$ so that
$$ \hat{Y}_\omega = \sum_{i \in A_R} w_{1i} \omega_{2i} y_i $$
can be used to estimate $Y$.

By incorporating the two-phase sampling framework, we can apply the calibration method of two-phase sampling introduced in Section 11.4 to solve our problem. Thus, using the generalized entropy function $G( \cdot)$, we can minimize 
$$ Q( \bm \omega_2 ) = \sum_{i \in A_R} w_{1i} G \left( \omega_{2i} \right) c_i $$
subject to 
\begin{equation}
\sum_{i \in A_R}  w_{1i} \omega_{2i} \bx_{i1} =\sum_{i \in A}  w_{1i}  \bx_{i1}
\label{eq:12-16} 
\end{equation}
and  
\begin{equation}
\sum_{i \in A_R } w_{1i} \omega_{2i} \left( \hat{g}_i c_i \right)  = \sum_{i \in A} w_{1i} \left(  \hat{g}_i c_i \right) 
\label{eq:12-17}
\end{equation}
where $\hat{g}_i = g\left( \hat{p}_i^{-1} \right)$ with $g( \omega) = d G( \omega)/ d \omega$, and  
 $\hat{p}_i = p(\bx_i; \hat{\bm \phi})$. Note that $\hat{\bm \phi}$ solves (\ref{eq:12-4b}).

 Now, for a given $\hat{\bm \phi}$, the solution to the above optimization problem can be written as 
 \begin{equation} \hat{\omega}_{2i} = g^{-1} \left( \bx_{i1}^{\top} \hat{\bm \lambda}_1 + \hat{g}_i c_i \hat{\lambda}_2  \right):= g^{-1} \left( \hat{\bz}_i^\top \hat{\bm \lambda} \right), 
 \label{eq:12-19}
 \end{equation}
 where $ \hat{\bm \lambda}^\top = ( \hat{\bm \lambda}_1^\top, \hat{\lambda}_2 )$ solves 
\begin{equation}
\hat{U}_2 \left(  \hat{\bm \phi}, \bm \lambda  \right) \equiv \sum_{i \in A} w_{1i} \left\{ \delta_i g^{-1} ( \hat{\bz}_i^{\top} \bm \lambda ) -1 \right\} \hat{\bz}_i = \mathbf{0} ,
\label{eq:lambda}
\end{equation}
 and $\hat{\bz}_i^\top= (\bx_i^{\top}, \hat{g}_i c_i )$.  The final generalized entropy calibration PS estimator is defined as \begin{equation}\widehat{Y}_{\rm gec, ps} = \sum_{i \in A_R} w_{1i} \hat{\omega}_{2i} y_i  , 
 \label{eq:12-20}
 \end{equation}
 where $\hat{\omega}_{2i}$ is defined in (\ref{eq:12-19}).

 The following theorem presents the asymptotic properties of the generalized entropy calibration  PS estimator in (\ref{eq:12-20}). 

\begin{theorem}
Assume that the response model  (\ref{four}) is correctly specified. Under some regularity conditions, the calibration PS estimator in (\ref{eq:12-20}) satisfies \begin{equation}
\widehat{Y}_{\rm gec, ps} = \widehat{Y}_{\rm gec, ps, \ell} + o_p (n^{-1/2} N ) ,
\label{eq:12-22}
\end{equation}
where 
\begin{eqnarray*}
 \widehat{Y}_{\rm gec, ps, \ell} 
 &=&\sum_{i \in A} w_{1i} \left\{       \mathbf{b}_i^\top \mathbf{B}_1^* + \bz_{i0}^{\top} \mathbf{B}_2^*  +   \frac{\delta_i}{p_i}   \left( y_i    -   \mathbf{b}_i^{\top} \mathbf{B}_1^* - \mathbf{z}_{i0}^\top \mathbf{B}_{2}^*  \right) \right\}, 
 \end{eqnarray*}
\begin{equation}
\mathbf{B}_1^* = \left( 
\sum_{i=1}^N p_i^{-1} (1-p_i) \mathbf{h}_i \mathbf{b}_i^\top \right)^{-1}  \sum_{i=1}^N p_i^{-1} (1-p_i) \mathbf{h}_i \left( y_i- \mathbf{z}_i^{\top} \mathbf{B}_2^*  \right) , 
\end{equation}
  \begin{equation}
\mathbf{B}_2^* =  \left\{
\sum_{i=1}^N \frac{p_i}{g' ( p_i^{-1} ) c_i  }  \bz_{i0} \bz_{i0}^\top  \right\}^{-1} \sum_{i=1}^N \frac{p_i}{g' ( p_i^{-1} ) c_i } 
\bz_{i0}  y_i 
\label{eq:12-24} 
\end{equation}
$p_i=p(\bx_i; \bm \phi_0)$ and  $ \bz_{i0}^\top  = \left(  \bx_i^\top, g( p_i^{-1} ) c_i \right)$. 
\end{theorem}
\begin{proof}
The generalized entropy calibration PS estimator is constructed in two-steps. In the first step, $\hat{\bm \phi}$ is computed from (\ref{eq:12-4b}). In the second step, for a given $\hat{\bm \phi}$, $\hat{\bm \lambda}$ is computed from (\ref{eq:lambda}). Thus, we can express $\hat{\bm \lambda}= \hat{\bm \lambda} (\hat{\bm \phi})$
as 
$\hat{\bm \lambda}$ is a function of $\hat{\bm \phi}$. To properly reflect the uncertainty of the two estimated parameters, we apply Taylor expansion separately in the two-step estimation. 

First, for a given $\hat{\bm \phi}$, the generalized entropy calibration estimator can be expressed as  
$$ \widehat{Y}_{\rm gec, ps} = \widehat{Y}_{\rm gec, ps} (  \hat{\bm \lambda} ), $$
where $\hat{\bm \lambda}$ solves  (\ref{eq:lambda}). In this case, we can apply Theorem \ref{thm:11-1} directly to get 
\begin{equation}
\widehat{Y}_{\rm gec, ps}
= \widehat{Y}_{\rm gec, ps, \ell 2}  + o_p \left(n^{-1/2} N \right) 
\label{eq:12-res1}
\end{equation}
where 
 \begin{equation*}
    \widehat Y_{\rm  gec, ps,  \ell 2}  =\sum_{i \in A_1} w_{1i} \left\{  \hat{\bz}_i^\top  {\bm \gamma}^*    +   \frac{\delta_i}{\hat{p}_{i}}  \left(   y_i - \hat{\bz}_i^\top   {\bm \gamma}^*  \right) \right\} , 
    \end{equation*}
 $\hat{\bz}_i^\top  =\left(\bx_i^\top , \hat{g}_i c_i\right) $ and ${\bm \gamma}^*$ is the probability limit of $\hat{\bm \gamma}$ given by      \begin{equation*} 
\hat{\bm \gamma} 
= \left( 
\sum_{i \in A_2} w_{1i}   \frac{1}{g'(\hat{p}_{i}^{-1} ) c_i }  \hat{\bz}_i \hat{\bz}_i^\top  \right)^{-1} \sum_{i \in A_2} w_{1i} \frac{1}{g'(\hat{p}_{i}^{-1}) c_i }\hat{\bz}_i y_i , 
\end{equation*}
where $\hat{p}_{i} = p ( \bx_i; \hat{\bm \phi}) $.
Now, we can show that $\hat{\bm \gamma}$ converges in probability to $\mathbf{B}_2^*$ in (\ref{eq:12-24}). 
Thus, we can express 
 \begin{equation}
    \widehat Y_{\rm  gec, ps,  \ell 2}  =\sum_{i \in A_1} w_{1i} \left\{  \hat{\bz}_i^\top  \mathbf{B}_2^*    +   \frac{\delta_i}{\hat{p}_{i}}  \left(   y_i - \hat{\bz}_i^\top   \mathbf{B}_2^*  \right) \right\} , 
 \label{eq:12-25} 
    \end{equation}
    where $\mathbf{B}_2^*$ is defined in (\ref{eq:12-24}). Note that we can express $\widehat{Y}_{\rm gec, ps, \ell2}$ in (\ref{eq:12-25}) as 
$ \widehat{Y}_{\rm gec, ps, \ell 2} =\widehat{Y}_{\rm gec, ps, \ell 2} ( \hat{\bm \phi} ) 
$ to emphasize its dependency on $\hat{\bm \phi}$.

Now, it remains to apply Taylor expansion with respect to $\bm \phi$. We can apply a similar argument for Theorem 12.1 to obtain the linearization. 
Define 
$$ \widehat{Y}_{\rm gec, ps, \ell} (\bm \phi, \mathbf{B}_1 ) = 
\widehat{Y}_{\rm gec, ps, \ell 2} 
( {\bm \phi}  )   - \hat{U}_b ( \bm \phi )^\top \mathbf{B}_1 , $$
where $\hat{U}_b (\bm \phi)$ is defined in (\ref{eq:12-4b}), and note that 
\begin{equation}
 \widehat{Y}_{\rm gec, ps, \ell} (\hat{\bm \phi},  \mathbf{B}_1 ) = \widehat{Y}_{\rm gec, ps, \ell 2} (\hat{\bm \phi})
 \label{eq:12-res2}
 \end{equation}
for all $\mathbf{B}_1$. 
Our goal is to find $\mathbf{B}_1^*$  such that 
 \begin{equation}
\widehat{Y}_{\rm gec, ps, \ell} \left( \hat{\bm \phi}, \mathbf{B}_1^*   \right) 
= \widehat{Y}_{\rm gec, ps, \ell} \left( {\bm \phi}^*,  \mathbf{B}_1^* \right) + o_p \left(n^{-1/2} N \right) 
\label{eq:12-res3}\end{equation}
 where $\bm \phi^* =  p \lim \hat{\bm \phi}$. 

 A sufficient condition for (\ref{eq:12-res3}) is 
\begin{equation}
 E\left\{ \nabla_{\phi} \widehat{Y}_{\rm gec, ps, \ell} ( \bm \phi^*, \mathbf{B}_1^* ) \right\}=0 
 \label{eq:randles3}
 \end{equation}
which is the Randles' condition applied to 
$\widehat{Y}_{\rm gec, ps, \ell} ( \bm \phi, \mathbf{B}_1 )$. Now, using 
\begin{eqnarray*} 
E\left\{ \frac{\partial}{\partial \bm \phi} 
\nabla_{\phi} \widehat{Y}_{\rm gec, ps, \ell} ( \bm \phi^*, \mathbf{B}_1  )\right\}
&=& - \sum_{i=1}^N p_i^{-1} (1-p_i) \mathbf{h}_i \left( y_i- \mathbf{z}_i^{\top} \mathbf{B}_2^* \right) \\
&& + \sum_{i=1}^N p_i^{-1} (1-p_i) \mathbf{h}_i \mathbf{b}_i^\top   \mathbf{B}_1 ,
\end{eqnarray*} 
we obtain 
$$
\mathbf{B}_1^* = \left( 
\sum_{i=1}^N p_i^{-1} (1-p_i) \mathbf{h}_i \mathbf{b}_i^\top \right)^{-1}  \sum_{i=1}^N p_i^{-1} (1-p_i) \mathbf{h}_i \left( y_i- \mathbf{z}_i^{\top} \mathbf{B}_2^*  \right) $$
as the solution to (\ref{eq:randles3}). 
Therefore, combining (\ref{eq:12-res1}), (\ref{eq:12-res2}), and (\ref{eq:12-res3}), we can establish (\ref{eq:12-22}).  
\end{proof}
By Theorem 12.2, we  show that $\hat{Y}_{\rm gec, ps}$ is asymptotically unbiased under the assumption that the response model (\ref{four}) is specified correctly. Therefore, double robustness of the regression estimator $\hat{Y}_{\rm gec, ps}$ can be established.

%% file: chapters/chapter13.tex
\chapter{Imputation}
\section{Introduction}

In this chapter, we consider the imputation which is a commonly used method for handling item nonresponse in survey sampling. The primary goal of imputation is to provide consistent point estimates by filling in missing data so that different users analyzing the data will achieve consistent results. If the missing data were left for users to analyze on their own, different users could derive different results for the same parameter, violating the One Number Principle, which is one of the fundamental principles of official statistics. This principle emphasizes that only one estimate should be produced for a given parameter. Failure to adhere to the One Number Principle could result in multiple estimates for a single parameter, causing confusion among users of the statistics and potentially undermining the credibility of the statistics themselves.

So, what is a good method for imputation? To help with understanding, consider the following example.

\begin{example}
Suppose we have a finite population following a bivariate normal distribution model, and a simple random sample is drawn from this population. In this case, the $n$ sample data points $(x_i, y_i)$ will follow the bivariate normal distribution model:
\begin{equation}
\left( \begin{array}{l} x_i 
\\
y_i \end{array} \right) \iid N\left[ \left(\begin{array}{l} \mu_x \\
\mu_y \end{array} \right), \left( \begin{array}{ll} \sigma_{xx} & \sigma_{xy} \\
\sigma_{xy} & \sigma_{yy} \end{array} \right) \right]. \label{biv1}
\end{equation}
Assume that all $x_i$ values are observed, but $y_i$ is observed only for the first $r (<n)$ observations. If we assume that the response probability of $y_i$ depends on $x_i$ but not on $y_i$, the model can be expressed as:
\begin{eqnarray}
x_i &\iid & N\left( \mu_x, \sigma_{xx} \right) \label{biv2} \\
y_i \mid x_i & \iid & N\left( \beta_{0} +\beta_{ 1} x_i, \sigma_{ee}
\right) \notag
\end{eqnarray}
In this case, $\beta_{0}=\mu_y -\beta_{ 1} \mu_x $, $\beta_{ 1} = \sigma_{xx}^{-1} \sigma_{xy} $, and $\sigma_{ee} = \sigma_{yy} - \beta_{1}^2 \sigma_{xx}= \sigma_{yy} \left( 1- \rho^2 \right) $. Under this model, the optimal predictor for the missing data used for imputation is given by:
\begin{equation}
\hat{y}_i = \bar{y}_r +\left( x_i - \bar{x}_r \right) \hat{\beta}_1
\label{imputed}
\end{equation}
Here, $(\bar{x}_r, \bar{y}_r)$ are the sample means of $(x_i, y_i)$ obtained using only the first $r$ responses. Using the imputed data to estimate $\mu_y$, we get:
\begin{equation}
\hat{\mu}_{yI} = \frac{1}{n} \left\{ \sum_{i=1}^r y_i +
\sum_{i=r+1}^n \hat{y}_i \right\}
\end{equation}
This is identical to the maximum likelihood estimator (MLE) of $\mu_y$:
\begin{equation}
\hat{\mu}_y = \bar{y}_r +\left( \bar{x}_n - \bar{x}_r \right)
\hat{\beta}_1 \label{mle}
\end{equation}
as shown by \cite{anderson57}. The variance in this case is:
\begin{equation}
V\left( \hat{\mu}_y \right) = \frac{\sigma_{xx} {\beta}_1^2}{n} +
\frac{\sigma_{ee}}{r}= \frac{\sigma_{yy}}{r}\left[ 1 -\left( 1 -
\frac{r}{n} \right)\rho^2 \right] \label{mlevar}
\end{equation}
Thus, $1 -\left( 1 - n^{-1} r \right)\rho^2$ represents the efficiency gain obtained by regression imputation using the $x$ values observed in the entire sample. Since the imputed values in (\ref{imputed}) result in the MLE for $\mu_y$, this method achieves  efficient estimation.

However, this imputation method does not provide an unbiased estimator for $\sigma_{yy}$. That is, 
\begin{equation*}
\hat{\sigma}_{yyI} = \frac{1}{n} \left\{ \sum_{i=1}^r y_i^2 +
\sum_{i=r+1}^n \hat{y}_i^2 \right\} - \hat{\mu}_{yI}^2. 
\end{equation*}
The imputed estimator of $\sigma_{yy}$, defined as above, approximately satisfies:

$$E \left( \hat{\sigma}_{yy I} \right) \doteq \frac{n-r}{n}\sigma_{yy}\left( 1- \rho^2\right) < \sigma_{yy}. $$
Thus, while this imputation method is optimal for $\mu_y$, it results in an underestimation for $\sigma_{yy}$.
\end{example}

Here, we will only consider  estimation of $Y = \sum_{i=1}^N y_i$. When the missing value for the target variable $y_i$ is imputed as $y_i^*$, the estimator using these imputed values can be expressed as follows:
\begin{equation}
\hat{Y}_I = \sum_{i \in A} \frac{1}{\pi_i} \left\{ \delta_i y_i + (1-\delta_i) y_i^* \right\}
\label{12.4}
\end{equation}
where $\delta_i$ is an indicator variable denoting the response status. Under what conditions is the imputation estimator (\ref{12.4}) unbiased for $Y$? The following lemma provides a sufficient condition.

\begin{lemma}
If $y_i^*$ satisfies the following condition
\begin{equation}
E\left( y_i^* \mid \delta_i=0 \right) = E\left( y_i \mid \delta_i=0 \right),
\label{12-5}
\end{equation}
then the imputation estimator (\ref{12.4}) is unbiased for $Y$.
\end{lemma}

\begin{proof}
First, define $\hat{Y}_{\rm HT} = \sum_{i \in A} \pi_i^{-1} y_i$. Then,
$$ \hat{Y}_I - \hat{Y}_{\rm HT} = \sum_{i \in A} \frac{1}{\pi_i} (1-\delta_i) \left( y_i^* - y_i \right)
    $$
holds. 
Thus,
\begin{equation}
E \left( \hat{Y}_I - \hat{Y}_{\rm HT} \mid \delta_1, \cdots, \delta_N , y_1, \cdots, y_N \right) = \sum_{i=1}^N (1-\delta_i) ( y_i^* - y_i)
\end{equation}
Taking the expectation with respect to the sampling mechanism, we get
\begin{eqnarray*}
E\{ (1-\delta_i) ( y_i^* - y_i) \} &=& E \left[ E\{ (1-\delta_i) ( y_i^* - y_i)\mid \delta_i \} \right] \\
&=& E \left[ (1-\delta_i)E\{ ( y_i^* - y_i)\mid \delta_i \} \right] \\
&=& E \left[ (1-\delta_i)E\{ ( y_i^* - y_i)\mid \delta_i=0 \} \right]. 
\end{eqnarray*}
By condition (\ref{12-5}), the right-hand side is 0, thus: $$ E \left( \hat{Y}_I - \hat{Y}_{\rm HT}  \right) =  \sum_{i=1}^N E\{ (1-\delta_i) ( y_i^* - y_i) \} = 0 .$$
Since $\hat{Y}_{\rm HT}$ is unbiased for $Y$, $\hat{Y}_I$ is also unbiased. 
\end{proof}

So, how can we implement a nonresponse imputation method that satisfies (\ref{12-5})? Assume that we can observe an auxiliary variable $X$ from the entire sample data and that the following condition holds for any measurable set $B$:
\begin{equation}
P(Y \in B \mid X, \delta=1) = P(Y \in B \mid X, \delta=0).
\label{mar}
\end{equation}
This condition is commonly known as Missing At Random (MAR), a concept first proposed by \cite{rubin1976}. In (\ref{mar}), the MAR assumption is made at the population level, not at the sample level.

Under the MAR condition, if we compute the imputed value $y^*$ from $E(Y \mid X, \delta=1)$, we can show that it satisfies (\ref{12-5}). The following lemma summarizes this result.

\begin{lemma}
If $y^*$ is obtained such that
\begin{equation}
E\left( y_i^* \mid \mathbf{x}_i, \delta_i=1 \right) = E\left( y_i \mid \mathbf{x}_i, \delta_i=1 \right)
\label{12-8}
\end{equation}
and the MAR condition (\ref{mar}) holds, then the imputation estimator expressed by (\ref{12.4}) is unbiased.
\end{lemma}

\begin{proof}
Under the MAR condition, the following holds:
\begin{eqnarray*}
E\left( y_i \mid \delta_i=0 \right) &=& E \left\{
E\left( y_i \mid X_i, \delta_i=0 \right) \mid \delta_i=0
\right\} \\ &=& E \left\{ E \left( y_i \mid X_i, \delta_i=1 \right)
\mid \delta_i =0 \right\}.
\end{eqnarray*}
Therefore, if $y^*$ is obtained to satisfy (\ref{12-8}), then (\ref{12-5}) holds, and by Lemma 13.1, the imputation estimator is unbiased.
\end{proof}
Therefore, according to the above lemma, if we estimate the model $f(y \mid x, \delta=1)$ from the response sample and use this to compute the imputed values, we implement an imputation method that satisfies unbiasedness. For the MAR condition in equation (\ref{mar}) to hold, the following must be true:
\begin{equation}
P (\delta=1 \mid X, Y) = P( \delta=1 \mid X)
\label{mar2}
\end{equation}
That is, MAR occurs when $Y$ and $\delta$ are conditionally independent given $X$.

   \section{Regression imputation}


Nonresponse imputation methods can be classified in several ways. First, they can be divided into deterministic imputation and stochastic imputation. \index{deterministic imputation}\index{stochastic imputation} Deterministic imputation refers to cases where the imputed value does not change even if the imputation process is repeated multiple times, as seen in (\ref{imputed}). In contrast, stochastic imputation involves a random component in determining the imputed values, so the imputed values may vary with each repetition of the imputation process. For example, in the previous example using a regression model, the imputed value can be set as:
$$y_i^* = \hat{y}_i + e_i^*$$where $\hat{y}i$ is the deterministic part (as in \eqref{imputed}) and $e_i^*$ is the stochastic part, which can be drawn from $N\left( 0, \hat{\sigma}_{ee} \right)$ or randomly sampled from the residuals of the regression model.

Another way to classify nonresponse imputation methods is into parametric imputation and nonparametric or semiparametric imputation. Parametric imputation\index{parametric imputation} assumes a parametric probability distribution, such as the normal distribution, to derive the imputed values. Nonparametric imputation\index{nonparametric imputation}\index{semiparametric imputation} does not assume any probability distribution for the imputation. Parametric imputation has the advantage of providing consistent estimators for all parameters under the model, but it is not traditionally well accepted in survey sampling and is rarely used for imputation in survey data. Semiparametric imputation assumes models for the mean, variance, and other aspects, such as:
\begin{eqnarray}
E_\zeta\left( y_i \right) &=& \bx_i^\top  \bm \beta \label{regmodel}\\
V_\zeta \left( y_i \right) &=& \sigma^2 \notag
\end{eqnarray}
A commonly accepted imputation method for models like (\ref{regmodel}) is as follows:
\begin{enumerate}
\item Divide the entire sample into $G$ cells using key categorical variables such as gender, age group, and region.
\item Apply a regression model to the entire response data to obtain $\hat{\bm \beta}$ and calculate the residuals $\hat{e}_i = y_i - \bx_i^\top  \hat{\bm \beta}$.
\item Determine the final imputed value as
$$y_i^* = \bx_i^\top  \hat{\bm \beta} + e_i^*$$
where $e_i^*$ is randomly sampled from the residuals within the corresponding cell.
\end{enumerate}
This method is called stochastic regression imputation\index{stochastic regression imputation}. Similar principles can be used to implement stochastic ratio imputation or stochastic cell mean imputation (also known as hot deck imputation). The advantages of stochastic regression imputation include: (1) eliminating nonresponse bias in the estimation of $\bm \beta$, (2) achieving consistent estimation of $\sigma_{yy}$, (3) preserving the correlation between $\bx$ and $y$, and (4) providing a robust imputation method by complementing the residuals' cell means derived from the model (\ref{regmodel}).

\begin{example}
Consider the following Horvitz-Thompson (HT) estimator: 
$$ \hat{Y}_{\rm HT} = \sum_{i \in A} \frac{1}{\pi_i} y_i. $$
When $y_i$ is not observed and only the auxiliary variable $\mathbf{x}_i$ is observed, we often use the regression model (\ref{regmodel}) to obtain imputed values. Under this model, if deterministic imputation is used, the imputation estimator is expressed as:
\begin{equation}
\widehat{Y}_{Id} = \sum_{i \in A} \frac{1}{\pi_i} \left\{ \delta_i y_i + \left( 1- \delta_i \right) \mathbf{x}_i^{\top} \hat{\bm \beta} \right\}
\label{regimp}
\end{equation}
where $\hat{\bm \beta}$ is the solution to
$$ U\left( \bm \beta \right)= \sum_{i \in A} \frac{1}{\pi_i} \delta_i \mathbf{x}_i \left( y_i - \mathbf{x}_i^{\top} {\bm \beta} \right) =
\mathbf{0}. $$
If $\mathbf{x}_i$ includes an intercept term, we have 
$$ \sum_{i \in A} \frac{\delta_i}{\pi_i}   \left( y_i - \bx_i^\top  \hat{\bm \beta} \right) =0 $$ 
and $\hat{Y}_{Id}$ in (\ref{regimp}) can be written as 
\begin{equation}
\widehat{Y}_{Id} = \sum_{i \in A} \frac{1}{\pi_i} \bx_i^{\top}  \hat{\bm \beta}.   \label{regimp2}
\end{equation}
Using a Taylor series expansion used in Lemma \ref{lem:8-1}, we can show that 
$$ \hat{\bm \beta} = \bm \beta +
\left(\sum_{i=1}^N \delta_i \mathbf{x}_i \mathbf{x}_i^\top   \right)^{-1}
\sum_{i \in A} \frac{1}{\pi_i} \delta_i \mathbf{x}_i \left( y_i -
\mathbf{x}_i^\top  \bm \beta  \right) + o_p \left( n^{-1/2}
\right). $$ 

Substituting this into (\ref{regimp2}), we get:
\begin{eqnarray} 
\widehat{Y}_{Id} &=&  \sum_{i \in A} \frac{1}{\pi_i} \bx_i^\top \bm \beta \notag \\
& +&  \left( \sum_{i \in A} \frac{1}{\pi_i} \bx_i \right)^\top \left(\sum_{i=1}^N  \delta_i \mathbf{x}_i \mathbf{x}_i^\top  \right)^{-1} \sum_{i \in A} \frac{1}{\pi_i} \delta_i \bx_i e_i + o_p \left(n^{-1/2} N \right) \notag 
\\
&=& \sum_{i \in A} \frac{1}{\pi_i}
\left\{\mathbf{x}_i^\top  {\bm \beta} + \delta_i  {\mathbf{c}}^\top  \mathbf{x}_i e_i \right\} + o_p \left( n^{-1/2} N \right)
\label{implinear2}
\end{eqnarray}
where
\begin{eqnarray*}
{\mathbf{c}}  &=&\left(\sum_{i=1}^N  \delta_i \mathbf{x}_i \mathbf{x}_i^\top  \right)^{-1} \sum_{i \in A} \pi_i^{-1}  \mathbf{x}_i, \\
{e}_i &=& y_i - \mathbf{x}_i^\top  {\bm \beta}.
\end{eqnarray*}

Defining $\eta_i = \mathbf{x}_i^\top  {\bm \beta} + \delta_i \mathbf{c}^\top  \mathbf{x}_i e_i$, we find that $\hat{Y}_{Id}$ is asymptotically equivalent to $\hat{\eta}_{\rm HT} = \sum_{i \in A} \pi_i^{-1} \eta_i$. 
To study variance estimation, we use the reverse approach of \cite{fay92} and \cite{shao99}. Similarly to (\ref{totalvar}), we apply the reverse approach to $V(\hat{\eta}_{\rm HT})$ to get the following decomposition: 
$$ V \left( \hat{\eta}_{\rm HT} \right) = E \left\{ V \left( \hat{\eta}_{\rm HT} \mid \mathcal{R}_N, \mathcal{F}_N \right) \right\} + V \left\{ E \left( \hat{\eta}_{\rm HT} \mid \mathcal{R}_N, \mathcal{F}_N  \right) \right\}:= V_1 +V_2, $$
where $\mathcal{R}_N = \{ \delta_1, \ldots, \delta_N \}$.

The reference distribution in the conditional expectation given $\mathcal{R}_N$ and $\mathcal{F}_N$ is the sampling mechanism treating $\mathcal{R}_N$ as fixed. 
The first term $V_1$ is estimated by
\begin{equation}
\hat{V}_1  = \sum_{i \in A} \sum_{j \in A} \frac{\pi_{ij} - \pi_i \pi_j }{\pi_{ij} } \frac{\hat{\eta}_i}{\pi_i} \frac{\hat{\eta}_j}{\pi_j}
\label{impvar}
\end{equation}
where 
$$ \hat{\eta}_i =\mathbf{x}_i^\top 
\hat{\bm \beta}  + \delta_i  \hat{\mathbf{c}}^\top  \mathbf{x}_i
\hat{e}_i
$$
and 
\begin{eqnarray*}
\hat{\mathbf{c}} &=&\left(\sum_{i \in A}\pi_i^{-1}  \delta_i \mathbf{x}_i
\mathbf{x}_i^\top 
\right)^{-1} \sum_{i \in A} \pi_i^{-1}   \mathbf{x}_i \\
\hat{e}_i &=& y_i - \mathbf{x}_i^\top \hat{\bm \beta}.
\end{eqnarray*}
Regarding the second term $V_2$, note that 
$$ V_2 = V \left\{ \sum_{i=1}^N \eta_i \right\}= \sum_{i=1}^N \delta_i \left( \mathbf{c}^\top  \bx_i \right)^2  V \left( e_i \mid \bx_i \right). $$
Thus, we can use 
$$ \hat{V}_2= \sum_{i \in A} \frac{\delta_i}{\pi_i} \left( \hat{\mathbf{c}}^\top  \bx_i \right)^2  \hat{e}_i^2 .$$
The second term, $\hat{V}_2$ is of smaller order than the first term is $n/N$ is asymptotically negligible.

If stochastic imputation is used instead of deterministic imputation, where $y_i^*$ is a stochastic imputed value for $y_i$, and
\begin{equation}
y_i^* \mid \hat{\bm \beta} \indep \left( \mathbf{x}_i^\top  \hat{\bm \beta}, \hat{\sigma}_{ee} \right),
\label{imputed2}
\end{equation}
the imputation estimator is
\begin{equation}
\hat{Y}_{Is} = \sum_{i \in A} \frac{1}{\pi_i} \left\{ \delta_i y_i + \left( 1- \delta_i \right) y_i^* \right\}.
\end{equation}
Since $\hat{Y}_{Is} = \hat{Y}_{Id} + \left( \hat{Y}_{Is} - \hat{Y}_{Id} \right)$ and using 
$Cov \left(\hat{Y}_{Id} , \hat{Y}_{Is} - \hat{Y}_{Id} \right) = 0$, the variance estimator can be derived as:
\begin{equation}
V \left( \hat{Y}_{Is}\right) = V \left( \hat{Y}_{Id} \right) + V\left( \hat{Y}_{Is} - \hat{Y}_{Id} \right).
\end{equation}The first term on the right-hand side is estimated using (\ref{impvar}), and the second term is estimated as:
$$\hat{Y}_{Is} - \hat{Y}_{Id} = \sum_{i \in A}\frac{1}{\pi_i} \left( 1- \delta_i \right)
\left( y_i^* - \mathbf{x}_i^\top  \hat{\bm \beta} \right).$$ If each imputed value is independently drawn as in (\ref{imputed2}), then

$$\hat{V} \left( \hat{Y}_{Is} - \hat{Y}_{Id} \right)= \sum_{i \in A}\frac{1}{\pi_i^2} \left( 1- \delta_i \right)
\left( y_i^* - \mathbf{x}_i^\top \hat{\bm \beta} \right)^2 $$can be easily implemented. More details on this can be found in \cite{kim2009}.\end{example}

\section{Model-based imputation}

The regression imputation introduced in the previous section uses the linear regression model to construct the imputed values. More generally, we can use a statistical model to develop a model-based imputation.

To explain the idea, let’s consider that $\mathbf{x}_i$ is observed for the entire sample data, but $y_i$ is only observed for respondents. In such cases, the most natural approach is to find the conditional distribution $f(y \mid \mathbf{x})$ and use this conditional distribution to obtain the imputed values. In sample surveys, it is crucial to distinguish between two models: one is the distribution of the population data before sampling (superpopulation model), and the other is the model followed by the sample data (sample model). The sample model may differ from the population model because, using Bayes' theorem, we can express
$$ f_s \left( y \mid x \right) = f_p \left( y \mid x \right) \frac{P \left( I=1 \mid x, y\right) }{ P \left( I=1 \mid x\right) }, $$
where $ f_s( \cdot ) $ is the probability density function of the sample model, and $ f_p( \cdot) $ is the probability density function of the population model. If
\begin{equation}
P \left( I=1 \mid x, y\right) = P \left( I=1 \mid x\right),
\label{noninformative}
\end{equation}
then the two models are the same. The case where equation (\ref{noninformative}) holds is called noninformative sampling \citep{sudgen1984}. In general, the distribution of the sample data and the population data are different, and equation (\ref{noninformative}) will not hold. \index{noninformative sampling}

Generally speaking, we are  interested in the parameters of the population model. We assume the MAR in the population level: 
\begin{equation}
f \left( y \mid x,  \delta = 1 \right) = f \left( y \mid x,  \delta = 0 \right).
\label{pmar}
\end{equation}
Condition (\ref{pmar}), called the population MAR (PMAR), is the classical MAR condition in survey sampling literature. \index{PMAR} See \cite{berg2016imputation} for further discussion on this topic.

When $x_i$ is always observed and PMAR holds, then the imputed value of $y_i$ can be generated from $f\left( y_i \mid \bx_i; \bm \theta \right)$, the population model of $y_i$ given $x_i$. When estimating the parameters in the population model, we need to use the sampling weights because the sampling design can be informative. That is, we maximize 
\begin{equation}
\ell_p (\bm \theta) = \sum_{i \in A} w_i \delta_i \log f(y_i \mid \bx_i; \bm \theta) 
\label{8-4-3}
\end{equation}
to estimate $\bm \theta$ in $f(y \mid \bx; \bm \theta)$, where $w_i$ is the sampling weight of unit $i$ such that $\sum_{i \in A} w_i y_i$ is a design-consistent estimator of $Y$.

The model-based imputation can be implemented by the following two-steps. 
\begin{description}
\item{[Step 1]} Specify a conditional distribution $f( y \mid \bx; \bm \theta)$ and estimate $\bm \theta$ by finding the maximizer of (\ref{8-4-3}) with respect to $\bm \theta$. 
\item{[Step 2]} For each unit $i$ with $\delta_i=0$, generate Monte Carlo samples from $f( y \mid \bx_i; \hat{\bm \theta})$ to obtain the imputed value(s) for unit $i$. 
\end{description}

If PMAR assumption is in question, one can also employ a non-MAR response model and develop  imputation from 
$$ f( y_i \mid \bx_i, \delta_i=0) = \frac{ f( y_i \mid \bx_i) P( \delta_i=0 \mid \bx_i, y_i) }{ \int f( y\mid \bx_i) P( \delta_i=0 \mid \bx_i, y) dy }. $$
See Chapter 8 of \cite{kimshao21} for more details of handling non-MAR missingness.

\section{Multiple imputation}


\cite{rubin1987} developed the multiple imputation method, which applies a Bayesian perspective to the problem of nonresponse imputation. Suppose the estimator
\begin{equation}
\hat{\theta}_n = \sum_{i \in A} w_i y_i \label{thetan}
\end{equation}
is the one that would be used in the absence of nonresponse. Then, the estimator using imputed values can be expressed as:
$$ \hat{\theta}_I = \sum_{i \in A}  w_i \left[ \delta_i y_i + \left( 1- \delta_i \right) y_i^* \right] . $$

When the parametric distribution of the missing values $y_i$ is denoted as $f\left( y_i \mid \bm \beta \right)$, if we knew the parameter $\bm \beta$, we could generate the imputed values $y^*$. In practice, since $\bm \beta$ is unknown, we use the Bayesian approach to generate imputed values from the posterior predictive distribution:
\begin{equation}
y_i^* \sim f \left( y_i \mid \mathbf{y}_{\rm obs} \right),
\label{bayesian}
\end{equation}
where $\mathbf{y}_{\rm obs}$ denotes the observed $y$ values. The posterior predictive distribution is given by:
$$f \left( y_i \mid \mathbf{y}_{obs} \right) = \int f\left( y_i \mid \bm \beta \right) p \left( \bm \beta \mid \mathbf{y}_{obs} \right)
d \bm \beta
$$
where $p \left( \bm \beta \mid \mathbf{y}_{obs} \right)$ represents the posterior distribution of the parameter $\bm \beta$.

If $M$ parameter values $\bm \beta_{(k)}^*, k=1,2, \cdots, M$ are drawn from this posterior distribution, the posterior predictive distribution can be approximated as:
$$\int f\left( y_i \mid \bm \beta \right) p \left( \bm \beta \mid \mathbf{y}_{obs} \right)
d \bm \beta = p\lim_{M \rightarrow \infty} \frac{1}{M} \sum_{k=1}^M f
\left(y_i \mid \bm \beta_{(k)}^* \right)
$$This approximation becomes more accurate as 
$M$ approaches infinity.

For informative sampling designs, the sample likelihood function is unknown and the posterior distribution of $\bm \beta$ cannot be computed easily. \cite{kim2017b} 
 propose an alternative algorithm for generating imputated values from (\ref{bayesian})  using a modificiation of data augmentation algorithm under informative sampling.

When the point estimator $\hat{\theta}_n$ and its variance estimator $\hat{V}_n$ for the case without nonresponse are given, Rubin's multiple imputation method can be broadly carried out through the following steps:
\begin{description}
\item[Step 1] Using the posterior predictive distribution in equation (\ref{bayesian}), independently generate $M (>1)$ imputed values ($y_{(1)}^*, \cdots, y_{(M)}^*$) for each missing value ($y_i$), resulting in $M$ sets of imputed data.

\item[Step 2] For each $k$-th ($k=1,2,\cdots, M$) set of imputed data, calculate $\hat{\theta}_{I(k)}$ and $\hat{V}_{I(k)}$ by applying the estimators $\hat{\theta}_n$ and $\hat{V}_n$.

\item[Step 3] Use the point estimator $\hat{ \theta}_{\rm MI}$ as:
\begin{equation}
\hat{\theta}_{\rm MI} = \frac{1}{M} \sum_{k=1}^M \hat{\theta}_{I(k)},
\label{mitheta}
\end{equation}
and the variance estimator as:
\begin{equation}
\hat{V}_{\rm MI} = U_M + \left( 1 + \frac{1}{M} \right)B_M,
\label{thetavar}
\end{equation}
where
\begin{eqnarray*}
U_M &=&  \frac{1}{M} \sum_{k=1}^M \hat{V}_{I(k)}, \\
B_M &=&  \frac{1}{M-1} \sum_{k=1}^M \left( \hat{\theta}_{I(k)} - \hat{\theta}_{MI} \right)^2.
\end{eqnarray*}
\end{description}

Multiple imputation has the advantage of increasing the efficiency of the point estimator by using multiple imputed values for a single missing value, and it also facilitates the estimation of variance. However, the consistency or asymptotic unbiasedness of the variance estimator in (\ref{thetavar}) is not always guaranteed. For further details, refer to \cite{kim2006} or \cite{yang2016}.

\section{Fractional imputation}

Fractional imputation, first proposed by \cite{kalton1984}, and later developed by \cite{kim2004}, \cite{fuller2005}, and \cite{kim2011}, is an alternative to multiple imputation.

For the full sample estimator in (\ref{thetan}), fractionally imputed estimator is constructed such that  
$$ \hat{Y}_{\rm FI} = \sum_{i \in A}  w_i \left\{  \delta_i y_i + \left( 1- \delta_i \right) \hat{y}_i \right\} $$
where 
$$ \hat{y}_i = \sum_{j=1}^m w_{ij}^* y_i^{*(j)}$$
and $w_{ij}^*$ is the fractional weight associated with $y_{ij}^{*}$, the $j$-th imputed value of $y_i$. The fractional weights are constructed such that 
$$ \sum_{j=1}^m w_{ij}^* = 1$$
and 
$$ \sum_{j=1}^m w_{ij}^* y_i^{*(j)} \cong \hat{E}( Y \mid \bx_i, \delta_i=0) .$$

 Let $y_{i1}^{*}, \ldots, y_{im}^{*}$ be $m$ imputed values  generated from a proposal distribution $f_0(y \mid x)$.
  The choice of the proposal distribution is somewhat arbitrary. The parametric model for $Y$ is assumed to follow with density  $f( y \mid \bx; \bm \theta)$.     
  If we do not have a good guess about $\bm \theta$, we may use
$$ f_0(y \mid x)=\hat{f}( y \mid \delta=1)=\frac{ \sum_{i \in A} w_i \delta_i I (y_i=y) }{\sum_{i \in A} w_i \delta_i  }, $$
which estimates the marginal distribution of $y_i$ using the set of respondents. If $x$ is categorical, then we can use
 $$ f_0(y \mid x)=\frac{ \sum_{i \in A} w_i \delta_i I (x_i = x, y_i=y) }{\sum_{i \in A} w_i \delta_i I(x_i=x) }. $$
For continuous $x$, we may use a kernel-type  proposal distribution
 $$ f_0(y \mid x)=\frac{  w_i \delta_i  K_h(x_i, x) K_h(y_i, y) }{\sum_{i \in A} w_i \delta_i K_h(x_i,x) }. $$

The fractional weight associated with $y_{ij}^{*}$ is computed as
$$ w_{ij0}^*  = \frac{ f ( y_{ij}^{*} \mid x_i ; \hat{\bm \theta})/f_0(y_{ij}^{*} \mid x_i)
}{ \sum_{k=1}^m f ( y_{ik}^{*} \mid x_i ; \hat{\bm \theta})/f_0(y_{ik}^{*} \mid x_i) } .
$$
When $m$ is small, the fractional weights can be further modified in the calibration step. The proposed calibration equation for improving the fractional weights in this case is
\begin{equation}
 \sum_{i \in A} \sum_{j=1}^m w_i(1-\delta_i) w_{ij}^* S(\hat{\bm \theta}; x_i, y_{ij}^* ) =0
 \label{8-4-8}
\end{equation}
and $\sum_{j=1}^m w_{ij}^* = 1$ for each $i$ with $\delta_i=0$, where $\hat{\bm \theta}$ is computed from (\ref{8-4-3}). Using the idea of regression weighting, the final calibration fractional weights can be  computed by
 \begin{equation}
 w_{ij}^{*} = w_{ij0}^{*} + w_{ij0}^{*} {\Delta}    \left(S_{ij}^{*}- \bar{S}_{i \cdot}^{*} \right),
 \label{6-22-b}
 \end{equation}
 where  $ S_{ij}^{*}= S ( \hat{\bm \theta}; x_i, \by_{ij}^{*}  )$, $\bar{S}_{i \cdot}^{*}=  \sum_{j=1}^m w_{ij0}^{*} S_{ij}^{*}$, and
$$ {\Delta}=-  \left\{\sum_{i \in A} w_i (1-\delta_i) \sum_{j=1}^m  w_{ij0}^{*} S_{ij}^{*} \right\}' \left[ \sum_{i \in A} w_i (1-\delta_i) \sum_{j=1}^m   w_{ij0}^{*}  \left(S_{ij}^{*}- \bar{S}_{i \cdot}^{*} \right)^{\otimes 2} \right]^{-1}.
$$
See \cite{fuller2005} for more details. 

The calibration condition (\ref{8-4-8}) guarantees that the imputed score equation leads to the same $\hat{\bm \theta}$, computed from (\ref{8-4-3}).
Once the FI data are created,  the fractionally imputed estimator of $Y=\sum_{i=1}^N y_i $ is obtained by
$$ \hat{Y}_{\rm FI} = \sum_{i\in A} w_i \left\{ \delta_i y_i + \left( 1- \delta_i \right) \sum_{j=1}^m w_{ij}^* y_{ij}^{*} \right\}. $$

We now consider the \emph{fractional hot deck imputation} in which $m$ imputed values are taken from the set of respondents.  
Let $\left\{y_1, \ldots, y_r\right\}$ be the set of respondents and, for unit $i$ with $\delta_i=0$, let  
$$ y_{ij}^*= y_j, \ \ j=1, \ldots, r$$
be the $j$-th imputed value for unit $i$. Thus, we  consider the special case of $m=r$. Let $w_{ij}^*$ be the fractional weights assigned to $y_{ij}^{*}=y_j$ for $j=1,2,\ldots,r$.
 In this case, the fractional weight represents the point mass assigned to each  responding $y_i$. Thus, under MAR, it is desirable to compute the fractional weights $w_{i1}^* , \ldots, w_{ir}^*$ such that
$ \sum_{j=1}^r w_{ij}^*=1 $ and
\begin{equation}
 \sum_{j=1}^r w_{ij}^* I (y_j < y) \cong P\left( y_i< y \mid x_i  \right).
 \label{107}
 \end{equation}

If we can assume a parametric model $f(y \mid x ; \bm \theta) $ for the conditional distribution of $y$ on $x$ , then the fractional weights satisfying (\ref{107}) are given by
\begin{eqnarray*}
 w_{ij}^*  &\propto & \frac{ f ( y_j \mid \bx_i ; \hat{\bm \theta} )}{f(y_j \mid \delta_j=1)}
\end{eqnarray*}
and 
$$ \sum_{j=1}^r w_{ij}^*=1. $$
Since
\begin{eqnarray*}
 f \left( y \mid \delta=1 \right) & = & \int  f \left( y_j \mid x \right) f ( x \mid \delta=1) dx \\
 & \propto &  \sum_{i \in A}  w_i \delta_i  f\left( y \mid x_i \right),
 \end{eqnarray*}
 we can compute
 \begin{equation}
 w_{ij}^* \propto   \frac{ f ( y_j \mid x_i, \hat{\bm \theta} ) }{
 \sum_{k \in A} w_k \delta_k   f ( y_j \mid x_k ; \hat{\bm \theta} ) }
 \label{fwgt_hd}
\end{equation}
with $\sum_{j; \delta_j =1 } w_{ij}^*=1$.  \index{Fractional hot deck imputation} The fractional weights in (\ref{fwgt_hd}) lead to robust estimation in the sense that a certain level of misspecification in $f( y \mid x)$ can still provide  consistent estimates. See \cite{kimyang14} for more details. 
  The fractional weights can be further adjusted to satisfy
\begin{equation}
\sum_{i=1}^{n}\{\delta_{i}S(\hat{\bm \theta};x_{i},y_{i})+(1-\delta_{i})\sum_{j;\delta_{j}=1}w_{ij}^{*}S(\hat{\bm \theta};x_{i},y_{j})\}=0,
\label{eq:constraint1}
\end{equation}
where $S({\bm \theta};x_{i},y_{i})$ is the score function of $\bm \theta$,  and $\hat{\bm \theta}$ is the pseudo MLE of $\bm \theta$, computed from (\ref{8-4-3}).

\section{Mass imputation for two-phase sampling}

\index{Mass imputation}

Missing data can sometimes occur naturally, but in some cases, it is intentionally created to reduce costs. For example, in two-phase sampling, $X$ is observed in the first-phase sample, but $Y$ is not. In this scenario, $Y$ is observed only in the subset, the second-phase sample, to reduce costs, making it a type of planned missing data.

Furthermore, if there are two independent data sets where one observes $(X, Y)$ and the other observes only $X$, this can also be considered a type of two-phase sampling with a non-nested data structure and classified as planned missing data. For instance, in the first-phase sample, health-related items are measured through a questionnaire, while in a small second-phase sample, health indicators are measured using health examination equipment.

When there is planned missing data, treating it as nonresponse and applying the nonresponse imputation methods described in the previous section is called synthetic data imputation. The term synthetic data implies that the data values are entirely created. In other words, even though $Y$ is not observed in the data, $\hat{Y}$ is generated for all elements in the data using other information. This method is called synthetic data imputation because it generates a large number of imputed values, also known as mass imputation. \index{mass imputation} \index{synthetic data imputation}

First, let's discuss how to perform mass imputation as part of the estimation methods for two-phase sampling data covered in Chapter 10. Suppose that in the first-phase sample $A_1$,  $\bx_i$ is observed, and in the second-phase sample $A_2$, both $\bx_i$ and $y_i$ are observed. Let $Y=\sum_{i=1}^N y_i$
  be the parameter of interest. If the weights for the first-phase sample are denoted as $w_i$, the regression imputation estimator of  $Y$ can be expressed as follows:
\begin{equation}
\widehat{Y}_{\rm reg,I} = \sum_{i \in A_2} w_i y_i + \sum_{i \in A_2^c \cap A_1} w_i \mathbf{x}_i^\top  \hat{\bm \beta}_q
\label{12.25}
\end{equation}
where $\hat{\bm \beta}_q$
  is the regression coefficient  obtained from the second-phase sample observations, calculated as:  
\begin{equation}
 \hat{\bm \beta}_q = \left( \sum_{i \in A_2} w_i  \mathbf{x}_i \mathbf{x}_i^\top / q_i  \right)^{-1} \sum_{i \in A_2} w_i  \mathbf{x}_i y_i/ q_i, 
 \label{eq:13-34}
 \end{equation}
 where $q_i=q( \bx_i)$ is a known function of $\bx_i$. 
Writing $\hat{Y}_{\rm HT}= \sum_{i \in A_1} w_i y_i$, the bias of $\widehat{Y}_{\rm reg, I}$
  can be calculated as: 
\begin{eqnarray*}
 E( \widehat{Y}_{\rm reg,I} ) - Y  & = & E\{ \hat{Y}_{\rm reg,I} - \hat{Y}_{\rm HT} \} \\
 &=& - E\left\{  \sum_{i \in A_2^c  \cap A_1} w_i \left( y_i - \mathbf{x}_i^{\top} \hat{\bm \beta}_q \right)
\right\} \\
&=& - E\left\{  \sum_{i \in A_2 } w_i \left(  \frac{1}{\pi_{i2 \mid 1} } -1 \right)  \left( y_i - \mathbf{x}_i^{\top} \hat{\bm \beta}_q  \right)
\right\}. 
\end{eqnarray*}
Therefore, if $\hat{y}_i= \mathbf{x}_i^{\top} \hat{\bm \beta}_q$  satisfies the following equation, the mass imputation estimator in (\ref{12.25}) is approximately unbiased:
\begin{equation}
\sum_{i \in A_2}  w_i \left(  \frac{1}{\pi_{i2 \mid 1} } -1 \right)  \left( y_i - \mathbf{x}_i^{\top} \hat{\bm \beta}_q  \right) = 0 .
\label{12-39}
\end{equation}
Note that condition  (\ref{12-39}) implies 
\begin{equation}
\hat{Y}_{\rm reg, I} = \sum_{i \in A_1} w_{i} \bx_i^{\top} \hat{\bm \beta}_q 
+ \sum_{i \in A_2} w_{i} \pi_{2i \mid 1}^{-1} \left\{y_i -\bx_i^{\top} \hat{\bm \beta}_q  \right\},
\label{8-6-3}
\end{equation}
which takes the form of the two-phase regression estimator in (\ref{9.15a}). Therefore, as long as (\ref{8-6-3}), the mass imputation estimator in (\ref{12.25}) is design consistent. 
If $\hat{\bm \beta}_q$ is calculated as in (\ref{eq:13-34}), then to satisfy (\ref{12-39}), there must exist a 
$\blambda$ such that 
\begin{equation}
 q_i \left( \pi_{i2 \mid 1}^{-1}-1 \right)  = \mathbf{x}_i^{\top} \bm \lambda
\label{12-40}
\end{equation}
for all $i$. A simple method to satisfy condition (\ref{12-40}) is to include $q_i ( \pi_{2i \mid 1}^{-1}-1 )$ in $\bx_i$, assuming that it can be calculated for all $i \in A_1$. This mass imputation is particularly useful for small area estimation because it applies the prediction for they values to the entire sample in $A_1 \cap A_2^c$. When mass imputation is used to implement small area estimation, the resulting small area estimator is called a synthetic estimator.  \index{synthetic estimation}

Next, let's consider the variance estimation of $\widehat{Y}_{\rm reg, I}$. Variance estimation can be simply calculated using the replication method. Similar to what was introduced in Section 10.3, by replacing $w_i$ 
  with the replication weight $w_i^{(k)}$ 
  and calculating $\widehat{Y}_{\rm reg, I}^{(k)}$ 
  in the same way as (\ref{12.25}), we can estimate the variance based on these replicates. That is, we compute
  $$
  \widehat{Y}_{\rm reg,I}^{(k)} = \sum_{i \in A_2} w_i^{(k)} y_i + \sum_{i \in A_2^c} w_i^{(k)} \mathbf{x}_i^\top  \hat{\bm \beta}_q^{(k)}, 
$$
where  
$$ \hat{\bm \beta}_q^{(k)} = \left( \sum_{i \in A_2} w_i^{(k)}  \mathbf{x}_i \mathbf{x}_i^{\top} /q_i  \right)^{-1} \sum_{i \in A_2} w_i^{(k)}  \mathbf{x}_i y_i /q_i . $$
Then, the replication variance estimator can be calculated as 
$$ \hat{V}_{\rm rep}  = \sum_{k=1}^L c_k \left( \widehat{Y}_{\rm reg,I}^{(k)} - \widehat{Y}_{\rm reg,I} \right)^2 . $$
\cite{rao1995} first proposed this variance estimation methodology for cases where the first-phase sample is a simple random sample. \cite{kimrao12} addressed the use of mass imputation in cases where common auxiliary variables 
$x$ are observed in two independent samples. See \cite{park2019} for more theoretical details on mass imputation for two-phase sampling.

%% file: chapters/chapter14.tex
\chapter{Analytic Inference}
 \section{Introduction}


In this chapter, we will discuss statistical analysis using statistical models from sample data obtained through probability sampling. A statistical model serves as a framework for learning from data  by describing the relationships among key variables observed in the data using probabilistic relationships. It is a tool used to understand the data and make more accurate predictions. Such statistical models should reflect the structure of the data and be set up to answer questions in the form of hypothesis testing that the data observation seeks to address. A good model provides a more accurate description of the data structure, thereby reducing model error and increasing the power of hypothesis tests.

First, consider a  population data set that follows a parametric model \( f(y; \bm \theta) \), where $\bm \theta \in \Omega \subset \mathbb{R}^p$.  Assume that there exists a true parameter value \(\bm \theta_0\) such that the population data \((y_1, \cdots, y_N)\) are randomly obtained from \( f( y ; \bm \theta_0) \). In this case, we can consider the concept of a census estimator for \(\bm \theta_0\), which is approached from the perspective of estimating \(\bm \theta_0\) if \( A \) were a census (i.e., if \( A = U \)).

In such cases, as the census data is an IID sample,  we can think about the value of \(\bm \theta\) that maximizes
$$ \ell_N ( \bm \theta) = N^{-1} \sum_{i=1}^N \log f( y_i ; \bm \theta), $$
and denote this value by \(\bm \theta_N\). Assume that this \(\bm \theta_N\) is unique. This census estimator \(\bm \theta_N\) is a consistent estimator of \(\bm \theta_0\), meaning it converges in probability to \(\bm \theta_0\) as \(N \rightarrow \infty\).

Therefore, from the observed data in a given sample \(A\), we need to obtain a consistent estimator \(\hat{\bm \theta}\) for \(\bm \theta_N\). Define
\begin{equation}
\hat{\ell}_{w}( \bm \theta) =N^{-1}  \sum_{i \in A} w_i \log f( y_i ; \bm \theta), \label{14-1}
\end{equation}
where \( w_i = 1/\pi_i \). Then, \(\hat{\ell}_{w}( \bm \theta)\) converges in probability to \( \ell_N (\bm \theta) \). If the estimator \(\hat{\bm \theta}\) that maximizes \(\hat{\ell}_{w}( \bm \theta)\) is unique, under some regularity conditions, it will converge in probability to \(\bm \theta_N\), and thus also to \(\bm \theta_0\). The estimator that maximizes (\ref{14-1}) is called the pseudo maximum likelihood estimator (PMLE). Conditions ensuring consistency and asymptotic normality of the pseudo maximum likelihood estimator (PMLE) are established, for example, in \cite{Binder1983},  \cite{rubin-bleuer2005}, and \cite{han2021}.  

\index{pseudo maximum likelihood estimator}

Differentiating equation (\ref{14-1}), the pseudo maximum likelihood estimator (PMLE) can also be understood as the solution to 
\begin{equation}
\hat{S}_{w} ( \bm \theta) \equiv N^{-1} \sum_{i \in A} w_i S( \bm \theta ; y_i ) = 0, 
\label{eq:pmle}
\end{equation}
where \( S(\bm \theta; y) = \partial \log f( y ; \bm \theta) / \partial \bm \theta \). Using a Taylor expansion, we can approximate the following:
\begin{eqnarray*}
0 = \hat{S}_w ( \hat{\bm \theta}) &\cong & \hat{S}_w ( {\bm \theta}_0 ) + E\left\{ \frac{\partial}{\partial \bm \theta'}  \hat{S}_w ( \bm \theta_0 ) \right\} ( \hat{\bm \theta} - \bm \theta_0 ) .
\end{eqnarray*}
Defining 
$$ I(\bm \theta; y) = - \frac{\partial}{ \partial \bm \theta^\top} S( \bm \theta; y), $$
we get
\begin{equation}
\hat{\bm \theta} - \bm \theta_0 \cong \left\{ \frac{1}{N} \sum_{i=1}^N I ( \bm \theta_0; y_i)  \right\}^{-1} \hat{S}_w ( {\bm \theta}_0 ).
\label{14-1b}
\end{equation}
Thus, under some regularity conditions, we can obtain
\begin{equation}
 \sqrt{n} \hat{S}_w ( {\bm \theta}_0 ) \stackrel{\mathcal{L}}{ \longrightarrow} N\left[ 0, n  V \{  \hat{S}_w ( {\bm \theta}_0 ) \}  \right]. 
 \label{score3}
 \end{equation}
 
Now, using equation (\ref{14-1b}), we obtain
\begin{equation}
 \sqrt{n} \left( \hat{\bm \theta} - \bm \theta_0 \right) \stackrel{\mathcal{L}}{ \longrightarrow} N\left[ 0, n V ( \hat{\bm \theta}  )  \right],
 \label{14-2}
\end{equation}
where
\begin{equation}
 V ( \hat{\bm \theta}  ) =  \left\{ N^{-1} \sum_{i=1}^N I ( \bm \theta_0; y_i)  \right\}^{-1} V \{  \hat{S}_w ( {\bm \theta}_0 ) \} \left\{ N^{-1} \sum_{i=1}^N I ( \bm \theta_0; y_i)^\top   \right\}^{-1}.
 \label{14-3}
 \end{equation}
This is commonly referred to as the sandwich formula.  \index{sandwich formula}

For the variance estimation expressed in equation (\ref{14-3}), we can use:
\begin{equation}
\hat{V} ( \hat{\bm \theta} ) = \left\{ N^{-1} \sum_{i \in A} w_i I ( \hat{\bm \theta} ; y_i) \right\}^{-1} \hat{V} \left\{ \hat{S}_w ( \bm \theta_0 ) \right\} \left\{ N^{-1} \sum_{i \in A} w_i I ( \hat{\bm \theta}; y_i)^\top  \right\}^{-1  },
\label{14-4}
\end{equation}
where \(\hat{V} \left\{ \hat{S}_w ( \bm \theta_0 ) \right\}\) is calculated as:
\begin{eqnarray*}
\hat{V} \left\{ \hat{S}_w ( \bm \theta_0 ) \right\} &=& \frac{1}{N^2}  \sum_{i \in A} \sum_{j \in A} \frac{ \pi_{ij} - \pi_i \pi_j }{ \pi_{ij} } w_i w_j S( \hat{\bm \theta}; y_i ) S( \hat{\bm \theta}; y_j )^\top  \\
&& + \frac{1}{N^2} \sum_{i \in A} w_i S( \hat{\bm \theta}; y_i ) S( \hat{\bm \theta}; y_i )^\top.
\end{eqnarray*}

The first term estimates \(E_\zeta \left[ V \left\{ \hat{S}_w ( \bm \theta_0 ) \mid \mathcal{F}_N \right\} \right]\), while the second term estimates
\[
V_\zeta \left[ E \left\{ \hat{S}_w ( \bm \theta_0 ) \mid \mathcal{F}_N \right\} \right] = V_\zeta \left\{ \sum_{i=1}^N S ( \bm \theta_0 ; y_i ) \right\}.
\]

If \(n/N = o(1)\), the second term becomes of smaller order than the first term and can be safely ignored.

\begin{example}Consider a sample data set where \((x_i, y_i)\) are observed, and \(y_i\) is an indicator variable that takes values of 0 or 1. We will set up a logistic regression model with \(X\) as the independent variable and \(Y\) as the dependent variable. The logistic regression model is expressed as:
$$ P( Y=1 \mid x) = \frac{ \exp ( \beta_0 + \beta_1 x)}{ 1+ \exp ( \beta_0 + \beta_1 x)} := p(x; \bm \beta), $$
where we aim to estimate \(\bm \beta = (\beta_0, \beta_1)^\top\) from the sample data.

The score function for \(\bm \beta\) is given by
\begin{equation}
 S(\bm \beta ; x,y) = \left\{ y - p(x; \bm \beta) \right\} (1, x_i)^\top,
 \label{14-5}
 \end{equation}
which leads to the equation
\begin{equation}
 \sum_{i \in A} w_i \left\{ y - p(x; \bm \beta) \right\} (1, x_i)^\top =(0,0)^\top.
 \label{14-6}
 \end{equation}
Solving this equation yields the PMLE. To solve (\ref{14-6}), we let \(\mathbf{x}_i = (1, x_i)'\) and use the Fisher scoring method:
$$ \bm \beta^{(t+1)} = \bm \beta^{(t)} + \left\{ \sum_{i \in A} w_i p_i^{(t)} ( 1- p_i^{(t)}) \mathbf{x}_i \mathbf{x}_i^{\top} \right\}^{-1} \sum_{i \in A} w_i (y_i - p_i^{(t)} ) \mathbf{x}_i, $$
where \(p_i^{(t)} = p( \bx; \bm \beta^{(t)} )\).

To estimate the variance of the pseudo maximum likelihood estimator \(\hat{\bm \beta}\), we use (\ref{14-5}) and
$$ I(\bm \beta; y) = p(\bx;\bm  \beta)\{ 1- p(\bx; \bm \beta) \} \mathbf{x}_i \mathbf{x}_i^{\top}, $$
applying these in (\ref{14-4}) to calculate the variance.

\end{example}

Now, by the second-order Taylor expansion, we  obtain
\begin{eqnarray}
\ell_w ( \bm \theta_0 ) &=& \ell_w ( \hat{\bm \theta} ) + \hat{S}_w^\top   ( \hat{\bm \theta} ) ( \bm \theta_0- \hat{\bm \theta}) - \frac{1}{2} ( \hat{\bm \theta} - \bm \theta_0  )^\top  \hat{I}_w ( \hat{\bm \theta}) ( \hat{\bm \theta} - \bm \theta_0 ) + o_p (n^{-1} ) \notag \\
&=& \ell_w ( \hat{\bm \theta} )  - \frac{1}{2} ( \hat{\bm \theta} - \bm \theta_0)^\top  \hat{I}_w ( \hat{\bm \theta}) (\hat{\bm \theta} - \bm \theta_0 ) + o_p (n^{-1} )  ,
\label{eqw3}
\end{eqnarray}
where
$
\hat{I}_w (\bm \theta) = N^{-1} \sum_{ i \in A} w_i  I( \bm \theta; y_i).
$
Define
\begin{equation}
W (\bm \theta_0) = - 2 n \{ \ell_w ( \bm \theta_0 ) - \ell_w ( \hat{\bm \theta})
\}.
\label{eqw4}
\end{equation}
 By  (\ref{eqw3}), we obtain
$$ W ( \bm \theta_0) = n ( \hat{\bm \theta} - \bm \theta_0   )^\top  \hat{I}_w ( \hat{\bm \theta}) ( \hat{\bm \theta} - \bm \theta_0  )  + o_p (1) . $$
Thus, by (\ref{14-2}),  we  have
\begin{equation}
W ( \bm \theta_0)   \longrightarrow \mathcal{G} = \sum_{i=1}^p c_i Z_i^2
\label{eqw5}
\end{equation}
in distribution as $n\to\infty$, 
where $c_1, \ldots, c_p$ are the eigenvalues of $V ( \hat{\bm \theta} ) I(\bm \theta_0)  $, and $Z_1, \ldots, Z_p$ are $p$ independent random variables from the standard normal distribution. Result (\ref{eqw5}) was  established by  \citet{rao1984chi} and \citet{lumley2014tests}, and it  can be regarded as  a version of the Wilks' theorem  for  survey sampling.  Unless the sampling design is simple random sampling and the sampling fraction is negligible, the limiting distribution does not reduce to the standard chi-squared distribution with $p$ degrees of freedom. 

\cite{kim2023} develop a bootstrap approach to approximate the limiting distribution in (\ref{eqw5}). Suppose that  the bootstrap weights are created such that the solution $\hat{\bm \theta}^*$ to 
$$N^{-1} \sum_{i \in A} w_i^* (\bm \theta; y_i )= 0 $$
satisfies 
\begin{equation}
	\sqrt{n} ( \hat{\bm \theta}^* - \hat{\bm \theta})\mid A \,{\longrightarrow }\,  N( 0, \hat{V} ( \hat{\bm \theta}_w) )
	\label{lemma1}
	\end{equation}
	in distribution
	as $n \rightarrow \infty$, and 
	the reference distribution on the left side of (\ref{lemma1}) is the bootstrap sampling distribution conditional on the sample $A$.
To establish a bootstrap version of Wilks' Theorem in (\ref{eqw5}), we use
\begin{equation*}
W^* ( \hat{\bm \theta} ) = -2 n \{ \ell_w^*(\hat{\bm \theta}) - \ell_w^* ( \hat{\bm \theta}^* ) \}
\label{eqw8}
\end{equation*}
as the bootstrap version of $W ( \bm \theta_0) $ in (\ref{eqw4}).
Under the assumptions for (\ref{eqw8}), we can show that 
\begin{equation}
	W^* ( \hat{\bm \theta} ) \mid A   {\,\longrightarrow\, }   \mathcal{G}   \label{8b}
\end{equation} in distribution
	as $n \rightarrow \infty$, and 
	the reference distribution is the bootstrap sampling distribution conditional on  the   sample $A$,  where $\mathcal{G}$ is the limiting distribution (\ref{eqw5}) of the quasi-likelihood ratio test statistic $W ( \bm \theta_0)$ in (\ref{eqw4}) under the null hypothesis.

By (\ref{8b}), we conclude that the corresponding test statistic generated by the proposed bootstrap method has the same asymptotic distribution as in (\ref{eqw4}). Thus, we can use the bootstrap distribution of $W^* ( \hat{\bm \theta} )$ to approximate the sampling distribution of $W ({\bm \theta}_0 )$. See \cite{kim2023} for more details of the bootstrap approach to hypothesis testing in survey sampling.

\section{Regression analysis} 
When conducting regression analysis based on sample survey data, we make the following assumptions about the population elements:
\begin{equation}
y_i = \mathbf{x}_i^{\top} \bm \beta + e_i,
\label{14-8}
\end{equation}
where \(E(e_i \mid \mathbf{x}_i) = 0\) and \(V(e_i \mid \mathbf{x}_i) = \sigma^2\). The census estimator for this parameter \(\bm \beta\) can be found by minimizing
$$ Q_N( \bm \beta) = \sum_{i=1}^N ( y_i - \mathbf{x}_i^{\top} \bm \beta)^2. $$

Thus, we use a design unbiased  estimator for \(Q_N( \bm \beta)\),
\begin{equation}
\hat{Q}_w ( \beta) = \sum_{i \in A} w_i ( y_i - \mathbf{x}_i^{\top} \bm \beta)^2,
\label{14-9}
\end{equation}
and find \(\hat{\bm \beta}_w\) that minimizes this. Under appropriate conditions, this estimator converges in probability to \(\bm \beta\) in (\ref{14-8}). To estimate the variance of this estimator, we use
$$ \hat{U}_w ( \bm \beta) = \sum_{i \in A} w_i ( y_i - \mathbf{x}_i^{\top} \bm \beta)\mathbf{x}_i $$
and apply the sandwich formula using Taylor series expansion:
\begin{equation}
\hat{V} ( \hat{\bm \beta}_w) = \left( \sum_{i \in A} w_i \mathbf{x}_i \mathbf{x}_i^{\top}  \right)^{-1} \hat{V} \{  \hat{U}_w ( \bm \beta) \} \left( \sum_{i \in A} w_i \mathbf{x}_i \mathbf{x}_i^{\top}  \right)^{-1},
\label{14-10}
\end{equation}
where
$$ \hat{V} \{ \hat{U}_w ( \bm \beta) \} = \sum_{i \in A} \sum_{j \in A} \frac{ \pi_{ij} - \pi_i \pi_j }{ \pi_{ij} } w_i w_j \hat{e}_i \hat{e}_j \mathbf{x}_i \mathbf{x}_j^\top + \sum_{i \in A} w_i \hat{e}_i^2 \mathbf{x}_i \mathbf{x}_i^{\top} $$
and \(\hat{e}_i = y_i - \mathbf{x}_i^{\top} \hat{\bm \beta}_w\).

Under appropriate conditions, we can show that
\begin{equation}
\left\{ \hat{V} ( \hat{\bm \beta}_w) \right\}^{-1/2} \left( \hat{\bm \beta}_w - \bm \beta \right) \stackrel{\mathcal{L}}{\longrightarrow} N(0, I),
\label{14-11}
\end{equation}
which can be used to construct confidence intervals for the regression coefficients or perform hypothesis testing.

The above methodology estimates based on design-based estimation, reflecting the sample design. However, using weights often results in larger variance estimates, which reduces the efficiency of the estimation. If condition (\ref{noninformative}) holds, making the sample design noninformative, then the regression model in (\ref{14-8}) also holds for the sample. In this case, we can obtain the ordinary least squares (OLS) estimator \(\hat{\bm \beta}_{ols}\) by minimizing 
$$ \sum_{i \in A} ( y_i - \mathbf{x}_i^{\top} \bm \beta )^2, $$
without using weights, and the variance estimation can also be calculated without considering weights. But how can we determine whether the given sample design is noninformative?

A method to verify whether the sample design is noninformative was suggested by \cite{dumouchel1983} through a hypothesis testing approach. The null hypothesis to be tested is:
\begin{equation}
H_0 : E( \hat{\bm \beta}_{obs}) = \bm \beta.
\label{14-12}
\end{equation}
To test this, consider the following extended regression model:
\begin{equation}
y_i = \mathbf{x}_i^{\top} \bm \beta + \mathbf{z}_i^\top  \bm \gamma + a_i,
\label{14-13}
\end{equation}
where \(E(a_i \mid \mathbf{x}_i, \mathbf{z}_i) = 0\) and \(\mathbf{z}_i = w_i \mathbf{x}_i\).

Under this model, \( E( \hat{\bm \beta}_{obs}) = \bm \beta + (X'X)^{-1} X' Z \bm \gamma \), so testing (\ref{14-12}) is equivalent to testing \(H_0: \bm \gamma=\mathbf{0}\) in equation (\ref{14-13}). The extended model in (\ref{14-13}) includes the weights as explanatory variables, allowing for the condition of noninformative sample design to hold. Thus, the OLS method can be used to obtain the estimator without weights, and testing the hypothesis \(H_0: \bgamma=0\) becomes straightforward.

Generally speaking, if weights are included as explanatory variables in the model setting stage, the sample design becomes noninformative when the model is properly fitted. This can be easily understood through the following lemma.
\begin{lemma}
In a finite population where $(X, Y, \pi)$ are obtained and a subset is selected as a sample, if the sample inclusion indicator variable \(I\) satisfies
\begin{equation}
Pr( I=1 \mid X, Y, \pi) = Pr ( I=1 \mid \pi),
\label{14-14}
\end{equation}
then
\begin{equation}
f ( Y \mid X, \pi, I=1) = f( Y \mid X, \pi ).
\label{14-15}
\end{equation}
\end{lemma}
\begin{proof}
Using Bayes' theorem,
$$ f ( Y \mid X, \pi, I=1) = f( Y \mid X, \pi ) \frac{  Pr( I=1 \mid X, Y, \pi) }{ Pr ( I=1 \mid X, \pi)}, $$
and applying (\ref{14-14}), we obtain (\ref{14-15}).
\end{proof}

Next, let's consider how to improve the efficiency of estimators when the noninformative sample design condition does not hold. Based on the regression model (\ref{14-8}), consider the following estimating equation:
\begin{equation}
\hat{U}_q ( \bm \beta) \equiv \sum_{i \in A} w_i q( \mathbf{x}_i) ( y_i - \mathbf{x}_i^{\top} \bm \beta ) \mathbf{x}_i = 0.
\label{14-16}
\end{equation}
Here, \(q( \mathbf{x})\) is an arbitrary function determined by \(\mathbf{x}\). Taking the expectation with respect to the sampling mechanism,
$$ E\{ \hat{U}_q ( \bm \beta) \mid \mathcal{F}_N \} = \sum_{i=1}^N q( \mathbf{x}_i) ( y_i - \mathbf{x}_i^{\top} \bm \beta ) \mathbf{x}_i, $$
and defining this as \(U_q ( \bm \beta)\), we have \(E\{ U_q ( \bm \beta)\}=0\) under model (\ref{14-8}). Thus, the solution to the estimating equation (\ref{14-16}),
\begin{equation}
\hat{\bm \beta}_q = \left( \sum_{i \in A} w_i q_i \bx_i \bx_i^{\top} \right)^{-1} \sum_{ i \in A} w_i q_i \bx_i y_i,
\label{14-17}
\end{equation}
is approximately unbiased regardless of the value of \(q_i = q(\mathbf{x}_i)\).

Then, what values of \(q_i\) can minimize the variance of the estimator defined by equation (\ref{14-17})? To determine this, we use the sandwich formula to calculate the variance of \(\hat{\bm \beta}_q\), which is given by:
\begin{equation}
V( \hat{\bm \beta}_q) \doteq \left( \sum_{i=1}^N q_i \bx_i \bx_i^{\top} \right)^{-1} V \left\{ \sum_{i \in A} w_i q_i \bx_i e_i \right\} \left( \sum_{i=1}^N q_i \bx_i \bx_i^{\top} \right)^{-1}.
\label{14-18}
\end{equation}
Here, since \(w_i = 1/\pi_i\) and assuming that \(e_i\) are independent,
\begin{eqnarray*}
V \left\{ \sum_{i \in A} w_i q_i \bx_i e_i \right\} &=& E\left\{ \sum_{i=1}^N \sum_{j=1}^N ( \pi_{ij} - \pi_i \pi_j ) \frac{ q_i e_i}{\pi_i} \frac{q_j e_j}{\pi_j} \bx_i \bx_j^\top  \right\} + V\left\{ \sum_{i=1}^N q_i \mathbf{x}_i e_i \right\} \\
&=& E\left\{ \sum_{i=1}^N \frac{1}{\pi_i} q_i^2 e_i^2 \mathbf{x}_i \mathbf{x}_i^{\top} \right\},
\end{eqnarray*}
where the second equality follows from the IID assumption.  
Thus, equation (\ref{14-18}) can be expressed as:
\begin{equation}
\left( \sum_{i=1}^N q_i \bx_i \bx_i^{\top} \right)^{-1} \sum_{i=1}^N q_i^2 v_i \mathbf{x}_i \mathbf{x}_i^{\top} \left( \sum_{i=1}^N q_i \bx_i \bx_i^{\top} \right)^{-1},
\end{equation}
where \(v_i = E(\pi_i^{-1} e_i^2 \mid \mathbf{x}_i)\).

Minimizing this expression with respect to \(q_i\) yields:
\begin{equation}
q_i^* = v_i^{-1} = \left\{ E( w_i e_i^2 \mid \mathbf{x}_i) \right\}^{-1}.
\label{14-19}
\end{equation}

Using the optimal \(q_i^*\) values from (\ref{14-19}) in the estimation of regression coefficients in (\ref{14-17}) yields an estimator equivalent to the generalized least squares (GLS) estimator, which minimizes
$$ Q^*( \bm \beta) = \sum_{i \in A} \frac{ w_i \left( y_i - \mathbf{x}_i^{\top} \bm \beta \right)^2 }{ E( w_i e_i^2 \mid \mathbf{x}_i) }.
$$
\index{generalized least squares estimator} To implement this GLS estimator, one must first estimate the \(q_i^*\) values defined in (\ref{14-19}) and then use these to compute the GLS estimator, known as the Estimated GLS (EGLS) method. The following two-step estimation procedure is used:

\begin{description}
\item{[Step 1]} First, set \(q_i=1\) and compute the \(\hat{\bm \beta}\) value from equation (\ref{14-17}). Let this be \(\hat{\bm \beta}^{(1)}\).
\item{[Step 2]} Using \(\hat{\bm \beta}^{(1)}\), calculate \(\hat{e}_i = y_i - \mathbf{x}_i^{\top} \hat{\bm \beta}^{(1)}\), and use these to fit a nonlinear regression model with \(w_i \hat{e}_i^2\) as the dependent variable and \(\mathbf{x}_i\) as the independent variable:
\begin{equation}
\label{14-20}
w_i \hat{e}_i^2 = q( \mathbf{x}_i; \gamma) + a_i.
\end{equation}
Estimate the regression coefficients \(\gamma\) and obtain \(\hat{q}_i^* = q( \mathbf{x}_i; \hat{\gamma})\). Then use (\ref{14-17}) to compute the EGLS estimator for \(\bm \beta\).
\end{description}

When estimating the regression coefficients in the nonlinear regression model (\ref{14-20}), it is advisable to use weights. The variance of the resulting EGLS estimator can be obtained using equation (\ref{14-10}), but replacing \(w_i\) with \(w_i \hat{q}_i\). The effect of using the estimated \(\hat{\gamma}\) can be approximately ignored. \cite{kim2013} considered an extension of this optimal estimation to generalized linear models (GLM). \index{generalized linear models} Besides the nonlinear regression model in equation (\ref{14-20}), nonparametric regression methodologies can also be applied.

\section{Likelihood-based approach}

In this section, we discuss likelihood-based approaches in survey sampling. 
Suppose that the finite population of $(x_i, y_i)$ is an independent and
identically distributed (IID) realization of the superpopulation model with
density $f(y \mid x; \bm \theta) g(x)$, where $\bm \theta$ is the parameter of interest
and the marginal density $g( \cdot)$ is completely unspecified. From the finite
population, we obtain a probability sample $A$ with a known first-order
inclusion probability $\pi_i$. We observe $x_i$ throughout the finite population but observe 
$(x_i, y_i)$ only in the sample. We are
interested in estimating the model parameter $\bm \theta$ from the complex sample,
which is  the main problem in the area of analytic inference in survey
sampling. See  \citet[Ch. 6]{fuller2009} for comprehensive overviews of analytic inference in survey sampling.

Let $I_i$ be the sampling indicator of unit $i$. Since we assume that $x_i$ are observed throughout the finite population, $I_i$ can be treated as the response indicator function of $y_i$ and we can apply the missing data analysis technique to compute the observed likelihood for $\bm \theta$. Let $\tilde{\pi}(x, y)= P( I=1 \mid x, y)$ be the conditional inclusion probability. If $\tilde{\pi}(x, y)$ is known, the observed likelihood for $\bm \theta$ can be expressed as  
\begin{equation}
 L_{\rm obs} (\bm \theta) = \prod_{i=1}^N \left[  f \left(  y_i \mid \mathbf{x}_i; \bm \theta \right) \tilde{\pi} \left( \mathbf{x}_i, y_i \right) \right]^{I_i } \left[ \int f \left(  y \mid \mathbf{x}_i; \bm \theta \right)
\{ 1- \tilde{\pi} (x_i, y) \}  d \mu(y) \right]^{1-I_i},
\label{6-1-1}
\end{equation}
where $\mu(\cdot)$ is the dominating measure. 
The observed score equation that solves 
$$ \frac{\partial}{ \partial \bm \theta} \log L_{\rm obs} (\bm \theta) = 0 $$
can be expressed as 
\begin{equation}
\sum_{i=1}^N \left[ I_i S( \bm \theta; x_i, y_i) + (1-I_i) E \left\{ S( \bm \theta; x_i, Y) \mid x_i, I_i=0; \bm \theta \right\} \right] = 0 , 
\label{14-24}
\end{equation}
where $S(\bm \theta; x, y) = \partial \log f(y \mid x; \bm \theta)/ \partial \bm \theta$ and 
\begin{equation}
 E \left\{ S( \bm \theta; x_i, Y) \mid x_i, I_i=0; \bm \theta \right\} = \frac{ \int S(\bm \theta; x_i, y)   \{ 1- \tilde{\pi} (x_i, y) \}f( y \mid x_i; \bm \theta) d \mu(y)}{\int    \{ 1- \tilde{\pi} (x_i, y) \}f( y \mid x_i; \bm \theta) d \mu(y)
}.
\label{14-25}
\end{equation}
As long as $\tilde{\pi}(x, y)$ is known and the model is identifiable, the actual computation for  solving (\ref{14-24}) can be implemented using the EM algorithm \citep{dempster77}. In the E-step, the conditional expectation in (\ref{14-25}) is evaluated at the current parameter value $\hat{\bm \theta}^{(t)}$. In the M-step, the parameter is updated by solving \begin{equation*}
\sum_{i=1}^N \left[ I_i S( \bm \theta; x_i, y_i) + (1-I_i) E \left\{ S( \bm \theta; x_i, Y) \mid x_i, I_i=0; \hat{\bm \theta}^{(t)} \right\} \right] = \mathbf{0} .
\end{equation*}

Now, if the sampling mechanism is non-informative in the sense that $\tilde{\pi}(x, y) = \tilde{\pi}(x)$, then the conditional expectation in (\ref{14-25}) reduces to 
\begin{eqnarray}
 E \left\{ S( \bm \theta; x_i, Y) \mid x_i, I_i=0; \bm \theta \right\} &=&  \frac{ \int S(\bm \theta; x_i, y)   \{ 1- \tilde{\pi} (x_i) \}f( y \mid x_i; \bm \theta) d \mu(y)}{\int    \{ 1- \tilde{\pi} (x_i) \}f( y \mid x_i; \bm \theta) d \mu(y)
} \notag  \\
&=&  \frac{ \int S(\bm \theta; x_i, y)   f( y \mid x_i; \bm \theta) d \mu(y)}{\int    f( y \mid x_i; \bm \theta) d \mu(y)
} \notag  \\
&=& E\{ S( \bm \theta; x, Y) \mid x; \bm \theta \} \notag \\
&=& \mathbf{0}, 
\label{barlett}
\end{eqnarray}
where the last equality follows from the basic property of the score function, which is sometime called the Bartlett identity. \index{Bartlett identity}

Another likelihood-based approach is to use the conditional likelihood. The conditional log-likelihood is based on the sample distribution: 
\begin{equation}
\ell_c (\bm \theta) = \sum_{i \in A}  \log \{ f(y_i \mid \bx_i, I_i=1; \bm \theta) \}
 \label{com-lik}
 \end{equation}
 where 
$$ f(y \mid \bx, I=1; \bm \theta) = \frac{ f(y \mid \bx; \bm \theta) \tilde{\pi} (\bx, y)  }{\int f(y \mid \bx; \bm \theta) \tilde{\pi} (\bx, y)  \mu (y) }. $$
Unlike the observed likelihood (\ref{6-1-1}),
the conditional likelihood can be used even when $\bx_i$'s associated with the missing $y$-values are not observed. 

The conditional score equation that solves 
$$\frac{ \partial }{ \partial \bm \theta} \ell_c (\bm \theta)= 0 
$$ 
can be expressed as 
\begin{eqnarray*}
  \sum_{i \in A}  \left[
S (\bm \theta; x_i, y_i) - E \left\{ S( \bm \theta; x_i, Y) \mid x_i, I_i=1; \bm \theta \right\} 
\right] =0 , \notag
\end{eqnarray*}
where 
$$E \left\{ S( \bm \theta; x_i, Y) \mid x_i, I_i=1; \bm \theta \right\} = \frac{ \int S(\bm \theta; x_i, y)   \tilde{\pi} (x_i, y) f( y \mid x_i; \bm \theta) d \mu(y)}{\int    \tilde{\pi} (x_i, y) f( y \mid x_i; \bm \theta) d \mu(y)
}.$$
The second term $E\left\{ S_i (\bm \theta)  \mid \mathbf{x}_i, I_i=1; \bm \theta \right\}$ can be understood as a bias term of the unweighted score function. 
If the sampling mechanism is noninformative such that $ \tilde{\pi}(x, y)= \tilde{\pi}(x)$, then the bias adjustment term is zero because of the Bartlett identify in (\ref{barlett}). 

Assuming that $\tilde{\pi}_i=\tilde{\pi}(x_i, y_i)$ is known and  writing 
$$ S_c ( \bm \theta; x, y)=S (\bm \theta; x, y) - E \left\{ S( \bm \theta; x, Y) \mid x, I=1; \bm \theta \right\},  $$
the Fisher-scoring method for obtaining the MLE from the conditional likelihood is given by
 $$ \hat{\bm \theta}^{(t+1)} = \hat{\bm \theta}^{(t)} + \left\{ \mathcal{I}_c\left(\hat{\bm \theta}^{(t)}\right) \right\}^{-1}
S_c\left(\hat{\bm \theta}^{(t)}\right),$$
where 
$$ S_c( \bm \theta ) = n^{-1} \sum_{i \in A} S_c( \bm \theta; x_i, y_i) $$
and
\begin{eqnarray}
\mathcal{I}_c(\bm \theta) &=&  - E\left\{ \frac{\partial }{\partial \bm \theta'} S_c (\bm \theta)  \right\} \notag \\ &=& 
  n^{-1} \sum_{i=1}^n  \left[ E\left\{ \tilde{\pi}(x_i,Y) S (\bm \theta; x_i, Y)^{\otimes 2}   \mid {x}_i; \bm \theta \right\}  -   \frac{ \left\{ E\left( S(\bm \theta; x_i, Y) \tilde{\pi}(x_i, Y)  \mid {x}_i; \bm \theta \right) \right\}^{\otimes 2} }{E\left(  \tilde{\pi}_i \mid \mathbf{x}_i; \bm \theta \right)  } \right] \notag 
\end{eqnarray}
 with $B^{\otimes 2} = B B^\top $. 
  \cite{wang2021c}  established the optimality of the maximum conditional likelihood estimator in the context of informative subsampling for big data analysis.

In practice, the conditional inclusion probability $\tilde{\pi}_i=\tilde{\pi}(x_i,y_i)$ is generally unknown.
    To compute the conditional inclusion probability,  we can use the following result, which was first established by \cite{Pfeffermann1999}. We assume that $(x_i, y_i, \pi_i)$ for element in $i \in U$ is a realization of the superpopualtion model. 
    \begin{lemma}
    Let $\tilde{\pi}(x,y)=E( \pi \mid x, y)$.  
\begin{equation} 
 E( \pi \mid x, y) = \frac{1}{E( w \mid x, y, I=1) },  \label{eq:14-29}
 \end{equation}
 where $w=\pi^{-1}$. 
 \end{lemma}
 \begin{proof}
The density of the conditional distribution of $w$ given $(x, y)$ and $I=1$ is 
\begin{eqnarray*} 
 f( w \mid x, y, I=1) &=& \frac{ f( w \mid x, y) P( I=1 \mid x, y, w) }{ \int f( w \mid x, y) P( I=1 \mid x, y, w) d \mu (w) } 
 \\
 &=& \frac{ f( w \mid x, y) w^{-1}  }{ \int f( w \mid x, y) w^{-1}  d \mu (w) } . \end{eqnarray*}
 Thus, 
 \begin{eqnarray*}
E \left( w \mid x, y, I=1 \right) &=& \int w f\left( w \mid x, y, I=1 \right) d \mu(w) \\
&=& \frac{ \int w f( w \mid x, y) w^{-1} d \mu(w)  }{ \int f( w \mid x, y) w^{-1}  d \mu (w) } \\
&=& \frac{1}{ \int f( w \mid x, y) w^{-1}  d \mu (w)} \\
&=& \frac{1}{ E( \pi \mid x, y) } .\end{eqnarray*} 
 \end{proof}


Therefore, regression of the weight on $x$ and $y$ among the sample data can be used to construct $\tilde{w}(x, y)= \{ \tilde{\pi}(x, y) \}^{-1}$, which is often called the weight smoothing.  The weight smoothing can reduce the variability of the sampling weight $w_i=\pi_i^{-1}$ in estimating parameters and thus can lead to more efficient estimation, as discussed by \cite{beaumont2008} and \cite{kim2013}. However, correct specification of the weight model in computing $E\{ w \mid x, y, I=1\}$ can be challenging. \cite{kim2023b} developed a score test for checking the validity of the weight model in developing the conditional maximum likelihood estimator. 


\section{Semiparametric model}

We now consider the case when the parameter of interest $\bm \theta$ is defined through $E\{ U( \bm \theta; X,Y) \}=0$. As in the previous section, we observe $(x_i, y_i)$ in sample $A$ but observe only $x_i$ outside the sample. A class of consistent estimators of $\bm \theta$ can be expressed as the solution to 
\begin{equation}
 \hat{U}_b ( \bm \theta) = 0 
 \label{eq:14-29}
 \end{equation}
where 
$$ \hat{U}_b (\bm \theta) = \sum_{i=1}^N b(\bm \theta ; x_i) + \sum_{i \in A} \frac{1}{\pi_i} \left\{ U( \bm \theta ; x_i, y_i) - b( \bm \theta; x_i)  \right\} 
$$
and $b(\bm \theta; x_i)$ is any measurable function of $x_i$. To emphasize its dependency on $b( \bm \theta; x)$, we use subscript $b$ in $\hat{U}_b( \bm \theta)$. 
 Note that 
$$ E \left\{ \hat{U}_b (\bm \theta) \mid \mathcal{F}_N \right\} = \sum_{i=1}^N U(\bm \theta; x_i, y_i). $$
Thus, the solution to (\ref{eq:14-29}) is approximately unbiased regardless of the choice of $b$-function in (\ref{eq:14-29}). The total variance is 
$$
V \left\{ \hat{U}_b( \bm \theta) \right\} = V \left\{ \sum_{i=1}^N U(\bm \theta; x_i, y_i ) \right\} + E \left\{ \sum_{i=1}^N \sum_{j=1}^N \frac{\pi_{ij}- \pi_i \pi_j}{\pi_i \pi_j}  e_i(\bm \theta) e_j(\bm \theta)  \right\}$$
where $e_i (\bm \theta) = U(\bm \theta; x_i, y_i) - b(\bm \theta; x_i)$. Under the IID setup, the second term of the variance is equal to 
$$E \left\{ \sum_{i=1}^N \sum_{j=1}^N \frac{\pi_{ij}- \pi_i \pi_j}{\pi_i \pi_j}  e_i(\bm \theta) e_j(\bm \theta)  \right\} = E\left[  \sum_{i=1}^N \left( w_i -1 \right) \left\{U(\bm \theta; x_i, y_i) - b(\bm \theta; x_i)\right\}^2   \right],  $$
where $w_i=\pi_i^{-1}$. 
Thus, the total variance is minimized at 
\begin{equation}
 b^* ( \bm \theta; x) = \frac{ E\left\{ (W-1) U(\bm \theta; x, Y) \mid x \right\} }{E\left\{ (W-1)  \mid x \right\},
} 
\label{optb}
\end{equation}
where $W=\pi^{-1}$ is the sampling weight obtained by the inverse of the first-order inclusion probability. \cite{morikawa2024} obtained $b^*(\bm \theta; x)$ in (\ref{optb}) using the projection technique in Hilbert space of the influence functions and proposed an adaptive estimator of $\bm \theta$.  

\section{Model selection}

Let $f$ be the true model that generate data $y$ and $\{ g ( y; \bm \theta), \bm \theta  \in \Omega\}$ be the  candidate model for $f$. 
Let $\hat{\bm \theta}$ be the MLE of $\bm \theta$ under the model $g( y; \bm \theta) $ computed from one realized sample. 
The Kullback-Leibler divergence of  $g(y; \bm \theta)$ from the true density  $f(y)$  is defined by 
$$KL (g \parallel f ) = \min_{\bm \theta} E_f \left\{ \log \left( \frac{f(Y)}{ g( Y; {\bm \theta}) }  \right)  \right\}. $$
Since we never know $f$, we cannot compute $KL(f, g)$ for given $g$. But, for different choices of $g$, only the second term in 
$$ 
KL (f, g) = E_f \left\{ \log f(Y)  \right\} - \max_{\bm \theta} E_f \left\{ \log g( Y; {\bm \theta}) \right\} $$
changes. Thus, finding $g$ that minimizes $KL(f,g)$ is equivalent to finding $g$ that maximizes 
$$- \max_{\bm \theta} E_f \left\{ \log g( Y; {\bm \theta}) \right\} =- E_f \left\{ \log g( Y; \hat{\bm \theta}) \right\}$$
where $\hat{\bm \theta}$ is the MLE of $\bm \theta$ obtained from a random sample from $f$. That is, 
\begin{equation}
T \equiv E_f \left\{ \log g( Y; \hat{\bm \theta}) \right\} 
= \int \int  \log g(y; \hat{\bm \theta}) f_{ \hat{\theta}} (\hat{\bm \theta}) f (y)   d \hat{\bm \theta} dy ,
\label{neq0}
\end{equation}
where $f_{ \hat{\theta}} ( \cdot)$ is the density for the sampling distribution of $\hat{\bm \theta}$. 
Thus, if we have two independent random sample from $f$,  say $x_1, \cdots, x_n$ and $y_1, \cdots, y_n$, 
we use the first set to obtain $\hat{\bm \theta}=\hat{\bm \theta} ( x_1, \cdots, x_n)$ and then use the second set to compute 
\begin{equation}
 \hat{T}= n^{-1} \sum_{i=1}^n \log g(y_i; \hat{\bm \theta}) 
 \label{neq1}
 \end{equation}
 as an estimate for $T$ in  (\ref{neq0}). Minimizing the KL distance is equivalent to  maximizing $T$ and so we can find $g$ that maximizes $\hat{T}$.    This is the basic idea of model selection using cross validation. 
 
 Now, we consider an alternative approach that estimate (\ref{neq0}) without using independent validation sample. Let $\hat{\bm \theta}$ be the PMLE obtained by maximizing 
 $$\ell_g ( \bm \theta) =N^{-1}  \sum_{i \in A} w_i \log g( y_i; \bm \theta) $$ where $y_1, \cdots, y_N$ are generated from $f$. Let
 $\bm \theta^* = p \lim \hat{\bm \theta}$ and 
  taking a Taylor expansion of $\ell_g ( {\bm \theta}^*)$ around $\bm \theta=\hat{\bm \theta}$, we have 
$$  \ell_g ( {\bm \theta}^* ) \cong \ell_g (   \hat{\bm \theta} ) + ( \hat{\bm \theta} - \bm \theta^*)^\top 
\underbrace{\dot{\ell}_g ( \hat{\bm \theta})}_{=0}  + \frac{1}{2}  ( \hat{\bm \theta} - \bm \theta^*)^\top   \ddot{\ell}_g ( \hat{\bm \theta}) ( \hat{\bm \theta} - \bm \theta^*) $$  
where $\dot{\ell}_g ( \bm \theta) = \partial l_g( \bm \theta) / \partial \bm \theta^\top  $ and $\ddot{\ell}_g( \bm \theta) = \partial^2 l_g ( \bm \theta) / \partial \bm \theta  \partial \bm \theta^\top  $. Thus, ignoring the smaller order terms, we have 
\begin{eqnarray}
E\left\{ \ell_g( \bm \theta^*) \right\} &=& E\left\{ \ell_g( \hat{\bm \theta} ) \right\} + \frac{1}{2} E \left\{
( \hat{\bm \theta} - \bm \theta^*)^\top    \ddot{\ell}_g ( \hat{\bm \theta}) ( \hat{\bm \theta} - \bm \theta^*)
\right\} \notag \\
&=& E\left\{ \ell_g( \hat{\bm \theta} ) \right\} - \frac{1}{2} \mbox{tr} \left\{J( \hat{\bm \theta}) V( \hat{\bm \theta})  \right\}
\label{eq:14-5-3}
\end{eqnarray}
where $$ J(\bm \theta) \equiv \ddot{l}_w (\bm \theta) = -  \frac{1}{N}  \sum_{i \in A} w_i \frac{\partial^2}{ \partial \bm \theta \partial \bm \theta^\top } \log g( y_i ; \bm \theta) $$

Now, note that 
\begin{equation}
 E\left\{ \ell_g( \bm \theta^*)  \right\} = E\left\{ \log g(Y; \bm \theta^*) \right\}
 \label{eq:14-5-4}
 \end{equation}
and, by the Taylor expansion again, 
\begin{eqnarray*}
\log g(Y ; \hat{\bm \theta} ) &\cong&  \log g(Y; \bm \theta^*) +  ( \hat{\bm \theta} - \bm \theta^*)^\top 
 { \nabla_\theta } \log g(Y;   { \bm \theta}^*) \\
 && + \frac{1}{2}  ( \hat{\bm \theta} - \bm \theta^*)^\top  \left\{ \nabla_{\theta \theta} \log g(Y;   {\bm \theta}^*) \right\}( \hat{\bm \theta} - \bm \theta^*),
\end{eqnarray*}
we can obtain 
\begin{equation}
 E \left\{ \log g(Y ; \hat{\bm \theta} ) \right\} \cong E\left\{ \log g(Y; \bm \theta^*)  \right\} -\frac{1}{2} \mbox{tr} \left\{J( \hat{\bm \theta}) V( \hat{\bm \theta})  \right\}.
 \label{eq:14-5-5}
 \end{equation}
Therefore, combining (\ref{eq:14-5-3})-(\ref{eq:14-5-5}), we obtain 
\begin{equation*}
 E\left\{ \log g(Y; \hat{\bm \theta} ) \right\} = E\left\{ \ell_g( \hat{\bm \theta} ) \right\} -  \mbox{tr} \left\{J( \hat{\bm \theta}) V( \hat{\bm \theta})  \right\} 
 \end{equation*}
 and we can use 
 \begin{equation}
  \hat{T} = \ell_g (\hat{\bm \theta}) - \mbox{tr} \left\{ J( \hat{\bm \theta}) \hat{V}( \hat{\bm \theta}) \right\}
  \label{tic}
  \end{equation}
 to estimate $T$ in (\ref{neq0}). The adjustment in (\ref{tic}) was first developed by \cite{takeuchi1976}. 
Thus, we can consider minimizing $AIC_w  = -2n \hat{T}_{w1}$ as discussed by \cite{lumley2015}. 

\section{Bayesian inference}

Bayesian approaches are widely used to handle complex problems. For the
Bayesian approach, a sample distribution is specified for the data, a prior distribution is specified
for parameters, and inferences are  based on the posterior distribution for the parameters.
However, when applied to survey sampling, Bayesian approaches implicitly assume that  the sampling design is non-informative.  Ignoring the sampling design can cause selection biases
(\citealp{Pfeffermann1999}), which
raises the question of how to take  sampling information into account in the Bayesian approach.
\cite{little2012calibrated} 
advocated using frequentist methods for specifying an analysis model,
and using Bayesian methods for inference under this model, leading to the
so-called calibrated Bayesian  approach. In the calibrated Bayes, the posterior means and credible intervals are calibrated to design-based estimates
and confidence intervals in this approach.

In the classical Baysian inference, the posterior values of parameter $\bm \theta$ are generated from
	\begin{equation}
 p(\bm \theta\mid  \mbox{data})\propto L(\bm \theta)\pi(\bm \theta),\label{eq:14-6-1}
	\end{equation}
	where   $L(\bm \theta)$ is the likelihood function of $\bm \theta$, and $\pi(\bm \theta)$ is a prior distribution.  Under non-informative sampling, $L(\bm \theta)=\prod_{i=1}^{n}f(y_{i} ;\bm \theta)$.

We consider an alternative that does not use the
likelihood $L(\bm \theta)$ in (\ref{eq:14-6-1}). If $\tilde{\pi}(y)=P( I =1 \mid  y)$ is known, then we may use the conditional likelihood in (\ref{com-lik}). Thus, the posterior distribution is 
$$p(\bm \theta\mid  \mbox{data})\propto L_c(\bm \theta)\pi(\bm \theta)$$
where 
$$ L_c( \bm \theta) = \prod_{i \in A} \frac{ f(y_i; \bm \theta) \tilde{\pi}(y_i)}{ \int  f(y ; \bm \theta) \tilde{\pi}( y) d\mu(y) } .
$$
If $\tilde{\pi}(y)$ is unknown, we may use (\ref{eq:14-29}) to estimate $\tilde{\pi}(y)$ from the sample. In this case, the uncertainly of the estimated parameter in $\tilde{\pi}(y)$ should be incorporated into the Bayesian framework. However, the conditional likelihood approach is known to be sensitive to model misspecification. A more flexible modeling approach can be considered to protect against model misspecification.    

Another approach is to use an approximation Bayesian computation technique. 
Let $\hat{\bm \theta}$ be the PMLE of $\bm \theta$, which is the solution to (\ref{eq:pmle}). By (\ref{14-2}), we can obtain the approximate distribution of $\hat{\bm \theta}$ given $\bm \theta$ as 
\begin{equation}
 \hat{\bm \theta}  \mid \bm \theta \sim N \left[ \bm \theta, \hat{V}( \hat{\bm \theta} \mid \bm \theta) \right]. 
 \label{eq:14-6-2}
 \end{equation}
\cite{wang2018} proposed using the sampling
distribution in (\ref{eq:14-6-2})  to obtain an approximate posterior distribution.
That is, instead of using (\ref{eq:14-6-1}), one may use 
\begin{equation}
\bm \theta^{*}\sim p(\bm \theta\mid\hat{\bm \theta})\propto g_{1}(\hat{\bm \theta}\mid\bm \theta)\pi(\bm \theta),\label{eq:14-6-3}
\end{equation}
where $p(\bm \theta\mid\hat{\bm \theta})$ is the posterior distribution of $\bm \theta$ give $\hat{\bm \theta}$, and $g_{1}(\hat{\bm \theta}\mid\bm \theta)$ is the approximate sampling distribution
of $\hat{\bm \theta}$ in (\ref{eq:14-6-2}), which can be expressed as 
$$ g_1( \hat{\bm \theta} \mid \bm \theta ) = \left( 2 \pi \right)^{-p/2} \mbox{det} \left\{ \hat{V}(\bm \theta)\right\}^{-1/2} \exp \left\{ - 0.5 ( \hat{\bm \theta} - \bm \theta )^\top  \left\{ \hat{V}( \bm \theta ) \right\}^{-1} ( \hat{\bm \theta} - \bm \theta ) \right\} $$
where $\hat{V}(\bm \theta) = \hat{V}( \hat{\bm \theta} \mid \bm \theta )$. 
\cite{wang2018} showed that 
the proposed Bayesian method is valid in the sense that the coverage properties hold for the resulting posterior credible set.

%% file: chapters/chapter15.tex
\chapter{Analysis of voluntary  samples}
 \section{Introduction}

 Non-probability samples are common, even though they do not guarantee an accurate representation of the target population. Nowadays, collecting a strict probability sample has become more challenging due to unavoidable issues such as frame undercoverage and increasing nonresponse rates. Additionally, the cost of strict probability sampling is rising. However, non-probability samples are prone to selection biases and often fail to represent the target population accurately. A popular approach to address these biases in non-probability samples is calibration weighting, which incorporates auxiliary information observed throughout the finite population. This method relies on the assumption that the sampling mechanism for the non-probability sample is ignorable after adjusting for the auxiliary variables used in the calibration weighting. This assumption is essentially the ``missing at random'' (MAR) assumption described by \cite{rubin1976}.

To formally discuss how to analyze non-probability samples, let's first introduce some notation. 
Let $U=\{1,\cdots, N\}$ be  the index set of the finite population of size $N$. Let $S \subset U$ be the index set of the sample.  
Let $\delta_i$ be the sample selection indicator for unit $i$, where  $\delta_i =1$ if $i \in S$ and $\delta_i=0$ otherwise.  
We observe $y_i$ only when $\delta_i=1$. We assume that the vectors of auxiliary variables $\bx_i$ are available  throughout the finite population. Our goal is to estimate the population total  $Y= \sum_{i=1}^N y_i$ from the sample.

The implicit model for regression calibration is the linear regression model 
\begin{equation}
 y_i = \bx_i^{\top}  \bm \beta+ e_i 
 \label{eq:15-1-1}
 \end{equation}
where $e_i$ is mean zero, independent of $\bx_i$, and the sampling mechanism is non-informative  under model (\ref{eq:15-1-1}).  As long as the weights are constructed to satisfy 
\begin{equation}
 \sum_{i \in S} \omega_i \bx_i = \sum_{i=1}^N \bx_i ,
 \label{eq:15-1-2}
 \end{equation}
the resulting calibration estimator $\hat{Y}_{\rm cal} = \sum_{i \in A} \omega_i y_i$ is unbiased for $Y=\sum_{i=1}^N y_i$. To see this, 
$$ \hat{Y}_{\rm cal} - Y =  \underbrace{\left( \sum_{i \in S} \omega_i \bx_i - \sum_{i=1}^N \bx_i\right)^\top }_{= 0 } \bm \beta +\sum_{i \in S} \omega_i e_i - \sum_{i=1}^N e_i$$
and the second term has zero expectation. 
Note that the justification is based on the regression model assumption in (\ref{eq:15-1-1}).  Calibration weighting methods for non-probability samples have been discussed in \cite{dever2016}, \cite{Kott2010},  and \cite{elliott2017inference}, among others.

Another justification for the calibration estimator lies within the quasi-randomization approach, where the reference distribution is the randomization distribution under the sampling mechanism. Unlike probability sampling, the sampling mechanism for voluntary mechanism is unknown. 
In this case, the first-order inclusion probability  is also  unknown,  and the theory in the previous chapters cannot be directly applied. 

Consider  the regression estimator defined by \index{Regression estimator}
\index{Regression weighting | \see{Regression estimator}}
\begin{equation}
\widehat{Y}_{\rm reg} = \sum_{i \in S}   \hat{\omega}_i y_i = \sum_{i=1}^N \bx_i^\top \hat{\bm \beta},
\label{eq:15-1-4}
\end{equation}
where
$$
\hat{\omega}_i =\left(  \sum_{i=1}^N \mathbf{x}_i \right)^\top  \left( \sum_{i \in S}  \mathbf{x}_i \mathbf{x}_i^{\top}    \right)^{-1} \mathbf{x}_i ^{-1}   $$
and 
$$ \hat{\bm \beta}= \left( \sum_{i \in S} \mathbf{x}_i \mathbf{x}_i^{\top}  \right)^{-1} \sum_{i \in S}   \mathbf{x}_i y_i ^{-1} . $$
 Note that the regression weight in (\ref{eq:15-1-4}) satisfies the calibration equation in (\ref{eq:15-1-2}).  

Let $\pi_i$ be the (unknown) first-order inclusion probability of sample $S$. The regression estimator in (\ref{eq:15-1-4}) is internally bias calibrated if it satisfies 
\begin{equation}
\sum_{i \in S} \frac{1}{\pi_i} \left( y_i - \bx_i^\top \hat{\bm \beta} \right) = 0 .
\label{eq:15-ibc}
\end{equation}
See equation (\ref{ibc}) for the IBC condition of the model-based projection estimator. As long as the condition in (\ref{eq:15-ibc}) is satisfied, the resulting regression estimator is design consistent under the quasi-randomization approach. 

Since $\hat{\bm \beta}$ satisfies 
$$ \sum_{i \in S} \left( y_i - \bx_i^\top \hat{\bm \beta} \right) \bx_i ^{-1} = \mathbf{0}, 
$$
by construction. A sufficient condition for the IBC condition in (\ref{eq:15-ibc}) is that    the unknown first-order inclusion probability satisfies
\begin{equation}
\frac{1}{\pi_i} = \mathbf{x}_i^{\top}  \bm \lambda, \ \  
\label{eq:15-1-3}
\end{equation}
for some  $\bm \lambda$.

Therefore, the regression estimator can be justified using  two different approaches: the superpopulation model approach and the quasi-randomization approach. Because the regression estimator is justified under two different models, it enjoys the double robustness property as discuss in \cite{bang05} and \cite{kimhaziza12}.

The key assumption under the quasi-randomization approach is the condition in (\ref{eq:15-1-3}). This condition is essentially the internal bias calibration (IBC) condition \citep{firth1998} of the projection estimator.
To satisfy this condition, one may build a model for $\pi_i$ and then include $ \hat{\pi}_i^{-1}$ into the auxiliary variables in the regression estimation. We explore this approach in more detail in the next section.

\section{Calibration weighting under missing at random}

Calibration weighting under  voluntary sampling is essentially the same as the calibration weighting for nonresponse adjustment. Thus, we can apply the ideas in Section 12.4 directly to our problems. 

Our goal is to construct the final weight $\omega_i$ in $S$ so that
$$ \widehat{Y}_\omega = \sum_{i \in S} \omega_i y_i $$
can be used to estimate $Y$.  
To construct the  calibration weights satisfying to the calibration constraint in (\ref{eq:15-1-2}), we use 
\begin{equation}
 Q(\omega) = \sum_{i \in S} G \left( {\omega_i} \right)  
 \label{eq:15-2-2}
 \end{equation}
as the objective function for calibration, where  $G: \mathcal V \to \mathbb R$ is the convex function for generalized entropy calibration (GEC) that was introduced in Section 9.4.

    Using the Lagrange multiplier method, we find the minimizer of 
    $$ \mathcal{L} ( \bomega, \bm \lambda) =  \sum_{i \in S} G( \omega_i)  - \bm \lambda^\top \left( \sum_{i \in S} \omega_i \bx_i - \sum_{i=1}^N \bx_i \right) $$
    with respect to $\bm \lambda$ and $\bomega$. 
    By setting 
    $ \frac{ \partial}{\omega_i} \mathcal{L} = 0$
    and solving for $\omega_i$, we obtain 
    $$ \hat{\omega}_i ( \bm \lambda) = g^{-1} \left( \bx_i^\top \bm \lambda  \right) , $$
    where $g(\omega) = d G( \omega) / d \omega$. Thus, by plugging $\hat{\omega}_i ( \bm \lambda)$ into $\mathcal{L}$,    
    we can formulate a  dual optimization problem: 
 \begin{equation}
 \hat{\bm \lambda} = \mbox{arg} \min_{\lambda} \left[  \sum_{i=1}^N \bx_i^\top \bm \lambda - \sum_{i=1}^N \delta_i \rho \left( \bx_i^\top \bm \lambda   \right)  \right] ,  
 \label{dual}
 \end{equation}
 where 
 $ \rho \left( \nu \right) $
 is the convex conjugate function of $G$, which is defined by 
 $$ \rho \left( \nu \right) = \nu \cdot g^{-1} ( \nu) - G \{ g^{-1} ( \nu) \}$$
 where $g(\omega) = d G( \omega) / d \omega$. See Table \ref{tab:15-1} for the examples of the generalized entropy functions and their convex conjugate functions. 

\begin{table}[ht]
\begin{center}
\caption{Examples of generalized entropies, $G(\omega)$, and the corresponding convex conjugate functions, and their derivatives } \begin{tabular}{ccccc}
\hline
Generalized Entropy                                             & $G(\omega)$                 & $\rho(\nu)$     & $\rho^{(1)} ( \nu)$      \\ \hline
Squared loss                                       & $\omega^2/2$                  & $\nu^2/2$              & $\nu$                   \\  
Kullback-Leibler & $\omega \log (\omega)$ & $\exp ( \nu -1)$ &  $\exp ( \nu -1 )$ \\ 
Shifted KL & $(\omega-1) \{ \log (\omega-1)- 1\}$ & $
\nu + \exp ( \nu )$ &  $1+ \exp ( \nu )$ \\
Empirical likelihood & $- \log ( \omega)$ & $-1-\log ( - \nu)$  & $1/ \nu$ \\
Squared Hellinger & $( \sqrt{\omega} -1 )^2$ & $\nu/(\nu-1)$ & $-(\nu-1)^{-2}$ \\
R\'enyi entropy($\alpha \neq 0, -1$) & $ \frac{1}{\alpha+1} \omega^{\alpha + 1}$   & $ \frac{\alpha}{\alpha+1} \nu^{\frac{\alpha+1}{\alpha}}$ &  $\nu^{1/\alpha}$\\
\hline
\end{tabular}
\end{center}
\label{tab:15-1}
\end{table}

 By construction, we have 
\begin{equation}
\sum_{i=1}^N \delta_i \rho^{(1)} \left( \bx_i^\top \hat{\bm \lambda}   \right) \bx_i =  \sum_{i=1}^N \bx_i. 
\label{15-cal2}
\end{equation}  
 Once $\hat{\bm \lambda}$ is obtained, we can obtain 
 $$ \widehat{Y}_{\rm GEC} = \sum_{i=1}^N \delta_i \rho^{(1)} \left( \bx_i^{\top} \hat{\bm \lambda}   \right) y_i $$
as the proposed calibration estimator, where $\rho^{(1)} (\nu)  = d \rho(\nu)/ d \nu$. 
 
 To emphasize its dependency on $\hat{\bm \lambda}$, we can write $\widehat{Y}_{\rm GEC} = \widehat{Y}_{\rm GEC} ( \hat{\bm \lambda} ) $. Now, define 
 \begin{equation}
 \widehat{Y}_{\ell} \left( \bm \lambda , \bm \gamma \right) = \hat{Y}_{\rm GEC} ( \bm \lambda) + \left\{ \sum_{i=1}^N \bx_i  - \sum_{i=1}^N \delta_i \rho^{(1)} \left( \bm \lambda^\top \bx_i   \right) \bx_i \right\}^\top \bm \gamma . 
 \label{lin} 
 \end{equation}
Note that $\hat{Y}_{\ell} \left( \bm \lambda, \bm \gamma \right)$ satisfies the following two properties. 
\begin{enumerate}
\item $\widehat{Y}_{\ell} ( \hat{\bm \lambda}, \bm \gamma ) = \hat{Y}_{\rm GEC}$ for all $\bm \gamma$. 
\item Let $\bm \lambda^*$ be the probability limit of $\hat{\bm \lambda}$.  If $\bm \gamma^*$ is  the solution to 
\begin{equation}
 E \left\{ \nabla_{\lambda} \widehat{Y}_{\ell} \left( \bm \lambda^*, \bm \gamma \right) \right\}= \mathbf{0}, 
 \label{eq:15-2-4}
 \end{equation}
then $\widehat{Y}_{\ell} ( \hat{\bm \lambda}, \bm \gamma^*)$ is asymptotically equivalent to $\widehat{Y}_{\ell} ( {\bm \lambda}^*, \bm \gamma^*)$ in the sense that 
$$ \widehat{Y}_{\ell} ( \hat{\bm \lambda}, \bm \gamma^*) = \widehat{Y}_{\ell} ( {\bm \lambda}^*, \bm \gamma^*) + o_p \left(n^{-1/2} N \right) 
$$
\end{enumerate}
Condition (\ref{eq:15-2-4}) is often called Randles' condition \citep{randles1982aysmptotic}. For $\widehat{Y}_{\ell}$ in (\ref{lin}), the  equation for $\bm \gamma^*$ is 
\begin{equation}
 E \left\{ \sum_{i=1}^N \delta_i \rho^{(2)} \left( \bx_i^{\top} \bm \lambda^*   \right) \left( y_i - \bx_i^{\top} \bm \gamma^* \right) \bx_i  \right\} = \mathbf{0} , 
 \label{randles2}
 \end{equation}
where 
$ \rho^{(2)} (\nu) = d^2  \rho ( \nu) / d \nu^2. $ Thus, $\bm \gamma^*$ is consistently estimated by 
$$ \hat{\bm \gamma} = \left\{ \sum_{i \in S}\rho^{(2)} \left( \bx_i^{\top} \hat{\bm \lambda}   \right)\bx_i \bx_i^{\top}  \right\}^{-1} \sum_{i \in S} \rho^{(2)} \left( \bx_i^{\top} \hat{\bm \lambda}   \right)
\bx_i y_i. $$

\begin{theorem}

Let $\hat{\omega}_i$ be obtained by minimizing (\ref{eq:15-2-2}) subject to (\ref{eq:15-1-2}).  Under some regularity conditions, the resulting calibration estimator 
$ \widehat{Y}_{\rm GEC} = \sum_{i \in S}  \hat{\omega}_i y_i $ 
satisfies 
\begin{equation}
 \widehat{Y}_{\rm GEC}  = \sum_{i=1}^N \eta_i + o_p \left( n^{-1/2} N \right) 
\label{eq:15-2-5}
\end{equation}
where 
$$ \eta_i = \bx_i^\top \bm \gamma^*   + \delta_i \rho^{(1)} \left( \bx_i^\top \bm \lambda^*   \right) \left( y_i - \bx_i^{\top} \bm \gamma^* \right), $$
$$\bm \gamma^* = \left( \sum_{i=1}^N \delta_i \rho^{(2)} \left( \bx_i^\top \bm \lambda^*  \right) \bx_i \bx_i^{\top}  \right)^{-1} \sum_{i=1}^N \delta_i \rho^{(2)} \left( \bx_i^\top \bm \lambda^*  \right) \bx_i y_i  $$
and $\bm \lambda^* = p \lim \hat{\bm \lambda}$. 
\end{theorem}

Result (\ref{eq:15-2-5}) presents the asymptotic equivalence between the GEC estimator and its dual form of the regression estimator.  That is, we can obtain  
\begin{equation}
 \hat{Y}_{\rm GEC}  = \hat{Y}_{\rm GREG}+ o_p \left( n^{-1/2} N \right) 
\label{eq:15-2-6}
\end{equation}
where 
$$\hat{Y}_{\rm GREG} 
= \sum_{i=1}^N \bx_i^{\top} \hat{\bm \gamma} + \sum_{i \in S} \hat{\omega}_i \left( y_i - \bx_i^{\top} \hat{\bm \gamma} \right) 
$$
with $\hat{\omega}_i = \rho^{(1)} ( \bx_i^\top \hat{\bm \lambda} )$. 
Note that the linearization in (\ref{eq:15-2-5}) does not use any model assumption. The consistency of $\hat{Y}_{\rm GEC}$ can be established under one of the two model assumptions.
\begin{enumerate}
\item Outcome regression (OR) model in (\ref{eq:15-1-1}). 
\item Propensity score (PS) model given by 
\begin{equation}
 P \left( \delta_i =1 \mid \bx_i \right) = \left\{ \rho^{(1)} \left( \bx_i^\top \bm \phi_0 \right) \right\}^{-1} .
\label{eq:15-2-7}
\end{equation}
for some $\bm \phi_0$.
\end{enumerate}

Under the linear regression model in (\ref{eq:15-1-1}), we have $\bm \gamma^* = \bm \beta$ and 
$$ \eta_i = y_i+  \left\{ \delta_i \rho^{(1)} \left( \bx_i^\top \bm \lambda^* \right) -1 \right\} e_i $$
where $e_i = y_i - \bx_i^\top \bm \beta$. Therefore, under (\ref{eq:15-1-1}) and MAR assumption,  we have 
$$ E\left( \eta_i - y_i\mid \delta_i, \bx_i \right) = 0 $$
and 
$$ V\left( \eta_i - y_i\mid \delta_i \bx_i \right) =\left\{ \delta_i \rho^{(1)} \left( \bx_i^\top \bm \lambda^* \right) -1 \right\}^2 V( y_i \mid \bx_i) $$
Therefore, $\widehat{Y}_{\rm GEC}$ is asymptotically unbiased with its asymptotic variance equal to 
$$ V \left( \widehat{Y}_{\rm GEC} - Y \right) \cong E \left[ \sum_{i=1}^N 
\left\{ \delta_i \rho^{(1)} \left( \bx_i^\top \bm \lambda^* \right) -1 \right\}^2 V( y_i \mid \bx_i)\right] .
$$


On the other hand, under the PS model in (\ref{eq:15-2-7}), we can obtain $\bm \lambda^*= \phi_0$ and establish 
$$ E \left( \eta_i \mid \bx_i, y_i \right) = y_i $$
where the conditional distribution is with respect to the probability law in $[ \delta \mid \bx, y]$. Thus, under (\ref{eq:15-2-7}), ignoring the smaller order terms, 
$$ 
V \left( \widehat{Y}_{\rm GEC} - Y \right) \cong E \left\{ \sum_{i=1}^N \left( \pi_i^{-1} -1 \right) \left( y_i - \bx_i^\top \bm \gamma^* \right)^2 \right\},
$$
where $\pi_i = \{ \rho^{(1)} \left( \bx_i^\top \bm \lambda^* \right)\}^{-1}$. 

For variance estimation, we can use 
$$ \hat{V} = \frac{N}{N-1} \sum_{i=1}^N \left( \hat{\eta}_i - \bar{\eta}_N \right)^2 $$
where 
$$ \hat{\eta}_i = \bx_i^{\top} \hat{\bm \gamma} + \delta_i \rho^{(1)} \left( \bx_i^\top \hat{\bm \lambda}  \right) \left( y_i - \bx_i^{\top} \hat{\bm \gamma}  \right) $$
 and $\bar{\eta}_N= N^{-1} \sum_{i=1}^N \hat{\eta}_i$.  The above variance estimator is doubly robust in the sense that it is justified under the assumption that either one of the two models is correctly specified.

\section{Calibration  weighting under missing not at random}

We now consider more challenging case of non-MAR sampling mechanism. 
 Let $\pi (\bx,y) =P(\delta =1 \mid \bx, y)$ be the propensity score (PS) function for the sampling mechanism.   In the voluntary sampling, we do not know the propensity score function $\pi(\bx,y)$. Instead, we may use a model on  $\pi(\bx,y)$, say $\pi(\bx,y)=\pi(\bx,y; \bm \phi_0)$ for some $\bm \phi_0$, and develop methods for bias adjustment under the model. We first discuss how to estimate $\bm \phi_0$ from the voluntary sample and then discuss estimation of $Y=\sum_{i=1}^N y_i$. 

  We assume that the propensity score function follows a parametric model such that 
    $\pi(\bx,y)=\pi(\bx,y; \bm \phi_0)$ for some $\bm \phi_0 \in \Phi \subset \mathbb{R}^q$. 
 The observed likelihood function of $\phi$ derived from the marginal density of $(\delta_i, \delta_i y_i )$ given $\bx_i$ can be written as 
    \begin{eqnarray}
L_{\rm obs} ( \phi) 
&=& \prod_{i=1}^N   \left\{ f( y_i \mid \bx_i, \delta_i=1) \right\}^{\delta_i} \times \prod_{i=1}^N \{ \tilde{\pi} (\bx_i; \phi)\}^{\delta_i} \{ 1- \tilde{\pi} (\bx_i; \phi)\}^{1-\delta_i} ,    \notag  \end{eqnarray}
where 
$$
\tilde{\pi}(\bx; \bm \phi) = \mathbb{E} \left\{ \pi(\bx, Y; \bm \phi) \mid \bx \right\}. 
$$

Note that the first component of the observed likelihood 
is about a regression model assumption for $f( y \mid \bx, \delta=1)$.  As $(\bx_i, y_i)$ are observed for $\delta_i=1$,  we can apply model diagnostic tools to verify the model. Thus, we can safely assume that $f_1( y \mid \bx) \equiv f(y \mid \bx, \delta=1)$ is correctly specified. In this case, we can use the following identity which was originally proved by \cite{Pfeffermann1999}: 
 \begin{equation} \label{eq:pfeffermann}
    \tilde{\pi}(\bx;\bm \phi) = \left[\int \{ \pi(\bx,y;\bm \phi)\}^{-1} f_1( y \mid \bx) dy \right]^{-1}.
\end{equation}

Thus, to estimate $\bm \phi$, we can maximize 
\begin{equation} 
    \ell_{\rm obs} (\bm \phi) = \sum_{i=1}^{N} \left\{ \delta_{i} \log \hat{\pi}(\bx_{i};\bm \phi) + (1-\delta_{i}) \log \left( 1-\hat{\pi}(\bx_{i};\bm \phi) \right) \right\},
    \label{15-obs2} 
\end{equation}
where 
\begin{equation}
\hat{\pi}(\bx;\bm \phi) = \left[\int \{ \pi(\bx,y;\bm \phi)\}^{-1} \hat{f}_1( y \mid \bx) dy \right]^{-1}
\label{hatpi}
\end{equation}
and $\hat{f}_1( y \mid \bx)$ is a consistent estimator of $f_1( y \mid \bx)$.  If we define 
$ \omega( \bx, y; \bm \phi) = \{ \pi(\bx, y; \bm \phi) \}^{-1} $
and $\widehat{\omega} (\bx; \bm \phi) = \{ \hat{\pi}(\bx;\bm  \phi) \}^{-1} $, then (\ref{hatpi}) can be expressed as 
\begin{equation}
\widehat{\omega}(\bx;\bm \phi) = \int  \omega(\bx,y;\bm \phi) \hat{f}_1( y \mid \bx) dy .
\label{hatpi2}
\end{equation}
 The propensity weight in (\ref{hatpi2}) can be called the smoothed propensity weight \citep{kim2013}. To achieve model identifiability, we assume that the mapping
$ \bm \phi \mapsto \hat{\pi}(\bx; \bm \phi)  $ 
is one-to-one, almost everywhere \citep{morikawa2018}. A sufficient condition is to satisfy the nonresponse instrumental variable assumption \citep{wang2014}.

Given $\hat{f}_1( y \mid \bx)$, we can find  the maximizer of $\ell_{\rm obs} ( \bm \phi)$ in (\ref{15-obs2}) to obtain $\hat{\bm \phi}$. Using $\hat{\bm \phi}$, we can construct $\hat{\pi}_i=\pi(\bx_i, y_i; \hat{\bm \phi}) $ and apply the generalized entropy calibration method. That is, we find the minimizer of $Q(\omega)$ in 
(\ref{eq:15-2-2}) subject to (\ref{eq:15-1-2}) and 
\begin{equation}
\sum_{i=1}^N \delta_i \omega_i g\left( \hat{\pi}_i^{-1} \right) = \sum_{i=1}^N g\left( \hat{\pi}_i^{-1} \right). \label{eq:15-2-1}
\end{equation} 
However, we cannot use (\ref{eq:15-2-1}) because we cannot compute $\hat{\pi}_i =\pi( \bx_i, y_i; \hat{\bm \phi})$ outside the sample as $y_i$ are not observed for $\delta_i=0$. Therefore, instead of using (\ref{eq:15-2-1}), we may use 
\begin{equation}
\sum_{i \in S} \omega_i g_i = \sum_{i=1}^N \tilde{g}_i  
\label{15-3-3}
\end{equation}
where $g_i= g \left( \hat{\pi}_i^{-1} \right)$ and 
$ \tilde{g}_i$ is a consistent estimator of $E \left( g_i \mid \bx_i \right)$. Since we have assumed that $\hat{f}_1( y \mid \bx)$ is consistent for $f_1( y \mid \bx) = f( y \mid \bx, \delta=1)$, we may use 
\begin{eqnarray*}  E \left( g_i \mid \bx_i\right) &=& \int g \left[ \{\pi (\bx_i, y; \hat{\bm \phi} )\}^{-1} \right] f( y \mid \bx_i) d \mu(y)  \\
&=& \frac{ \int g \left[ \{\pi (\bx_i, y; \hat{\bm \phi} )\}^{-1} \right] f_1( y \mid \bx_i)\{\pi (\bx_i, y; \hat{\phi} )\}^{-1} d\mu(y)}{\int  f_1( y \mid \bx_i)\{\pi (\bx_i, y; \hat{\phi} )\}^{-1} d\mu(y)}
\end{eqnarray*} 
to compute $\tilde{g}_i$.

\cite{kim2023c} considered the special case of  $G(\omega)=-\log (\omega)$. In this case, the debiasing calibration constraint in (\ref{15-3-3}) is changed to 
\begin{equation}
 \sum_{i \in S} \omega_i {\pi} (\bx_i, y_i ; \hat{\bm \phi} ) = \sum_{i=1}^N \hat{\pi} (\bx_i; \hat{\bm \phi} ).   
 \label{eq:15-3-4}
 \end{equation}
The final weights can be obtained by maximizing $\sum_{i \in S} \log \left( \omega_i \right)$ subject to (\ref{eq:15-1-2}) and (\ref{eq:15-3-4}). According to \cite{kim2023c}, 
the resulting EL  estimator $ \widehat{Y}_{\rm EL} = \sum_{ i \in S} \hat{\omega}_i y_i $ satisfies 
\begin{equation}
 \widehat{Y}_{\rm EL} = \sum_{i=1}^N \left\{ \hat{y}_i^{(0)} + \frac{\delta_i}{ \hat{\pi}_i } \left( y_i - \hat{y}_i^{(1)} \right) \right\} + o_p(n^{-1/2} N ), 
\label{eq:15-3-5}
\end{equation}
where $\hat{\pi}_i = \pi(\bx_i, y_i; \hat{\bm \phi})$, $\hat{y}_i^{(0)}  = \hat{\pi} (\bx_i; \hat{\bm \phi} ) \hat{\beta}_1 + \tilde{\bx}_i^{\top}  \hat{\bm \beta}_2 $, $\hat{y}_i^{(1)}  = {\pi} (\bx_i, y_i; \hat{\bm \phi} ) \hat{\beta}_1 + {\bx}_i^{\top}  \hat{\bm \beta}_2 $ and 
 $$
 \begin{pmatrix}
\hat{\beta}_1 \\
\hat{\bm \beta}_2 \\
\end{pmatrix}
= \left\{ 
\sum_{i \in S} \hat{\pi}_i^{-2} \begin{pmatrix} 
\hat{\pi}_i - \hat{W} \\
{\bx}_i - \bar{\mathbf{X}}_N  \end{pmatrix} \begin{pmatrix} 
\hat{\pi}_i - \hat{W} \\
{\bx}_i - \bar{\mathbf{X}}_N
\end{pmatrix}^\top  \right\}^{-1}\sum_{i \in S} \hat{\pi}_i^{-2} \begin{pmatrix} 
\hat{\pi}_i - \hat{W} \\
{\bx}_i - \bar{\mathbf{X}}_N
\end{pmatrix} y_i . $$ 
 Here, $\hat{\pi}_i = \pi(\bx_i, y_i ; \hat{\bm \phi})$ and $\hat{W}=N^{-1} \sum_{i=1}^N{\pi} (\bx_i; \hat{\bm \phi} ) $.

\section{Data Integration approach}

We now consider the setup where $\bx_i$ are not available throughout the finite population. Instead, we assume that there is an independent probability sample, selected from the same finite population. In this case, we can utilize the auxiliary information in the probability sample to construct the calibration weights in the non-probability sample.

 We use the set-up considered in \cite{yang2020}
where sample A is a probability sample observing $\bx$ and sample B is the nonprobability sample observing $(\bx, y)$.  
Table \ref{table:15-2} presents the general setup of the two sample structures for data integration.  
As indicated in Table \ref{table:15-2}, sample $B$  is not representative of the target population. 

\begin{table}[h!]
\begin{center} 
\caption{Data Structure for  Data integration and Data fusion}
\label{table:15-2}\par
\begin{tabular}{c|c|ccc}
\hline 
\multicolumn{5}{c}{
Data Integration} \\
\hline 
\hline
Sample & Type & $X$ & $Y$ & Representative? \\
\hline 
$A$ & Probability Sample & \checkmark &   & Yes \\
$B$ & Non-probability Sample & \checkmark & \checkmark & No \\ 
\hline 
\end{tabular}
\end{center} 
\end{table}

The formulation is somewhat similar to the two-phase sampling:  
\begin{enumerate}
\item The first-phase sample $S_1 \equiv A \cup B$ is selected from $U$ and  $\bx_i$ is observed for all units in $S_1$ 
\item The second-phase sampling  $S_2=B$ is selected from $S_1$ and  $y_i$  is observed for all units in $S_2$. 
\end{enumerate} 
Unlike classical two-phase sampling, we do not know the first-order inclusion probability of $S_1$. Instead, we only know the first-order inclusion probability of the sample $A$. That is,  $\pi_i^{(A)}= P( i \in A \mid i \in U )$ is the (known) first-order inclusion probability of sample $A$. 

Let $\pi_i^{(B)} = P( i \in B \mid i\in U )$ be the (unknown) first-order inclusion probability of sample $B$. We assume that  $\pi_i^{(B)}= P( i \in B \mid i \in U)$ can be parametrically modelled as 
\begin{equation}
\pi_i^{(B)} = \pi_B( \bx_i; \bm \phi)
\label{eq:15-4-3}
\end{equation}
for some parameter $\bm \phi$. 
\cite{chen2020doubly} used  the pseudo maximum likelihood method  to estimate $\bm \phi$ in this setup.

Note that the first-order inclusion probability of $S_1$ can be written as 
\begin{eqnarray}
P( i \in S_1 \mid i \in U) &=& P( i \in A \cup B \mid i \in U)\notag  \\
&=& \pi_i^{(A)} + \pi_i^{(B)} - \pi_i^{(A)} \pi_i^{(B)} \label{0}
\end{eqnarray}
where the last equality follows from the independence of two samples.  Thus,  we can express the conditional inclusion probability for the second-phase sample as  
\begin{equation}
P( i \in S_2 \mid i \in S_1) 
= \frac{P( i \in B \mid i \in U )  }{P( i \in A \cup B \mid i \in U) }= \frac{\pi_i^{(B)} }{\pi_i^{(A)} + \pi_i^{(B)}  - \pi_i^{(A)} \cdot \pi_i^{(B)} }. 
\label{eq:15-4-1}
\end{equation}

Now, using  the PS model in (\ref{eq:15-4-3}),    we can obtain 
\begin{equation}
\frac{\pi_i^{(B)}( \bm \phi) }{\pi_i^{(A)} + \pi_i^{(B)} (\bm \phi)- \pi_i^{(A)} \cdot \pi_i^{(B)} ( \bm \phi)  } := {p}(\bx_i; {\bm \phi}) 
\label{eq:15-4-4}
\end{equation}
and compute the conditional maximum likelihood estimator of $\bm \phi$.  That is, 
 $$ 
 \hat{\bm \phi} = \mbox{arg} \max_{\phi} \sum_{i \in S_1} \left[\delta_i\log p(\bx_i; \bm \phi) + \left( 1- \delta_i\right) \log \{ 1- p(\bx_i;\bm  \phi) \} \right]  ,
 $$
 where $\delta_i = \mathbb{I} (i \in B)$ is the indicator function of the event   $i \in B$. 
 The conditional MLE of $\bm \phi$ is more efficient than the pseudo maximum likelihood estimator.

  In practice, we cannot use (\ref{eq:15-4-4}) directly as the first-order inclusion probabilities are unknown outside the sample.  One way to handle this problem is to estimate $w_i^{(A)}=1/\pi_i^{(A)}$ by 
\begin{equation}
\tilde{w}_i^{(A)} = E\{w_i^{(A)}  \mid \bx_i , I_i^{(A)} =1 \}
\label{sw}
\end{equation}
following the result of \cite{Pfeffermann1999}.
We can fit a unweighted regression of $w_i^{(A)}$ on $\bx_i$ in the sample to obtain $\hat{w}_i^{(A)}$. We can use $\hat{\pi}_i^{(A)} = 1/ w_i^{(A)}$ instead of $\pi_i^{(A)}$ in (\ref{eq:15-4-4}) to construct  $p( \bx_i; \bm \phi)$ and apply the conditional ML estimation method to estimate  $\bm \phi$.

 Once $\hat{\bm \phi}$ is obtained, we can construct the debiased calibration weights for sample B. Specifically,  we find the minimizer of $Q(\bm \omega)= \sum_{i \in B} G( \omega_i) $ subject to 
 \begin{equation}
\sum_{i \in B} \omega_i \bx_i = \sum_{i \in A} \frac{1}{\pi_i^{(A)} } \bx_i
\label{eq:15-4-5}
 \end{equation}
 and 
\begin{equation}
\sum_{i \in B} \omega_i g\left( 1/ \hat{\pi}_i^{(B)} \right) = \sum_{i \in A} \frac{1}{ \pi_i^{(A)}} g\left( 1/ \hat{\pi}_i^{(B)}  \right) 
\label{eq:15-4-6}
\end{equation} 
where $g( \omega) = d G( \omega)/ d \omega$ and $\hat{\pi}_i^{(B)}=\pi_B \left( \bx_i; \hat{\bm \phi}\right)$.
Constraint (\ref{eq:15-4-5}) is the usual calibration constraint using the linear outcome  regression model. Constraint (\ref{eq:15-4-6}) is the debiasing calibration constraint using the PS model in (\ref{eq:15-4-3}). Once $\hat{\omega}_i$ are obtained from this debiasing calibration, we can compute $\widehat{Y}_{\rm cal} = \sum_{i \in B} \hat{\omega}_i y_i$ to estimate $Y=\sum_{i=1}^N y_i$. 


\section{Multiple bias calibration}

The calibration weighting method discussed in Section 15.3 is based on the assumption that the propensity score (PS) model is correctly specified.  Since the nonignorable PS model depends on the missing study variable, it is difficult to verify the model assumption from the observed sample.  To protect bias against model misspecification, we now consider multiple PS  models and develop a consistent estimation method that gives valid inference results as long as one of the multiple PS  models is correctly specified. Specifically, we wish  to extend the EL estimator in the previous section to multiple PS models, such that it is $\sqrt{n}$-consistent  as long as one of the working PS models is correctly specified.


Suppose that  multiple working PS models are  expressed as $\{\pi_{k}(\bx,y;\phi_{k}): k = 1, \ldots, m\}$.  We assume that one of the working PS model is correctly specified. If all the PS models are MAR  in the sense that $\pi_k ( \bx, y)=\pi_k(\bx)$, then the debiasing constraint can be constructed easily as 
\begin{equation}
\sum_{i \in S} \omega_i \pi_k ( \bx_i) = \sum_{i=1}^N \pi_k( \bx_i) , \ \ k=1, \ldots, m  
\label{eq:15-debiase2}
\end{equation}
for the case of $G(\omega) =- \log (\omega) $. Such debiasing constraint was considered by \cite{han2014} in the context of multiply robust estimation under MAR assumptions under missing data. If some of the PS models is not MAR, then we cannot compute $\pi(\bx_i, y_i)$ for units with $\delta_i=0$. In this case, instead of using (\ref{eq:15-debiase2}), one may consider \begin{equation}
\sum_{i \in S} \omega_i \pi_k ( \bx_i,y_i) = \sum_{i=1}^N \tilde{\pi}_k( \bx_i) , \ \ k=1, \ldots, m  
\label{eq:15-debiase3}
\end{equation}
as a debiasing constraint where 
$ \tilde{\pi}_k( \bx_i)
= E \left\{ \pi_k( \bx_i, Y) \mid \bx_i \right\}. 
$ 
Since $\pi_k(\bx, y)$ is not necessarily correct, we cannot apply (\ref{eq:pfeffermann}) to compute $\tilde{\pi}_k(\bx)$. Instead, we use  
$$ \tilde{\pi}_k( \bx) = \int \pi_k(\bx, y) f( y \mid \bx) d \mu(y)  $$
to calculate the right hand side of (\ref{eq:15-debiase3}). Therefore, we need to estimate $f( y \mid \bx)$ first in order to apply the debiasing constraint in (\ref{eq:15-debiase3}). 

To motivate simultaneous estimation of the model parameter in the outcome model $f( y \mid \bx)$ and the parameters in the multiple PS models, let's assume for now that  the true PS function $\pi(\bx, y)=P( \delta=1 \mid \bx, y)$ is known. In this case,  if  we are interested in estimating $\bm \beta$ in $f( y \mid \bx; \bm \beta)$, then the maximum likelihood estimator of $\bm \beta$ can be obtained by solving 
\begin{equation}
\sum_{i=1}^N \left[ \delta_i S (\bm \beta ; \bx_i, y_i) + (1- \delta_i) E_0\left\{ S (\bm \beta; \bx_i, Y) \mid \bx_i ; \bm  \beta\right\}\right] = 0, 
\label{eq:15-5-1}
\end{equation}
where 
$$E_0\left\{ S  (\bm \beta; \bx_i, Y) \mid \bx_i ;  \bm \beta\right\} =  \frac{ \int S (\bm \beta; \bx_i, y) f( y \mid \bx_i ; \bm \beta) \{ 1- \pi(\bx_i, y) \} dy }{\int  f( y \mid \bx_i ; \bm \beta) \{ 1- \pi(\bx_i, y) \} dy} $$
and $S  (\bm \beta; \bx, y)= \partial \log f( y \mid \bx; \bm \beta) / \partial \bm \beta$. 
The actual implementation of (\ref{eq:15-5-1}) can be done using EM algorithm.  
According to Lemma 4.1 of \cite{morikawa2018}, (\ref{eq:15-5-1}) is asymptotically equivalent to solving 
\begin{equation}
\sum_{i=1}^N \left[ \frac{ \delta_i}{ \pi(\bx_i, y_i) } S (\bm \beta; \bx_i, y_i) + \left(1-\frac{\delta_i}{ \pi(\bx_i, y_i)} \right) E^\star\{ S (\bm \beta; \bx_i, Y) \mid \bx_i; \bm \beta\} \right] = 0,
\label{eq:15-5-2}
\end{equation}
where 
$$E^\star\{ S (\bm \beta; \bx_i, Y) \mid \bx_i; \bm \beta\} =  \frac{ \int S (\bm \beta; \bx_i, y) f( y \mid \bx_i ; \bm \beta) \{ 1/\pi(\bx_i, y)-1 \} dy }{\int  f( y \mid \bx_i ; \bm  \beta) \{ 1/\pi(\bx_i, y)-1 \} dy}. $$

Now, using the theory of \cite{qin02}, another way of implementing (\ref{eq:15-5-2})  is through the  empirical likelihood: Maximize 
$$ \ell_{p} ( \omega) = \sum_{i  \in S}    \log (\omega_i) $$
subject to $\sum_{i \in S}   \omega_i=1$, 
\begin{equation}
 \sum_{i \in S}  \omega_i \pi(\bx_i, y_i) = N^{-1} \sum_{i=1}^N \tilde{\pi} (\bx_i; \bm \beta) 
 \label{eq:15-5-3}
 \end{equation}
and 
\begin{equation}
 \sum_{i \in S}  \omega_i E^{\star} \{ S(\bm \beta; \bx_i, Y) \mid \bx_i; \bm \beta\} = N^{-1} \sum_{i=1}^N E^{\star} \{ S(\bm \beta; \bx_i, Y) \mid \bx_i; \bm \beta\},
 \label{eq:15-5-4}
 \end{equation}
 and 
 \begin{equation}
  \sum_{i \in S}  \omega_i S(\bm \beta; \bx_i, y_i) = 0 , 
  \label{eq:15-5-5}
  \end{equation}
where 
$$ \tilde{\pi} (\bx_i; \bm \beta) = \int \pi(\bx_i, y) f( y \mid \bx_i; \bm \beta) d y . $$
The above equivalence is not accurate, but it should be a reasonable approximation. 

Now, when $\pi(\bx, y)$ is unknown, we consider  multiple working PS models  $\{\pi_{k}(\bx,y;\bm \phi_{k}): k = 1, \ldots, m\}$ and assume that one of the working PS models is correctly specified.  In this case, constraints in (\ref{eq:15-5-3}) and (\ref{eq:15-5-4}) can be replaced by 
\begin{equation}
\sum_{i \in S} \omega_i \pi_k(\bx_i, y_i; \bm {\phi}_k ) = N^{-1} \sum_{i=1}^N \tilde{\pi} (\bx_i; \bm {\phi}_k,  \bm \beta ) , \ \ \ k=1, \ldots, m 
\label{eq:15-5-6}
\end{equation}
and 
\begin{equation}
 \sum_{i \in S} \omega_i E^{\star} \left\{ S ( \bm \beta; \bx_i, Y) \mid \bx_i; \bm \beta, \bm  \phi_k \right\}   = N^{-1} \sum_{i =1}^N E^{\star} \left\{ S( \bm \beta; \bx_i, Y) \mid \bx_i; \bm \beta, \bm \phi_k \right\}
, \ \ \ k=1, \ldots, m
\label{eq:15-5-7}
\end{equation}
respectively,  where 
$$ \tilde{\pi} (\bx_i; \phi_k, \bm \beta) = \int \pi(\bx_i, y; \bm \phi_k) f( y \mid \bx_i; \bm \beta) d \mu(y) . $$

Now, to include estimation of $\bm \phi_k$, we need additional constraints:
\begin{equation}
\sum_{i \in S} \omega_i E^{\star} \left\{ S_k \left( \bm \phi_k; \bx_i, Y, \delta_i \right) \mid \bx_i \right\} = N^{-1} \sum_{i=1}^N E^{\star} \left\{ S_k \left( \bm \phi_k; \bx_i, Y, \delta_i  \right) \mid \bx_i \right\},  \label{eq:15-5-8}
\end{equation}
and 
\begin{equation}
\sum_{i \in S} \omega_i S_k (\bm  \phi_k ; \bx_i, y_i,\delta_i) = 0, \label{eq:15-5-9}
\end{equation}
for $k=1, \ldots, m$, 
 where  
 $$E^\star\{ S_k  (\bm \phi_k; \bx_i, Y, \delta_i) \mid \bx_i\} =  \frac{ \int S_k (\bm \phi_k; \bx_i, y, \delta_i) f( y \mid \bx_i ; \bm \beta) \{ 1/\pi_k (\bx_i, y; \bm \phi_k )-1 \} dy }{\int  f( y \mid \bx_i ; \bm \beta) \{ 1/\pi_k(\bx_i, y; \bm \phi_k )-1 \} dy} $$
 $$ S_k (\bm \phi_k; \bx, y, \delta) = \frac{\partial}{\partial \bm \phi} \log f_k (\delta \mid \bx, y; \bm \phi_k )    $$
 and 
 $$f_k (\delta \mid \bx, y; \bm \phi_k) = \{ \pi_k(\bx, y; \bm \phi_k) \}^{\delta} \{ 1 - \pi_k(\bx, y; \bm \phi_k) \}^{1-\delta}.   $$
 Constraint (\ref{eq:15-5-6}) is the internal bias calibration \citep{firth1998} constraint incorporating the multiple PS models. Since the bias calibration is based on the multiple PS models, it can be called the multiple bias calibration constraint  \citep{cho2024}. Constraints  (\ref{eq:15-5-5}) and (\ref{eq:15-5-7}) are included to estimate $\beta$. Also, constrains (\ref{eq:15-5-8}) and (\ref{eq:15-5-9}) are included to estimate $\phi_k$.